\newif\ifsubmit
\newif\ifeprint
\newif\ifconference
\newif\ifproceedings

\newif\ifstoc
\stocfalse 
\submittrue 
\eprinttrue

\conferencefalse

\proceedingsfalse

\documentclass[11pt]{article}
\usepackage[utf8]{inputenc}

\ifproceedings
\usepackage[T1]{fontenc}
\fi

\ifeprint
\usepackage[pdftex,left=1in,top=1in,bottom=1in,right=1in]{geometry}
\usepackage{amsthm}
\fi

\usepackage{amsmath,amssymb,dsfont,mathrsfs}

\usepackage{braket}
\usepackage{physics}
\usepackage{mathtools}

\usepackage{xcolor}
\usepackage[framemethod=tikz]{mdframed}
\usepackage{graphicx}
\usepackage{caption}
\usepackage{bm}

\usepackage{verbatim} 
\usepackage{enumitem}

\usepackage[colorlinks=true,citecolor=blue,linkcolor=blue]{hyperref}
\usepackage[nameinlink, noabbrev, capitalize]{cleveref}

\hypersetup{
    colorlinks,
    linkcolor={red!50!black},
    citecolor={blue!50!black},
    urlcolor={blue!80!black}
}

\ifeprint
\usepackage{tocloft}

\fi

\definecolor{darkgreen}{rgb}{0,0.5,0}
\definecolor{darkblue}{rgb}{0,0,0.6}

\allowdisplaybreaks

\usepackage{tabularx}

\DeclareMathAlphabet{\mathpzc}{OT1}{pzc}{m}{it}

\newcommand{\varr}[1]{\mathrm{{#1}}}  \newcommand{\sch}[1]{\mathsf{#1}} \newcommand{\algo}[1]{\mathsf{#1}} \newcommand{\gam}[1]{\mathsf{#1}} 
\newcommand{\samp}{\leftarrow}

\newcommand{\U}{\mathcal{U}}

\newcommand{\zo}{\{0,1\}}
\newcommand{\poly}{\mathsf{poly}}

\newcommand{\regi}{\mathsf{R}} \newcommand{\regis}[1]{\regi_{\mathsf{#1}}}

\newcommand{\ora}{\mathcal{O}}

\newcommand{\pke}{\sch{PKE}}

\newcommand{\prf}{\sch{PRF}}
\newcommand{\prs}{\sch{PRS}}

\newcommand{\pk}{\varr{pk}}
\newcommand{\sk}{\varr{sk}}

\newcommand{\ct}{\varr{ct}}

\newcommand{\enc}{\algo{Enc}}
\newcommand{\dec}{\algo{Dec}}

\newcommand{\ver}{\algo{Ver}}
\newcommand{\ceval}{\algo{Eval}}
\newcommand{\gen}{\algo{Gen}}

\newcommand{\genkey}{\algo{GenKey}}

\newcommand{\sign}{\algo{Sign}}

\newcommand{\negl}{\mathsf{negl}}

\ifeprint
\newtheorem{theorem}{Theorem}
 
 \newtheorem{claim}{Claim}
 \newtheorem{lemma}{Lemma}
 
 \newtheorem{corollary}{Corollary}
 
 \newtheorem{definition}{Definition}

\fi

\usepackage{needspace}
\usepackage{etoolbox}

\newcommand{\PTIME}{\texorpdfstring{\ensuremath{\mathsf{P}}}{P}}
\newcommand{\PP}{\texorpdfstring{\ensuremath{\mathsf{PP}}}{PP}}
\newcommand{\NP}{\texorpdfstring{\ensuremath{\mathsf{NP}}}{NP}}
\newcommand{\BPP}{\texorpdfstring{\ensuremath{\mathsf{BPP}}}{BPP}}
\newcommand{\BQP}{\texorpdfstring{\ensuremath{\mathsf{BQP}}}{BQP}}
\newcommand{\QMA}{\texorpdfstring{\ensuremath{\mathsf{QMA}}}{QMA}}
\newcommand{\QMAH}{\texorpdfstring{\ensuremath{\mathsf{QMAH}}}{QMAH}}
\newcommand{\QCMA}{\texorpdfstring{\ensuremath{\mathsf{QCMA}}}{QCMA}}
\newcommand{\QCMAH}{\texorpdfstring{\ensuremath{\mathsf{QCMAH}}}{QCMAH}}
\newcommand{\QCIP}{\texorpdfstring{\ensuremath{\mathsf{QCIP}}}{QCIP}}

\newcommand{\PH}{\texorpdfstring{\ensuremath{\mathsf{PH}}}{PH}}

\newcommand{\AM}{\texorpdfstring{\ensuremath{\mathsf{AM}}}{AM}}
\newcommand{\QPT}{\texorpdfstring{\ensuremath{\mathsf{QPT}}}{QPT}}
\newcommand{\PPT}{\texorpdfstring{\ensuremath{\mathsf{PPT}}}{PPT}}
\newcommand{\PROMISEP}{\texorpdfstring{\ensuremath{\mathsf{PromiseP}}}{PromiseP}}
\newcommand{\PROMISEBQP}{\texorpdfstring{\ensuremath{\mathsf{PromiseBQP}}}{PromiseBQP}}
\newcommand{\PROMISEQCMA}{\texorpdfstring{\ensuremath{\mathsf{PromiseQCMA}}}{PromiseQCMA}}
\newcommand{\PROMISEQCMAH}{\texorpdfstring{\ensuremath{\mathsf{PromiseQCMAH}}}{PromiseQCMAH}}

\newcommand{\QMAbb}{\texorpdfstring{\ensuremath{\mathsf{BB}\text{-}\QMA}}{BB-QMA}}
\newcommand{\QCMAbb}{\texorpdfstring{\ensuremath{\mathsf{BB}\text{-}\QCMA}}{BB-QCMA}}
\newcommand{\QCMAHbb}{\texorpdfstring{\ensuremath{\mathsf{BB}\text{-}\QCMAH}}{BB-QCMAH}}
\newcommand{\QCIPbb}{\texorpdfstring{\ensuremath{\mathsf{BB}\text{-}\QCIP}}{BB-QCIP}}

\newcommand{\PTIMEbb}{\texorpdfstring{\ensuremath{\mathsf{BB}\text{-}\PTIME}}{BB-P}}
\newcommand{\coNPbb}{\texorpdfstring{\ensuremath{\mathsf{BB}\text{-}\mathsf{coNP}}}{BB-coNP}}

\newcommand{\CHS}{\texorpdfstring{\ensuremath{\mathsf{CHS}}}{CHS}}
\newcommand{\CHN}{\texorpdfstring{\ensuremath{\mathsf{CHN}}}{CHN}}
\newcommand{\Forrelation}{\texorpdfstring{\ensuremath{\mathsf{Forrelation}}}{Forrelation}}

\newcommand{\Decode}{\algo{Decode}}
\newcommand{\Prepare}{\algo{Prepare}}
\newcommand{\PrepTables}{\mathcal{D}_{\mathrm{prep}}}
\newcommand{\EvalHash}{\algo{EvalHash}}
\newcommand{\EvalCoset}{\algo{EvalCoset}}
\newcommand{\EvalKey}{\algo{EvalKey}}

\newcommand{\BiasYZ}{\algo{BiasYZ}}
\newcommand{\Answer}{\algo{Answer}}

\newcommand{\Switch}{\algo{Switch}}
\newcommand{\Reduce}{\algo{Reduce}}
\newcommand{\Replace}{\algo{Replace}}
\newcommand{\CSig}{\sch{CSig}}
\newcommand{\QSig}{\sch{QSig}}
\newcommand{\Piracy}{\gam{Piracy}}
\newcommand{\Forge}{\gam{Forge}}
\newcommand{\EqTest}{\algo{Eq}}
\newcommand{\KeyGen}{\algo{KeyGen}}
\newcommand{\StateGen}{\algo{StateGen}}
\newcommand{\VKGen}{\algo{VKGen}}
\newcommand{\PKE}{\mathsf{PKE}}
\newcommand{\PKEGen}{\algo{PKE.Gen}}
\newcommand{\PKEEnc}{\algo{PKE.Enc}}
\newcommand{\PKEDec}{\algo{PKE.Dec}}
\newcommand{\PRFGen}{\algo{PRF.Gen}}
\newcommand{\PRFEval}{\algo{PRF.Eval}}
\newcommand{\CSigGen}{\algo{CSig.Gen}}
\newcommand{\CSigSign}{\algo{CSig.Sign}}
\newcommand{\CSigVer}{\algo{CSig.Ver}}
\newcommand{\QSigGen}{\algo{QSig.Gen}}
\newcommand{\QSigSign}{\algo{QSig.Sign}}
\newcommand{\QSigVer}{\algo{QSig.Ver}}
\newcommand{\QSigVKGen}{\algo{QSig.VKGen}}

\newcommand{\Good}{\mathsf{Good}}
\newcommand{\Sim}{\mathsf{Sim}}

\newcommand{\Learn}{\mathsf{Learn}}
\newcommand{\YES}{\mathsf{YES}}
\newcommand{\NO}{\mathsf{NO}}

\newcommand{\vk}{\mathsf{vk}}
\newcommand{\oradist}{\mathsf{O}}
\newcommand{\oranoise}{\ora^{\mathsf{noise}}}
\newcommand{\oraplant}{\ora^{\mathsf{plant}}}
\newcommand{\WitnessReg}{\mathsf{W}}
\newcommand{\SigScheme}{\mathsf{Sig}}

\newcommand{\orabase}{\ora_{\mathsf{base}}}
\newcommand{\oraQCMA}{\ora_{\QCMA}}
\newcommand{\cA}{\mathcal{A}}
\newcommand{\cC}{\mathcal{C}}
\newcommand{\cE}{\mathcal{E}}

\newcommand{\cG}{\mathcal{G}}
\newcommand{\cH}{\mathcal{H}}
\newcommand{\cI}{\mathcal{I}}
\newcommand{\cM}{\mathcal{M}}
\newcommand{\cP}{\mathcal{P}}
\newcommand{\cS}{\mathcal{S}}
\newcommand{\cX}{\mathcal{X}}
\newcommand{\cY}{\mathcal{Y}}
\newcommand{\cZ}{\mathcal{Z}}
\newcommand{\bbE}{\mathbb{E}}
\newcommand{\bbF}{\mathbb{F}}

\newcommand{\indicator}{\mathbf{1}}
\newcommand{\Hadamard}{\mathrm{H}}
\newcommand{\HaarMeasure}{\mathrm{Haar}}

\newcommand{\Bernoulli}{\operatorname{Bernoulli}}
\newcommand{\DTdepth}{\operatorname{DT}}
\newcommand{\Clip}{\operatorname{clip}}
\newcommand{\Comp}{\operatorname{Comp}}
\newcommand{\supp}{\operatorname{supp}}
\newcommand{\dom}{\operatorname{dom}}
\newcommand{\trdist}[2]{\tfrac12\left\|#1-#2\right\|_1}
\providecommand{\ceil}[1]{\left\lceil #1\right\rceil}

\newenvironment{boxedalgorithm}[1]{\begin{mdframed}[skipabove=8pt,skipbelow=8pt,nobreak=true]
  \noindent\textbf{#1}\par\smallskip\normalfont
}{\end{mdframed}}

\ifstoc
  \AtBeginDocument{\pagenumbering{gobble}\hypersetup{pageanchor=false}}
  \AddToHook{file/intro.tex/before}{\clearpage
    \pagenumbering{arabic}\hypersetup{pageanchor=true}}
\fi

\newcommand{\paperpart}[1]{\clearpage\part{#1}}
\newcounter{appendixpart}
\renewcommand{\theappendixpart}{\Roman{appendixpart}}
\makeatletter
\ifeprint
\patchcmd{\NR@part}{\raggedright}{\centering}{}{\PackageWarning{paper}{Could not center numbered part headings}}
\patchcmd{\NR@spart}{\raggedright}{\centering}{}{\PackageWarning{paper}{Could not center unnumbered part headings}}
\renewcommand{\l@part}[2]{\ifnum\c@tocdepth>\m@ne
    \addpenalty\@secpenalty
    \addvspace{\cftbeforepartskip}\begingroup
      \centering\cftpartfont #1\par
    \endgroup
    \nobreak
  \fi}
\fi
\newlength{\cftbeforeappendixgroupskip}
\newcommand{\toclevel@paperappendices}{-1}
\newcommand{\l@paperappendices}[2]{\ifnum\c@tocdepth>\m@ne
    \addpenalty\@secpenalty
    \addvspace{\cftbeforeappendixgroupskip}\begingroup
      \centering
      \rule{.7\linewidth}{.3pt}\par\nobreak
      \vskip.4\baselineskip\nobreak
      \cftpartfont #1\par
    \endgroup
    \nobreak
  \fi}
\newcommand{\toclevel@appendixpart}{0}
\newcommand{\l@appendixpart}[2]{\begingroup
    \renewcommand{\cftpartfont}{\large\bfseries}\setlength{\cftbeforepartskip}{\cftbeforeappendixgroupskip}\l@part{#1}{#2}\endgroup}
\makeatother
\newcommand{\paperappendices}{\clearpage\appendix
  \setcounter{appendixpart}{0}\part*{Appendices}\addcontentsline{toc}{paperappendices}{Appendices}}
\newcommand{\paperappendixpart}[1]{\ifnum\value{appendixpart}>0\relax\clearpage\fi
  \stepcounter{appendixpart}\part*{Appendix \theappendixpart: #1}\addcontentsline{toc}{appendixpart}{Appendix \theappendixpart: #1}}

\ifeprint
\renewcommand{\cftpartfont}{\fontsize{14}{16}\selectfont\bfseries}
\cftpagenumbersoff{part}

\fi

\setlist[enumerate,2]{label=\arabic{enumi}.\arabic*.,ref=\arabic{enumi}.\arabic*,leftmargin=*}
\setlist[enumerate,3]{label=\arabic{enumi}.\arabic{enumii}.\arabic*.,ref=\arabic{enumi}.\arabic{enumii}.\arabic*,leftmargin=*}

\newcounter{overviewstep}[subsubsection]

\crefname{overviewstep}{Section}{Sections}
\Crefname{overviewstep}{Section}{Sections}

\title{Classical Oracle Separation of $\QMA$ from the $\QCMA$ Hierarchy\protect\\
and Quantum Cryptography Even When $\PTIME=\QCMA$}

\ifproceedings
\fi

\date{}

\ifproceedings
\author{Alper Cakan\orcidID{0000-0003-3567-1704}}

\authorrunning{A. Cakan}

\institute{Carnegie Mellon University, Pittsburgh PA, USA \email{alpercakan98@gmail.com}}
\fi

\ifeprint
\author{Alper Cakan\thanks{Carnegie Mellon University. \texttt{alpercakan98@gmail.com}.}}
\fi

\begin{document}
\maketitle
\begin{mdframed}[linewidth=0.5pt,innertopmargin=6pt,innerbottommargin=6pt,nobreak=true]
\textbf{This work is in submission at STOC'27 and thus the full manuscript has been publicly posted on the conference website since November 3rd, 2026.}
\end{mdframed}

\begin{abstract}

  Inspired by the  \emph{Codeword-Hash Structure} problem recently introduced by Cakan (arXiV, 2026) to separate $\QMA$ from $\QCIP$, 
  we introduce a new problem called \textbf{\emph{Code\-word-Hash pairs in Noise}} (\CHN{}) and prove that, in the query complexity model, it is in \QMA{} but hard for the \QCMA{} hierarchy $\QCMAH = \bigcup_{d\geq1}\allowbreak\QCMA^{\QCMA^{\cdots^{\QCMA}}}$, the classical-witness quantum-verifier analogue of $\PH=\bigcup_{d\ge1}\NP^{\NP^{\cdots^{\NP}}}$. 

  Using this problem, we show that there exists a classical oracle $\ora$ such that 
  \[
   \PTIME^\ora=\BQP^\ora=\QCMA^\ora
   =\QCMAH^\ora \subset \QMA^\ora
   \qquad\text{but}\qquad
   \QMA^\ora\not\subseteq\QCMAH^\ora,
  \]
  thus showing that \emph{quantum state} or \emph{quantum proof advantage} could exist even if classical computers can efficiently solve any problem that can be quantumly verified with a classical proof.
  \\ \mbox{}\\
  Further, using our problem, we show that various quantum cryptographic primitives can exist even when $\PTIME = \QCMA$ and thus neither quantum-computable one-way functions nor (verifiable) quantum \emph{computational} advantage exist. In more detail, we show that there exists a classical oracle $\ora$ such that $\PTIME^\ora = \BQP^\ora =\QCMA^\ora=\QCMAH^\ora$ but the following primitives exist relative to $\ora$, answering an open question of Kretschmer, Qian, and Tal (STOC 2025).

  \begin{itemize}
    \item Pseudorandom states (PRS) with classical keys.
    \item Pseudorandom functions (PRF) with (copy-protected) quantum secret keys, and thus also pseudorandom unitaries (PRU) with quantum secret keys,
    \item Public-key encryption (PKE) with a (copy-protected) quantum decryption key, classical public keys and ciphertexts,
    \item Signature schemes with quantum signing keys and either classical verification keys and quantum signatures or quantum verification keys and classical signatures,
    \item Collusion-resistant public-key quantum money (PKQM) with classical verification keys.
  \end{itemize}

  \ifproceedings
  \fi
\end{abstract}
\ifeprint
\vfill\par\textbf{Keywords:} Quantum complexity, quantum cryptography, MicroCrypt.
\fi

\newpage\tableofcontents
\setcounter{tocdepth}{2}

\newpage\section{Introduction}\label{sec:introduction}

The $\QMA$ versus $\QCMA$ question asks whether quantum proofs are
more powerful than classical ones, and has been a central question in quantum complexity since the introduction of these classes~\cite{AN02}. Aaronson and Kuperberg~\cite{AK07} gave the first evidence for
such an advantage for quantum proofs by separating these classes relative to a quantum unitary
oracle. After many attempts~\cite{FK18,NN24,ITCS:LLPY24,BK24,ITCS:Zhandry25,STOC:LiuMutYue25,QFire}, recently the breakthrough work of Bostanci,
Haferkamp, Nirkhe, and
Zhandry~\cite{BHNZ} finally showed the first \emph{classical} oracle separation. Later, a simpler classical oracle separation was shown by Bostanci, Huang, and Vaikuntanathan~\cite{BHV}.

However, a fundamental question still remains open: whether $\QMA$ is contained in the $\QCMA$ hierarchy 
\[
 \QCMAH=\bigcup_{d\ge1}
 \underbrace{\QCMA^{\QCMA^{\cdots^{\QCMA}}}}_{d\text{ levels}},
\]
the classical-witness quantum-verifier analogue of the polynomial hierarchy $\PH = \bigcup_{d\ge1}\NP^{\NP^{\cdots^{\NP}}}$.

Even more importantly, whether $\QMA$ is larger even when $\BQP=\QCMA$. In other terms, can \textbf{\emph{quantum state advantage}} or \textbf{\emph{quantum proof advantage}} exist even if every problem
that admits a classical proof can also be efficiently solved? 
Thus, we ask\footnote{Separating $\QMA^\ora$ from $\BQP^\ora = \QCMA^\ora$ is stronger than just separating $\QMA^\ora$ from $\QCMAH^\ora$ since $\BQP^\ora = \QCMA^\ora$ implies $\BQP^\ora = \QCMAH^\ora$ for any oracle $\ora$.}:

\begin{mdframed}
\begin{quote}
\centering
\emph{Is $\QMA$ in the $\QCMA$ hierarchy? Can quantum state/proof advantage exist ($\QMA \not\subseteq \QCMA$) even if $\BQP = \QCMA$, or even if $\PTIME = \QCMA$?}
\end{quote}
\end{mdframed}

The languages\footnote{Spectral Forrelation problem studied in~\cite{BHNZ} is not necessarily in $\AM$, although the language used in their oracle separation lies in $\AM$ because of the version of Spectral Forrelation  needed in their $\QCMA$ lower-bound proof. In contrast, the problem in~\cite{BHV} itself lies in $\AM$.} $L \in \QMA$ constructed in~\cite{BHNZ} and~\cite{BHV} both lie in $\AM$, which is contained in the second level $\QCMA^{\QCMA}$ of the $\QCMA$ hierarchy. So, they do not obtain a $\QMA$ separation from $\QCMAH$, or $\QMA$ advantage when $\BQP = \QCMA$.

Thus, this question was left open and indeed answering it requires a substantial strengthening of their results: Drawing a parallel to \textbf{\emph{quantum computational advantage}}\footnote{We mean efficiently verifiable quantum advantage without quantum memory.}, we note that an oracle separation of $\BQP$ from $\NP$ has been known since
2000~\cite{FOCS:Watrous00}, but a separation of $\BQP$ from $\PH$, and separation of $\BQP$ from $\PTIME$ when $\PTIME=\NP$ was obtained much later by the celebrated works of
Raz and Tal~\cite{RT} and Aaronson, Ingram, Kretschmer~\cite{AIK}, respectively.

\paragraph{Cryptography}
Aside from its complexity-theoretic interest, the question above also
has an important connection to quantum cryptography. Initiated by Kretschmer~\cite{KretschmerHaar24}, the study of \emph{MicroCrypt}, a fruitful major line of research in quantum cryptography~\cite{C:AnaQiaYue22,MY22,ITCS:BraCanQia23,KQST,KT,MY24OneWayness,C:ChuGolGra24,FOCS:BatJai24,CountCrypt,KQT,BCN25,BMMMY25}, has shown evidence for the tantalizing possibility that quantum cryptography can exist even if there is no classical cryptography. 

While earlier oracle separations in this area, including \cite{KretschmerHaar24}, used quantum unitary oracles, Kretschmer, Qian, Sinha, and Tal~\cite{KQST} finally showed a classical oracle relative to which $\PTIME=\NP$ yet single-copy pseudorandom states (PRS) exist. Later, Kretschmer, Qian, and Tal~\cite{KQT} showed that there is a classical oracle relative to which $\PTIME=\NP$ but quantum-computable one-way functions (OWFs), public-key encryption (PKE), and multi-copy PRS can still exist, and left open the question of cryptography when $\PTIME=\QCMA$.

Further, we note that the results of~\cite{KQT} necessarily have that {quantum computational advantage} exists relative to their oracle:  Indeed, observe that existence of quantum-computable one-way functions implies $\BQP \neq \QCMA$, and since they also have $\PTIME = \NP$, their oracle world has $\PTIME \neq \BQP$. Thus, we ask:

\begin{mdframed}
\begin{quote}
\centering
\emph{Can quantum cryptography exist even if $\PTIME = \QCMA$, that is, without quantum-computable one-way functions (OWF) or quantum computational advantage?}
\end{quote}
\end{mdframed}

We believe this is an important question given that there have been many examples of quantum algorithms later being de-quantized~\cite{STOC:Tang19,STOC:CGLLTW20,KOW26}, whereas there is more concrete evidence for quantum state advantage in the standard model~\cite{STOC:GKKRW07,STOC:RegKla11,FOCS:Mahadev18a,FOCS:CCHL21}.

\subsection{Our results}
In this work we answer both of the questions affirmatively, through a new problem called \textbf{\emph{Codeword-Hash pairs in Noise}} (\CHN{}) that we introduce. Briefly, the problem is to decide if the constant-$1$ function have been planted  on top of a string of sparse Bernoulli noise, at locations that are codewords and their symbol-wise hashes (i.e. Yamakawa--Zhandry hashes~\cite{YZ}).

\subsubsection{Complexity}
We use the prefix $\mathsf{BB}\text{-}$ (for black-box) to distinguish classes in the query complexity model from the usual time-bounded classes.
We first prove the following query complexity result for \CHN{}\footnote{We define the problem \CHN{} through two distributions (planted versus noise) over instances with overlapping supports, which means technically it is not even a promise problem. However, discarding the negligible sets of planted and noise samples on which the $\QMA$ verifier’s respective acceptance and rejection guarantees fail yields an actual promise problem with the stated result.}.
\begin{theorem}[Informal, \cref{thm:qma-verification,thm:hierarchy}]\label{thm:intro-chn}
In the query complexity model,
\[
 \CHN\in\QMAbb\setminus\QCMAHbb.
\]
\end{theorem}

Then, with a small extension of our query complexity argument, we construct a classical oracle relative to which $\QCMAH$ collapses to $\PTIME$ while $\QMA$ remains strictly larger.
\begin{theorem}[Informal, \cref{thm:boolean-collapse}]\label{thm:intro-complexity}
There exist a classical oracle $\ora:\zo^*\to\zo$ and a unary language
$L$ such that
\[
 \PTIME^\ora=\QCMA^\ora=\QCMAH^\ora\subset\QMA^\ora,
 \qquad
 L\in\QMA^\ora\setminus\QCMAH^\ora.
\]
\end{theorem}

We note that our cryptography theorem below (with small caveats) subsumes this oracle separation theorem. In fact, we show the stronger separation where
$
 \PROMISEP^{\ora}
 =\PROMISEQCMA^{\ora}
 =\PROMISEQCMAH^{\ora},
$
but there is a language in $\QMA^\ora$ that is hard even on average for $\QCMAH^\ora$ machines
(\cref{cor:pke-average-qma}). However, note that they do not subsume the query complexity result (\cref{thm:intro-chn}) regarding \CHN{}.

\paragraph{Quantum State/Proof Advantage}
Our result shows that \emph{quantum state advantage}, or \emph{quantum proof advantage},
can exist even when classical computers can efficiently solve every problem
that admits quantum verification with a classical witness ($\PTIME=\QCMA$).
While our problem itself has no direct connection to the \Forrelation{}
problem~\cite{AaronsonPH,STOC:AarAmb15}, in some sense it can be thought
of as a quantum-state analogue of it, since \Forrelation{} has been used to
show that \emph{quantum computational advantage} can exist even when
$\PTIME=\NP$~\cite{RT,AIK}.

\subsubsection{Cryptography}

We use \CHN{} to construct quantum cryptographic primitives in a
world where $\PTIME=\QCMA=\QCMAH$, and hence there is no quantum computational advantage, quantum-computable
one-way functions or even efficiently verifiable one-way puzzles~\cite{C:ChuGolGra24}; answering an open question of Kretschmer,
Qian, and Tal~\cite{KQT}.

\Needspace{12\baselineskip}
\begin{theorem}[Informal, \cref{thm:fixed-oracle-crypto}]
\label{thm:intro-crypto}
There exists a classical oracle $\ora:\zo^*\to\zo$ such that
\[
 \PTIME^\ora=\QCMA^\ora=\QCMAH^\ora,
\]
and relative to $\ora$ the following primitives exist with correctness and security:
\begin{itemize}
 \item Pseudorandom states (PRS) with classical keys, secure for
 any polynomial number of copies.
 \item Copy-protected public-key encryption (PKE) with quantum
 decryption keys and classical public keys and ciphertexts.
 \item Copy-protected pseudorandom functions (PRF) with quantum secret
 keys, pseudorandom against quantum evaluation queries.
 \item Signature schemes with quantum signing keys and either
 classical verification keys and quantum signatures or quantum
 verification keys and classical signatures.
\end{itemize}
\end{theorem}
We emphasize that, unlike~\cite{KQT}, our $\PKE$ construction also satisfies \emph{copy-protection}, a stronger security requirement that has been extensively studied~\cite{AaronsonCP,ALLLZ,GZ20,CLLZ,CG25CopyProtection}.

\paragraph{Corollaries} Through the known black-box constructions in cryptography, we also obtain that the following exist relative to the same oracle:
\begin{itemize}
    \item Collusion-resistant public-key
    quantum money (PKQM) with classical verification keys~\cite{AC13Money},
    \item Bounded-collusion-resistant copy-protected PKE and PRF schemes~\cite{CGKNY26MultiCopy},
    \item Pseudorandom unitaries with a quantum key~\cite{MH25},
    \item (Inefficient-verifier) one-way puzzles~\cite{KT},
    \item One-way state generators with a classical key~\cite{MY22},
    \item Signature scheme with quantum verification keys
    and classical signing keys and signatures, secure against a bounded number of signing queries~\cite{MY24OneWayness}.
\end{itemize}
We refer the reader to the Microcrypt Zoo~\cite{MicrocryptZoo} for more implications of these primitives.

\paragraph{Quantum State Advantage in Cryptography} It is not hard to see that when $\PTIME = \QCMA$, there is no quantum \emph{computational} advantage: Any problem that can be solved efficiently by a quantum computer can also be solved efficiently classically, and any efficient quantum computational (even interactive) process that holds no quantum memory can be efficiently simulated by a classical computer\footnote{There can be some efficient quantum (sampling) processes that can be \emph{inefficiently} distinguished~\cite{AaronsonSampling14,STOC:MorShiYam25}.}.  Thus, we obtain cryptography from quantum state advantage even if there is no quantum computational advantage. Note that this is in contrast to the results of~\cite{KQST,KQT}, which, as discussed, necessarily require quantum computational advantage ($\PTIME \neq \BQP$).

Aside from \cite{KretschmerHaar24,KQST,KQT}, very recently, Gulati, Khurana, and Tomer~\cite{GKT26PRU} ruled out fully black-box constructions of quantum-computable one-way
functions from PRUs. However, they do not show that PRUs can exist
even if quantum-computable one-way functions do not exist, or if
$\PTIME=\QCMA$.\footnote{See~\cite{AC:BaeBrzFis13,GKT26PRU} for a discussion on the difference between fully black-box separations and oracle separations.}

\subsection{Organization}
In \cref{sec:overview}, we give a technical overview of the paper. See \cref{sec:more-organization} for the detailed organization and \cref{sec:ai-methodology} for the AI methodology.

In \cref{sec:future-directions} we discuss some future directions and in \cref{sec:more-related-work} we discuss some more related work.

\ifstoc
\else
\newpage
\fi

\section{Technical Overview}\label{sec:overview}
In this section, we give a technical overview of our work, explaining our techniques and providing some intuition. Formal definitions, statements, and proofs can be found in the relevant sections. 

\paragraph{Complexity} In this overview, we focus on our query complexity\footnote{Also called \emph{black-box}~\cite{FOCS:BBCMW98,ITCS:AarAmb11,RT} or \emph{oracle-input} \cite{ITCS:Zhandry25} model.} result about \CHN{} to highlight the main techniques that also underlie our cryptographic results. Through (somewhat) routine techniques, such query complexity results can be converted to oracle separations (see, e.g.,~\cite{RT,AIK,ITCS:Zhandry25,BHNZ,BHV}). Indeed, our oracle separation showing that \emph{$\QMA^\ora$ is strictly larger than $\PTIME^\ora = \QCMAH^\ora$ relative to some fixed oracle $\ora$} (\cref{thm:boolean-collapse}), follows from this with small additions (which we provide for reference in \cref{app:fixed-oracles}). As mentioned before, our cryptography results actually subsume this oracle separation, but not the query complexity result for \CHN{}.

In \cref{sec:overview-model}, we briefly recall the model. Then \cref{sec:overview-chn} we introduce our problem \CHN{} and state our query complexity result. We then give an overview of its proof in \cref{sec:overview-proof}.

\paragraph{Cryptography} We introduce our function encoding, PKE construction and collapse oracle in \cref{sec:overview-crypto}. We then give the proof overview in \cref{sec:overview-crypto-simulation}, first assuming that the required states are supplied and then explaining their preparation using classical oracles.

\paragraph{Detailed Organization and Further Discussion} We give a detailed map of the paper in \cref{sec:more-organization}. Then we discuss future directions in \cref{sec:future-directions}, and review further related work in \cref{sec:more-related-work}.

\subsection{Model}\label{sec:overview-model}
{}{}

We now briefly recall the query complexity model, the familiar readers may skip. See \cref{sec:complexity-preliminaries} for formal definitions.

\paragraph{Algorithms and Machines}
An algorithm is a classical or quantum procedure whose output may be
probabilistic, whereas a machine specifies a partial decision function
with values in $\{0,1,\bot\}$. We call an algorithm or machine
\emph{query-efficient} if it makes polynomially many queries in total
to its oracles, each on an \emph{address} (also called a \emph{query input})
of polynomial length. We allow unbounded computation outside oracle queries. Usually, we will use the word algorithm for query-efficient ones.

\paragraph{Query Complexity Model}

An algorithm or machine accesses its input as a finite Boolean oracle
$\ora:\zo^{p(\lambda)}\to\zo$, where $p$ is a fixed polynomial
and $N=2^{p(\lambda)}$ is size of the oracle's truth table.
Resource bounds, such as the number of queries and the witness size,
are measured in $\lambda$ rather than in $N$, so \emph{polynomial}
means polynomial in $\lambda$. We also consider oracles with domain
$\cX_\lambda$, writing $\ora:\cX_\lambda\to\zo$.
Besides its input oracle, an algorithm or machine may have oracle
access to another machine and receive a classical auxiliary input $z$ {},
supplied directly rather than through oracle access.\footnote{The auxiliary
input allows a higher-level machine to query a lower-level machine
on different strings while both access the same input oracle.}
Unless otherwise specified, witnesses and auxiliary inputs have
polynomial size.
We use the prefix $\mathsf{BB}\text{-}$ (for \emph{black-box}), as in $\QMAbb$ and $\QCMAHbb$, to distinguish these query-model classes from their usual time-bounded counterparts.

\paragraph{$\QCMAbb$ and $\QCMAHbb$ in Query Complexity Model}
A \QCMAbb{} machine $M$ is specified by a query-efficient quantum
algorithm $V$, called the \emph{verifier} that additionally receives
\emph{classical witness} $w$, and is defined as
\[
 M^\ora(z)=
 \begin{cases}
  1,&\exists w:\Pr[V^\ora(z,w)=1]\ge2/3,\\
  0,&\forall w:\Pr[V^\ora(z,w)=1]\le1/3,\\
  \bot,&\text{otherwise}.
 \end{cases}
\]

Following Agarwal and Ben-David~\cite{AB}, we define the \QCMAbb{}
hierarchy using \emph{completions} of partial functions.\footnote{See~\cite{AB} for a discussion on the choice of this definition for the $\QCMAbb$ hierarchy.}
For $f:\zo^*\to\{0,1,\bot\}$, we define its set of Boolean completions by
\[
 \Comp(f)=
 \left\{
 \widetilde f:\zo^*\to\zo
 \;\middle|\;
 \forall z,\ f(z)\ne\bot\implies\widetilde f(z)=f(z)
 \right\}.
\]

Starting with $\QCMAHbb_1=\QCMAbb$, for $d\ge2$ a $\QCMAHbb_d$ machine $M_d$
is specified by a query-efficient quantum verifier $V_d$ and a
$\QCMAHbb_{d-1}$ machine $M_{d-1}$, and is defined as
\[
 M_d^\ora(z)=
 \begin{cases}
  1,&\forall\widetilde f\in\Comp(M_{d-1}^\ora)\ \exists w:
     \Pr[V_d^{\ora,\widetilde f}(z,w)=1]\ge2/3,\\[2pt]
  0,&\forall\widetilde f\in\Comp(M_{d-1}^\ora)\ \forall w:
     \Pr[V_d^{\ora,\widetilde f}(z,w)=1]\le1/3,\\
  \bot,&\text{otherwise}.
 \end{cases}
\]
We define the $\QCMAbb$ hierarchy $\QCMAHbb$ as union of constant levels $d$,
\[
 \QCMAHbb=\bigcup_{d\ge1}\QCMAHbb_d.
\]

\subsection{The \CHN{} Problem and Our Query Complexity Result}\label{sec:overview-chn}
\paragraph{Starting point: \CHS{} Problem}
Our starting point is the \emph{Codeword-Hash Structure} (\CHS{}) problem recently introduced by Cakan \cite{CakanQCIP26}, to separate $\QMAbb$ from $\QCIPbb$, which in turn is based on the \emph{Code Intersection Subset Size} problem introduced by the elegant classical-oracle $\QMA$--$\QCMA$ separation of Bostanci, Huang and Vaikuntanathan \cite{BHV}. Both problems build on the Yamakawa-Zhandry code intersection problem~\cite{YZ} where we consider the symbol-wise hash function $H(a) = H_1(a_1) || \dots || H_q(a_q)$ and a  \emph{suitable} linear code $C \in \Sigma^q$, and the task is to, for a given image $y \in \zo^q$, find a codeword $v \in C$ such that $H(v) = y$. Indeed, many $\QMA$--$\QCMA$  separations rely on this same problem \cite{ITCS:LLPY24,BK24,ClonableQMA}. In particular, \cite{BHV,ClonableQMA,CakanQCIP26} use the \emph{biased} version introduced in~\cite{STOC:GGJL25}.

More precisely, \CHS{} is defined as follows. Let $C_\lambda\subseteq\Sigma_\lambda^{q_\lambda}$ be
a suitable linear code, $q(\lambda) = \poly(\lambda)$ and we sample each value of the predicates
$H_i:\Sigma_\lambda\to\zo$ independently as
\[
 H_i(a)\leftarrow\Bernoulli(\lambda^{-4})
 \qquad(i\in[q_\lambda],\ a\in\Sigma_\lambda),
\]
 We then define the \emph{codeword-hash relation}
over $\cX_\lambda=\Sigma_\lambda^{q_\lambda}\times\zo^\lambda$ by
\begin{equation}\label{eq:overview-code-pairs}
 \cP_H=\bigl\{(v,y)\in\cX_\lambda:v\in C_\lambda,
 H(v)=H_1(v_1)\|\cdots\|H_{q_\lambda}(v_{q_\lambda})
      =y\|0^{q_\lambda-\lambda}\bigr\}.
\end{equation}
Then, the \CHS{} problem is to distinguish the \emph{codeword-hash structure}
distribution $\ora=\indicator_{\cP_H}$ (i.e. $\ora(v,y)=1$ iff $v\in C_\lambda$ and
$H(v)=y\|0^{q_\lambda-\lambda}$),
from the Bernoulli noise oracle $\ora=B$, where
$
 B(v,y)\leftarrow\Bernoulli(\delta)$ independently for  $(v,y)\in C_\lambda\times\zo^\lambda,
 \delta=2^{-\lceil\lambda/12\rceil}.
$

\paragraph{Quantum versus classical witnesses.}
As observed by \cite{EC:Liu23}, the code intersection problem satisfies a useful \emph{equivocation} property: Using only the \emph{predicate-preimage state} $\ket{\Psi_H}\propto \bigotimes_{i=1}^{q_\lambda}\left(\left(\sum_{a\in\Sigma_\lambda:H_i(a)=0}\ket a\right)\otimes\left(\sum_{a\in\Sigma_\lambda:H_i(a)=1}\ket a\right)\right)$,
the decoding algorithm of \cite{YZ} can find a codeword $v \in C$ with
$H(v) = y$ for any $y$, without querying $H$. Crucially, the same state $\ket{\Psi_H}$ works for any $y$ value. Based on this idea and a suitable choice of a code; \cite{BHV} shows that, using $\ket{\Psi_H}$ as the witness, we can check the codeword-hash structure in an input (such as $\ora=\indicator_{\cP_H}$ above), simply by sampling a random $y\leftarrow\zo^\lambda$ and running the decoder of \cite{YZ} for  $y\|0^{q_\lambda-\lambda}$, and checking if the output $v$ satisfies $\ora(v,y)=1$. However, since a classical fixed witness $w$ can only encode a small amount of pairs $(v,y)$, they show that if the code satisfies sufficiently strong combinatorial properties, such structure cannot be checked reliably with classical witnesses, proving their $\QMA$--$\QCMA$ separation. As we discuss later on, \cite{CakanQCIP26} defines \emph{contractive simultaneous incompressibility} property for oracles that codifies this notion, and extends this proof idea beyond classical witnesses/$\QCMAbb$ to \emph{classical-interactive proofs with quantum verifiers} ($\QCIPbb$).

\paragraph{$\coNPbb$ Attack for \CHS{}}
While \CHS{} separates $\QMAbb$ from $\QCIPbb$, unfortunately it is in $\coNPbb\subseteq\mathsf{BB}\text{-}(\NP^{\NP})$: To recognize the structured case, $\coNPbb$ can simply check that no codeword
$v$ is accepted for two distinct targets $y\ne y'$.
Since $H(v)$ determines $y$, this holds for $\indicator_{\cP_H}$,
whereas $B$ accepts exponentially many targets for any fixed codeword
with overwhelming probability.
In fact, small tweaks to this problem will not help: If a noticeable relative gap
remains in the number of accepted targets
$N_\ora(v):=|\{y\in\zo^\lambda:\ora(v,y)=1\}|$ for a fixed codeword $v$, we can use  Stockmeyer's approximate-counting algorithm~\cite{STOC:Stockmeyer83}
to approximate this count and decide the problem in
$\mathsf{BB}\text{-}(\BPP^{\NP})\subseteq\QCMAHbb_2$.

\paragraph{Our Problem: \CHN{}}
To remove this counting gap, we use noise in both cases, introducing the problem \emph{Codeword-Hash pairs in Noise} (\CHN{}):

\begin{mdframed}[nobreak=true,innertopmargin=9pt,innerbottommargin=9pt]
\noindent\textbf{The \CHN{} problem. $\beta=2^{-\lceil\sqrt\lambda\rceil}$}
\begin{equation*}
\begin{array}{@{}p{.5\linewidth}@{}p{.5\linewidth}@{}}
 \underline{Planted (YES) distribution: $\oradist_{\mathsf{plant},\lambda}$}
 &\underline{Noise (NO) distribution: $\oradist_{\mathsf{noise},\lambda}$}\\[11pt]
 $B(v,y)\leftarrow\Bernoulli(\beta)\quad(v,y)\in\cX_\lambda$
 &$B(v,y)\leftarrow\Bernoulli(\beta)\quad(v,y)\in\cX_\lambda$\\[6pt]
 $H_i(u)\leftarrow\Bernoulli(\lambda^{-4})\quad{i\in[q_\lambda], u\in\Sigma_\lambda}$
 &\\[11pt]
 $\displaystyle\oraplant_{H,B}(v,y):=
 \begin{cases}
 1,&(v,y)\in\cP_H,\\
 B(v,y),&\text{otherwise}
 \end{cases}$
 &$\oranoise_B(v,y):=B(v,y).$
\end{array}
\end{equation*}
\end{mdframed}
We emphasize that the algorithms access only the sampled oracle $\oraplant_{H,B}$ or $\oranoise_B$, with no separate access to $H$ or $B$. With overwhelming probability, the noise contributes exponentially many
accepted targets for a fixed codeword $v$ in either case, whereas
$\cP_H$ adds at most one. Thus, the additional pair has only an
exponentially small relative effect on $N_\ora(v)$, eliminating the
noticeable gap used by the counting attack. Of course, merely eliminating the counting attack does not prove $\QCMAHbb$ hardness. However, looking ahead, having noise in both cases will allow us to use a \emph{switching lemma} \cite{Hastad86,Kelley} to remove witnesses one by one from depth-$d$ $\QCMAHbb$ machines and reducing its depth, akin to some classical oracle separations regarding $\PH$~\cite{Hastad86,FOCS:RosSerTan15}. But first, to make the switching lemmas from classical complexity applicable, we will \emph{classicalize} our quantum verifiers. More precisely, we will use the sparsity of the noise ($\beta=2^{-\lceil\sqrt\lambda\rceil}$) and the properties of the codeword-hash relation to show \emph{contractive simultaneous incompressibility}, and then use this to show how to simulate quantum verifiers using \emph{deterministic-classical queries}. Finally, we note that even with our addition of the noise to the YES distribution, it is not too hard to prove that $\QMAbb$ completeness and soundness still holds with the same verifier as \cite{BHV,CakanQCIP26}.

\paragraph{Our result.}
We obtain the following query complexity result for \CHN{}.

\begin{theorem}[Informal, \cref{thm:qma-verification,thm:hierarchy}]
\label{thm:overview-chn}
There is a \QMAbb{} machine that accepts planted samples and rejects noise samples,
except on a negligible fraction from each distribution.
For every constant $d$ and query-efficient $\QCMAHbb_d$ machine $M_d$
satisfying $M_d^\ora(1^\lambda)\ne\bot$ almost surely under both distributions,
\begin{equation}\label{eq:overview-target}
 \left|
 \Pr_{\ora\leftarrow\oradist_{\mathsf{plant},\lambda}}[M_d^\ora(1^\lambda)=1]
 -\Pr_{\ora\leftarrow\oradist_{\mathsf{noise},\lambda}}[M_d^\ora(1^\lambda)=1]
 \right|\le\negl(\lambda).
\end{equation}
\end{theorem}
Although the distributions have overlapping supports, as mentioned beofre, removing the
negligible exceptional sets where $\QMAbb$ acceptance (for planted distribution) and rejection (for noise distribution) guarantee fails yields a promise problem in
$\QMAbb\setminus\QCMAHbb$ in the query complexity model (\cref{thm:hierarchy}).

\subsection{Complexity Proof Overview}\label{sec:overview-proof}
Throughout the whole proof, we fix the prefix predicates $(H_1,\ldots,H_\lambda)$ to any fixed one that satisfies the following: $H$ is a \emph{good hash} (we can invert all $y$) with overwhelming probability over the sampling of the suffix predicates $H_{\lambda+1},\dots,H_{q}$ (\cref{lem:code-hash}). Indeed, the sampling of the prefix in both $\oradist_{\mathsf{plant},\lambda}, \oradist_{\mathsf{noise},\lambda}$ already satisfies this property with overwhelming probability. We implicitly allow all algorithms to depend on this choice.

Our proof of  \cref{thm:overview-chn} consists of three main steps.
\begin{enumerate}
\item \textbf{Deterministic-classical-query simulation of $\QCMAbb$ Verifier:} First, using the \emph{contractive simultaneous incompressibility} property of our oracle distributions, we show the following: With a query-efficient simulator algorithm $\Sim_V^\ora(z,w)$ that makes \emph{deterministic-classical} queries (with randomized post-processing), we can simulate (for most oracles under either the planted or noise distribution) a $\QCMAbb$ verifier's execution $V^\ora(z,w)$ on any witness $w$ (\cref{sec:classical-simulation}).
\item \textbf{Removing the witness: DNF represention of the simulator and switching}   
Then, thanks to the fact that its queries are deterministic, classical and polynomially many, we represent the
simulator's execution and the existential quantification of an accepting witness in a \emph{small/polynomial width} disjunctive
normal form (DNF) function of the oracle's bits. Note that even a single quantum query can require exponential-width DNF. Small-width DNF allows us to apply a \emph{switching lemma}~\cite{Hastad86,Kelley} to remove the existential quantification over witnesses. Thus, Step 1 then 2 \emph{morally} turns the $\QCMAbb$ machine into a $\PTIMEbb$ machine, with a caveat that this machine will depend on some sections of the oracle. (\cref{sec:classical-switching}). 
\item \textbf{(Repeated) Depth reduction} Finally, this allows us to reduce the level/depth of a $\QCMAHbb$ machine by one, by replacing the \emph{deepest} $\QCMAbb$ machine with a \emph{witness-less} algorithm (\cref{sec:hierarchy-depth}). Then repeatedly applying the same steps (again, with caveats) completes our proof of \cref{thm:overview-chn}, since any $\QCMAHbb$ machine has constant number of levels (\cref{sec:hierarchy-separation}). 
\end{enumerate}
We note that switching lemmas are commonly used to remove witnesses or reduce depth~\cite{Hastad86,FOCS:RosSerTan15,Impagliazzo11,HiraharaNanashima22,AIK,AB}. However, apart from this high-level resemblance, our proof has no direct connection to the arguments in these works (see \cref{sec:more-related-work}).

\subsubsection{Deterministic-Classical-Query Simulation of $\QCMAbb$ Verifiers}\label{sec:overview-simulation}
\label{sec:overview-simulation-proof}

To bring ourselves to a setting where we can utilize classical complexity techniques, we first show how to simulate a $\QCMAbb$ verifier $V$ using deterministic-classical queries. Our simulator will work simultaneously for all witnesses and for both oracle distributions. In fact, our simulation will work even when the description of $V$ depends on full descriptions of some suffix predicates $(H_i)_{i \in K}$ (where $K \subseteq \{\lambda+1,\dots,q\}$). As we see later on, this is crucial since our next step (\emph{switching}) will introduce dependence of some of the suffix predicates, but for repeated depth reduction, we will need to apply simulation again even with such dependence.

Now we introduce some notation. For nonempty
$I\subseteq\{\lambda+1,\ldots,q_\lambda\}$, write
\[
 H_I=(H_i)_{i\in I},\qquad
 \cP_I=\{(v,y)\in\cX_\lambda:v\in C_\lambda,\
 H_i(v_i)=y_i\ (i\le\lambda),\
 H_i(v_i)=0\ (i\in I)\}.
\]
Set $\cP_\varnothing=\cX_\lambda$ and
$\overline K=\{\lambda+1,\ldots,q_\lambda\}\setminus K$.

We will prove the following simulation lemma\footnote{ We note that a classical-query simulation lemma also appears in \cite{CakanQCIP26} for the noise distribution and the \emph{structure} distribution of \CHS{}, but in a weaker form. In particular, it uses randomized queries (whereas deterministic queries is crucial for us) and it does not allow dependence of $V$ on sections of oracles (e.g. $H_K$), which is again crucial for us. Separately, a similar argument to our query-database size bounding approach used in our proof also appears in \cite{BHV} (and in a weaker form in \cite{QFire}).}.

\begin{lemma}[Informal, \cref{thm:classical-simulation}]
\label{lem:overview-simulation}
Let $K\subseteq\{\lambda+1,\ldots,q_\lambda\}$ satisfy
$|\overline K|\ge\lfloor(q_\lambda-\lambda)/d\rfloor$ for a constant $d\ge1$. Fix any  $H_K$ and $B|_{\cX_\lambda\setminus\cP_K}$.
For every verifier algorithm $V_{H_K,\,B|_{\cX_\lambda\setminus\cP_K}}$,
there is a simulator $\Sim_{V_{H_K,\,B|_{\cX_\lambda\setminus\cP_K}}}$
making polynomially many deterministic-classical queries followed by
randomized oracle-free computation, such that, with overwhelming
probability over $H_{\overline K}$ and $B|_{\cP_K}$,
\begin{equation}\label{eq:overview-simulation}
 \left|\Pr[V_{H_K,\,B|_{\cX_\lambda\setminus\cP_K}}^\ora(z,w)=1]-\Pr[\Sim_{V_{H_K,\,B|_{\cX_\lambda\setminus\cP_K}}}^\ora(z,w)=1]\right|
 \le1/100
\end{equation}
simultaneously for every $z,w$ and both
$\ora\in\{\oraplant_{H,B},\oranoise_B\}$.
\end{lemma}
The full theorem (\cref{thm:classical-simulation}) is more general, which is needed for our cryptography results later on.

\paragraph{Contractive simultaneous incompressibility.}
To prove our simulation lemma, we first recall the notion of \emph{contractive simultaneous
incompressibility}, from \cite{CakanQCIP26}.\footnote{A similar notion is
used implicitly in~\cite{BHV}, and a weaker version appears in \cite{QFire}.}
Let $F$ be a set sampled according to some distribution and consider a $c(\lambda)$-bit classical witness $w$ that may
depend arbitrarily on it (i.e. chosen by an unbounded sender who sees $F$). Then, given $w$, an unbounded receiver with no access
to $F$ tries to output elements of $F$, and we allow the sender and this receiver to pre-share randomness independent of $F$.
While the general version can have different parameters, in our version, we say $F$ satisfies \emph{contractive simultaneous incompressibility}
if every such sender and receiver, the probability that the receiver outputs
\[
 t_{\rm inc}(\lambda,c(\lambda))
 =\ceil{\frac{2(c(\lambda)+8\lambda^2)}{\sqrt\lambda}}
\]
distinct elements of $F$ is at most $2^{-8\lambda^2}$,
for every polynomial $c(\lambda)$ and all sufficiently large $\lambda$.

\paragraph{Incompressibility in the planted and noise cases.} Using a suitable choice of the code $C$, we show that the set
\begin{equation}\label{eq:overview-sparse-support}
 F_K:=\cP_H\cup\{(v,y)\in\cP_K:B(v,y)=1\}
\end{equation}
satisfies the above contractive simultaneous incompressibility bound,
even after fixing $H_K$ and $B|_{\cX_\lambda\setminus\cP_K}$
(\cref{cor:code-noise-incompressibility}). This covers both distributions for our argument below, since $F_K$ contains every input in
$\cP_K$ at which either $\oraplant_{H,B}$ or $\oranoise_B$ answers $1$.

\paragraph{Our simulator.}
Write $V=V_{H_K,\,B|_{\cX_\lambda\setminus\cP_K}}$, let $Q=\poly(\lambda)$
denote its number of queries, and let $t,m$ be polynomials (chosen later).  For every set $E\subseteq\cP_K$, define
 \[
 \ora_E(u)=
 \begin{cases}
 B(u),&u\notin\cP_K,\\
 \indicator_E(u),&u\in\cP_K.
 \end{cases}
\]
For $j\in[Q]$, let $\pi^{(j)}_{z,w,E}$ be the distribution defined as measuring the query input register for the $j$-th query during (purified) execution of $V^{\ora_E}(z,w)$. Now independently sample $u^{(j,\ell)}_{z,w,E} \samp \pi^{(j)}_{z,w,E}$ for $\ell \in [m], j \in [Q]$ and for every $z,w,E$ with $|E|<t$; and set $\textbf u^{(j)}_{z,w,E} = (u^{(j,1)}_{z,w,E}, \dots, u^{(j,m)}_{z,w,E})$. Thus, each list contains $m$ samples from the corresponding query,
obtained from independent executions of $V^{\ora_E}(z,w)$.
We write $\mathbf U$ for the full collection of these samples.

Essentially, our simulator will try to estimate a database $D$ that records the addresses of the non-zero bits of our oracle, \emph{that $V$ could query}. Intuitvely, we will be able to do so due to the simultaneous incompressibility property above, since the queries of $V$ with non-zero output will be concentrated in a small list.

\begin{boxedalgorithm}{$\underline{\Sim_V^\ora(z,w)}$}
\begin{enumerate}[leftmargin=*,itemsep=4pt]
 \item Set $D=\varnothing$.
 \item While $|D|<t$:
 \begin{enumerate}[leftmargin=*,itemsep=3pt]
  \item For $j=1,\ldots,Q$, consider $(u^{(j,\ell)}_{z,w,D})_{\ell \in [m]}$ in order, and query $\ora(u)$ for each one that has $u\in\cP_K\setminus D$. As soon as a query returns $1$, add its
  input $u$ to $D$ and go back to Step~2.
  \item If all $Q$ lists have been completed without adding to $D$,
  go to Step~3.
 \end{enumerate}
 \item Without further oracle queries, compute $p_D$, value of $\Pr[V^{\ora_D}(z,w)=1]$ with additive error at most $\lambda^{-2}$, and output
 $\Bernoulli(p_D)$.
\end{enumerate}
\end{boxedalgorithm}
Note that $\Sim$ makes at most $tQm$ oracle queries.

\paragraph{Showing $|D| < t$ by incompressibility.}
We first show that the probability that the simulator reaches $|D|=t$ for any $(z,w)$ is negligible in total. To show this, we willl use our simulator as an encoding scheme for the incompressibility game. Consider the incompressibility game for
$F=F_K$, with $(\textbf u^{(j)}_{z,w,D})_{\substack{j\in [Q],z,w,D:\\ |D| < t}}$ as shared randomness, which is independent of
$F$. Given $F$, the sender tries every $(z,w)$ and subset $S\subseteq F$,
running the simulator while answering its queries with $\ora = \indicator_S$.
If a run reaches $|D|=t$, it sends the $c(\lambda)$-bit message
\[
 (z,w,k_1,\ldots,k_t),\qquad
 c(\lambda)=|z|+|w|+t\ceil{\log_2(tQm)},
\]
where $k_i\le tQm$ is the query number of $\Sim$ at which the $i$th element
was added to $D$ during the execution. Now the receiver
runs the simulator, answering $1$ exactly at
these query numbers and also records the query input as $x_i$. It answers $0$ elsewhere. Finally, the receiver outputs $x_1, \dots, x_t$. Note that all $x_i$ lie in $F$ and are distinct, since inputs
added to $D$ are never queried again. Thus, choosing polynomial $t$ with
$t_{\rm inc}(\lambda,c(\lambda))\le t$, incompressibility bounds the
probability of reaching $|D|=t$ by $2^{-8\lambda^2}$.

\paragraph{Simulation Accuracy.} We now argue simulation accuracy as follows: Any non-zero set of addresses of our oracle will either be eventually added to the database $D$, or the verifier's query weight on that address set is small enough that not having it in the database makes a small difference.
Let $D_{\rm fin}$ be the final database during execution of $\Sim$, and assume $|D_{\rm fin}|<t$ as above.
Write $W(E)$ for the total query weight of $V^{\ora_E}(z,w)$ on inputs
where $\ora_E$ and $\ora$ differ, summed over all $Q$ queries.
We claim that, with probability $1-\negl(\lambda)$ over $\mathbf U$,
we have $
 W(D_{\rm fin})\le\frac{1}{\lambda^2Q},
$
simultaneously for all $z,w$ and both oracles.

Then, by the hybrid bound (\cite{BBBV}), replacing $\ora_{D_{\rm fin}}$ with $\ora$
changes the verifier's acceptance probability by $O(1/\lambda)$.
Our simulator outputs $\Bernoulli(p_{D_{\rm fin}})$, where
$p_{D_{\rm fin}}$ approximates the acceptance probability of
$V^{\ora_{D_{\rm fin}}}(z,w)$ within $\lambda^{-2}$, so its simulation error is
$O(1/\lambda)+\lambda^{-2}<1/100$ for sufficiently large $\lambda$.

Now to prove our claim, fix $z,w,\ora$ and any database $E$ with $|E|<t$. Now observe that
each sample $u^{(j,\ell)}_{z,w,E}$ is obtained by measuring the query input
register at query $j$ in an independent execution of $V^{\ora_E}(z,w)$.
Thus, any input $u$ with query weight $a$ at this step is sampled with
probability $a$. By independence, the probability that all $Qm$ samples avoid the inputs
where $\ora_E$ and $\ora$ differ is at most $e^{-mW(E)}$.

Now note that for the simulator to terminate with $D_{\rm fin}=E$ for some $E$,
it must be the case that during the final loop, all $Qm$ values $(u^{(j,\ell)}_{z,w,E})_{j \in [Q], \ell \in [m]}$ avoid the inputs where $\ora_E$ and $\ora$ differ (since we have already
ruled out the other termination condition $|D_{\rm fin}|=t$).
Otherwise, $\Sim$ would query an input $u\notin E$
with $\ora(u)=1$ and add it to $D$, and hence that would not be the last loop (recall that we only stop when there is no new additions).
Thus, for each fixed $E$ with $|E|<t$, we have
$\Pr_{\mathbf U}[D_{\rm fin}=E]\le e^{-mW(E)}$.
A union bound over all such $E$ with $W(E)>1/(\lambda^2Q)$,
all $z,w$, and both oracles proves our claim about $W(D_{\rm fin})$, for sufficiently large
polynomial $m$.

\paragraph{Making $\Sim$ deterministic query.} We now fix a deterministic choice of $\mathbf U=(\mathbf u^{(j)}_{z,w,D})_{z,w,D,j}$. For any $\mathbf U$,  let $\varepsilon(\mathbf U)$ be the probability over
$H_{\overline K},B|_{\cP_K}$ that acceptance error of $\Sim$ exceeds $1/100$
for some $z,w$ and either oracle. Choose the lexicographically first $\mathbf U^\star$ such that
$
 \varepsilon(\mathbf U^\star)
 \le\mathbb E_{\mathbf U}[\varepsilon(\mathbf U)]
 \le\negl(\lambda).
$ Now we hardwire\footnote{Recall that in our query complexity model, the only restriction is number of queries, so we can have such exponential length description.} this into our $\Sim$. Note that $\mathbf U^\star$ depends only on $V,H_K,B|_{\cX_\lambda\setminus\cP_K}$ that are already fixed; beyond that, it only depends on the distributions of our oracles and not actual sampled oracles. 

\Needspace{6\baselineskip}

\subsubsection{Removing the witness: From deterministic-classical queries to DNFs and Switching}\label{sec:overview-switching}
\paragraph{Small-Width DNF Represention}
Now observe that since $\Sim_{V_{H_K,\,B|_{\cX_\lambda\setminus\cP_K}}}$ (\cref{lem:overview-simulation})
makes at most $k=\poly(\lambda)$ \emph{deterministic-classical} queries, for fixed $z, w$  its output depends only on $k$ oracle bits (beyond dependence on ${{H_K,\,B|_{\cX_\lambda\setminus\cP_K}}}$). Since this simulator is adaptive, the addresses of the bits can depend on the earlier queried bits, but still, for any auxiliary input $z$ and witness $w$, whether $\Sim_{V_{H_K,\,B|_{\cX_\lambda\setminus\cP_K}}}^\ora(z,w)$ accepts (with some probability\footnote{Due to randomized post-processing} $p$) can be written as a disjunction of width-$k$ conjunctions of the oracle bits: Simply check if there is a \emph{query transcript}\footnote{Simulator's adaptive queries and their answers} that matches the oracle and such that the simulator accepts with probability $p$. Note that this is in stark contrast to quantum-query algorithms, since even a single deterministic quantum query can depend on all of the oracle bits.

Further, the existence of an accepting witness can be written as another layer of disjunction.  More precisely, for some query outputs $a_1, \dots, a_k$, let $p_{z,w,a}$ be the acceptance probability of $\Sim_{V_{H_K,\,B|_{\cX_\lambda\setminus\cP_K}}}(z,w)$ and denote its query inputs during this execution as $u^{z,w,a}_1, \dots, u^{z,w,a}_k$. Then we have

\begin{equation}\label{eq:overview-transcript-dnf}
      f_{z,{{H_K,\,B|_{\cX_\lambda\setminus\cP_K}}}}(\ora):=\mathbf 1\!\left[
        \exists w:\Pr[\Sim_{V_{H_K,\,B|_{\cX_\lambda\setminus\cP_K}}}^\ora(z,w)=1]\ge\tfrac12
      \right]
      =\bigvee_{\substack{w,a:\\p_{z,w,a}\ge1/2}}
        \;\bigwedge_{j=1}^{k}[\ora(u^{z,w,a}_j)=a_j].
     \end{equation}

\paragraph{Removing the witness by switching.}
While $f_{z,{{H_K,\,B|_{\cX_\lambda\setminus\cP_K}}}}$ has polynomial \emph{width}, it still depends on exponentially many bits of the oracle $\ora$. However, since it is a DNF, switching
lemmas~\cite{Hastad86,Kelley} can help us compute $f_{z,{{H_K,\,B|_{\cX_\lambda\setminus\cP_K}}}}$ using a polynomial number of classical-queries, after the  oracle bits at all positions but a random small fraction are fixed (to independent random values). Note that here \emph{small} means as a fraction of whole oracle table size $N$, but still superpolynomial in $\lambda$, so this is indeed non-trivial.

More precisely, Kelley's switching lemma~\cite{Kelley} applies to
any fixed width-$k$ DNF and a random set $S$ of \emph{free positions},
provided that it satisfies a \emph{joint sparsity condition}: For an integer $Q\ge1$ and $0<p\le1/(16k)$,
every fixed set $A$ of at most $k+Q$ positions satisfies
\[
 \Pr_S[A\subseteq S]\le p^{|A|}.
\]
After independently and uniformly fixing the oracle bits outside our \emph{free positions} $S$, with probability at least $1-2(8kp)^Q$, a classical-query deterministic
algorithm whose description depends only on the position set $S$ and the fixed outside bits can compute the DNF value using fewer than $Q$ queries (for any oracle that has our fixed values outside $S$).

Since $\Sim_{V_{H_K,\,B|_{\cX_\lambda\setminus\cP_K}}}$ and thus $f_{z,{{H_K,\,B|_{\cX_\lambda\setminus\cP_K}}}}$  depend on $H_K$ and $B|_{\cX_\lambda\setminus\cP_K}$, we will consider $H_{\overline{K}}$ and  $B|_{\cP_K}$  for our switching lemma. Let $J\subseteq\{\lambda+1,\ldots,q_\lambda\}\subseteq \overline{K}$ be such that $|J|\ge\lfloor(q_\lambda-\lambda)/d\rfloor$ for a constant $d\ge1$. These will be our \emph{still fresh suffix indices} that we will use for $S$ in the switching lemma. We consider any such $J$ instead of simply $\overline{K}$ so that we are only \emph{spending} some of the remaining fresh suffix predicates: Observe that our switching lemma will introduce dependence on $H_J$, but in the depth reduction step later on we will need to apply it multiple times. Now observe that our oracles have the form needed for switching. $\oranoise_B=B$ and
$\oraplant_{H,B}=B\vee\indicator_{\cP_H}$ agree with $B$ on
$\cP_K\setminus\cP_{K\cup J}$, and the fresh predicates $H_J$ give
\begin{equation}\label{eq:overview-switching-sparsity}
 \Pr_{H_J}[u_1,\ldots,u_t\in\cP_{K\cup J}]
 \le2^{-\Omega(\lambda t)}
 \qquad\text{for distinct }u_1,\ldots,u_t\in\cP_K,\quad t\le2^\lambda,
\end{equation}
So, in the switching theorem above\footnote{Technically, Kelley's lemma in its original form only applies to independent uniform random fixing outside $S$, whereas our $B$ has biased bits. But the biased version of their lemma is an easy extension (see \cref{thm:biased-switching}).}, we can use $S = \cP_{K\cup J}$ as our free positions and randomly \emph{fix} outputs of the other (i.e. $\cP_K\setminus\cP_{K\cup J}$) positions in $\cP_K$ to these common $B$-values, recalling that the common answers on $\cP_K\setminus\cP_{K\cup J}$ are independent $\Bernoulli(\beta)$ bits in both the planted and noise distributions. Thus, we obtain a classical-query deterministic algorithm $T$ with the same description for both the planted and noise oracles. This is precisely why adding noise to both the YES and NO distributions helps, as
mentioned before.

\begin{lemma}[Informal, \cref{thm:biased-switching}]
\label{lem:overview-switching}
Let $K,J\subseteq\{\lambda+1,\ldots,q_\lambda\}$ be disjoint, with
$|J|\ge\lfloor(q_\lambda-\lambda)/d\rfloor$ for a constant $d\ge1$.
Fix any $H_K$ and $B|_{\cX_\lambda\setminus\cP_K}$, and fix the
simulator $\Sim_{V_{H_K,\,B|_{\cX_\lambda\setminus\cP_K}}}$ from
\cref{lem:overview-simulation}.

For any polynomially bounded integer $Q\ge1$ there is a classical-query deterministic algorithm $T_{H_K,\,H_J,\,B|_{\cX_\lambda\setminus\cP_{K\cup J}}}$
taking no witness and making fewer than $Q$ queries, all in
$\cP_{K\cup J}$. 
For each fixed auxiliary input $z$, with probability at least
$1-2^{-\Omega(Q\sqrt\lambda)}$ over $H_J$ and $B|_{\cP_K\setminus\cP_{K\cup J}}$,
\[
\begin{aligned}
 T_{H_K,\,H_J,\,B|_{\cX_\lambda\setminus\cP_{K\cup J}}}^\ora(z)&=f_z(\ora)\\
 &=\mathbf1\!\left[\exists w:\Pr[\Sim_{V_{H_K,\,B|_{\cX_\lambda\setminus\cP_K}}}^\ora(z,w)=1]\ge\tfrac12\right]
\end{aligned}
\]
for both $\ora\in\{\oraplant_{H,B},\oranoise_B\}$.

We emphasize that unlike the simulator $\Sim_{V_{H_K,\,B|_{\cX_\lambda\setminus\cP_K}}}$, $T_{H_K,\,H_J,\,B|_{\cX_\lambda\setminus\cP_{K\cup J}}}$ receives no witness. 
\end{lemma}
The full theorem (\cref{thm:biased-switching}) is more general, which is needed for our cryptography results later on.

\subsubsection{(Repeated) Depth reduction}\label{sec:overview-depth}
\paragraph{Warm-up: $\QCMAbb$ Hardness}
Observe that for the first level of $\QCMAHbb$, that is, for a fixed $\QCMAbb$ machine $M_1$ with verifier $V$, we are already done. Since $M_1$ does not depend on the oracle, \cref{lem:overview-simulation} with $K=\varnothing$ gives, except with negligible probability,
$f_z(\ora)=\mathbf1\!\left[
  \exists w:\Pr[\Sim_V^{\ora}(z,w)=1]\ge\tfrac12
\right]=M_1^\ora(z)$ for both oracles $\ora\in\{\oranoise_B,\oraplant_{H,B}\}$ whenever $M_1^\ora(z)\ne\bot$. Using $J = \{\lambda + 1, \dots, q\}$ in \cref{lem:overview-switching}
with $Q=1$ and $z=1^\lambda$ then gives a deterministic classical algorithm
$T=T_{H,\,B|_{\cX_\lambda\setminus\cP_H}}$
that makes no queries and has the same description for both
oracles and satisfies, except with negligible probability,
$M_1^\ora(1^\lambda)\ne\bot\implies T^\ora(1^\lambda)=M_1^\ora(1^\lambda)$
for both oracles. Therefore, $M_1$ cannot distinguish the oracles and we are done. For a general $\QCMAHbb_d$ machine, we will repeatedly do the same argument starting with the deepest $\QCMAbb$ machine $M_1$, but more carefully: Note that already for a $\QCMAHbb_2$ machine defined by $V_2, M_1$, the above argument would be an issue since our replacement for $M_1$ depends on all of $H$, so we cannot apply another step of simulation or switching.

\paragraph{General case ($\QCMAHbb_d$)}
We now sketch the proof for the general case (see \cref{thm:hierarchy}
and the full proofs in \cref{sec:hierarchy-depth,sec:hierarchy-query-proof}).
Consider a $\QCMAHbb_d$ machine $M_d$ with verifiers $V_1,\ldots,V_d$,
where the deepest/bottom verifier $V_1$ queries only $\ora$ and each $V_j$, for $j>1$, also
queries a Boolean completion of the lower machine $M_{j-1}^\ora$
(defined by $V_1,\ldots,V_{j-1}$). Starting with the deepest level $M_1$, we will replace each machine by a classical-query deterministic algorithm $T_1$
taking no witness, then repeat for the next level. At each step,
the replacement will agree with the non-$\bot$ outputs of the machine being replaced (except with negligible probability), which also preserves the higher
levels' non-$\bot$ outputs by the Boolean-completion definition
(\cref{def:qcmah-machine}).

To leave fresh predicates for each step, partition the suffix indices
into $d$ groups $J_1,\ldots,J_d$ whose sizes differ by at most one,
and write
\[
 S_i=\bigcap_{j\le i}\cP_{J_j},\qquad
 \cX_\lambda=S_0\supseteq S_1\supseteq\cdots\supseteq S_d=\cP_H.
\]
At proof step $i$, $H_K$ and $B|_{\cX_\lambda\setminus S_{i-1}}$ are already fixed. Let $K=J_1\cup\cdots\cup J_{i-1}$. The
remaining verifiers, from top to bottom, are $V_d,\ldots,V_i$.
The bottom verifier $V_i$ queries $\ora$ and, for $i>1$, the
previously constructed algorithm
$T_{i-1,\,H_K,\,B|_{\cX_\lambda\setminus S_{i-1}}}^\ora$
(for $i=1$, it queries only $\ora$).
We now remove $V_i$ and its witness. First, we incorporate $T_{i-1}$ into its code so that it only queries $\ora$.
Simulation (\cref{lem:overview-simulation}) gives us
$\Sim_{i,\,H_K,\,B|_{\cX_\lambda\setminus S_{i-1}}}$.
For $i<d$, we then apply switching (\cref{lem:overview-switching})
with $J=J_i$, choosing $Q=\poly(\lambda)$ large enough to union bound
over all $2^{\poly(\lambda)}$ auxiliary inputs $z$.
This gives us a classical-query deterministic algorithm
$T_i=T_{i,\,H_K,\,H_{J_i},\,B|_{\cX_\lambda\setminus S_i}}$
with the same description for both oracles, as guaranteed by
\cref{lem:overview-switching}. It takes no witness, makes fewer
than $Q$ queries in $S_i$, and satisfies, except with negligible probability over $H_{\overline K}$ and $B|_{S_{i-1}}$,
for all such $z$ and each $\ora\in\{\oraplant_{H,B},\oranoise_B\}$,
\[
\begin{aligned}
 \exists w:\Pr[V_i^{\ora,T_{i-1}^{\ora}}(z,w)=1]\ge\tfrac23
 &\quad\Longrightarrow\quad T_i^{\ora}(z)=1,\\
 \forall w:\Pr[V_i^{\ora,T_{i-1}^{\ora}}(z,w)=1]\le\tfrac13
 &\quad\Longrightarrow\quad T_i^{\ora}(z)=0.
\end{aligned}
\]
So, $T_i$ agrees with a Boolean completion of the current level-$i$
machine on every possible query, and so we can remove $V_i$ and its witness
and continue with $V_{i+1}^{\ora,T_i^\ora}$ without changing the
hierarchy's non-$\bot$ outputs (\cref{def:qcmah-machine}).

Finally, for the last step $i = d$, take $Q=1$, where only $z=1^\lambda$ matters.
The resulting $T_d$ makes no queries and has the same description
for both oracles, so it outputs the same bit for both. Since our
replacements preserve $M_d^\ora(1^\lambda)$ whenever it is not
$\bot$ (except with negligible probability), this proves
\cref{thm:overview-chn}.

\subsection{Cryptographic Oracle and Constructions}
\label{sec:overview-crypto}\label{sec:overview-oracle-details}
We focus on our PKE result in this overview and refer the reader to the relevant sections for our other cryptography results.

We construct an oracle $\ora$ such that $\PTIME^\ora=\QCMA^\ora$ and
there is a scheme $\PKE^\ora$ with classical ciphertexts and public
keys that satisfies the following strong guarantee called
\emph{copy-protection}: a QPT adversary $\cA_0$ tries to split the
quantum decryption key $\ket K$ into two registers $\regi_1,\regi_2$
for two noncommunicating adversaries $\cA_1,\cA_2$ that receive
independent ciphertext challenges, and we require that their probability
of both guessing their challenge bits is at most $1/2+\negl(\lambda)$
(\cref{def:pke-piracy}).

\paragraph{Intuition}
On a very high level, our main idea is to use our problem \CHN{} to
encode a function $\mathfrak F$ (that provides us e.g. a copy-protected PKE construction) into an oracle $E$, and use our
\QMAbb{} witness state $\ket{\Psi_H}$ to \emph{unlock} it and achieve
\emph{functionality/correctness}. Then, using the tools underlying the
$\QCMAHbb$ hardness of \CHN{}, we will reduce attacks in presence of an oracle that collapses $\QCMAH$ to $P$, to QPT attacks with direct
$\mathfrak F$ access. Indeed, \cite{KQT}
{} uses the same idea, but instead with the Forrelation
problem.

\paragraph{Our Function Encoding and Decoding}
To encode a function $\mathfrak F$, we modify \CHN{} as follows:
instead of planting the constant-$1$ function at codeword-hash pairs,
we plant the outputs of $\mathfrak F$. More precisely, let
$C_\kappa\subseteq\Sigma_\kappa^{q_\kappa}$ be our code and
$\mathfrak F:\zo^\kappa\to\zo$ be the function to encode.\footnote{Here
$\kappa=\poly(\lambda)$ is the \CHN{} parameter, chosen large enough
to encode each function input in $\kappa$ bits.} We sample $H$ as in
\CHN{} (\cref{def:hash-distribution}) and define
$E=E_{H,B,\mathfrak F}$ on
$(v,y)\in\Sigma_\kappa^{q_\kappa}\times\zo^\kappa$ by
\begin{equation}\label{eq:overview-function-encoding}
 E(v,y)=
 \begin{cases}
 \mathfrak F(y),&v\in C_\kappa\text{ and }H(v)=y\|0^{q_\kappa-\kappa},\\
 B(v,y),&\text{otherwise},
 \end{cases}
 \qquad
 \begin{gathered}
 H\leftarrow\cH_\kappa,\\
 B(v,y)\overset{\mathrm{ind}}\leftarrow\Bernoulli(1/2).
 \end{gathered}
\end{equation}
Now recall that the witness state $\ket{\Psi_H}$ lets us find, for any
chosen $y\in\zo^\kappa$, a codeword $v\in C_\kappa$ satisfying
$H(v)=y\|0^{q_\kappa-\kappa}$ with overwhelming probability. Since
$E(v,y)$ has the same output for every such $v$, we can obtain
$\mathfrak F(y)$ for any $y$ without damaging $\ket{\Psi_H}$, up to
negligible error (\cref{sec:quantum-evaluation}). We note that a similar
encoding was used recently by~\cite{ClonableQMA}, where they encode
$H_i$ itself to achieve a $\mathsf{ClonableQMA}$--$\QCMA$ separation.

For cryptography, we will use many independent instances of our
problem, indexed by a serial number $s\in\zo^\lambda$, and write
$E_s=E_{H_s,B_s,\mathfrak F_s}$. For now, let us assume that our
constructions can \emph{magically} obtain copies of
$\ket\Omega=2^{-\lambda/2}\sum_s\ket s\ket{\Psi_{H_s}}$.
Measuring its serial register gives the witness state for a random
$s$, and later we will show how to prepare this superposition with
$\Prepare$ using classical Boolean oracles. A similar step was also
used in~\cite{CountCrypt}, but with quantum unitary oracles and
Haar-random states.

\paragraph{Constructing PKE}
Since we are trying to achieve a copy-protected PKE scheme, we start
with the copy-protection construction of~\cite{ALLLZ,CLLZ}, which uses
\emph{coset states} as decryption keys. For each serial number
$s\in\zo^\lambda$, sample a uniform $2\lambda$-dimensional subspace
$A_s\le\bbF_2^{4\lambda}$, uniform shifts
$a_s,a'_s\in\bbF_2^{4\lambda}$ and uniform Boolean functions
$F_{s,0},F_{s,1}:\zo^\lambda\to\zo$, and set
$F_s=F_{s,0}\oplus F_{s,1}$. Then consider the function
\[
 C_s(x,c,\xi)=
 \begin{cases}
 F_{s,0}(x),&c=0\text{ and }\xi\in A_s+a_s,\\
 F_{s,1}(x),&c=1\text{ and }\xi\in A_s^\perp+a'_s,\\
 \bot,&\text{otherwise}.
 \end{cases}
\]
As shown by~\cite{ALLLZ,CLLZ}, using
$\ket{\chi_s}=2^{-\lambda}\sum_{\xi\in A_s}
(-1)^{\langle a'_s,\xi\rangle}\ket{\xi+a_s}$, one can obtain
$F_{s,c}(x)$ for any $c,x$ without damaging $\ket{\chi_s}$, and this
can be used to decrypt ciphertexts $(u,m\oplus F_s(u))$.

The coset security argument of~\cite{ALLLZ} gives copy-protection
with direct oracle access to $C_s$, and Behera, Cakan, and
Goyal~\cite{BCG26} give a simpler proof. For our goal, we will encode $C_s$
into $E_s$ and show that attacks with $E,N,A$ access can be converted
to query-efficient attacks with direct $C_s$ access, at which point
we invoke the coset security proof. Thus, at a high level, the coset
state provides copy-protection security, while the \CHN{} encoding
lets us preserve it in the presence of the collapse oracle.

More precisely, we do the following.
\begin{itemize}
 \item Encode the coset functions $(C_s)_s$ into the oracle
 $E(s,v,y)=E_s(v,y)$ using our function encoding above. Here
 $\mathfrak F_s$ returns the requested validity or value bit of $C_s$,
 so that its output $\bot$ also has a Boolean representation.
 \item $\PKEGen$ samples a quantum key
 $\ket{K_s}=\ket{\Psi_{H_s}}\otimes\ket{\chi_s}$ with public key
 $s$ by measuring the serial register of
 $\ket{\Omega'}=2^{-\lambda/2}\sum_s\ket s\ket{K_s}$.
 \item $\PKEEnc$\footnote{This sampling is necessary: if encryption
 were pseudodeterministic given explicit randomness $(s,m;r)$, then
 $\PTIME=\QCMA$ would let us invert it and break security.}
 creates $\ket{\Gamma_s}=2^{-\lambda/2}\sum_u\ket u\ket{F_s(u)}$
 and measures it to obtain $(u,F_s(u))$ with uniform $u$, then encrypts
 the message as $(u,m\oplus F_s(u))$.
\end{itemize}
Again, for now assume that we can magically obtain $\ket{\Omega'}$
and $\ket{\Gamma_s}$, which we will later prepare with $\Prepare$.
Decryption correctness follows by combining our function decoding
with the coset-state decryption described above. The Boolean oracles
used by $\Prepare$, together with $E$, form $\orabase$
(\cref{def:pke-oracle}).

\paragraph{Adding the Collapse Oracle.}
To obtain $\PTIME^\ora=\QCMA^\ora$, we add two oracles $N,A$ to
$\orabase$. Morally, $N$ is an $\NP$ oracle, as in~\cite{KQT}, while
$A$ is a $\BQP$ oracle that lets us simulate quantum verifiers.
We build them from the ground up: suppose their answers on strings
of length at most $n$ have already been defined, and write
\[
 \ora_{\le n}=(\orabase|_{\le n},N_{\le n},A_{\le n}).
\]
We now define answers to queries of length $(n+2)^2$. Each such
query specifies a circuit $M$ and input $x$, and $M$ may query only
strings of length at most $n$, so $M^{\ora_{\le n}}$ uses only
answers already constructed:
\begin{enumerate}[leftmargin=*,itemsep=2pt]
 \item For $N$, interpret $M$ as a deterministic classical circuit
 and set
 $N(\langle M,x\rangle)=\mathbf1[\exists w\in\zo^{\le n}:
 M^{\ora_{\le n}}(x,w)=1]$.
 \item For $A$, interpret $M$ as a quantum circuit and sample
 \[
 A(\langle M,x\rangle,i)\overset{\mathrm{ind}}\leftarrow
 \Bernoulli\!\left(\frac{1+\Pr[M^{\ora_{\le n}}(x)=1]}3\right),
 \qquad i\in[n^4].
 \]
\end{enumerate}
Each sampled answer remains fixed, and we set
$\ora=(\orabase,N,A)$ (\cref{def:recursive-qcma}).

\paragraph{Showing $\PTIME^\ora=\QCMA^\ora$.}
Given a quantum verifier $V^\ora(x,w)$ for a problem in
$\QCMA^\ora$, define the deterministic classical circuit
\[
 M_{\rm test}^{A}(x,w)=\operatorname{Maj}_{i\in[n^4]}
 A(\langle V,(x,w)\rangle,i).
\]
This distinguishes the accepting and rejecting witnesses of $V^\ora$
with overwhelming probability, so a single longer query
$N(\langle M_{\rm test},x\rangle)$ decides whether there is an
accepting witness. Thus $\PTIME^\ora=\QCMA^\ora$ for almost every
sampled oracle (\cref{lem:helper-estimates}).

\subsection{Cryptography Proof Overview}
\label{sec:overview-crypto-simulation}\label{sec:overview-crypto-proof-details}
We now give a high level overview of our proof.

\begin{enumerate}
\item \textbf{Simulating the encoded function.}
First, assume that $\ket{\Psi_H}$ is supplied together with an auxiliary
state $\rho$, such as the coset state. In this setting, we show that
access to $E,N,A$ can be simulated using only $\mathfrak F$.
\begin{lemma}[Informal, \cref{thm:crypto-common-simulation}]
\label{lem:overview-crypto-simulation}
Fix a query-efficient quantum algorithm $\cA$ and Boolean oracle
$\ora_{\rm aux}$. There is a
query-efficient quantum simulator $\cS$ such that
\begin{equation}\label{eq:overview-coset-simulation}
 \cA^{\ora_{\rm aux},E_{H,B,\mathfrak F},N,A}(\ket{\Psi_H},\rho)
 \approx\cS^{\mathfrak F}(\rho).
\end{equation}
\end{lemma}

\item \textbf{Preparing the states using classical oracles.}
Next, we remove the assumption of supplied states by constructing
Boolean oracles that prepare the witness and coset states, as well as
the graph states used for encryption. To use the preceding simulation
in this setting, we also show that sampling the preparation oracles
for an exponentially close target remains undetectable, even with
access to $N,A$.
\begin{lemma}[Informal, \cref{lem:public-preparation,thm:preparation-replacement}]
\label{lem:overview-preparation}
For every $n$-qubit real-amplitude state $\ket\Omega$, there is a
distribution on Boolean oracles $R,S$ and a uniform QPT algorithm
$\Prepare$ such that, for sufficiently large $n$,
\[
 \bbE_{R,S}\,\tfrac12\|\Prepare^{R,S}(1^n)-\ket\Omega\!\bra\Omega\|_1
 \le2^{-n}.
\]
Write $\ora_\Omega=(\ora_{\rm aux},R,S,N,A)$ for the oracle obtained
by sampling $R,S$ for $\ket\Omega$ and then sampling $N,A$, with fixed
$\ora_{\rm aux}$. For another real-amplitude state
$\ket{\widetilde\Omega}$, $n\ge\lambda$ and every query-efficient
quantum algorithm $\cA$,
\[
 \|\ket\Omega-\ket{\widetilde\Omega}\|_2\le2^{-\Omega(\lambda)}
 \quad\Longrightarrow\quad
 \cA^{\ora_\Omega}(\rho)\approx\cA^{\ora_{\widetilde\Omega}}(\rho),
\]
where $\rho$ and the challenger's operations may depend on both
targets, but not on the sampled preparation or recursive answers.
\end{lemma}

\item \textbf{Completing the PKE proof.}
Finally, we keep the issued key and the rule for forming challenge
ciphertexts unchanged, but sample the preparation oracles for new
targets. This removes the preparation oracles' dependence on the
corresponding secret information. Each new target
is at distance $O(2^{-\lambda/2})$ from the original, so the preceding
guarantee justifies this change of oracles. We can then apply
encoded-function simulation and invoke the coset security proof
of~\cite{BCG26}, which bounds joint success on the independent
ciphertext challenges by $1/2+\negl(\lambda)$
(\cref{def:coset-experiments,lem:two-fresh-values}).
\end{enumerate}
\label{sec:overview-security-roadmap-end}

\subsubsection{Simulating Adversaries Using the Complexity Tools}
\label{sec:overview-crypto-simulation-proof}
To obtain the first guarantee, we simulate full-oracle queries by a classical-query deterministic
algorithm $T_\eta^E$, whose classical description $\eta$ is independent
of $\mathfrak F$, so that
\[
 \cA^{\ora_{\rm aux},E,N,A}(\ket{\Psi_H},\rho)
 \approx\cA^{T_\eta^E}(\ket{\Psi_H},\rho).
\]
The adversary may make quantum queries, so we evaluate $T_\eta^E$
coherently and leave its other operations unchanged. Thus it suffices
to construct $T_\eta^E$ and then answer its remaining $E$ queries
using $\mathfrak F$.

\paragraph{Classical-query simulation and switching.}
For this purpose, as in the complexity proof, partition the suffix predicates into
constantly many groups $J_1,\ldots,J_m$. The encoding differs from $B$
only on $\cP_H=\bigcap_j\cP_{J_j}$, so at each recursive level we can
use another group's incompressibility and switching guarantees.
Moreover, only constantly many levels can reach an encoding input,
because query lengths decrease by a square root at each level, whereas
the adversary's query length is polynomial and encoding inputs have
length at least $\lambda$. We can therefore choose enough groups to
leave one unused at the end.

With these groups fixed, we now construct the answering algorithms in increasing query length.
After substituting the algorithms for shorter queries into $M$, fix
the predicates and noise used so far and write $M^Y$ for the resulting
circuit, where $Y$ contains the remaining oracle answers. For an $N$
query, $M^Y$ is deterministic, so $\exists w:M^Y(x,w)=1$ has a
polynomial-width accepting-transcript DNF. The switching theorem
therefore gives a polynomial-query deterministic algorithm for this
condition (\cref{thm:biased-switching}).

For an $A$ query, however, $M^Y$ is quantum, so we first estimate
$p_M(Y)=\Pr[M^Y(x)=1]$ using quantitative classical-query simulation
(\cref{thm:classical-simulation}). We need exponentially small error,
which can require exponentially many simulation queries. However,
answers outside the chosen group's codeword-hash set are fixed, and
contractive simultaneous incompressibility bounds the number of
distinct queried inputs inside that set. We can therefore specify the
algorithm using only polynomially many queries to $Y$.
Crucially, both guarantees hold simultaneously for every choice of $Y$,
and, as in the complexity proof, the resulting descriptions may depend on the
predicates already used and the noise outside the current set, leaving
the other groups available for subsequent levels.

\paragraph{Keeping the recursive distribution exact.}
The estimate alone does not yet give the required simulation, because
replacing $p_M(Y)$ by $\widehat p_M(Y)$ in the sampling of $A$ could
change a later existential answer. Thus we must preserve $A$'s exact conditional
distribution throughout the recursion, and only afterwards argue
that the adversary cannot detect the remaining difference.

To preserve this distribution, observe that a bit with bias $(1+\widehat p_M(Y))/3$ can be sampled
using a fair coin with probability $2/3$, and a
$\Bernoulli(\widehat p_M(Y))$ coin otherwise. We correct the fair part on an intersection of new
independent random sets, chosen so that any input belongs to the
intersection with probability $\delta$, where
$\delta\ge\sup_Y|p_M(Y)-\widehat p_M(Y)|$ is exponentially small.
Outside the intersection we use a fair bit, and inside it we sample
with probability
\[
 \frac12+\frac{p_M(Y)-\widehat p_M(Y)}{2\delta}.
\]
Consequently, averaging over the intersection gives exactly the required
bias $(1+p_M(Y))/3$. Applying this sampling independently at each
address preserves $A$'s conditional distribution given all shorter
answers, while the algorithm for $N$ computes its answer exactly.

\paragraph{The quantum query hybrid.}
Now that the recursive answers have the correct distribution, we replace the auxiliary
correction functions by uniform Boolean functions and include them in
$\eta$, while leaving $E$ unchanged. To bound the effect, recall that each correction intersection has
one random set that the recursive simulation has not used. The
construction ensures that, after fixing $H,B,\mathfrak F,\eta$ and the
supplied state and challenger information, every fixed input still
belongs to such a remaining set with exponentially small probability.
Since these choices determine the query states in the new experiment,
the expected query weight on inputs where the answers differ is
exponentially small. The quantum query hybrid therefore proves the
desired simulation
(\cref{thm:split-recursive-preservation}).

\paragraph{Obtaining direct function access.}
It remains to answer the encoding queries. For fixed $\ora_{\rm aux}$,
$\eta$ depends only on $H,B$ and independent
coins, so the simulator can sample these objects and prepare
$\ket{\Psi_H}$ itself. It then runs $T_\eta^E$, answering an encoding
query with $\mathfrak F(y)$ on $\cP_H$ and with $B(v,y)$ elsewhere.
Moreover, using this one sample at all stages preserves noncommunication and
proves the encoded-function simulation guarantee
(\cref{lem:overview-crypto-simulation}). In particular, setting
$\rho=\ket{\chi_s}\!\bra{\chi_s}$ reduces the supplied-key experiment
to direct access to $C_s$, as required for the coset security proof.

\subsubsection{Preparing the States Using Classical Oracles}
\label{sec:overview-crypto-preparation}

So far, the states have been supplied directly. We now construct the
classical oracles used by $\Prepare$ to obtain the key superposition
$\ket{\Omega'}$ and graph state $\ket{\Gamma_s}$.
More generally, we want to prepare an arbitrary real-amplitude state
\[
 \ket\Omega=\sum_{x\in\{0,1\}^n}a_x\ket x
\]
using classical Boolean oracles. The target determines how the oracles
are sampled, while the preparation algorithm receives only quantum
query access to the sampled functions. In the security proof, we will
replace these oracles by ones sampled for a slightly different target,
so we also want close targets to give close probabilities for each
oracle bit. Merely preparing the right state does not ensure this
second property, as illustrated by a bit specifying whether $a_x\ge1/2$,
which can flip under an arbitrarily small change in $a_x$.

\paragraph{Obtaining amplitudes by averaging Boolean answers.}
A Boolean phase query multiplies each basis vector by $+1$ or $-1$,
so it does not directly produce arbitrary amplitudes. However,
averages of signs can approximate any number in $[-1,1]$, suggesting
that we obtain amplitudes by averaging Boolean answers.\footnote{A
related averaging step is used in the state-preparation method
of~\cite{SLSB19} to convert the fraction of inputs satisfying a Boolean
condition into an amplitude.}

To use this averaging step with signs, let $S(x,y)\in\{0,1\}$, where
$y\in\{0,1\}^{4n}$. Prepare $\ket+^{\otimes n}\ket+^{\otimes4n}$,
query $S$ in phase, and apply Hadamards to the last $4n$ qubits.
Every $y$ contributes equally to the measurement outcome $0^{4n}$,
so the remaining unnormalized vector on that outcome is
\[
 \ket{\eta_S}=2^{-n/2}\sum_x\overline\sigma_x\ket x,
 \qquad
 \overline\sigma_x=2^{-4n}\sum_y(-1)^{S(x,y)}.
\]
Thus the task becomes choosing the signs so that their averages
$\overline\sigma_x$ approximate the desired amplitudes.

At this point, we could assign signs using a deterministic threshold, but an individual
answer could then jump when the target changes, just as in the example
above. Instead, encode $a_x$ in the mean of a random sign, following
the sampling used in~\cite{KQST} to turn a distinguisher for quantum
states into one for distributions of Boolean functions. In our case,
we choose the sign means to equal the target amplitudes, sampling each
$S(x,y)$ independently with
\[
 \Pr[S(x,y)=1]=\frac{1-a_x}{2},
 \qquad \bbE[(-1)^{S(x,y)}]=a_x.
\]
Now changing $a_x$ by $\varepsilon$ changes its bit probability by only
$|\varepsilon|/2$. For example, a desired mean of $3/5$ gives
probabilities $4/5$ and $1/5$ for signs $+1$ and $-1$. The oracle is
sampled once and fixed, so the different $y$'s provide independent
signs, rather than fresh randomness on repeated queries to one input.

Since each average $\overline\sigma_x$ has mean $a_x$ and uses
$2^{4n}$ signs, we get
\[
 \bbE_S\ket{\eta_S}=2^{-n/2}\ket\Omega,
 \qquad
 \bbE_S\bigl\|\ket{\eta_S}-2^{-n/2}\ket\Omega\bigr\|_2^2
 \le2^{-4n}.
\]
Consequently, the root-mean-square error is exponentially smaller than the
desired vector's norm $2^{-n/2}$, so normalization gives a state
exponentially close to $\ket\Omega$ with overwhelming probability.
Although this gives the right state, preparation is still inefficient
because success has probability only about $2^{-n}$, the squared
norm of $2^{-n/2}\ket\Omega$.

\paragraph{Increasing the success probability.}
To increase the success probability to about $p_{\mathrm{succ}}$, we
would instead like the successful vector to be
$\sqrt{p_{\mathrm{succ}}}\,\ket\Omega$, which the same calculation
suggests obtaining by sampling
\[
 \Pr[S(x,y)=1]=\frac{1-\sqrt{2^n p_{\mathrm{succ}}}\,a_x}{2}.
\]
This is a valid probability only if
$\sqrt{2^n p_{\mathrm{succ}}}\,|a_x|\le1$ for every $x$.
Therefore, a target with small amplitudes permits larger
$p_{\mathrm{succ}}$, but for $\ket\Omega=\ket{0^n}$ we still cannot
exceed $p_{\mathrm{succ}}=2^{-n}$. We therefore need to spread the amplitudes by a
reversible transformation, prepare the transformed state, and then undo
the transformation.

However, Hadamards alone cannot do this for every target, since they
spread $\ket{0^n}$ uniformly but send $\ket+^{\otimes n}$ to $\ket{0^n}$.
A related difficulty arises in state synthesis, where states with
amplitudes $\pm2^{-n/2}$ are easy to prepare but can have exponentially
small overlap with the target. The construction of~\cite{GKW26}
addresses this using random signs and Hadamards, and we use the same
transformation to bound all amplitudes from above, as required by our
sampling probabilities. Specifically, sample a uniform Boolean function
$R:\{0,1\}^n\to\{0,1\}$ and define
\[
 U_R\ket z=(-1)^{R(z)}\ket z,
 \qquad
 \ket\alpha=\Hadamard^{\otimes n}U_R\ket\Omega
 =\sum_x\alpha_x\ket x.
\]
Indeed, $\alpha_x=2^{-n/2}\sum_z(-1)^{x\cdot z+R(z)}a_z$, so the
random signs prevent the contributions from systematically adding with
the same sign. Since $\sum_z a_z^2=1$, we have
$\bbE_R\alpha_x^2=2^{-n}$, and applying Hoeffding's inequality followed
by a union bound gives
\[
 \max_x|\alpha_x|\le\frac{5n}{2^{n/2+1}}
 \quad\text{except with probability }2^{-n^2}
 \text{ for sufficiently large }n.
\]
We can therefore take $p_{\mathrm{succ}}=1/(25n^2)$, making every sign
mean $\sqrt{2^n p_{\mathrm{succ}}}\,\alpha_x$ at most $1/2$ in absolute
value. The resulting bit probabilities lie in $[1/4,3/4]$, so each
answer can be sampled using a fair coin half the time, supplying the
uniform randomness needed for the later switching argument.\footnote{For
$q\in[1/4,3/4]$, a Bernoulli($q$) bit can be sampled using a fair coin
with probability $1/2$, and a Bernoulli($2q-1/2$) coin otherwise.}

\paragraph{The preparation algorithm.}
We can now put these ingredients together. For an $R$ satisfying the amplitude
bound above, sample the bits of $S$ independently with
\[
 p_{\Omega,R}(x):=\Pr[S(x,y)=1\mid R]
 =\frac{1-\sqrt{2^n p_{\mathrm{succ}}}\,\alpha_x}{2}
 \in[1/4,3/4].
\]
Run the averaging circuit with $S$ and, on outcome $0^{4n}$, apply
$\Hadamard^{\otimes n}$ and then $U_R$ to the remaining qubits, and
output them. The successful vector before these last operations has
mean $\sqrt{p_{\mathrm{succ}}}\,\ket\alpha$ and expected squared
error at most $2^{-4n}$, so, with overwhelming probability over $S$,
normalization and the inverse transformation give a state exponentially
close to $\ket\Omega$.

If the outcome is not $0^{4n}$, try again with fresh registers and the
same fixed oracles for at most $500n^3$ attempts in total, outputting
$\ket{0^n}$ if all fail. On the high-probability event just described,
each attempt succeeds with probability at least
$p_{\mathrm{succ}}/4=1/(100n^2)$, so the total failure probability is
at most $e^{-5n}$. These steps use only $n$ and quantum queries to $R,S$,
without an initial copy or a classical description of $\ket\Omega$.

Finally, for exceptional $R$, clip the displayed probability to $[1/4,3/4]$
when defining $p_{\Omega,R}(x)$. This makes the sampling well
defined for every $R$ without changing the argument on its
high-probability event. Including the exceptional choices and the
preparation error, the output $\rho_{R,S}$ satisfies
\[
 \bbE_{R,S}\frac12
 \bigl\|\rho_{R,S}-\ket\Omega\!\bra\Omega\bigr\|_1
 \le2^{-n}
\]
for sufficiently large $n$ (\cref{lem:public-preparation}).

\paragraph{Controlling the preparation-oracle probabilities.}
Having established correctness, we return to the security requirement
by bounding how much each bit's probability changes when the preparation
oracles are sampled for close targets, despite the factor
$\sqrt{2^n p_{\mathrm{succ}}}$ in the sign means. To show this, fix two targets
$\ket\Omega,\ket{\widetilde\Omega}$ at distance
$\delta=\|\ket\Omega-\ket{\widetilde\Omega}\|_2$ and sample the same
$R$ for both, since its distribution does not depend on the target.
Writing
$\widetilde\alpha_x=\bra x\Hadamard^{\otimes n}U_R\ket{\widetilde\Omega}$,
their transformed difference is
$\Hadamard^{\otimes n}U_R(\ket\Omega-\ket{\widetilde\Omega})$,
so applying the same concentration bound to this vector of norm
$\delta$ gives
\[
 \max_x|\alpha_x-\widetilde\alpha_x|
 \le\frac{5n\delta}{2^{n/2+1}}
 \quad\text{except with probability }2^{-n^2}.
\]
Thus the increase in the sign means cancels the decrease in each
amplitude difference, yielding
\[
 \max_x|p_{\Omega,R}(x)-p_{\widetilde\Omega,R}(x)|
 \le\frac{\sqrt{2^n p_{\mathrm{succ}}}}2
       \max_x|\alpha_x-\widetilde\alpha_x|
 \le\frac\delta4.
\]
Since clipping cannot increase this difference, the bound also applies to
the fully defined sampling. Notice that we applied concentration to
the difference of the targets, since bounding their amplitudes
separately would give no factor of $\delta$.

We have therefore obtained both efficient preparation and a bound on
every bit's change in probability when the preparation oracles are
sampled for the other target, with independent bits
conditional on $R$ (\cref{lem:public-preparation}). This is a bound on
individual probabilities, not on the statistical distance between the
entire sampled functions.

\paragraph{Adding $N,A$.}
We now use this bound to show that replacing the preparation oracles
by ones sampled for an exponentially close target remains undetectable
with access to $N,A$. Since each $p_{\Omega,R}(x)$ lies in
$[1/4,3/4]$, its Bernoulli distribution contains a fair bit with
probability $1/2$. As in the recursive simulation above, we put the
small change of bias into this fair part on intersections of independent
random sets, with exponentially small probability of containing any
fixed input. Classical recursive simulation and the quantum query
hybrid then bound the effect of this change.
Thus sampling the preparation oracles for the other target and sampling
$N,A$ accordingly is undetectable, even when the same supplied state
$\rho$ depends on both targets (\cref{thm:preparation-replacement}).

\subsubsection{Completing the PKE Proof}\label{sec:overview-pke-proof}
We now combine the two guarantees to prove security of the actual
scheme (\cref{sec:pke-security-proof}). First, preparation correctness
lets us give the challenger the ideal key $\ket{K_s}$ and ideal graph
samples, while the adversary still accesses the full oracle. Its two
challenge ciphertexts are $(u_i,m_{i,\beta_i}\oplus F_s(u_i))$, where
$u_1,u_2$ and $\beta_1,\beta_2$ are independently uniform and the
ciphertexts are given only after $\cA_0$ outputs its message pairs
and two systems.

Next, we change the public preparation oracles while keeping the
issued key and the rule for forming challenge ciphertexts unchanged. First sample the key-preparation
oracles for the target obtained from $\ket{\Omega'}$ by replacing
$\ket{K_s}$ in its $s$ term by $\ket0$.
Then resample $F_{s,0},F_{s,1}$ at $u_1,u_2$ to obtain $G_0,G_1$
and sample the $s$ graph-preparation oracles for
$2^{-\lambda/2}\sum_u\ket u\ket{G_0(u)\oplus G_1(u)}$.
If $u_1=u_2$, resample that input only once. The new targets differ
from the originals in one key term and at most two graph terms,
respectively, so their distances are $O(2^{-\lambda/2})$.
Consequently, the preceding guarantee justifies each change of
preparation oracles, with $N,A$ sampled accordingly, while the
challenger continues using the original key and $F_s,u_i,\beta_i$
to form the ciphertexts.

We can now apply encoded-function simulation. Condition on $G_0,G_1$ and include the public preparation
oracles, all $E_t$ with $t\ne s$, and the independent components for
the other primitives in $\ora_{\rm aux}$. These oracles are now
independent of the issued coset, $H_s,B_s$ and the original challenge
values, so the simulation gives direct $C_s$
access, reducing the adversary to the coset ciphertext experiment.
Finally, the coset security bound (\cref{lem:two-fresh-values}) gives joint
success at most $1/2+\negl(\lambda)$ and completes the proof.

\subsection{Detailed Organization}\label{sec:more-organization}
\subsubsection{Complexity}
\paragraph{Main Results} We define \CHN{} and prove \QMAbb{} verification in \cref{sec:oracle-problem}, then establish depth reduction for \QCMAHbb{} in \cref{sec:hierarchy-depth}. Finally, \cref{sec:hierarchy-separation} proves \QCMAHbb{} query hardness and constructs an oracle $\ora$ with $\PTIME^\ora=\QCMAH^\ora\subsetneq\QMA^\ora$.

\paragraph{Technical Tools} We introduce the codeword-hash relation and its incompressibility guarantees in \cref{sec:code-hash}, followed by deterministic-classical-query simulation of \QCMAbb{} verifiers in \cref{sec:classical-simulation} and witness removal by switching in \cref{sec:classical-switching}.

\subsubsection{Cryptography}

\paragraph{Main Results} We give our cryptographic constructions in \cref{sec:prf-construction,sec:pke-construction,sec:signatures,sec:pseudorandom-states}. We then prove that these primitives coexist relative to one oracle $\ora$ with $\PTIME^\ora=\QCMAH^\ora$ in \cref{sec:fixed-crypto}.

\paragraph{Technical Tools} We prove classical simulation of the recursive oracle in \cref{sec:recursive-oracle-tools}, then construct Boolean oracles for state preparation in \cref{sec:public-preparation} and bound the effect of changing their target with full $N,A$ access in \cref{sec:preparation-replacement}. Finally, \cref{sec:quantum-evaluation} gives our function encoding, evaluation and simulation guarantees.

\subsection{Future Directions}\label{sec:future-directions}
Our work leaves open some interesting questions.

\paragraph{Pseudorandom unitaries (PRU) with a classical key when $\PTIME = \QCMA$}
Our PRUs with quantum keys follow from our PRF with quantum keys through the construction of Ma and Huang~\cite{MH25}. However, unlike classical-key PRFs, classical-key PRUs are not ruled out by $\PTIME = \QCMA$, so one could seek a direct construction that does not go through PRFs. We leave this as an open problem, for which \CHN{} and our techniques may be useful.

\paragraph{Quantum Money and Copy-Protection without Obfuscation} A series of works have shown how to construct public-key quantum money (PKQM)~\cite{AC13Money,EC:Zhandry19b} and quantum copy-protection schemes for various cryptographic primitives and various classes of functionalities~\cite{ALLLZ,CLLZ,TCC:LLQZ22,STOC:ColGun24,C:AnaBeh24,TCC:CakGoy24,CG25CopyProtection,ABHKY26UPO,KNSY26Trapdoor,BCG26}. These constructions rely on indistinguishability obfuscation (iO), and obtaining such primitives without iO remains an important question in quantum cryptography. Since our PKQM and copy-protection constructions survive without classical cryptography, they may suggest ways to obtain plain-model constructions without iO.

\paragraph{Higher Levels of $\QCMA$ Hierarchy} The class $\QCMAH$ is the union of the constant levels of the $\QCMA$ hierarchy, which suffice to separate $\QMA$ from $\PTIME=\QCMA$. Our proof may also extend to $\lambda^c$ levels for some fixed $c>0$, and we leave this question for future work.

\subsection{Related work}\label{sec:more-related-work}

\paragraph{MicroCrypt separations.}
Oracle separations provide evidence that quantum cryptography can exist without classical one-way functions. Using quantum unitary oracles, Kretschmer~\cite{KretschmerHaar24} obtained pseudorandom unitaries with $\BQP=\QMA$, while \cite{CountCrypt} obtained commitments and multiparty noninteractive key exchange with classical communication when $\BQP=\QCMA$, and quantum lightning relative to a separate oracle with $\BQP=\QMA$. Bartusek and Goldin~\cite{BG26UE} also obtained reusable unclonable encryption relative to a quantum unitary oracle with $\BQP=\QMA$.

Kretschmer, Qian, and Tal~\cite{KQT} obtained quantum-computable, quantum-secure trapdoor one-way functions relative to a classical oracle with $\PTIME=\NP$. Their result also gives pseudorandom unitaries, as well as public-key encryption, signatures and oblivious transfer with classical communication. Their PKE does not provide copy-protection and has classical keys, which would be impossible under our stronger collapse $\PTIME=\QCMA$.

In a different direction, Gulati, Khurana, and Tomer~\cite{GKT26PRU} rule out uniform fully black-box constructions of quantum-computable one-way functions from PRUs. However, this does not establish an oracle where PRUs exist but quantum-computable one-way functions do not, or where $\PTIME=\QCMA$.

\paragraph{Switching lemmas and oracle separations.}
Switching lemmas have long been used to prove oracle separations by reducing circuit depth. H{\aa}stad~\cite{Hastad86} used his switching lemma to prove a circuit depth hierarchy, yielding an oracle relative to which $\PH$ is infinite. Rossman, Servedio, and Tan~\cite{FOCS:RosSerTan15} strengthened this to a random-oracle separation through an average-case depth hierarchy theorem, using random projections and a corresponding switching lemma. 

On the quantum side, Aaronson, Ingram, and Kretschmer~\cite{AIK} combine a random-restriction theorem for $\QMAbb$ query algorithms with classical switching to show that $\PP\not\subseteq\QMAH$ relative to a random oracle. Whereas their argument allows quantum witnesses, Agarwal and Ben-David~\cite{AB} consider hierarchies with classical witnesses, developing a switching lemma for quantum query algorithms to show that both QCPH and $\QCMAH$ are infinite relative to a random oracle.

\subsection{AI Methodology}\label{sec:ai-methodology}
We used ChatGPT-6 Astra Ultra as our sole AI tool. Our results were obtained by us without AI assistance except for the following: We used AI interactively to assist with obtaining our results regarding the classical-oracle state-preparation construction and its statistical analysis (\cref{sec:public-preparation}). After our results were completed, we also used it to help choose some parameters and check correctness. We also interactively used it under our close direction for drafting and editing throughout the paper (except \cref{sec:introduction,sec:overview} and the abstract), and we subsequently heavily edited and verified the text.
We verified the whole paper and the results and take full responsibility for the full paper.

\newpage\section{Preliminaries}\label{sec:preliminaries}
In this section, we fix notation and give the definitions used for
complexity, quantum information and cryptography.

\subsection{Notation and Conventions}\label{sec:notation}

We use standard notation from cryptography and quantum information,
referring to~\cite{Goldreich_2001,Goldreich_2004,NC10} for further details.
The following conventions apply unless stated otherwise.

\paragraph{Functions.}
The security parameter is $\lambda$. A function $p$ is polynomially
bounded if $p(\lambda)\le\lambda^c$ for some constant $c$ and all
sufficiently large $\lambda$, and we sometimes abbreviate
``polynomially bounded'' to ``polynomial.'' We write $\negl(\lambda)$
for an unspecified nonnegative negligible function, meaning that
\[
 \forall c>0\ \exists\lambda_0\ \forall\lambda\ge\lambda_0:
 \qquad \negl(\lambda)\le\lambda^{-c}.
\]
A function $f$ is superlogarithmic if $f(\lambda)=\omega(\log\lambda)$,
equivalently if $2^{-f(\lambda)}$ is negligible. Omitted parameter
arguments are understood to be $\lambda$.

\paragraph{Classical and quantum information.}
We write classical keys and signatures as $\pk,\sk,\vk,\sigma$
and use $\regi,\regis{sk},\regis{vk},\regis{sig}$ for quantum
registers, which may contain classical components and may be entangled
with other registers. We use ket notation for pure states and density
operators for mixed states, and $\regi\samp\rho$ means that $\regi$
is initialized in state $\rho$. State sums are implicitly normalized
unless explicitly described as unnormalized. We write $\HaarMeasure_n$
for the unitarily invariant distribution on pure $n$-qubit states.

The trace distance between states $\rho$ and $\sigma$ is
$\tfrac12\|\rho-\sigma\|_1$. In a trace-distance expression, an
algorithm denotes its output density operator and a classical variable
denotes its corresponding classical register, with all other randomness averaged. For a quantum channel $\Phi$
and any auxiliary system $W$, contractivity gives
\[
 \trdist{(\Phi\otimes\mathrm{id}_W)(\rho)}
          {(\Phi\otimes\mathrm{id}_W)(\sigma)}
 \le\trdist{\rho}{\sigma}.
\]

\paragraph{Distributions.}
We write $x\samp S$ for uniform sampling from a finite set,
$x\samp\mathcal D$ for sampling from a distribution, and
$x\samp A(a)$ for a classical output of a randomized or quantum
algorithm. We use the analogous notation $\regi\samp A(a)$ when
the output is a quantum register.

\paragraph{Algorithms and oracles.}
The abbreviations \PPT{} and \QPT{} denote probabilistic and quantum
polynomial-time algorithms. Cryptographic scheme algorithms are uniform
\QPT{}, and an algorithm using a quantum key returns its updated
register for later calls, which we leave implicit in the notation.
We write $A^\ora$ for an algorithm with access to the classical oracle
$\ora$, distinguishing classical and quantum access in the complexity
model below. In cryptographic experiments, every algorithm and adversary
has quantum access to the same oracle $\ora$. If $\ora$ is sampled,
its sampling is independent of the adversary's description and is
included in the experiment's probability. Cryptographic security is
against \QPT{} adversaries without additional quantum advice, with
nonuniform classical circuit descriptions allowed where stated.

\paragraph{Security games.}
A security game is an experiment between a challenger and an adversary,
whose output is a bit. We say that the adversary wins when the experiment
outputs $1$. A distinguishing experiment instead returns the adversary's
output bit, and its advantage is the absolute difference between the two
specified acceptance probabilities.

\subsection{Complexity}\label{sec:complexity-preliminaries}

The acceptance conditions for oracle machines are common to computational
and query complexity, while their resource bounds differ. We first define
these machines, including their behavior on inputs outside a promise, and
then specify the classes in each model.

\subsubsection{Oracle machines and the $\QCMA$ hierarchy}
\label{sec:hierarchy-definitions}

A quantum oracle verifier $V$ takes a classical input $z$ and a witness,
has access to a Boolean oracle $\ora$, and outputs one bit. Its oracle
queries act as $\ket u\ket b\mapsto\ket u\ket{b\oplus\ora(u)}$
(\cref{eq:quantum-oracle-access}). A hierarchy verifier may also query a
lower machine, as defined below. For now, we require finite computation
but impose no efficiency bound. The witness length $\ell$ may depend on
the input length or the parameter $\lambda$, as specified in each model.
The probabilities below are over the verifier's execution with its
input, witness, and oracles fixed.

We use the usual quantum and classical witness acceptance conditions
\cite{AN02}, written as partial functions with values in $\{0,1,\bot\}$.

\begin{definition}[\QMA{} machine]\label{def:qma-machine}
Let $V$ be a quantum oracle verifier, and let $\rho$ range over density
operators on $\ell$ qubits. We define its associated \QMA{} machine by
\[
 M^\ora(z)=
 \begin{cases}
  1,&\exists\rho:\Pr[V^\ora(z,\rho)=1]\ge 2/3,\\
  0,&\forall\rho:\Pr[V^\ora(z,\rho)=1]\le 1/3,\\
  \bot,&\text{otherwise}.
 \end{cases}
\]
\end{definition}

\begin{definition}[\QCMA{} machine]\label{def:qcma-machine}
Let $V$ be a quantum oracle verifier, and let $w$ range over
$\zo^\ell$. We define its associated \QCMA{} machine by
\[
 M^\ora(z)=
 \begin{cases}
  1,&\exists w:\Pr[V^\ora(z,w)=1]\ge 2/3,\\
  0,&\forall w:\Pr[V^\ora(z,w)=1]\le 1/3,\\
  \bot,&\text{otherwise}.
 \end{cases}
\]
\end{definition}

Thus $\bot$ denotes the remaining inputs, on which neither acceptance
condition holds. The witness may depend on $z$ and $\ora$ in both
definitions. To use such a partial machine as an oracle, we must specify
what its oracle answers can be on these remaining inputs.

\begin{definition}[Boolean completion]\label{def:boolean-completion}
A \emph{Boolean completion} of $N:\zo^*\to\{0,1,\bot\}$ is a total
function $\widetilde N:\zo^*\to\zo$ that agrees with every non-$\bot$
value of $N$. We write
\begin{equation}\label{eq:boolean-completions}
 \Comp(N)=
 \left\{\widetilde N:\zo^*\to\zo:
   \forall z,\ N(z)\ne\bot\ \Longrightarrow\ \widetilde N(z)=N(z)
 \right\}.
\end{equation}
\end{definition}

Each completion assigns arbitrary bits where $N(z)=\bot$, but it is
one fixed function throughout the computation, including coherent
queries. Following Agarwal and Ben-David~\cite{AB}, we require a hierarchy
machine's answer to agree for every completion of its lower oracle.

\begin{definition}[\QCMA{} hierarchy machine]\label{def:qcmah-machine}
A $\QCMAH_1$ machine is a \QCMA{} machine. For $d\ge2$, a
$\QCMAH_d$ machine $M_d$ consists of a quantum oracle verifier $V_d$,
a classical witness length $\ell$, and a $\QCMAH_{d-1}$ machine
$M_{d-1}$. With $w\in\zo^\ell$, we define
\begin{equation}\label{eq:hierarchy-machine}
 M_d^\ora(z)=
 \begin{cases}
 1,&\displaystyle
   \forall\widetilde M_{d-1}\in\Comp(M_{d-1}^\ora)\ \exists w:
       \Pr[V_d^{\ora,\widetilde M_{d-1}}(z,w)=1]\ge 2/3,\\[2pt]
 0,&\displaystyle
   \forall\widetilde M_{d-1}\in\Comp(M_{d-1}^\ora)\ \forall w:
       \Pr[V_d^{\ora,\widetilde M_{d-1}}(z,w)=1]\le 1/3,\\[2pt]
 \bot,&\text{otherwise}.
 \end{cases}
\end{equation}
\end{definition}

In particular, the accepting witness may depend on the completion,
whereas the answer must be the same for every completion. A depth-$d$
machine has verifiers $V_1,\ldots,V_d$, where $V_1$ queries only $\ora$
and $V_i$ also queries a completion of $M_{i-1}^\ora$ for $i>1$.
The lower machine receives the query string as its classical input
and has access to the same base oracle $\ora$.

For the replacement argument, it is useful to express the hierarchy
condition using the partial \QCMA{} machine $M$ associated with the
two-oracle verifier $V_d$ (\cref{def:qcma-machine}). For $b\in\zo$,
\begin{equation}\label{eq:hierarchy-qcma-form}
 M_d^\ora(z)=b
 \quad\Longleftrightarrow\quad
 \forall\widetilde M_{d-1}\in\Comp(M_{d-1}^\ora):
                 M^{\ora,\widetilde M_{d-1}}(z)=b.
\end{equation}

\subsubsection{Computational complexity}

A promise problem is a pair $\Pi=(\Pi_{\YES},\Pi_{\NO})$ of disjoint
subsets of $\zo^*$. An algorithm must accept inputs in $\Pi_{\YES}$
and reject inputs in $\Pi_{\NO}$, with no requirement outside their
union. A language $L\subseteq\zo^*$ corresponds to the total promise
$(L,\zo^*\setminus L)$.

In this model, the oracle $\ora$ is fixed and the explicit input is
$x\in\zo^n$. A uniform quantum polynomial-time verifier is a family
of polynomial-size quantum oracle circuits whose description on
inputs of length $n$ is generated by a deterministic polynomial-time
algorithm on $1^n$. The generator specifies the oracle gates without
access to $\ora$, and each oracle call counts as one step. Witness
length, workspace, and query length are all polynomial in $n$.

For a fixed oracle $\ora$, the classes $\PTIME^\ora$, $\BPP^\ora$
and $\BQP^\ora$ use uniform deterministic, randomized and quantum
polynomial-time algorithms, respectively, with error at most $1/3$
for the latter two. The class $\NP^\ora$ allows a polynomial-size
classical witness checked by a deterministic polynomial-time verifier.
We use $\PP$ for probabilistic polynomial time with a strict majority
acceptance threshold, and \PPT{} and \QPT{} for probabilistic and
quantum polynomial-time algorithms. Here ``efficient'' means
polynomial-time.

We now give the witness conditions explicitly for both quantum
verification classes~\cite{AN02}.

\begin{definition}[$\QMA$ and $\QCMA$]\label{def:qma-qcma-classes}
A promise problem $\Pi$ belongs to $\QMA^\ora$ if there are a
polynomial $p$ and a uniform \QPT{} verifier $V$ such that
\[
 \begin{aligned}
 x\in\Pi_{\YES}&\ \Longrightarrow\
  \exists\rho\text{ on }p(|x|)\text{ qubits}:\
       \Pr[V^\ora(x,\rho)=1]\ge2/3,\\
 x\in\Pi_{\NO}&\ \Longrightarrow\
  \forall\rho\text{ on }p(|x|)\text{ qubits}:\
       \Pr[V^\ora(x,\rho)=1]\le1/3.
 \end{aligned}
\]
It belongs to $\QCMA^\ora$ if there are a polynomial $p$ and a
uniform \QPT{} verifier $V$ such that
\[
 \begin{aligned}
 x\in\Pi_{\YES}&\ \Longrightarrow\
  \exists w\in\zo^{p(|x|)}:\Pr[V^\ora(x,w)=1]\ge2/3,\\
 x\in\Pi_{\NO}&\ \Longrightarrow\
  \forall w\in\zo^{p(|x|)}:\Pr[V^\ora(x,w)=1]\le1/3.
 \end{aligned}
\]
\end{definition}

The corresponding hierarchy uses these time and witness bounds at
each level, together with the completion rule above~\cite{AB}.

\begin{definition}[Computational $\QCMA$ hierarchy]
\label{def:computational-qcmah}
For a fixed depth $d\ge1$, a promise problem $\Pi$ belongs to
$\QCMAH_d^\ora$ if there is a hierarchy machine $M_d$ with uniform
\QPT{} verifiers $V_1,\ldots,V_d$ such that
\[
 x\in\Pi_{\YES}\ \Longrightarrow\ M_d^\ora(x)=1,
 \qquad
 x\in\Pi_{\NO}\ \Longrightarrow\ M_d^\ora(x)=0.
\]
For each level $i$, there is a polynomial $p_i$ bounding the running
time and witness length of $V_i$ on every explicit input $z$ by
$p_i(|z|)$. These resource bounds hold for every base oracle and
every total Boolean lower oracle, whereas the acceptance conditions
in \cref{eq:hierarchy-machine} quantify over completions of the
specified lower machine. We define
\[
 \QCMAH_1^\ora=\QCMA^\ora,
 \qquad
 \QCMAH^\ora=\bigcup_{d\ge1}\QCMAH_d^\ora,
\]
where $d$ is a constant independent of the input length.
\end{definition}

Without an oracle superscript, these classes have no base oracle.
We apply the same definitions to languages using their total promises,
and use the prefix $\mathsf{Promise}$ when emphasizing promise problems,
as in $\PROMISEQCMA^\ora$.

\subsubsection{Query complexity}\label{sec:query-complexity-model}

In the query model, the input itself is a Boolean oracle
$\ora:\cX_\lambda\to\zo$, where $\cX_\lambda$ is finite and its
elements have polynomial-length binary encodings. The size
$N=|\cX_\lambda|$ may be exponential, and resource bounds are measured
in $\lambda$, not $N$. A classical algorithm queries individual
inputs, whereas a quantum algorithm may query coherently
(\cref{eq:quantum-oracle-access}), with unrestricted finite computation
between queries~\cite{AB}.

An oracle promise problem is a family
$\Pi=(\Pi_{\YES,\lambda},\Pi_{\NO,\lambda})_{\lambda\ge1}$ of disjoint
sets of Boolean functions $\cX_\lambda\to\zo$. We use a verifier
family $(V_\lambda)_{\lambda\ge1}$, whose description may depend on
$\lambda$ but not on the input oracle. No uniformity or polynomial
description-size bound is required. The prefix $\mathsf{BB}\text{-}$
denotes these black-box classes, distinguishing them from the
time-bounded classes for a fixed oracle.

\begin{definition}[$\QMAbb$ and $\QCMAbb$]
\label{def:qma-qcma-query-classes}
An oracle promise problem $\Pi$ belongs to $\QMAbb$ if there are
polynomials $p,q$ and quantum query verifiers $V_\lambda$ using at
most $q(\lambda)$ queries on every oracle and witness, such that
\[
 \begin{aligned}
 \ora\in\Pi_{\YES,\lambda}&\ \Longrightarrow\
  \exists\rho\text{ on }p(\lambda)\text{ qubits}:\
       \Pr[V_\lambda^\ora(1^\lambda,\rho)=1]\ge2/3,\\
 \ora\in\Pi_{\NO,\lambda}&\ \Longrightarrow\
  \forall\rho\text{ on }p(\lambda)\text{ qubits}:\
       \Pr[V_\lambda^\ora(1^\lambda,\rho)=1]\le1/3.
 \end{aligned}
\]
It belongs to $\QCMAbb$ if there are such polynomials and verifiers
with classical witnesses satisfying
\[
 \begin{aligned}
 \ora\in\Pi_{\YES,\lambda}&\ \Longrightarrow\
  \exists w\in\zo^{p(\lambda)}:
       \Pr[V_\lambda^\ora(1^\lambda,w)=1]\ge2/3,\\
 \ora\in\Pi_{\NO,\lambda}&\ \Longrightarrow\
  \forall w\in\zo^{p(\lambda)}:
       \Pr[V_\lambda^\ora(1^\lambda,w)=1]\le1/3.
 \end{aligned}
\]
\end{definition}

We use the same completion rule to define the query hierarchy, but
bound queries and witnesses rather than running time~\cite{AB}.

\begin{definition}[Black-box $\QCMA$ hierarchy]
\label{def:query-qcmah}
For a fixed depth $d\ge1$, an oracle promise problem $\Pi$ belongs to
$\QCMAHbb_d$ if there is a family of hierarchy machines
$M_{d,\lambda}$, with verifiers $V_{1,\lambda},\ldots,V_{d,\lambda}$,
such that
\[
 \ora\in\Pi_{\YES,\lambda}\ \Longrightarrow\ M_{d,\lambda}^\ora(1^\lambda)=1,
 \qquad
 \ora\in\Pi_{\NO,\lambda}\ \Longrightarrow\ M_{d,\lambda}^\ora(1^\lambda)=0.
\]
There is a polynomial $p$ such that, starting from $1^\lambda$,
every verifier makes at most $p(\lambda)$ queries, and all classical
inputs to verifiers, witnesses, and oracle-query inputs have length
at most $p(\lambda)$. These bounds hold for every base oracle,
every total Boolean lower oracle, and every witness, while finite
local computation and description size are unrestricted. We define
\[
 \QCMAHbb_1=\QCMAbb,
 \qquad
 \QCMAHbb=\bigcup_{d\ge1}\QCMAHbb_d,
\]
where $d$ is independent of $\lambda$.
\end{definition}

The same prefix is used for other query-model classes. For
distributional results, we sample $\ora$ and abbreviate
$M_{d,\lambda}^\ora(1^\lambda)$ to $M_d^\ora$ when $\lambda$ is clear.
Our quantitative statements also allow classical auxiliary inputs $z$,
whose bounds are specified along with the query and witness bounds.
Every lower machine receives its parent's query string $z$ and
accesses the same base oracle $\ora$, with the resource bounds holding
even for total lower oracles that are not legal completions.

Unless a statement explicitly permits oracle-dependent descriptions,
the verifier family is fixed for all input oracles, while a witness
may depend on the oracle being verified. Bounds on additional advice
or initial states are stated separately. For a full oracle, only the
finite set of inputs accessible within the query-length bound matters.
These query bounds do not imply polynomial running time.

\subsection{Quantum Information}\label{sec:quantum-information}

We specify quantum access to a classical oracle and recall the two
query bounds used below, using the state and register notation fixed
in \cref{sec:notation}.

\subsubsection{Quantum oracle access and query bounds}
\label{sec:quantum-query-tools}

A Boolean oracle is a function $\ora:\cX\to\zo$, where
$\cX\subseteq\zo^*$ is finite for the distributional lemmas
and $\cX=\zo^*$ for a full oracle. Quantum access to $\ora$
means access to the unitary
\begin{equation}\label{eq:quantum-oracle-access}
 \ket u\ket b\longmapsto\ket u\ket{b\oplus \ora(u)}.
\end{equation}
Oracles with finite output alphabets, such as $\{\bot,0,1\}$, use
fixed encodings.\footnote{Encodings of tuples and output symbols
are implicit. They are polynomial-time computable, and malformed
queries have a fixed answer.}

As in the query model (\cref{sec:query-complexity-model}), query bounds
hold on every oracle.
The query lemmas impose no bound on local computation or description
length.

\begin{lemma}[Quantum query hybrid bound~\cite{BBBV}]
\label{lem:query-hybrid}
Let $\ora,\ora'$ be two classical oracles that agree outside a set $E$.
Fix a $Q$-query quantum algorithm $A$ and an initial state, which is
the same in both executions. Let $p_j$ be the probability that
measuring the input of query $j$ in the execution $A^\ora$ gives an
element of $E$. Then
\[
 \trdist{A^\ora}{A^{\ora'}}
 \le2\sum_{j=1}^Q\sqrt{p_j}
 \le2\sqrt{Q\sum_{j=1}^Qp_j}.
\]
The same bound holds for the joint output with any auxiliary system
$W$ on which $A$ does not act. If both executions use the same purification of the initial
state and the same intervening operations, their final pure states also
satisfy
\[
 \|\ket{\Psi_\ora}-\ket{\Psi_{\ora'}}\|_2^2
 \le4Q\sum_{j=1}^Qp_j.
\]
\end{lemma}

To substitute a classical algorithm for an oracle, we use the
following coherent implementation.

\begin{lemma}[Coherent classical computation]\label{lem:coherent-classical-query}
Let $T$ be a deterministic classical oracle algorithm that outputs one
bit and makes at most $h$ queries on every auxiliary input and oracle. There is
a quantum oracle algorithm using at most $2h$ queries that implements
\[
 \ket z\ket b\longmapsto\ket z\ket{b\oplus T^\ora(z)}
\]
exactly, including on coherent inputs. Its local computation and
description may be unrestricted.
\end{lemma}
The proof of \cref{lem:coherent-classical-query} appears in \cref{app:query-bounds}.

\subsection{Cryptography}\label{sec:crypto-definitions}

In this section, we recall pseudorandom functions, public-key encryption,
signatures, and pseudorandom states. We state explicitly the quantum-key
and repeated-use requirements used by our constructions.

\subsubsection{Pseudorandom Functions}

We use pseudorandomness against quantum queries~\cite{ZhandryQRF12}.
For our constructions, we allow a reusable quantum key and an optional
public serial supplied with each evaluation.
\begin{definition}[Pseudorandom Functions]\label{predef:prf}
A pseudorandom function (PRF) scheme is a family of functions
$F_k:\zo^{m(\lambda)}\to\zo^{n(\lambda)}$ along with the following
uniform \QPT{} algorithms.
\begin{itemize}
 \item $\gen^\ora(1^\lambda):$ Takes the security parameter and samples
 an index $k$, then outputs a key register $\regis{sk}$ in state
 $\rho_k$ and an optional public serial $j$. The challenger also knows $k$, which need not be public.
 \item $\ceval^\ora(j,\regis{sk},\regis{x},\regis{b}):$ Takes the
 serial, key, input and response registers, and evaluates $F_k$ in the
 response register, returning the updated key.
\end{itemize}
We require the following.

\paragraph{Correctness.}
The ideal evaluation applies
\[
 U_{F_k}:\ket{x,b}\longmapsto\ket{x,b\oplus F_k(x)}
\]
and leaves the key unchanged. For every polynomial number of calls,
the real and ideal executions have negligible trace distance, allowing
coherent inputs entangled with arbitrary auxiliary systems, and
intervening operations outside the key.

\paragraph{Security.}
Consider the following experiment between a challenger and a \QPT{}
adversary $\cA$, with $b\in\zo$ specifying the challenge oracle.
\paragraph{\underline{Experiment $\mathsf{PRF}^{\ora}_{\cA}(\lambda,b)$}}
\begin{enumerate}
 \item The challenger runs $\gen^\ora(1^\lambda)$, stores its index
 $k$, and samples an independent uniform function
 $R:\zo^{m(\lambda)}\to\zo^{n(\lambda)}$.
 \item The challenger submits only $j$ to $\cA$, with quantum access
 to $F_k$ if $b=0$ and to $R$ if $b=1$.
 \item The adversary outputs a bit, which the experiment returns.
\end{enumerate}
We require that for every \QPT{} adversary $\cA$,
\begin{equation}\label{eq:prf-privacy-game}
 \left|\Pr[\mathsf{PRF}^{\ora}_{\cA}(\lambda,0)=1]
       -\Pr[\mathsf{PRF}^{\ora}_{\cA}(\lambda,1)=1]\right|
 \le\negl(\lambda).
\end{equation}
For an ordinary classical-key PRF, the serial $j$ may be empty.
\end{definition}
Copy-protection for one-bit-output PRFs is defined in
\cref{def:prf-security}.

\subsubsection{Public-Key Encryption}

We recall chosen-plaintext security for classical messages and
ciphertexts, allowing the decryption key to be quantum and reused.
\begin{definition}[Public-Key Encryption]\label{predef:pke}
A public-key encryption scheme with message space $\cM_\lambda$
consists of the following uniform \QPT{} algorithms.
\begin{itemize}
 \item $\gen^\ora(1^\lambda):$ Takes the security parameter and outputs
 a classical public key $\pk$ and a secret-key register $\regis{sk}$.
 \item $\enc^\ora(\pk,m):$ Takes a public key $\pk$ and a message
 $m\in\cM_\lambda$, and outputs a classical ciphertext $\ct$.
 \item $\dec^\ora(\regis{sk},\ct):$ Takes a secret key and a ciphertext,
 and outputs a classical message $\widehat m$, returning the updated key.
\end{itemize}
We require the following.

\paragraph{Correctness.}
Over polynomially many honest encryptions and decryptions using one
generated key, the probability of any incorrect decryption is negligible.

\paragraph{Security.}
Consider the following game between the challenger and an adversary
$\cA=(\cA_0,\cA_1)$.
\paragraph{\underline{Experiment $\mathsf{CPA}^{\ora}_{\cA}(\lambda)$}}
\begin{enumerate}
 \item The challenger generates
 $(\pk,\regis{sk})\leftarrow\gen^\ora(1^\lambda)$ and submits $\pk$ to
 $\cA_0$.
 \item The adversary $\cA_0$ outputs $(m_0,m_1,\regi)$, where
 $m_0,m_1\in\cM_\lambda$ and $|m_0|=|m_1|$.
 \item The challenger samples $b\leftarrow\zo$ and
 $\ct\leftarrow\enc^\ora(\pk,m_b)$, and submits $\ct$ to $\cA_1$.
 \item The adversary outputs
 $\widehat b\leftarrow\cA_1^\ora(\ct,\regi)$.
 \item The challenger outputs one if and only if $\widehat b=b$.
\end{enumerate}
We require that for every \QPT{} adversary $\cA$,
\[
 \left|\Pr[\mathsf{CPA}^{\ora}_{\cA}(\lambda)=1]-\tfrac12\right|
 \le\negl(\lambda).
\]
\end{definition}
Copy-protection for bit encryption, with two independent ciphertext
challenges after the split, is defined in \cref{def:pke-piracy}.

\subsubsection{Digital Signature Schemes}

We use adaptive existential unforgeability under classical chosen-message
attacks, allowing quantum signing keys, verification keys, and signatures.
\begin{definition}[Digital Signature Schemes]\label{predef:digsig}\label{def:signature-security}
A digital signature scheme with message space $\zo^\lambda$ consists
of the following uniform \QPT{} algorithms.
\begin{itemize}
 \item $\gen^\ora(1^\lambda):$ Takes the security parameter and outputs
 a verification key $\regis{vk}$ and a signing key $\regis{sk}$.
 \item $\sign^\ora(\regis{sk},m):$ Takes a signing key and a message
 $m\in\zo^\lambda$, and outputs a signature register $\regis{sig}$,
 returning the updated signing key.
 \item $\ver^\ora(\regis{vk},m,\regis{sig}):$ Takes a verification key,
 a message and an alleged signature, and outputs an acceptance bit.
 \item $\VKGen^\ora(\regis{sk}):$ When verification keys are quantum,
 takes a signing key and outputs a fresh verification key, returning
 the updated signing key.
\end{itemize}
We require the following.

\paragraph{Correctness.}
Polynomially many honestly generated signatures are all accepted
except with negligible probability, including adaptive messages and
verification-key requests.

\paragraph{Security.}
Consider the following experiment between a challenger and an adversary
$\cA$.
\paragraph{\underline{Experiment $\Forge^\ora_{\SigScheme,\cA}(\lambda)$}}
\begin{enumerate}
 \item The challenger generates
 $(\regis{vk},\regis{sk})\leftarrow\gen^\ora(1^\lambda)$, submits
 $\regis{vk}$ to $\cA$, and initializes $\cM=\varnothing$.
 \item On each adaptive classical signing request $m$, the challenger
 runs $\sign^\ora(\regis{sk},m)$, keeps the updated signing key,
 returns only the signature and adds $m$ to $\cM$.
 For a fresh quantum verification key, it runs $\VKGen^\ora(\regis{sk})$,
 keeps the updated signing key and returns only the new verification key.
 \item The adversary outputs a message $m$ and an alleged signature
 $\regis{sig}$.
 \item The challenger outputs one if and only if
 $m\in\zo^\lambda\setminus\cM$ and
 $\ver^\ora(\regis{vk},m,\regis{sig})=1$, using a fresh honest
 verification key when it is quantum.
\end{enumerate}
We require that for every \QPT{} adversary $\cA$,
\[
 \Pr[\Forge^\ora_{\SigScheme,\cA}(\lambda)=1]\le\negl(\lambda).
\]
\end{definition}

\subsubsection{Pseudorandom Quantum States}

We recall pseudorandom quantum states, introduced by Ji, Liu and
Song~\cite{C:JiLiuSon18}, using the generator formulation of Brakerski
and Shmueli~\cite{C:BraShm20} for an output length $n(\lambda)$.
\begin{definition}[Pseudorandom Quantum States]\label{predef:prs}
A pseudorandom quantum state (PRS) generator consists of the following
uniform \QPT{} algorithms.
\begin{itemize}
 \item $\KeyGen^\ora(1^\lambda):$ Takes the security parameter and
 outputs a classical key $k$.
 \item $\StateGen^\ora(1^\lambda,k):$ Takes the security parameter
 and a classical key $k$ in the support of $\KeyGen^\ora(1^\lambda)$,
 and outputs an $n(\lambda)$-qubit state, which may be mixed.
\end{itemize}
We require the following.

\paragraph{Security.}
Consider the following experiment between a challenger and an adversary
$\cA$, for $b\in\zo$ and a polynomial number of copies $r=r(\lambda)$.
\paragraph{\underline{Experiment $\mathsf{PRS}^{\ora}_{\cA}(\lambda,b,r)$}}
\begin{enumerate}
 \item If $b=0$, the challenger samples
 $k\leftarrow\KeyGen^\ora(1^\lambda)$, runs
 $\StateGen^\ora(1^\lambda,k)$ independently $r$ times using this same
 key, and submits all $r$ outputs to $\cA$.
 If $b=1$, it samples one
 $\ket\psi\leftarrow\HaarMeasure_{n(\lambda)}$ independently of $\ora$
 and submits $\ket\psi^{\otimes r}$ to $\cA$.
 \item The adversary outputs a bit, which the experiment returns.
\end{enumerate}
We require that for every \QPT{} adversary $\cA$ and every polynomial $r$,
\[
 \left|\Pr[\mathsf{PRS}^{\ora}_{\cA}(\lambda,0,r)=1]
       -\Pr[\mathsf{PRS}^{\ora}_{\cA}(\lambda,1,r)=1]\right|
 \le\negl(\lambda).
\]
\end{definition}

\paragraph{Preparation accuracy of our construction.}
Our construction additionally has prescribed pure target states
$\ket{\psi_k}$. We measure its preparation error by
\begin{equation}\label{eq:prs-preparation-error}
 \epsilon_{\mathrm{prs}}(\lambda,\ora):={}
 \max_{k\in\supp(\KeyGen^\ora(1^\lambda))}
 \trdist{\StateGen^\ora(1^\lambda,k)}{\ket{\psi_k}\!\bra{\psi_k}}.
\end{equation}
The fixed-oracle theorem (\cref{thm:fixed-oracle-crypto}) gives
$\epsilon_{\mathrm{prs}}(\lambda,\ora)=\negl(\lambda)$ for almost every sampled
$\ora$. Writing $\rho_k$ for the output density matrix of
$\StateGen^\ora(1^\lambda,k)$, changing one tensor factor at a time gives
\[
 \frac12\left\|\rho_k^{\otimes r}
 -(\ket{\psi_k}\!\bra{\psi_k})^{\otimes r}\right\|_1
 \le r\epsilon_{\mathrm{prs}}(\lambda,\ora).
\]
Thus using these ideal tensor powers in the security experiment
changes any adversary's acceptance probability by a negligible amount.

\addtocontents{toc}{\protect\setcounter{tocdepth}{1}}
\paperpart{Technical Tools}
\label{part:technical-tools}
\section{Codeword-Hash Pairs, Noise, and Incompressibility}
\label{sec:code-hash}
In this section, we state the code and quantum decoding guarantees for
codeword-hash pairs (\cref{lem:code-hash}). We then define
contractive simultaneous incompressibility and prove it for codeword-hash pairs
and sparse noise (\cref{cor:code-noise-incompressibility}).

\subsection{Codeword-hash pairs and quantum decoding}
\label{sec:code-hash-decoding}

The codeword-hash relation introduced by \cite{YZ} underlies our
\CHN{} problem. We use its \emph{biased} version, introduced by
\cite{STOC:GGJL25} and later also used by \cite{BHV,CakanQCIP26}.

\begin{definition}[Codeword-hash pairs]\label{def:code-pairs}
Let $C_\lambda\subseteq\Sigma_\lambda^{q_\lambda}$ be a linear code of length $q_\lambda$, where
$q_\lambda>\lambda$, $q_\lambda=\poly(\lambda)$ and symbols have $b_\lambda=\poly(\lambda)$-bit
encodings. Put $\cX_\lambda=\Sigma_\lambda^{q_\lambda}\times\zo^\lambda$.
For $H=(H_1,\ldots,H_{q_\lambda})$ with predicates $H_i:\Sigma_\lambda\to\zo$, we define
the symbol-wise hash $H(v)$ and the set $\cP_H$ of codeword-hash pairs by
\[
 \begin{aligned}
 H(v)&=H_1(v_1)\|\cdots\|H_{q_\lambda}(v_{q_\lambda}),\\
 \cP_H&=\{(v,y)\in\cX_\lambda:v\in C_\lambda,\ H(v)=y\|0^{q_\lambda-\lambda}\}.
 \end{aligned}
\]
\end{definition}

We keep the prefix predicates fixed and sample the suffix predicates from the
following distribution.

\begin{definition}[Biased hash distribution]\label{def:hash-distribution}
For fixed prefix predicates $H_1,\ldots,H_\lambda$, we define the biased
hash distribution $\cH_\lambda$ by independently sampling
\[
 H_i(u)\leftarrow\Bernoulli(\lambda^{-4})
 \qquad(i>\lambda,\ u\in\Sigma_\lambda).
\]
The choice of the fixed prefix predicates will be specified below.
\end{definition}

We next define the predicate-preimage state and the decoding property
that we require of $H$.

\begin{definition}[Predicate-preimage state]\label{def:hash-witness}
Put $a_\lambda=2q_\lambda b_\lambda$. When $H_i^{-1}(b)$ is nonempty for every $i\in[q_\lambda]$ and $b\in\zo$,
we define the predicate-preimage state to be the normalized $a_\lambda$-qubit state
\begin{equation}\label{eq:hash-witness}
 \ket{\Psi_H}=
 \bigotimes_{i=1}^{q_\lambda}\bigotimes_{b\in\zo}
    \left(\sum_{u\in\Sigma_\lambda:H_i(u)=b}\ket u\right),
\end{equation}
with normalization factors suppressed. Otherwise, set
$\ket{\Psi_H}=\ket{0^{a_\lambda}}$.
\end{definition}
The decoder prepares the public uniform superposition over $C_\lambda$ itself.

\begin{definition}[Good hashes]\label{def:good-hashes}
For an oracle-free quantum decoder $\Decode$, we call a hash $H$
\emph{good} if all preimages $H_i^{-1}(b)$ are nonempty and
\[
 \forall y\in\zo^\lambda:\qquad
 \Pr_{v\leftarrow\Decode(y,\ket{\Psi_H})}
       [(v,y)\in\cP_H]\ge1-2^{1-\lambda}.
\]
We denote the set of good hashes by $\Good_\lambda$.
\end{definition}

To use fresh predicates at successive levels of the hierarchy, we group
the suffix constraints as follows.

\begin{definition}[Suffix-constraint sets]\label{def:suffix-constraints}
For an integer $d\ge1$, partition the suffix coordinates into $d$
groups of sizes differing by at most one:
\begin{equation}\label{eq:code-hash-groups}
 \{\lambda+1,\ldots,q_\lambda\}=J_1\sqcup\cdots\sqcup J_d.
\end{equation}
We define the suffix-constraint sets $P_1(H),\ldots,P_d(H)$ by
\begin{equation}\label{eq:code-hash-sparse-sets}
 P_j(H)=\{(v,y)\in\cX_\lambda:v\in C_\lambda,
 H_1(v_1)\|\cdots\|H_\lambda(v_\lambda)=y,
 \ H_i(v_i)=0\text{ for all }i\in J_j\}.
\end{equation}
\end{definition}

We now state the code and decoding guarantees used below.

\begin{lemma}[Existence of suitable codes]\label{lem:code-hash}
There is a uniformly constructible family of linear codes
$C_\lambda\subseteq\Sigma_\lambda^{q_\lambda}$ over efficiently represented
alphabets, with polynomial-time encoding and membership testing,
and a uniform polynomial-time oracle-free quantum decoder $\Decode$
returning a codeword or $\bot$, satisfying the following.
\begin{enumerate}
\item \emph{Independent suffix constraints.} For every constant $d\ge1$, all sufficiently large $\lambda$,
and every choice of the fixed predicates $H_1,\ldots,H_\lambda$, the sets
$P_1(H),\ldots,P_d(H)$ for $H\leftarrow\cH_\lambda$ are mutually
independent, satisfy $\cP_H=\bigcap_{j=1}^dP_j(H)$, and obey
\begin{equation}\label{eq:code-joint-inclusion}
 \Pr_{H\leftarrow\cH_\lambda}[u_1,\ldots,u_t\in P_j(H)]
 \le2^{-\lambda t/(8d)}
\end{equation}
for every $j\in[d]$ and all distinct $u_1,\ldots,u_t\in\cX_\lambda$
with $1\le t\le2^\lambda$.
\item \emph{Good hashes.} The predicates $H_1,\ldots,H_\lambda$ can be fixed so that,
for all sufficiently large $\lambda$,
\begin{equation}\label{eq:good-hash-probability}
 \Pr_{H\leftarrow\cH_\lambda}[H\in\Good_\lambda]\ge1-2^{-\lambda}.
\end{equation}
Neither this choice nor $\Decode$ depends on $d$.
\item \emph{Code-state preparation.} The normalized uniform superposition over
$C_\lambda$ can be prepared without oracle queries to trace-distance error
$\epsilon$ in time $\poly(\lambda,\log(1/\epsilon))$.
\end{enumerate}
\end{lemma}

The proof of \cref{lem:code-hash} appears in \cref{app:code-export},
using the code and decoder of~\cite{BHV}.
Henceforth, we fix this code family and the prefix predicates from
\cref{lem:code-hash}(2), unless another prefix is explicitly specified.
The decoder needs neither the prefix description nor access to $H$,
since its hash-dependent input is the state $\ket{\Psi_H}$.

\subsection{Contractive simultaneous incompressibility}
\label{sec:incompressibility-check}

We recall simultaneous incompressibility from~\cite{CakanQCIP26}.
Throughout this paper, we fix the error
$\varepsilon_{\rm inc}(\lambda)=2^{-8\lambda^2}$, independently of the message length.

\begin{definition}[Contractive simultaneous incompressibility]
\label{def:simultaneous-incompressibility}
Let $F_\lambda\subseteq\cX_\lambda$ be random, where $\cX_\lambda\subseteq\zo^*$ is finite.
Fix $\gamma:\mathbb N\to\mathbb R_{>0}$ with $\gamma(\lambda)\to\infty$ and define the recovery threshold
\[
 t_{\rm inc}(\lambda,c)=\ceil{\frac{c+8\lambda^2}{\gamma(\lambda)}}.
\]
The family $(F_\lambda)_\lambda$ satisfies \emph{$\gamma$-contractive simultaneous incompressibility}
if, for every polynomially bounded integer-valued
$c(\lambda)\ge0$ and all sufficiently large $\lambda$, the following holds.
An unbounded sender receives $F_\lambda$ and sends
$w\leftarrow\mathsf{Enc}(F_\lambda,r)\in\zo^{c(\lambda)}$ to an unbounded receiver $G(w,r)$,
where $r$ is their shared random tape. The shared tape and the receiver's
description, initial state and auxiliary oracles are jointly independent of $F_\lambda$,
and the receiver has no access to $F_\lambda$.
For every such sender and receiver,
\begin{equation}\label{eq:simultaneous-incompressibility}
 \Pr_{F_\lambda,r}\!\left[
 G(\mathsf{Enc}(F_\lambda,r),r)\text{ outputs }t_{\rm inc}(\lambda,c(\lambda))\text{ distinct elements of }F_\lambda
 \right]\le\varepsilon_{\rm inc}(\lambda).
\end{equation}
\end{definition}

The parameter $\gamma(\lambda)$ is independent of the message length.
Here, \emph{contractive} refers to the recovery bound, since $t_{\rm inc}(\lambda,c(\lambda))=o(c(\lambda)+\lambda^2)$.
For codeword-hash pairs, we obtain $\gamma(\lambda)=\Theta(\lambda)$,
whereas sparse noise and its union with codeword-hash pairs give
$\gamma(\lambda)=\Theta(\sqrt\lambda)$
(\cref{cor:code-noise-incompressibility}). We prove these bounds using
the following lemma, which verifies contractive simultaneous incompressibility from
joint inclusion and remains valid after adding independent noise.

\begin{lemma}[Contractive simultaneous incompressibility and independent noise]
\label{lem:simultaneous-incompressibility}
Let $F_\lambda\subseteq\cX_\lambda$ be random, where $\cX_\lambda\subseteq\zo^*$ is finite.
Suppose $\rho(\lambda)>0$ tends to zero and, for all sufficiently large
$\lambda$ and every $1\le r\le2^\lambda$,
\begin{equation}\label{eq:sim-joint-sparsity}
 \Pr_{F_\lambda}[u_1,\ldots,u_r\in F_\lambda]\le\rho(\lambda)^r
 \quad\text{for all distinct }u_1,\ldots,u_r\in\cX_\lambda.
\end{equation}
Then $(F_\lambda)_\lambda$ satisfies $\gamma$-contractive simultaneous incompressibility for
\[
 \gamma(\lambda)=\log_2(1/\rho(\lambda)).
\]
If $B_\lambda\leftarrow\Bernoulli(\beta(\lambda))^{\cX_\lambda}$ is sampled independently of $F_\lambda$
and $\rho(\lambda)+\beta(\lambda)\to0$,
the same statement holds for $F_\lambda\cup\supp(B_\lambda)$
with $\rho+\beta$ in place of $\rho$.
\end{lemma}

\begin{proof}
We suppress $\lambda$ in $F,B,\rho,\beta$. For $1\le t\le2^\lambda$ and
distinct inputs $u_1,\ldots,u_t$, the noise bits satisfy
$\Pr_B[u_1,\ldots,u_t\in\supp(B)]=\beta^t$.
Assigning each input either to $F$ or to $\supp(B)$ and using their
independence therefore gives
\[
 \Pr_{F,B}[u_1,\ldots,u_t\in F\cup\supp(B)]
 \le\sum_{J\subseteq[t]}\rho^{|J|}\beta^{t-|J|}
 =(\rho+\beta)^t.
\]
For either $F$ or $F\cup\supp(B)$, write $p=\rho$ or $p=\rho+\beta$,
respectively. Fix a polynomially bounded $c=c(\lambda)$ and put
$t=\ceil{(c+8\lambda^2)/\log_2(1/p)}$.
Since $p\le1/2$ for all sufficiently large $\lambda$, we have
$1\le t\le c+8\lambda^2+1\le2^\lambda$ eventually.
Fix the shared tape $r$ and a message $w\in\zo^c$.
The receiver's output is independent of the random set, so averaging
the corresponding joint-inclusion bound over that output gives success
probability at most $p^t$. Summing over the $2^c$ messages and averaging
over $r$ bounds every sender's success probability by
\[
 2^cp^t\le2^{-8\lambda^2}=\varepsilon_{\rm inc}(\lambda).
\]
The same calculation holds for every message length $c$ for which
$t_{\rm inc}(\lambda,c)\le2^\lambda$.
\end{proof}

We now apply these bounds to the codeword-hash pairs and noise.

\begin{corollary}[Contractive simultaneous incompressibility of codeword-hash pairs and noise]
\label{cor:code-noise-incompressibility}
Fix a constant $d\ge1$ and any prefix predicates $H_1,\ldots,H_\lambda$,
and independently sample
\[
 H\leftarrow\cH_\lambda,\qquad
 B\leftarrow\Bernoulli(\beta(\lambda))^{\cX_\lambda},\qquad
 \beta(\lambda)=2^{-\ceil{\sqrt\lambda}}.
\]
Put $\rho_d(\lambda)=2^{-\lambda/(8d)}$. For every $j\in[d]$, the families
\[
 P_j(H),\qquad\supp(B),\qquad P_j(H)\cup\supp(B)
\]
satisfy contractive simultaneous incompressibility with parameters
\[
 \gamma(\lambda)=\frac{\lambda}{8d},\qquad
 \ceil{\sqrt\lambda},\qquad
 \log_2\!\frac{1}{\rho_d(\lambda)+\beta(\lambda)},
\]
respectively, and common error $\varepsilon_{\rm inc}(\lambda)=2^{-8\lambda^2}$.
The same holds with $\cP_H$ in place of $P_j(H)$,
setting $d=1$ in these parameters.
\end{corollary}
\begin{proof}
For $1\le r\le2^\lambda$ and any $r$ distinct inputs,
\cref{lem:code-hash}(1) bounds their joint inclusion in $P_j(H)$ by $\rho_d(\lambda)^r$.
Independence of the noise bits gives the bound $\beta(\lambda)^r$ for $\supp(B)$.
Since $\rho_d(\lambda)+\beta(\lambda)\to0$,
\cref{lem:simultaneous-incompressibility} proves the three claims.
Applying \cref{lem:code-hash}(1) with $d=1$ gives the claims for $\cP_H$.
\end{proof}

\clearpage\section{Deterministic-Classical-Query Simulation of Quantum Verifiers}\label{sec:classical-simulation}

In this section, we prove that contractive simultaneous incompressibility gives
deterministic-classical-query simulation for quantum verifiers with classical witnesses
(\cref{thm:classical-simulation}).

To state the simulation theorem, we define oracles that agree with a
fixed function $C$ outside $F$.

\begin{definition}[Oracles agreeing outside a set]\label{def:oracle-modification}
For a Boolean function $C:\cX\to\zo$, a set $F\subseteq\cX$ and
$R:F\to\zo$, we define $\ora_F^R$, which agrees with $C$ outside $F$, by
\[
 \ora_F^R(u)=
 \begin{cases}
 R(u),&u\in F,\\
 C(u),&u\notin F.
 \end{cases}
\]
\end{definition}

We can now state the deterministic-classical-query simulation theorem.

\begin{theorem}[Deterministic-classical-query simulation from contractive simultaneous incompressibility]
\label{thm:classical-simulation}\label{thm:sublinear-classical-simulation}
\label{thm:quantitative-classical-simulation}\label{cor:sparse-simulation}
Let $\cX,\cZ\subseteq\zo^*$ be finite and nonempty. Fix $C:\cX\to\zo$
and a one-bit $Q$-query quantum verifier $V(z,w)$, where $Q\ge1$,
$z\in\cZ$ and $w\in\zo^\ell$, with a prescribed initial state for each $(z,w)$. For integers $t,L\ge1$ and $0<\eta<1$, put
\begin{equation}\label{eq:simulation-parameters}
 \begin{gathered}
 c=\ell+\ceil{\log_2|\cZ|},\qquad
 Q_{\mathrm{sim}}=t^2L,\\
 m=c+t\bigl(\ceil{\log_2Q_{\mathrm{sim}}}+1\bigr),\qquad
 \delta_{\mathrm{samp}}=2^c(|\cX|+1)^t t e^{-\eta^2 L/(16Q^2)}.
 \end{gathered}
\end{equation}
Suppose $(F_\lambda)_\lambda$ satisfies $\gamma$-contractive simultaneous incompressibility
(\cref{def:simultaneous-incompressibility}), where $F=F_\lambda\subseteq\cX$,
$m$ is polynomially bounded and $m+8\lambda^2\le\gamma(\lambda)t$.
There is a deterministic-classical-query simulator $\Sim$ making at most
$Q_{\mathrm{sim}}$ queries, followed by a randomized oracle-free computation
producing one bit. Its description and local computation may be unrestricted.
Writing $\mathcal Q_{z,w}^\ora$ for its queried inputs, for all sufficiently large $\lambda$,
\begin{equation}\label{eq:simulation-good-event}
 \Pr_F\!\left[
 \begin{gathered}
 \forall z\in\cZ,\ w\in\zo^\ell,\ R:F\to\zo:\\
 |\mathcal Q_{z,w}^{\ora_F^R}\cap F|<t,\\
 \left|\Pr[V^{\ora_F^R}(z,w)=1]-\Pr[\Sim^{\ora_F^R}(z,w)=1]\right|\le\eta
 \end{gathered}\right]
 \ge1-2^{-8\lambda^2}-\delta_{\mathrm{samp}}.
\end{equation}
\end{theorem}
The proof of \cref{thm:classical-simulation} appears in
\cref{sec:simulation-proof}.
The same query phase simulates quantum outputs in trace distance if
oracle-free quantum computation is allowed after the queries.

\subsection{Proof of deterministic-classical-query simulation}
\label{sec:simulation-proof}

We now prove \cref{thm:classical-simulation}. Put $\alpha=\eta^2/(16Q^2)$
and $N=|\cX|$. We may assume $L\ge2Q$,
since otherwise $\delta_{\mathrm{samp}}>1$ and the probability bound is trivial.
The simulator keeps a database $D$ of queried inputs whose answers differ
from $C$. Since the answers are Boolean, $D$ determines these answers.

\begin{definition}[Database oracle]\label{def:database-oracle}
For a database $D\subseteq\cX$, we define the database oracle by
\[
 C_D(u)=C(u)\oplus\indicator[u\in D].
\]
\end{definition}

For each fixed $E\subseteq\cX$ with $|E|<t$, let
$\pi^{(j)}_{z,w,E}$ be the distribution of the measured input to query
$j$ of $V^{C_E}(z,w)$, with output $\bot$ if the execution makes fewer
than $j$ queries. Thus its total query weight on a set $A\subseteq\cX$ is
\begin{equation}\label{eq:sim-query-distribution}
 W_{z,w,E}(A)=\sum_{j=1}^Q\pi^{(j)}_{z,w,E}(A).
\end{equation}
Following the query-sampling arguments of~\cite{AIK,CakanQCIP26}, put
$L_{\rm qry}=\lfloor L/Q\rfloor$ and independently sample
\begin{equation}\label{eq:sim-query-samples}
 \mathbf u^{(j)}_{z,w,E}\leftarrow
   (\pi^{(j)}_{z,w,E})^{\otimes tL_{\rm qry}}
 \qquad(z\in\cZ,\ w\in\zo^\ell,\ |E|<t,\ j\in[Q]).
\end{equation}
These distributions depend only on $V,C,z,w,E$, so the whole collection
is independent of $F$. Fixing it determines the learner below.

\begin{boxedalgorithm}{$\Learn^\ora(z,w)$}
Set $D=\varnothing$.
\begin{enumerate}[leftmargin=*,itemsep=3pt]
 \item For $j=1,\ldots,Q$, consider the inputs in
 $\mathbf u^{(j)}_{z,w,D}$ in order. For each $u\notin D\cup\{\bot\}$:
 \begin{enumerate}[leftmargin=*,itemsep=2pt]
  \item Query $\ora(u)$ to obtain $a$.
  \item If $a\ne C(u)$, add $u$ to $D$, then return $D$ if $|D|=t$
  and restart Step~1 otherwise.
 \end{enumerate}
 \item Return $D$ after completing all $Q$ sequences without adding an input.
\end{enumerate}
\end{boxedalgorithm}

Each restart adds an input to $D$, so the learner makes at most
$t\cdot QtL_{\rm qry}\le t^2L=Q_{\mathrm{sim}}$ queries on every oracle.
To produce its output bit, the simulator uses the rounded probabilities
\[
 b=\ceil{\log_2(4/\eta)},\qquad
 p_{z,w,D}=2^{-b}\left\lfloor2^b\Pr[V^{C_D}(z,w)=1]\right\rfloor,
\]
which have error at most $\eta/4$ and are stored in its classical description.
\begin{boxedalgorithm}{$\Sim^\ora(z,w)$}
\begin{enumerate}[leftmargin=*,itemsep=3pt]
 \item Run $\Learn^\ora(z,w)$ to obtain $D$.
 \item Output a bit sampled from $\Bernoulli(p_{z,w,D})$.
\end{enumerate}
\end{boxedalgorithm}

We first state the sampling estimate that will bound the simulation error.
It applies to every fixed set $F$, without an incompressibility assumption.

\begin{lemma}[Remaining query weight]\label{lem:sim-remaining-query-probability}
Fix $F,E\subseteq\cX$, $z\in\cZ$ and $w\in\zo^\ell$, with $|E|<t$.
For the samples in \cref{eq:sim-query-samples}, put
\[
 S_i=\bigcup_{j=1}^Q\{u^{(j)}_{z,w,E,1},\ldots,
                       u^{(j)}_{z,w,E,iL_{\rm qry}}\}\cap\cX
 \qquad(0\le i\le t).
\]
Then
\begin{equation}\label{eq:sim-remaining-query-probability}
 \Pr_{\mathbf u}\left[
 |S_t\cap F|<t\ \land\
 W_{z,w,E}\bigl(F\setminus(E\cup S_t)\bigr)>2Q\alpha
 \right]\le t e^{-\alpha L}.
\end{equation}
\end{lemma}
The proof of \cref{lem:sim-remaining-query-probability} appears in
\cref{sec:simulation-supporting-proofs}.

\subsubsection{Applying contractive simultaneous incompressibility}
\label{sec:sublinear-simulation-proof}

We now bound the number of distinct elements of $F$ that the learner queries,
using their query numbers and answers as an incompressibility message.

\begin{claim}[Few queries in the random set]
\label{clm:sublinear-few-queries}\label{lem:classical-query-incompressibility}
For every fixed choice of the query samples in
\cref{eq:sim-query-samples},
\begin{equation}\label{eq:sim-many-queries}
 \Pr_F\!\left[\exists z,w,R:\
 \Learn^{\ora_F^R}(z,w)\text{ queries }t\text{ distinct elements of }F
 \right]\le2^{-8\lambda^2}.
\end{equation}
\end{claim}
\begin{proof}
Fix the query samples. Given $F$, the sender finds $(z,w,R)$ for
which the learner queries $t$ distinct elements of $F$ and sends
\[
 (z,w,(k_i,b_i)_{i=1}^t),\qquad
 m=c+t\bigl(\ceil{\log_2Q_{\mathrm{sim}}}+1\bigr),
\]
where $k_i$ is the first query number of the $i$th distinct element
of $F$, and $b_i$ is its answer.
The receiver runs $\Learn(z,w)$ without oracle access, answering its
$k$th query at $u$ by $b_i$ if $k=k_i$ and by $C(u)$ otherwise.
After query $k_t$, it outputs the first $t_{\rm inc}(\lambda,m)\le t$
inputs queried at the marked positions, or $\bot$ if the learner stops earlier.

These answers reproduce the chosen execution through query $k_t$.
Indeed, each newly queried element of $F$ has its first query marked,
and a repeated query to such an element has answer $C(u)$, since the
learner never queries inputs already added to $D$.
Every other query lies outside $F$ and also has answer $C(u)$.
The receiver therefore recovers $t_{\rm inc}(\lambda,m)$ distinct
elements of $F$, so contractive simultaneous incompressibility gives
\cref{eq:sim-many-queries}.
\end{proof}

\paragraph{Fixing the query samples.}
There are at most $2^c(N+1)^t$ triples $(z,w,E)$ with $|E|<t$.
For every fixed $F$, the remaining-query-weight estimate
(\cref{lem:sim-remaining-query-probability}) and a union bound show that,
except with probability $\delta_{\mathrm{samp}}$ over the query samples,
\begin{equation}\label{eq:sim-sampling-event}
 \forall z,w,E:\quad
 |S_t\cap F|<t\ \Longrightarrow\
 W_{z,w,E}\bigl(F\setminus(E\cup S_t)\bigr)\le2Q\alpha.
\end{equation}
Together with \cref{clm:sublinear-few-queries}, this bounds the probability
of either failure over $F$ and the samples by
$2^{-8\lambda^2}+\delta_{\mathrm{samp}}$.
Consequently, some fixed collection of samples has failure probability
at most this bound over $F$ alone. We use that collection in $\Learn$,
so its queries are deterministic and its description depends only on
the distribution of $F$, not on the sampled set.

\paragraph{Bounding the simulation error.}
Fix $F$ outside the exceptional event, fix $z,w,R$, and let
$D=\Learn^{\ora_F^R}(z,w)$ be the final database.
Since the learner queries fewer than $t$ distinct elements of $F$ and
$D\subseteq F$, it completes all sequences indexed by this final $D$.
Each sampled input in $F$ either belongs to $D$ or is queried in this
last pass, so its sampled set $S_t$ satisfies $|S_t\cap F|<t$.
Moreover, $C_D$ and $\ora_F^R$ agree on $D\cup S_t$ and outside $F$.
The simultaneous sampling bound therefore gives
\begin{equation}\label{eq:sim-small-remaining-query-probability}
 W_{z,w,D}\bigl(\{u:C_D(u)\ne\ora_F^R(u)\}\bigr)\le2Q\alpha.
\end{equation}
Applying the quantum query hybrid bound (\cref{lem:query-hybrid})
to $V^{C_D}(z,w)$ now yields
\[
 \trdist{V^{\ora_F^R}(z,w)}{V^{C_D}(z,w)}
 \le2\sqrt{Q\cdot2Q\alpha}=\eta/\sqrt2.
\]
Since $\Sim$ outputs $\Bernoulli(p_{z,w,D})$, its simulation error
is at most $\eta/\sqrt2+\eta/4<\eta$.
This holds simultaneously for every $z,w,R$ and proves
\cref{eq:simulation-good-event}.

\paragraph{Polynomial query bounds.}
\label{sec:quantitative-simulation}
For codeword-hash pairs and noise, the recovery threshold has the form
\[
 t_{\rm inc}(\lambda,m)=\ceil{\frac{m+8\lambda^2}{\gamma(\lambda)}},
 \qquad \gamma(\lambda)=\Omega(\sqrt\lambda).
\]
Suppose $Q,c,\log N,1/\eta$ are polynomially bounded. Choose
\[
 t=c+\lambda^2,\qquad
 L=\ceil{\alpha^{-1}
 \bigl(c\ln2+t\ln(N+1)+\ln t+8\lambda^2\ln2\bigr)}.
\]
Then $Q_{\mathrm{sim}}$ is polynomial and $m=O(t\log\lambda)$, so
$m+8\lambda^2\le\gamma(\lambda)t$ for sufficiently large $\lambda$.
Thus $t_{\rm inc}(\lambda,m)\le t$ and
$\delta_{\mathrm{samp}}\le2^{-8\lambda^2}$, giving simulation error
at most $\eta$ except with probability $2^{1-8\lambda^2}$.

\subsection{Proof of the sampling estimate}
\label{sec:simulation-supporting-proofs}

We now prove \cref{lem:sim-remaining-query-probability}.
Group the samples into $t$ blocks, with block $i$ containing samples
$(i-1)L_{\rm qry}+1,\ldots,iL_{\rm qry}$ from every query distribution.
If $|S_t\cap F|<t$, some block adds no new element of $F$.
For such a block $i$, the high-query-weight event implies
\[
 W_{z,w,E}\bigl(F\setminus(E\cup S_{i-1})\bigr)>2Q\alpha,
\]
since $S_{i-1}\subseteq S_t$.
Conditional on the preceding blocks, put
$w_j=\pi^{(j)}_{z,w,E}(F\setminus(E\cup S_{i-1}))$.
The probability that this block misses the displayed set is
\[
 \prod_{j=1}^Q(1-w_j)^{L_{\rm qry}}
 \le\exp\left(-L_{\rm qry}\sum_{j=1}^Qw_j\right)
 \le e^{-\alpha L},
\]
where the last inequality uses $L_{\rm qry}\ge L/(2Q)$.
Summing over the $t$ possible blocks proves the lemma.

\clearpage\section{Removing Classical Witnesses by Switching}
\label{sec:classical-switching}
\label{sec:qcma-replacement}

In this section, we remove classical witnesses while preserving a quantum
verifier's non-$\bot$ answers on a noise oracle and an oracle that differs
only on a sparse random set (\cref{thm:qcma-replacement}).
The accepting-transcript identity and switching bound are also used in
recursive-oracle simulation (\cref{lem:accepting-dnf,thm:biased-switching}).

We first specify the joint sampling of these oracles.

\begin{definition}[Oracles agreeing outside a sparse random set]
\label{def:sparse-oracle-modification}
Fix $c>0$ and a finite nonempty domain $\cX\subseteq\zo^*$, and put
$b=\lceil\sqrt\lambda\rceil$, $\beta=2^{-b}$ and $\rho=2^{-c\lambda}$.
We define $\mu$ to be any distribution of oracles agreeing outside a sparse
random set, sampled as follows:
\begin{enumerate}[leftmargin=*,itemsep=2pt]
 \item Sample $P\subseteq\cX$ from any distribution satisfying
 \[
 \Pr_P[u_1,\ldots,u_t\in P]\le\rho^t
 \quad(1\le t\le2^\lambda)
 \]
 for all distinct $u_1,\ldots,u_t\in\cX$.
 \item Independently sample $B(u)\leftarrow\Bernoulli(\beta)$ for each $u\in\cX$.
 \item Sample $\ora:\cX\to\zo$ arbitrarily subject to
 $\ora|_{\cX\setminus P}=B|_{\cX\setminus P}$.
\end{enumerate}
\end{definition}

For $(P,B,\ora)\sim\mu$, we obtain the following replacement
of a \QCMAbb{} machine by a classical-query deterministic algorithm.

\begin{theorem}[Removing classical witnesses]\label{thm:qcma-replacement}
Let $\mu$ be a joint distribution of $(P,B,\ora)$ as specified
(\cref{def:sparse-oracle-modification}) for a fixed constant $c>0$. Let
$\varnothing\ne\cZ\subseteq\zo^*$ be finite and put $L=|\cZ|$.
Let $N_V$ be the partial \QCMAbb{} machine (\cref{def:qcma-machine})
associated with a fixed quantum query verifier $V(z,w)$, with auxiliary input
$z\in\cZ$ and witness $w\in\zo^\ell$. Suppose that $V$ makes at
most $Q\ge1$ queries on every oracle and that
$Q,\ell,\log|\cX|,\log L$ are polynomial in $\lambda$.

For every polynomially bounded integer $h\ge1$, there is a mapping
\[
 \Replace_{V,\mu,h}:
 (P,B|_{\cX\setminus P})\longmapsto T.
\]
For every auxiliary input and oracle, $T$ outputs a bit after fewer than
$h$ deterministic-classical queries, all in $P$, and it outputs zero
outside $\cZ$. Its description and local computation may be unrestricted.
Uniformly in $\mu$ and the fixed local computation of $V$, for all
sufficiently large $\lambda$,
\begin{equation}\label{eq:qcma-replacement}
 \Pr_{(P,B,\ora)\samp\mu}\left[
 \forall F\in\{B,\ora\},\ z\in\cZ:\quad
 N_V^F(z)\in\{T^F(z),\bot\}
 \right]
 \ge1-2^{-\lambda^2}-2L\,2^{-ch\sqrt\lambda/4}.
\end{equation}
\end{theorem}

\subsection{\texorpdfstring{Proof of \cref{thm:qcma-replacement}}{Proof of Theorem \getrefnumber{thm:qcma-replacement}}}

We combine deterministic-classical-query simulation with two lemmas:
acceptance of the simulator has a small-width DNF representation,
and switching turns this DNF into a short classical query algorithm.
We first state these two lemmas.

\begin{definition}[Accepting-transcript DNF]\label{def:accepting-dnf}
Let $\cX,\cZ\subseteq\zo^*$ be finite, let $\ell,m\ge0$, and
let $S(z,w)$ be an algorithm with auxiliary input $z\in\cZ$ and
witness $w\in\zo^\ell$. Suppose that, on every oracle
$\ora:\cX\to\zo$, $S$ first makes at most $m$ deterministic-classical queries and then produces a bit without further oracle queries.
For each $(z,w)$, let $\cA_{z,w}\subseteq\bigcup_{t=0}^m\zo^t$
consist of the answer sequences $a$ on which the query phase terminates
after exactly $|a|$ queries. For each such sequence $a$, denote query
$i$ by $u_i(z,w,a)$, which depends only on $z,w,a_{<i}$, and let
$p_{z,w,a}$ be the acceptance probability of the final oracle-free
computation. We define the accepting-transcript DNF $f_z$ by
\begin{equation}\label{eq:accepting-transcript-dnf}
 f_z(\ora)=
 \bigvee_{\substack{w\in\zo^\ell,\ a\in\cA_{z,w}\\p_{z,w,a}\ge1/2}}
 \ \bigwedge_{i=1}^{|a|}[\ora(u_i(z,w,a))=a_i].
\end{equation}
\end{definition}

\begin{lemma}[Accepting transcripts]\label{lem:accepting-dnf}
The accepting-transcript functions $f_z$ (\cref{def:accepting-dnf}) are DNFs of
width at most $m$ and satisfy, for every $z,\ora$,
\begin{equation}\label{eq:accepting-transcript-identity}
 f_z(\ora)=1
 \quad\Longleftrightarrow\quad
 \exists w\in\zo^\ell:\Pr[S^\ora(z,w)=1]\ge\tfrac12.
\end{equation}
Moreover, let $M_V$ be the partial \QCMA{} machine for a verifier $V$
with the same input and witness domains (\cref{def:qcma-machine}).
For every oracle $\ora$ satisfying
\begin{equation}\label{eq:dnf-simulation-assumption}
 \forall z\in\cZ,w\in\zo^\ell:\quad
 \left|\Pr[V^\ora(z,w)=1]-\Pr[S^\ora(z,w)=1]\right|
 \le\tfrac1{100},
\end{equation}
the DNFs preserve every non-$\bot$ answer:
\begin{equation}\label{eq:dnf-machine-agreement}
 \forall z\in\cZ:\quad
 M_V^\ora(z)\ne\bot\ \Longrightarrow\ f_z(\ora)=M_V^\ora(z).
\end{equation}
\end{lemma}
The proof of \cref{lem:accepting-dnf} appears in \cref{sec:accepting-transcripts}.

We next state the switching bound for these DNFs.

\begin{definition}[Restricted Boolean function]\label{def:restricted-function}
Let $\cX\subseteq\zo^*$ be finite, $f:\zo^{\cX}\to\zo$,
$P\subseteq\cX$, and $U:\cX\setminus P\to\zo$.
We define the restriction $g^{P,U}$ of $f$ obtained by setting the
answers outside $P$ to $U$ as follows: for $r:P\to\zo$, let
\[
 \ora_r(u)=
 \begin{cases}r(u),&u\in P,\\U(u),&u\notin P,\end{cases}
 \qquad g^{P,U}(r)=f(\ora_r).
\]
\end{definition}

Random restrictions are a standard tool for reducing classical circuit
depth~\cite{Hastad86}. Here the fixed answers are biased toward zero,
and the decision tree must depend only on $P$ and $U$. We derive the
following bound from Kelley's switching lemma~\cite{Kelley}, which is
independent of the number of DNF terms.

\begin{theorem}[Switching with sparse answers]\label{thm:biased-switching}
Fix integers $b,k,h\ge1$ and a DNF $f$ of width at most $k$ on a finite
domain $\cX\subseteq\zo^*$. There is a mapping
\[
 \Switch_{f,h}:(P,U)\longmapsto T_{P,U}
\]
to deterministic classical algorithms that output one bit using fewer
than $h$ queries to any oracle $r:P\to\zo$.

Suppose that $0<\rho^{1/b}\le1/(16bk)$ and that $P$ is random with
\[
 \Pr_P[u_1,\ldots,u_t\in P]\le\rho^t
 \qquad(1\le t\le bk+h)
\]
for distinct $u_1,\ldots,u_t\in\cX$. Independently of $P$, sample
\[
 B\leftarrow\Bernoulli(2^{-b})^{\cX},\qquad U=B|_{\cX\setminus P}.
\]
The mapping and algorithms may have unrestricted local computation
and description length. For the restricted function $g^{P,U}$
(\cref{def:restricted-function}),
\begin{equation}\label{eq:switching-guarantee}
 \Pr_{P,B}\left[
 \forall r:P\to\zo:\quad T_{P,U}^{r}=g^{P,U}(r)
 \right]\ge1-2(8bk\rho^{1/b})^h.
\end{equation}
\end{theorem}
The proof of \cref{thm:biased-switching} appears in \cref{sec:switching}.

We now apply these lemmas to the deterministic-classical-query simulator
to prove \cref{thm:qcma-replacement}.

\paragraph{Constructing the classical algorithm.}
Use the parameters $b=\lceil\sqrt\lambda\rceil$,
$\beta=2^{-b}$ and $\rho=2^{-c\lambda}$, and let
$E=\supp(B)\cup P$. Choose the simulator $S$ from
deterministic-classical-query simulation (\cref{thm:classical-simulation})
for the distribution of $E$, with
the fixed function $C=0$, accuracy $\eta=1/100$, and parameters
\[
 \begin{aligned}
 c_V&=\ell+\lceil\log_2L\rceil,\qquad t=c_V+\lambda^2,\\
 L_{\rm sim}&=\left\lceil16\cdot10^4Q^2
 \bigl(c_V\ln2+t\ln(|\cX|+1)+\ln t+8\lambda^2\ln2\bigr)\right\rceil.
 \end{aligned}
\]
We verify the required incompressibility below. Write
$k=t^2L_{\rm sim}=\poly(\lambda)$ for its query bound and $f_z$
for its accepting-transcript DNF (\cref{def:accepting-dnf}), both
determined by $V$ and $\mu$.

Given $P$ and $U=B|_{\cX\setminus P}$, construct one tree
$\Switch_{f_z,h}(P,U)$ for each $z\in\cZ$ and store these trees
in the description of $T$.
\begin{boxedalgorithm}{$T^F(z)$}
\begin{enumerate}[leftmargin=*,itemsep=3pt]
 \item If $z\notin\cZ$, output zero and terminate.
 \item Start at the root of the stored tree $\Switch_{f_z,h}(P,U)$.
 At each internal node with query input $u\in P$, query $F(u)$ to
 obtain $a$, then follow the edge labelled $a$.
 \item On reaching a leaf, return its bit and terminate.
\end{enumerate}
\end{boxedalgorithm}
This defines $\Replace_{V,\mu,h}$ using only $(P,U)$ from the sample.

\paragraph{Approximating the verifier.}
For distinct inputs, independence and the joint-inclusion bound give
\[
 \Pr[u_1,\ldots,u_t\in E]\le(\beta+\rho)^t.
\]
The message length $m_{\rm msg}=c_V+t(\ceil{\log_2 k}+1)$ is polynomial in $\lambda$.
Since $k$ is polynomial,
$t=c_V+\lambda^2$, and $\beta+\rho\le2^{-\sqrt\lambda/2}$, we have
\[
 m_{\rm msg}+8\lambda^2\le t\log_2(1/(\beta+\rho))
\]
for sufficiently large $\lambda$. Thus the incompressibility bound
(\cref{lem:simultaneous-incompressibility}) gives
$t_{\rm inc}(\lambda,m_{\rm msg})\le t$ for $E$.
The choice of $L_{\rm sim}$ also bounds the sampling error in
deterministic-classical-query simulation (\cref{thm:classical-simulation}) by $2^{-8\lambda^2}$.
Consequently, except with probability $2^{-4\lambda^2}$ over $E$,
$S$ has error at most $1/100$ simultaneously for every auxiliary input,
witness and oracle supported in $E$. Both $B$ and $\ora$ have
support in $E$, so \cref{lem:accepting-dnf} gives
\begin{equation}\label{eq:replacement-dnf-event}
 \forall F\in\{B,\ora\},z\in\cZ:\quad
 N_V^F(z)\ne\bot\ \Longrightarrow\ f_z(F)=N_V^F(z).
\end{equation}

\paragraph{Removing the witness.}
Let $g_z^{P,U}$ be the restriction of $f_z$ obtained by fixing
the answers outside $P$ to $U$ (\cref{def:restricted-function}).
\cref{thm:biased-switching} gives the query bound on every
sample. Its correctness guarantee applies for sufficiently large
$\lambda$, since $bk+h\le2^\lambda$ and $\rho^{1/b}\le1/(16bk)$.
A union bound over $z\in\cZ$ then shows that every returned
tree computes $g_z^{P,U}$ on all assignments to $P$, except with
probability $2L(8bk\rho^{1/b})^h$.

Since both $B$ and $\ora$ equal $U$ outside $P$, successful switching gives
\[
 \forall F\in\{B,\ora\},\ z\in\cZ:\qquad
 T^F(z)=g_z^{P,U}(F|_P)=f_z(F).
\]
Finally, $b\le2\sqrt\lambda$ and $k$ is polynomial, so for sufficiently large $\lambda$,
\[
 8bk\rho^{1/b}=8bk\,2^{-c\lambda/b}
 \le2^{-c\sqrt\lambda/4}.
\]
Combining this bound with the displayed equality,
\cref{eq:replacement-dnf-event} and
$2^{-4\lambda^2}\le2^{-\lambda^2}$ proves
\cref{eq:qcma-replacement}.

\subsection{Proofs of the supporting lemmas}

\subsubsection{Accepting transcripts}\label{sec:accepting-transcripts}

We prove \cref{lem:accepting-dnf} by identifying the conjunction
satisfied by each execution of the simulator.

\begin{proof}
For fixed $z,w,\ora$, the conjunction indexed by $a$ holds exactly when
$a$ is the answer sequence of the deterministic query phase, in
which case $p_{z,w,a}=\Pr[S^\ora(z,w)=1]$. The disjunction over witnesses
and sequences with $p_{z,w,a}\ge1/2$ therefore gives
\cref{eq:accepting-transcript-identity}, with at most $m$ literals
in each term.

To obtain agreement with $M_V$, apply the approximation
in~\cref{eq:dnf-simulation-assumption} at the two \QCMA{} thresholds:
\[
 \begin{aligned}
 M_V^\ora(z)=1&\ \Longrightarrow\quad
   \max_w\Pr[S^\ora(z,w)=1]\ge2/3-1/100>1/2,\\
 M_V^\ora(z)=0&\ \Longrightarrow\quad
   \max_w\Pr[S^\ora(z,w)=1]\le1/3+1/100<1/2.
 \end{aligned}
\]
\cref{eq:accepting-transcript-identity} now gives
\cref{eq:dnf-machine-agreement}.
\end{proof}

\subsubsection{Switching with sparse answers}\label{sec:switching}

We now prove switching with sparse answers (\cref{thm:biased-switching})
using Kelley's switching lemma.

\begin{lemma}[Kelley's switching lemma~\cite{Kelley}]
\label{lem:kelley-switching}
Let $w,h\ge1$ be integers, let $F$ be a width-$w$ DNF on a finite set
$\cY\subseteq\zo^*$, and let $0<p\le1/(16w)$. Suppose
a random set $P'\subseteq\cY$
satisfies
\[
 \Pr_{P'}[S\subseteq P']\le p^{|S|}
 \qquad(S\subseteq\cY,\ |S|\le w+h).
\]
Independently sample
$B'\leftarrow\Bernoulli(1/2)^{\cY}$ and put
$U'=B'|_{\cY\setminus P'}$. Then
\[
 \Pr_{P',B'}\!\left[
   \DTdepth(g^{P',U'})\ge h
 \right]\le2(8wp)^h,
\]
where $g^{P',U'}$ is the restriction of $F$
(\cref{def:restricted-function}) and $\DTdepth$ denotes
deterministic decision-tree depth.
\end{lemma}

We first construct the returned decision tree, then use Kelley's lemma
to bound the probability that it fails to compute the restriction.

Given $(P,U)$, the restricted function $g^{P,U}$ is determined
without access to the unknown assignment $r:P\to\zo$. Define the
mapping by the following finite search.
\begin{boxedalgorithm}{$\Switch_{f,h}(P,U)$}
Fix an ordering of all decision trees of depth less than $h$
whose query inputs belong to $P$.
\begin{enumerate}[nosep]
 \item For each tree in order, enumerate all assignments $r:P\to\zo$
 and compare its output with $f(\ora_r)$, where
 $\ora_r|_P=r$ and $\ora_r|_{\cX\setminus P}=U$.
 Return the first tree whose output equals $f(\ora_r)$ for every $r$.
 \item If the enumeration ends without finding such a tree,
 return the constant-zero tree.
\end{enumerate}
\end{boxedalgorithm}
The returned tree makes fewer than $h$ queries on every assignment.
It remains to bound the probability that the search finds no tree
computing $g^{P,U}$.

Sample a uniform Boolean function $B':\cX\times[b]\to\zo$
independently of $P$, and set
\[
 B(u)=\bigwedge_{j=1}^bB'(u,j),\qquad P'=P\times[b].
\]
Substituting this conjunction for each variable of $f$ and distributing
gives a DNF $f'$ of width at most $bk$: a positive literal contributes
$b$ literals, while a negative literal becomes a disjunction of $b$
negative literals. To apply Kelley's lemma, we also need a
joint-inclusion bound for $P'$. For
$S\subseteq\cX\times[b]$ with $|S|\le bk+h$, let
$\pi(S)=\{u:(u,j)\in S\text{ for some }j\}$. Since
$|\pi(S)|\ge|S|/b$, the joint-inclusion hypothesis gives
\[
 \Pr[S\subseteq P']=\Pr[\pi(S)\subseteq P]
 \le\rho^{|\pi(S)|}\le(\rho^{1/b})^{|S|}.
\]
\cref{lem:kelley-switching} therefore applies with $w=bk$ and
$p=\rho^{1/b}$, giving a decision tree of depth less than $h$ after
the variables outside $P'$ are fixed according to $B'$, except with
probability $2(8bk\rho^{1/b})^h$. Answering each query $(u,j)\in P\times[b]$ by $r(u)$ makes this tree compute $g^{P,U}(r)$ with fewer
than $h$ queries, since the conjunction of the $b$ copies of $r(u)$
is again $r(u)$.

The search finds a correct tree of depth less than $h$ whenever one
exists, so the bound above gives \cref{eq:switching-guarantee}.

When only one hierarchy level remains, we use the case $h=1$:
$T_{P,U}$ then makes no queries, so $g^{P,U}$ is constant on the
successful event.

\paperpart{$\QMA$ versus $\QCMAH$}
\label{part:hierarchy}
\section{Codeword-Hash Pairs in Noise (\CHN{}) Problem and $\QMAbb$ Verification}
\label{sec:oracle-problem}

In this section, we define codeword-hash pairs in noise (\CHN{}) and
prove that it has a \QMAbb{} verifier using only three classical oracle
queries (\cref{thm:qma-verification}).

\label{sec:oracle-distributions}

\begin{definition}[\CHN{} oracle distributions]\label{def:oracle-distributions}
Let $C_\lambda\subseteq\Sigma_\lambda^{q_\lambda}$ be the linear code used for
codeword-hash pairs (\cref{def:code-pairs}), so oracle queries are
pairs $(v,y)\in\cX_\lambda=\Sigma_\lambda^{q_\lambda}\times\zo^\lambda$.
Use the biased hash distribution $\cH_\lambda$
(\cref{def:hash-distribution}), with prefix predicates fixed by
\cref{lem:code-hash}(2).
We define the \CHN{} oracle distributions $\oradist_{\mathsf{noise},\lambda}$ and
$\oradist_{\mathsf{plant},\lambda}$ by independently sampling
\begin{equation}\label{eq:chn-oracle-sampling}
 H\samp\cH_\lambda,\qquad
 B(v,y)\overset{\mathrm{ind}}\leftarrow
 \Bernoulli(2^{-\ceil{\sqrt\lambda}})
 \quad((v,y)\in\cX_\lambda),
\end{equation}
and returning the following Boolean oracles on $\cX_\lambda$:
\begin{enumerate}
 \item \emph{Noise.} For $\oradist_{\mathsf{noise},\lambda}$, return
 the noise oracle $\oranoise_B(v,y):=B(v,y)$.
 \item \emph{Planted pairs.} For $\oradist_{\mathsf{plant},\lambda}$,
 return the planted oracle
 \begin{equation}\label{eq:chn-planted-oracle}
 \oraplant_{H,B}(v,y):=
 \begin{cases}
 1,&v\in C_\lambda\text{ and }H(v)=y\|0^{q_\lambda-\lambda},\\
 B(v,y),&\text{otherwise}
 \end{cases}.
 \end{equation}
\end{enumerate}
When an oracle is sampled from either distribution, algorithms receive
quantum query access to it without separate access to $H$ or $B$.
\end{definition}

We now give the verifier's acceptance bounds for the planted and noise distributions.

\begin{theorem}[\QMAbb{} verification]\label{thm:qma-verification}
Let $a=a_\lambda$ be the predicate-preimage state length
(\cref{def:hash-witness}), and use the planted and noise \CHN{} distributions
$\oradist_{\mathsf{plant},\lambda},\oradist_{\mathsf{noise},\lambda}$
(\cref{def:oracle-distributions}). There is a uniform \QPT{}
verifier $V_{\rm \QMA{}}$ using at most three classical oracle queries
and a $3a$-qubit witness $\sigma$,
such that, for all sufficiently large $\lambda$,
\begin{align*}
 \Pr_{\ora\leftarrow\oradist_{\mathsf{plant},\lambda}}
 \left[\exists\sigma:\Pr[V_{\rm \QMA{}}^\ora(1^\lambda,\sigma)=1]\ge\frac23\right]
 &\ge1-2^{-\lambda},\\
 \Pr_{\ora\leftarrow\oradist_{\mathsf{noise},\lambda}}
 \left[\forall\sigma:\Pr[V_{\rm \QMA{}}^\ora(1^\lambda,\sigma)=1]\le\frac13\right]
 &\ge1-6\sqrt{(a+1)2^{-\lambda}}.
\end{align*}
\end{theorem}

\subsection{Proof of \QMAbb{} verification}\label{sec:qma-verification}

We now prove \cref{thm:qma-verification}. The following bound will control
acceptance under the noise distribution, even for an oracle-dependent witness.

\begin{lemma}[A random Boolean function]\label{lem:random-table}
Let $\lambda,a\ge1$ be integers. Fix an
oracle-free quantum algorithm $A$ that receives an $a$-qubit density operator
$\sigma$ and a classical input $y\in\zo^\lambda$ and outputs an element of
$C_\lambda\cup\{\bot\}$ for the linear code
$C_\lambda\subseteq\Sigma_\lambda^{q_\lambda}$ (\cref{def:code-pairs}).
For a uniform Boolean function $B:\Sigma_\lambda^{q_\lambda}\times\zo^\lambda\to\zo$,
\[
 \Pr_B\left[
 \exists\sigma:\quad
 \Pr_{y\leftarrow\zo^\lambda,\,v\leftarrow A(\sigma,y)}
     [v\ne\bot\ \land\ B(v,y)=1]>\frac23
 \right]
 \le6\sqrt{(a+1)2^{-\lambda}}.
\]
\end{lemma}
The proof of \cref{lem:random-table} appears in \cref{app:query-bounds}.

\paragraph{The verifier.}
Using the oracle-free decoder $\Decode$ (\cref{def:good-hashes}),
define the following three-query verifier.
\begin{boxedalgorithm}{$V_{\rm \QMA{}}^\ora(1^\lambda,\WitnessReg_1,\WitnessReg_2,\WitnessReg_3)$}
The witness consists of three $a$-qubit registers.
\begin{enumerate}
\item For $j=1,2,3$:
  \begin{enumerate}
  \item Sample $y_j\leftarrow\zo^\lambda$ uniformly and run
        $v_j\leftarrow\Decode(y_j,\WitnessReg_j)$.
  \item If $v_j=\bot$ or $\ora(v_j,y_j)=0$, output $0$ and terminate.
  \end{enumerate}
\item Output $1$.
\end{enumerate}
\end{boxedalgorithm}

\paragraph{The planted distribution.}
For every good hash $H\in\Good_\lambda$ (\cref{def:good-hashes}),
\[
 \Pr\!\left[V_{\rm \QMA{}}^{\oraplant_{H,B}}
       (1^\lambda,\ket{\Psi_H}^{\otimes3})=1\right]
 \ge1-3\cdot2^{1-\lambda}.
\]
\cref{lem:code-hash}(2) gives
$\Pr_H[H\in\Good_\lambda]\ge1-2^{-\lambda}$, proving
the planted-distribution bound for sufficiently large $\lambda$.

\paragraph{The noise distribution.}
We couple the sparse noise oracle to a uniform Boolean oracle that is
pointwise at least as large. Put
$\beta=2^{-\lceil\sqrt\lambda\rceil}$ and independently sample
\[
 U(u)\leftarrow\Bernoulli(1/2),\qquad
 Z(u)\leftarrow\Bernoulli(2\beta)
 \quad(u\in\cX_\lambda),
\]
and set $B(u)=U(u)Z(u)$. This gives a sample
$B\samp\oradist_{\mathsf{noise},\lambda}$ with $B\le U$.
For an $a$-qubit state $\tau$, write
\[
 p_B(\tau)=\Pr_{y\leftarrow\zo^\lambda,\,
                  v\leftarrow\Decode(y,\tau)}
              [v\ne\bot\ \land\ B(v,y)=1].
\]
Because the decoder is oracle-free, $p_B(\tau)\le p_U(\tau)$ for
every $\tau$, and \cref{lem:random-table} gives
\[
 \Pr_B\left[\sup_\tau p_B(\tau)>2/3\right]
 \le\Pr_U\left[\sup_\tau p_U(\tau)>2/3\right]
 \le6\sqrt{(a+1)2^{-\lambda}}.
\]
To apply this one-check bound to all three checks, even when the witness
registers are entangled, fix $B$ with $\sup_\tau p_B(\tau)\le2/3$.
After conditioning on the preceding checks accepting, $\WitnessReg_j$
still contains an $a$-qubit state, while the fresh challenge $y_j$
remains independent and uniform.
Each conditional acceptance probability is therefore at
most $2/3$, so
\[
 \Pr[V_{\rm \QMA{}}^{\oranoise_B}(1^\lambda,\sigma)=1]\le(2/3)^3=8/27<1/3
\]
for every $\sigma$, including entangled witnesses, which proves the noise-distribution bound.

For $H\in\Good_\lambda$ (\cref{def:good-hashes}),
the predicate-preimage state $\ket{\Psi_H}$ (\cref{def:hash-witness})
produces a valid pair outside any fixed set of oracle inputs
$D\subseteq\cX_\lambda$ with probability
\[
 \Pr_{\substack{y\samp\zo^\lambda\\v\samp\Decode(y,\ket{\Psi_H})}}\left[
 (v,y)\in\cP_H\setminus D
 \right]
 \ge1-2^{1-\lambda}-\frac{|D|}{2^\lambda}.
\]
Indeed, at most $|D|$ challenges appear in $D$, and the decoder
succeeds on every challenge with probability at least $1-2^{1-\lambda}$.
The state $\ket{\Psi_H}$ depends on $H$, whereas
\cref{thm:classical-simulation} requires an oracle-independent
initial state for each fixed classical auxiliary input and witness.

\clearpage\section{Depth Reduction for \QCMAHbb{} Machines}
\label{sec:hierarchy-depth}

In this section, we prove that, except with negligible probability, a constant-depth
\QCMAHbb{} machine cannot give different non-$\bot$ answers on two random
oracles that differ only on the intersection of independent sparse sets
(\cref{thm:depth-reduction}).

We first specify a common resource bound for all levels of the hierarchy.

\begin{definition}[Hierarchy resource bounds]\label{def:hierarchy-resource-bounds}
A hierarchy has resource bound $p$ on a set $\cZ$ of auxiliary inputs to the top machine
if, starting from each $z\in\cZ$, every verifier makes at most
$p(\lambda)$ queries and all auxiliary inputs to verifiers, allowed witnesses, and
oracle query inputs have length at most $p(\lambda)$. These bounds
must hold on every base oracle and every choice of total Boolean
lower oracles.
\end{definition}

With this resource convention, we can state the depth-reduction theorem.

\begin{theorem}[Iterated depth reduction]\label{thm:depth-reduction}
Fix constants $d\ge1$, $c>0$, and a polynomial $p$.
Let $\varnothing\ne\cX\subseteq\zo^{\le p(\lambda)}$.
Sample mutually independent sets
$P_1,\ldots,P_d$ satisfying
\[
 \Pr_{P_i}[u_1,\ldots,u_t\in P_i]\le2^{-c\lambda t}
 \qquad(1\le t\le2^\lambda)
\]
for every $i\in[d]$ and all distinct $u_1,\ldots,u_t\in\cX$.
Independently of these sets, sample
\[
 B(u)\overset{\mathrm{ind}}\leftarrow
 \Bernoulli(2^{-\lceil\sqrt\lambda\rceil})
 \quad(u\in\cX),\qquad
 \Gamma=\bigcap_{i=1}^dP_i.
\]
Given $(P_1,\ldots,P_d,B)$, sample $\ora:\cX\to\zo$ arbitrarily
subject to
\[
 \ora|_{\cX\setminus\Gamma}=B|_{\cX\setminus\Gamma}
 \quad\text{almost surely}.
\]

Let $M_d$ be a fixed depth-$d$ hierarchy of quantum query verifiers,
with a fixed nonempty set $\cZ\subseteq\zo^*$ of at most
$p(\lambda)$ auxiliary inputs to the top machine and resource bound $p$ on $\cZ$
(\cref{def:hierarchy-resource-bounds}).
There is a mapping
\[
 (P_1,\ldots,P_d,B|_{\cX\setminus\Gamma})
       \longmapsto(c_z)_{z\in\cZ}\in\zo^{\cZ}
\]
whose outputs satisfy the following bound, uniformly in the joint
distribution, finite domain and fixed local computation of $M_d$:
\begin{equation}\label{eq:iterated-depth-reduction}
 \Pr_{P_1,\ldots,P_d,B,\ora}\left[
 \forall z\in\cZ, F\in\{B,\ora\}:\quad
       M_d^F(z)\in\{c_z,\bot\}
 \right]\ge1-\varepsilon_{p,d,c}(\lambda),
 \qquad
 \varepsilon_{p,d,c}(\lambda)=2^{-\Omega(\sqrt\lambda)}.
\end{equation}
\end{theorem}

\subsection{Proof of iterated depth reduction}

To prove \cref{thm:depth-reduction}, we first state two supporting lemmas.
The first lets us substitute a classical query algorithm for a partial
lower oracle while preserving every non-$\bot$ answer.

\begin{lemma}[Substituting a completion]\label{lem:substitute-completion}
Let $A(z,w)$ be a quantum query verifier with access to a base oracle
and a Boolean lower oracle, let $N$ be a partial Boolean oracle machine,
and let $M_{\mathrm{comp}}$ be their completion-based composition
(\cref{def:qcmah-machine}). Fix a base oracle $\ora$ and finite
sets $\U,\cZ\subseteq\zo^*$ such that, on auxiliary inputs $z\in\U$,
every lower query of $A$ belongs to $\cZ$, for every witness and
every choice of the two oracles. Suppose a classical-query deterministic algorithm $T$ outputs one bit on every auxiliary input and oracle and satisfies
\[
 \forall u\in\cZ:\qquad
 N^\ora(u)\ne\bot\ \Longrightarrow\ T^\ora(u)=N^\ora(u).
\]
Define $W^\ora(z,w)=A^{\ora,T^\ora}(z,w)$, and let $M_W$ be its
\QCMA{} machine. If $A$ and $T$ use at most $Q$ and $t$ queries
on every oracle, then $W$ uses at most $Q(1+2t)$ queries and the
same witness length as $A$, and
\[
 \forall z\in\U:\qquad
 M_{\mathrm{comp}}^\ora(z)\ne\bot\ \Longrightarrow\ M_W^\ora(z)=M_{\mathrm{comp}}^\ora(z).
\]
\end{lemma}
The proof of \cref{lem:substitute-completion} appears in
\cref{sec:depth-supporting-proofs}.
Observe that this agreement also preserves non-$\bot$ answers higher
in the hierarchy when all queries to the replaced machine lie in $\U$.

\cref{thm:qcma-replacement} supplies the classical algorithm
needed for this substitution. The resulting hierarchy has bounded
resources even when replacement fails, allowing us to iterate the
reduction without conditioning on success.

\begin{lemma}[One-level depth reduction]\label{lem:one-depth-reduction}
Fix a constant $d\ge2$, a constant $c>0$, and a polynomial $p$.
Let $\varnothing\ne\cX\subseteq\zo^{\le p(\lambda)}$ and
sample $(P,B,\ora)\leftarrow\mu$, where $\mu$ satisfies
\cref{def:sparse-oracle-modification} with this $c$ and domain.
Let $M_d$ be a fixed depth-$d$ hierarchy of quantum query verifiers,
with a finite set $\cZ\subseteq\zo^*$ of auxiliary inputs to the top machine and
resource bound $p$ on $\cZ$
(\cref{def:hierarchy-resource-bounds}).

There is a mapping
\[
 \Reduce_{M_d,\mu}:
   (P,B|_{\cX\setminus P})\longmapsto M'_{d-1}
\]
with unrestricted output description and local computation. Uniformly
in $\mu$, $\cX$ and the fixed local computations of $M_d$, it has
the following properties:
\begin{enumerate}
\item For every input to the mapping, $M'_{d-1}$ is a depth-$(d-1)$ hierarchy
of quantum query verifiers on domain $P$, with resource bound
$p'(\lambda)=5(p(\lambda)+1)^2$ on $\cZ$.
\item For all sufficiently large $\lambda$,
\begin{equation}\label{eq:one-depth-preservation}
 \Pr_{(P,B,\ora)\leftarrow\mu}\left[
 \begin{gathered}
 \forall z\in\cZ,\quad F\in\{B,\ora\}:\\
 M_d^F(z)\ne\bot\ \Longrightarrow\ (M'_{d-1})^{F|_P}(z)=M_d^F(z)
 \end{gathered}\right]
 \ge1-\delta_{p,c}(\lambda),
 \qquad \delta_{p,c}(\lambda)=2^{-\Omega(\sqrt\lambda)}.
\end{equation}
\end{enumerate}
\end{lemma}
The proof of \cref{lem:one-depth-reduction} appears in
\cref{sec:one-depth-proof}.

We now apply the one-level reduction repeatedly to prove
\cref{thm:depth-reduction}.

\paragraph{Constructing the output bits.}
We define the mapping by induction on $d$. For $d=1$, apply
\cref{thm:qcma-replacement} with $P=P_1$, auxiliary input set
$\cZ$, and $h=1$. Since the replacement makes no queries,
its output on $z$ is a bit $c_z$ determined by
$P_1,B|_{\cX\setminus P_1}$. The error is
$2^{-\Omega(\sqrt\lambda)}$ because $|\cZ|$ is polynomial.

For $d\ge2$, let $\mu$ be the joint distribution of $(P_1,B,\ora)$
and construct the bits in two steps.
\begin{enumerate}
\item Apply one-level depth reduction (\cref{lem:one-depth-reduction})
with $P=P_1$, which is valid because $\Gamma\subseteq P_1$. Set
\[
 e=(P_1,B|_{\cX\setminus P_1}),\qquad
 M'_{d-1,e}=\Reduce_{M_d,\mu}(e),\qquad
 P_i'=P_1\cap P_i\quad(2\le i\le d).
\]
\item If $P_1\ne\varnothing$, apply the inductive mapping to
$M'_{d-1,e}$ on domain $P_1$, using $(P_i')_{i=2}^d$ and
$B|_{P_1\setminus\Gamma}$, and return its bits $(c_z)_{z\in\cZ}$.
If $P_1=\varnothing$, return the non-$\bot$ answer of $M'_{d-1,e}$
on the empty oracle at each $z$, or zero when that answer is $\bot$.
\end{enumerate}
The recursion decreases the depth, and its supplied answers together
are exactly $B|_{\cX\setminus\Gamma}$. Thus the construction terminates
and has the required dependence on the sampled objects.

\paragraph{Applying the induction.}
We check the hypotheses for Step~2 under the conditional distribution
given $e$. This conditioning fixes
$M'_{d-1,e}$ and its domain $P_1$, and
\cref{lem:one-depth-reduction}(1) gives the resource bound
$p'=5(p+1)^2$ for every $e$.
The sets $P_2',\ldots,P_d'$ remain mutually independent and
independent of $B|_{P_1}$, since $e$ fixes $P_1$ and only the bits
of $B$ outside $P_1$. In particular,
\[
 B|_{P_1}\samp\Bernoulli
       (2^{-\lceil\sqrt\lambda\rceil})^{P_1},\qquad
 \Pr[u_1,\ldots,u_t\in P_i'\mid e]\le2^{-c\lambda t}
\]
for distinct $u_1,\ldots,u_t\in P_1$, $1\le t\le2^\lambda$,
and $2\le i\le d$. Moreover,
$\ora|_{P_1}=B|_{P_1}$ outside $\bigcap_{i=2}^dP_i'$,
which supplies the remaining induction hypothesis.
For $P_1=\varnothing$, the two reduced executions have the same
empty oracle, so the chosen bits satisfy the conclusion exactly.

\paragraph{Combining the errors.}
Because the conditional error bound holds for every $e$ without
conditioning on successful replacement, we may average over $e$
and add the first replacement's failure probability to obtain
\[
 \begin{aligned}
 &\Pr[\exists z\in\cZ,\ F\in\{B,\ora\}:M_d^F(z)\notin\{c_z,\bot\}]\\
 &\quad\le\delta_{p,c}(\lambda)
   +\bbE_e\Pr[\exists z,F:
         (M'_{d-1,e})^{F|_{P_1}}(z)\notin\{c_z,\bot\}\mid e]\\
 &\quad\le\delta_{p,c}(\lambda)
       +\varepsilon_{p',d-1,c}(\lambda)
       =2^{-\Omega(\sqrt\lambda)}.
 \end{aligned}
\]
The last equality uses that $d$ is constant, completing the proof.

\subsection{Proofs of the supporting lemmas}
\label{sec:depth-supporting-proofs}

We now prove \cref{lem:substitute-completion}, which justifies replacing
a partial lower oracle by a classical query algorithm.

\begin{proof}[Proof of \cref{lem:substitute-completion}]
Construct $W(z,w)$ by running $A(z,w)$, forwarding its queries to $\ora$ and implementing each lower query with the coherent
version of $T^\ora$ from \cref{lem:coherent-classical-query}.
This keeps the witness unchanged and uses at most
$Q(1+2t)$ queries to $\ora$.

For correctness, extend $T^\ora|_{\cZ}$ to a completion
$\widetilde N\in\Comp(N^\ora)$ by setting
$\widetilde N(u)=N^\ora(u)$ at every non-$\bot$ input outside
$\cZ$ and setting it to zero at the remaining inputs.
Every lower query of $A$ belongs to $\cZ$, so
\[
 W^\ora(z,w)=A^{\ora,\widetilde N}(z,w)
 \qquad(z\in\U).
\]
If $M_{\mathrm{comp}}^\ora(z)\ne\bot$, it is the answer for every completion by
\cref{eq:hierarchy-qcma-form}, and hence also for $\widetilde N$.
This gives $M_W^\ora(z)=M_{\mathrm{comp}}^\ora(z)$.

Since $M_W^\ora$ preserves every non-$\bot$ answer of $M_{\mathrm{comp}}^\ora$ on $\U$,
the restriction of any completion of $M_W^\ora$ to $\U$ extends
to a Boolean completion of $M_{\mathrm{comp}}^\ora$
(\cref{eq:boolean-completions}).
Consequently, if a higher verifier queries $M_{\mathrm{comp}}$ only on
$\U$, every completion it receives after substitution is also a
permitted completion before substitution on those inputs. Repeated
application of~\cref{eq:hierarchy-qcma-form} therefore preserves
the higher machines' non-$\bot$ answers as well.
\end{proof}

\subsubsection{Removing one level}\label{sec:one-depth-proof}

We now prove \cref{lem:one-depth-reduction} by constructing the reduced
hierarchy and checking its resource and correctness guarantees.

\paragraph{Constructing the reduced hierarchy.}
Let $V$ be the deepest verifier and $N_V$ its \QCMAbb{} machine, with
witnesses padded to length $p(\lambda)$ by ignoring extra bits.
Let $\cZ_{\rm lower}\subseteq\zo^{\le p(\lambda)}$ contain
every query string that can reach $N_V$ from $\cZ$ under any base
oracle, total Boolean lower oracles, and allowed witnesses, choosing
this set independently of the sample. To absorb a union bound over
$\cZ_{\rm lower}$, put $h=1$ if this set is empty and otherwise
\[
 h=\left\lceil\log_2(2|\cZ_{\rm lower}|)\right\rceil
 \le p(\lambda)+2.
\]
\par\Needspace{14\baselineskip}
Write $U=B|_{\cX\setminus P}$ and construct $M'_{d-1}$ as follows.
\begin{boxedalgorithm}{$\Reduce_{M_d,\mu}(P,U)$}
\begin{enumerate}
\item If $\cZ_{\rm lower}$ is empty, set $T=0$.
      Otherwise construct
      $T=\Replace_{V,\mu,h}(P,U)$ for the deepest verifier $V$.
\item Let $A$ be the parent verifier of $N_V$. Replace it by
      $W^F(z,w)=A^{F,T^F}(z,w)$, with the same classical witness.
      Remove $N_V$ from the hierarchy.
\item In every remaining verifier, answer a query $F(u)$ by $U(u)$
      if $u\notin P$, and by a query to the given oracle on $P$ otherwise.
      Output this depth-$(d-1)$ hierarchy.
\end{enumerate}
\end{boxedalgorithm}

\paragraph{Bounding the resources.}
By \cref{lem:coherent-classical-query}, each query to $T$ uses at
most $2h$ queries to $F$, so the modified parent uses at most
\[
 p(1+2h)\le p(2p+5)\le5(p+1)^2
\]
queries to $F$. Step~3 answers queries outside $P$ locally and changes neither
witnesses nor query inputs, whose lengths remain at most $p(\lambda)$.
Thus $p'=5(p+1)^2$ bounds all stated query and length resources for
every $(P,U)$, including samples on which replacement is incorrect.

\paragraph{Preserving the answers.}
If $\cZ_{\rm lower}$ is empty, no call to $N_V$ occurs and
the substitution is exact. Otherwise,
\cref{thm:qcma-replacement} bounds the probability that $T$
fails to preserve a non-$\bot$ answer of $N_V$ on either $B$ or $\ora$ by
\[
 \begin{aligned}
 2^{-\lambda^2}
 +2|\cZ_{\rm lower}|\,2^{-ch\sqrt\lambda/4}
 &\le2^{-\lambda^2}+2^{-h(c\sqrt\lambda/4-1)}\\
 &=2^{-\Omega(\sqrt\lambda)},
 \end{aligned}
\]
where the first inequality uses $2|\cZ_{\rm lower}|\le2^h$.

Whenever $T$ preserves these non-$\bot$ answers,
\cref{lem:substitute-completion} preserves all non-$\bot$ answers
in Step~2. Step~3 implements queries to $B$ and $\ora$ exactly, including
on superpositions, because both equal $U$ outside $P$. This proves
\cref{eq:one-depth-preservation}.

\clearpage\section[\QCMAHbb{} Query Lower Bounds and \QMA{}--\QCMAH{} Oracle Separation]{\QCMAHbb{} Query Lower Bounds and\texorpdfstring{\\}{ }\QMA{}--\QCMAH{} Oracle Separation}
\label{sec:hierarchy-separation}

In this section, we prove that \CHN{} is hard for \QCMAHbb{} machines
(\cref{thm:hierarchy}) and state the corresponding separation relative
to a fixed classical oracle. Recall that \CHN{} compares a random noise
oracle $\oranoise_B=B$ with the planted oracle
$\oraplant_{H,B}=B\vee\indicator_{\cP_H}$
(\cref{def:oracle-distributions}).

\begin{theorem}[\QCMAHbb{} indistinguishability]\label{thm:hierarchy}
Let $d\ge1$ be constant and let $M_d$ be a $\QCMAHbb_d$ machine
with a polynomial resource bound on $\{1^\lambda\}$
(\cref{def:hierarchy-resource-bounds}), evaluated on auxiliary input $1^\lambda$.
For every fixed choice of prefix predicates $H_1,\ldots,H_\lambda$ in the
biased hash distribution $\cH_\lambda$ (\cref{def:hash-distribution}),
independently sample
\[
 H\leftarrow\cH_\lambda,\qquad
 B(u)\overset{\mathrm{ind}}\leftarrow
 \Bernoulli(2^{-\lceil\sqrt\lambda\rceil})
 \quad(u\in\cX_\lambda),
\]
and put $\oranoise_B=B$ and $\oraplant_{H,B}=B\vee\indicator_{\cP_H}$.
There is a mapping
\[
 (H,B|_{\cX_\lambda\setminus\cP_H})\longmapsto c\in\zo
\]
such that
\begin{equation}\label{eq:common-hierarchy-answer}
 \Pr_{H,B}\left[M_d^{\oranoise_B},M_d^{\oraplant_{H,B}}\in\{c,\bot\}\right]
 \ge1-\varepsilon_{M_d}(\lambda),
 \qquad \varepsilon_{M_d}(\lambda)=2^{-\Omega(\sqrt\lambda)}.
\end{equation}
The bound is uniform in the prefix and in fixed auxiliary oracle
answers independent of $H,B$. Therefore, for the planted and noise
\CHN{} distributions $\oradist_{\mathsf{plant},\lambda},\oradist_{\mathsf{noise},\lambda}$
(\cref{def:oracle-distributions}),
\begin{equation}\label{eq:hierarchy-promise-bound}
 \Pr_{\ora\leftarrow\oradist_{\mathsf{plant},\lambda}}[M_d^\ora=1]
 +\Pr_{\ora\leftarrow\oradist_{\mathsf{noise},\lambda}}[M_d^\ora=0]
 \le1+\varepsilon_{M_d}(\lambda).
\end{equation}
If $M_d$ outputs a bit with probability $1$ under both distributions, then
\begin{equation}\label{eq:hierarchy-indistinguishability}
 \left|\Pr_{\ora\leftarrow\oradist_{\mathsf{plant},\lambda}}[M_d^\ora=1]
       -\Pr_{\ora\leftarrow\oradist_{\mathsf{noise},\lambda}}[M_d^\ora=1]\right|
 \le\varepsilon_{M_d}(\lambda).
\end{equation}
\end{theorem}
The proof of \cref{thm:hierarchy} appears in \cref{sec:hierarchy-query-proof}.

We also obtain a separation relative to a fixed classical oracle,
including the promise collapse.

\begin{theorem}[\QCMA{} collapse and \QMA{} separation]\label{thm:boolean-collapse}
There are a classical Boolean oracle $\ora_{\mathsf{sep}}$ and a unary language $L$ such that
\[
 \begin{gathered}
 \PTIME^{\ora_{\mathsf{sep}}}=\BQP^{\ora_{\mathsf{sep}}}=\QCMA^{\ora_{\mathsf{sep}}}=\QCMAH^{\ora_{\mathsf{sep}}}\subsetneq\QMA^{\ora_{\mathsf{sep}}},
 \qquad L\in\QMA^{\ora_{\mathsf{sep}}}\setminus\QCMAH^{\ora_{\mathsf{sep}}},\\
 \PROMISEP^{\ora_{\mathsf{sep}}}=\PROMISEQCMAH^{\ora_{\mathsf{sep}}}.
 \end{gathered}
\]
\end{theorem}
The proof of \cref{thm:boolean-collapse} appears in \cref{app:boolean-collapse}.
Note that the cryptographic oracle also gives a promise problem in $\QMA$ that is average-case hard for $\QCMAH$ (\cref{cor:pke-average-qma}).

\subsection{Proof of \QCMAHbb{} hardness}
\label{sec:hierarchy-query-proof}

We now prove \cref{thm:hierarchy} by applying depth reduction to the
independent groups of suffix predicates.

Apply the codeword-hash bound (\cref{lem:code-hash}(1)) with $d$ groups to the suffix-constraint
sets $P_1(H),\ldots,P_d(H)$ (\cref{def:suffix-constraints}).
They are mutually independent, independent of $B$, intersect in
$\cP_H$, and satisfy the joint-inclusion bound
$2^{-\lambda t/(8d)}$. Thus \cref{thm:depth-reduction} applies with
sparsity exponent $1/(8d)$, intersection $\Gamma=\cP_H$ and
auxiliary input set $\{1^\lambda\}$ for the top machine,
after incorporating the fixed auxiliary oracle answers into the
verifiers' local computation. Its conclusion is
\[
 \Pr_{H,B}[M_d^{\oranoise_B},M_d^{\oraplant_{H,B}}\in\{c,\bot\}]
 \ge1-2^{-\Omega(\sqrt\lambda)},
\]
where the output bit $c$ depends only on the sets $P_i(H)$ and
$B|_{\cX_\lambda\setminus\cP_H}$. Since $H$ determines those
sets, this is precisely \cref{eq:common-hierarchy-answer}.

To derive the probability bounds, let $\cE$ be the event
in~\cref{eq:common-hierarchy-answer}. For every sample,
\[
 \indicator[M_d^{\oraplant_{H,B}}=1]+\indicator[M_d^{\oranoise_B}=0]
 \le1+\indicator[\neg\cE],
\]
so taking expectations gives
\cref{eq:hierarchy-promise-bound}, and when both answers lie in $\zo$
almost surely, their agreement on $\cE$ also gives
\cref{eq:hierarchy-indistinguishability}.

\paperpart{Technical Tools for Cryptography}
\label{part:crypto-tools}
\section{Simulating the Recursive Oracle}
\label{sec:recursive-oracle-tools}

We simulate the recursive oracle with classical queries to its Boolean
function $T$, while preserving auxiliary information about $T$.
\begin{definition}[The recursive oracle]\label{def:recursive-qcma}
For a Boolean oracle $\orabase$ and functions $h_u:[0,1]\to[0,1]$,
define the \emph{recursive oracle} $\ora=(\orabase,N,A)$ by increasing
query length, with $h_u(p)=(1+p)/3$ unless specified otherwise.
For $a=\langle M,x\rangle\in\zo^n$, where $M$ has at most $n$ gates,
witness bits and query length, the length-$(n+2)^2$ queries
$\langle N,a\rangle$ and $u=\langle A,a,i\rangle$, $i\in[n^4]$, receive
\begin{equation}\label{eq:recursive-helper-distribution}
\begin{aligned}
 N(\langle M,x\rangle)&=\indicator[\exists w:M^\ora(x,w)=1],\\
 A(\langle M,x\rangle,i)&\overset{\mathrm{ind}}\leftarrow
 \Bernoulli\!\left(h_u(\Pr[M^\ora(x)=1])\right).
\end{aligned}
\end{equation}
Here $M$ is deterministic for $N$ and quantum for $A$, and runs with
the full oracle $\ora$. Its queries have length at most $n$, so their
answers have already been defined. Conditional on them, sample each
$A$ bit once. Invalid queries receive zero, and
$\oraQCMA=(N,A)$.
\end{definition}

\begin{theorem}[Classical simulation of the recursive oracle]
\label{thm:helper-implementation}\label{thm:split-recursive-preservation}
For every polynomial $L\ge\lambda$ there is a constant $m\ge1$
such that the following holds for $c_0>0$ and sufficiently small $c>0$.
Fix $\cX\subseteq\bigcup_{\ell=\lambda}^{L}\zo^\ell$, a Boolean
oracle $\ora_{\rm fix}$ on $\cX^c$, and functions $h_u$ satisfying
\[
 2^{-c\lambda}\le h_u(v)\le1-2^{-c\lambda},\qquad
 |h_u(v)-h_u(w)|\le|v-w|\quad(v,w\in[0,1]).
\]
Independently sample $G^{(1)},\ldots,G^{(m)}\subseteq\cX$ and
$B\leftarrow\zo^{\cX}$ with
\[
 \Pr[S\subseteq G^{(i)}]\le2^{-c_0\lambda|S|}\quad(|S|\le2^\lambda).
\]
Then sample any $(T,Z)$, where $Z$ is classical and $T:\cX\to\zo$ satisfies
\begin{equation}\label{def:helper-data}
 T|_{(\bigcap_iG^{(i)})^c}=B|_{(\bigcap_iG^{(i)})^c},\qquad
 \orabase(u)=\begin{cases}T(u),&u\in\cX,\\
 \ora_{\rm fix}(u),&u\notin\cX.\end{cases}
\end{equation}
Sample the recursive oracle $\ora=(\orabase,N,A)$ with means $h_u$
(\cref{def:recursive-qcma}).
Any supported recursive prefix below $\lambda^2$ may be fixed.

For every quantum protocol $\cA_Z$ with $Z$-dependent input and
operations, a common polynomial query bound and query length at most $L$,
there is a classical-query deterministic algorithm $\Answer_E^T$ with
\begin{equation}\label{eq:base-exposure-record}
 E=\Phi(G^{(1)},\ldots,G^{(m-1)},B,R),\qquad
 R\perp(G^{(1)},\ldots,G^{(m)},B,T,Z).
\end{equation}
On inputs of length at most $L$, it answers $\orabase$ queries exactly
and makes polynomially many queries for every $T$. Uniformly over the fixed choices, its averaged
joint output satisfies
\[
 \trdist
 {\bbE[\ket{Z}\!\bra{Z}\otimes\cA_Z^\ora]}
 {\bbE[\ket{Z}\!\bra{Z}\otimes\cA_Z^{\Answer_E^T}]}
 \le2^{-\Omega(\lambda)}.
\]
\end{theorem}

\subsection{Proof}\label{sec:helper-preservation-proof}

We now prove \cref{thm:split-recursive-preservation}. Include
$(G^{(1)},\ldots,G^{(m)},B,T)$ in $Z$, since discarding these classical
registers gives the stated conclusion. We first express $T$ using
independent uniform Boolean functions, then construct its recursive
extension while keeping one sparse set independent of the simulation.

Fix $d=m-1$ large enough that $L<\lambda^{2^{d+1}}$, and choose
\begin{equation}\label{eq:helper-interval-parameters}
 \rho_0=2^{-c_0\lambda},\qquad
 \nu=\rho_0^{(256m)^{-d}},\qquad
 0<c<\frac{c_0}{256(256m)^d},\qquad b=2^{-c\lambda}.
\end{equation}
Thus $\nu^{1/256}\le b\le1/4$ for sufficiently large $\lambda$.

\paragraph{Coupling.}
Sample independent uniform Boolean functions
$B^{(1)},\ldots,B^{(m)}$. Define
\[
 G_{<m}=\bigcap_{i<m}G^{(i)},\qquad
 j(u)=\min\{i<m:u\notin G^{(i)}\},\qquad
 B(u)=\begin{cases}B^{(j(u))}(u),&u\notin G_{<m},\\B^{(m)}(u),&u\in G_{<m}.\end{cases}
\]
Here $G_{<m}=\cX$ when $m=1$. Conditional on the groups, $B$ is uniform,
so sample the original $T,Z$ from their prescribed conditional
distribution given the groups and $B$, and define
\[
 T_0(u)=\begin{cases}T(u),&u\in G_{<m},\\B^{(m)}(u),&u\notin G_{<m}.\end{cases}
\]
Thus $T(u)=B^{(j(u))}(u)$ when $u\notin G_{<m}$ and
$T(u)=T_0(u)$ otherwise. This sampling preserves the original
joint distribution, including $Z$.

Let $I$ specify the values of $\ora_{\rm fix}$ through length $L$
and any fixed recursive prefix, and put
$E_{\mathsf{base}}=(I,(G^{(i)},B^{(i)})_{i<m})$.
The following lemma constructs the recursive answers through functions
$T_j$ that equal independent uniform functions $B_j$ outside sparse
sets $G_j$. We will then remove these differences by a query hybrid.

\begin{lemma}[Recursive answer construction]\label{lem:recursive-answer-construction}
The coupled sampling above extends to $E_d$ and functions
$(G_j,B_j,T_j)_{j=0}^d$, with $(G_0,B_0)=(G^{(m)},B^{(m)})$,
and a deterministic algorithm $\mathsf{Rec}_{E_d}^{\mathbf T}$,
where $\mathbf T=(T_0,\ldots,T_d)$, such that
\[
 \Pr[\mathsf{Bad}(E_d)]\le3d\nu^4,\qquad
 \neg\mathsf{Bad}(E_d)\ \Longrightarrow\
 \mathsf{Rec}_{E_d}^{\mathbf T}(u)=\ora(u)\quad(|u|\le L).
\]
The algorithm makes polynomially many queries for every $\mathbf T$
and returns zero on failure. The sampling preserves the original
$(Z,\ora)$, and $E_d$ depends only on
$E_{\mathsf{base}}$ and independent coins. Conditional on successful
$E_d$, the sets $(G_j)_{j\ge1}$ and functions $(B_j)_{j\ge1}$ are mutually
independent and independent of $(Z,B^{(m)})$,
with uniform $B_j$ and
\[
 T_j|_{G_j^c}=B_j|_{G_j^c},\qquad
 \Pr[u\in G_j\mid E_d]\le\nu\quad(j\ge1).
\]
\end{lemma}
The proof of \cref{lem:recursive-answer-construction} appears in
\cref{sec:recursive-answer-construction}.

\Needspace{10\baselineskip}
We first use this lemma to prove \cref{thm:split-recursive-preservation},
then give the construction.
\paragraph{Replacing $T_j$ by $B_j$ for $j\ge1$.}

Let $\mathsf H_0$ run the original protocol. Define $\mathsf H_1$ by
answering $\orabase$ queries with $T$ or $\ora_{\rm fix}$, and $N,A$
queries with $\mathsf{Rec}_{E_d}^{\mathbf T}$. The latter algorithm
returns zero on $\mathsf{Bad}$, so recursive answer construction
(\cref{lem:recursive-answer-construction}) gives
\[
 \trdist{\mathsf H_0}{\mathsf H_1}\le3d\nu^4.
\]
Here and below the outputs include the original classical variable
$Z$ and all quantum output systems. The same failure rule defines
$\mathsf H_2$, which instead uses $B_j$ for $j\ge1$ and leaves
$T_0$ unchanged.

Conditional on successful $E_d$, $Z$, $B^{(m)}$ and the
uniform Boolean functions $(B_j)_{j\ge1}$, each new set $G_j$ still satisfies
$\Pr[u\in G_j\mid E_d,Z,B^{(m)},(B_\ell)_{\ell\ge1}]\le\nu$.
In $\mathsf H_2$, these quantities determine every oracle answer:
$T_0$ is determined by the original variables in $Z$ and $B^{(m)}$,
and the other functions are the displayed $B_j$. They therefore
determine the query weights.
Writing $w(j,u)$ for the weight on address $(j,u)$ in an internal query,
we have
\[
 \bbE\sum_{j\ge1,u}w(j,u)\indicator[u\in G_j]
 \le\nu\,\bbE\sum_{j\ge1,u}w(j,u)\le\nu.
\]
For $Q$ original queries and at most $t_{\mathsf{sim}}$ classical
queries per answer, coherent evaluation and uncomputation make
at most $2Qt_{\mathsf{sim}}$ internal queries. Use $\mathsf H_2$
for the prefix at each changed query and apply the quantum query
bound (\cref{lem:query-hybrid}), which gives
\[
 \trdist{\mathsf H_1}{\mathsf H_2}
 \le4Qt_{\mathsf{sim}}\sqrt\nu.
\]
The two experiments coincide on failure. Their input and other
operations may depend on $Z$, since the new sets are independent
of that complete original sample under the stated conditioning.

We now express every remaining oracle call using $T$. By construction,
\[
 T_0(u)=\begin{cases}T(u),&u\in G_{<m},\\B^{(m)}(u),&u\notin G_{<m}.\end{cases}
\]
Add $B^{(m)}$ and $(B_j)_{j\ge1}$ to the classical description.
On inputs longer than $L$, $\Answer_E^T$ returns zero. On a recursive
query of length at most $L$, it runs $\mathsf{Rec}_{E_d}$
with oracles $T_0,B_1,\ldots,B_d$, answering each $T_0$ query
by the displayed formula. On a query to $\orabase$, it returns $T(u)$
for $u\in\cX$ and $\ora_{\rm fix}(u)$ otherwise. For fixed original groups and $B$, the
conditional distribution of $(B^{(i)})_{i\in[m]}$ depends only on
$G^{(1)},\ldots,G^{(m-1)},B$: at each $u$, its selected bit must equal
$B(u)$, and the other bits are sampled independently from $\zo$. The sampled
$T,Z$ do not affect this conditional distribution. All further
coins in the construction are independent of the enlarged variable $Z$.
The resulting description therefore has precisely the form
$E=\Phi(G^{(1)},\ldots,G^{(m-1)},B,R)$ with $R\perp Z$.

Only oracle queries were replaced, so the protocol's other operations
and party separation are preserved. Finally,
\[
 \bbE_Z\trdist
 {\bbE_{\ora\mid Z}\cA_Z^\ora}
 {\bbE_R\cA_Z^{\Answer_E^T}}
 \le3d\nu^4+4Qt_{\mathsf{sim}}\sqrt\nu
 =2^{-\Omega(\lambda)},
\]
which proves the theorem.

\subsubsection{One random restriction}

Independently sample a random set $P\subseteq\cX$ and a uniform Boolean
function $C:\cX\to\zo$, and write $C_P^Y(u)=Y(u)$ on $P$ and $C(u)$
elsewhere.
Assume
\[
 \Pr_P[S\subseteq P]\le\rho^{|S|}\quad(|S|\le2^\lambda),\qquad
 \rho=2^{-c_{\rm res}\lambda},\qquad \log(|\cX|+1)=\poly(\lambda)
\]
for a constant $c_{\rm res}>0$.
The next lemma constructs classical query algorithms that work for
every $Y$. This simultaneous guarantee allows the later recursive
construction to choose its remaining answers after the algorithms
have been specified.

\begin{lemma}[One random restriction]\label{lem:recursive-one-restriction}
Fix quantum circuits $W_u$ and deterministic verifiers
$M_u(w)$ with polynomial query, index and witness bounds, and sample
$(P,C)$ as above.
For $\rho^{1/40}\le\epsilon\le1/4$, there are classical-query deterministic algorithms $\tau_u,T_u$ and numbers
$v_{u,\tau}\in[0,1]$, depending on $(P,C)$, where $\tau_u(Y)$ returns
the inputs queried in $Y$ and their answers, or $\bot$. Both algorithms make
polynomially many queries on every oracle, and except with probability
$3\rho^4$ over $(P,C)$,
\[
 \forall u,Y:\qquad
 |v_{u,\tau_u(Y)}-\Pr[W_u^{C_P^Y}=1]|\le\epsilon,\qquad
 T_u^Y=\indicator[\exists w:M_u^{C_P^Y}(w)=1].
\]
\end{lemma}

\Needspace{5\baselineskip}
We now prove \cref{lem:recursive-one-restriction}, first estimating
quantum acceptance and then computing the existential answers.
\paragraph{Acceptance estimates.}
If $\cX=\varnothing$, evaluate the tests and existential predicates
without queries, so assume $\cX\ne\varnothing$.
For quantum query bound $Q\ge1$ and index length $s$, set
$t=s+\lambda^2$ and put
\[
 M_{\mathrm{trial}}=\lceil\rho^{-1/10}\rceil,
 \qquad Q_{\mathrm{sim}}=t^2M_{\mathrm{trial}}.
\]
Apply deterministic-classical-query simulation (\cref{thm:classical-simulation})
using its simulator $\Sim$ with $F=P$, fixed function $C$,
$V(u)=W_u$, no witness, and accuracy $\eta=\epsilon$. Its query phase is
deterministic, so we implement that phase with access only to $Y$:
\begin{enumerate}[leftmargin=*,itemsep=2pt]
 \item Initialize the partial function $\tau=\varnothing$ on $P$.
 \item On a request $x\in P\setminus\dom\tau$, return $\bot$
 if $|\dom\tau|=t-1$, and otherwise query $Y(x)$ and set $\tau(x)$ to
 its answer. For every request $x$, supply
 $C(x)$ if $x\notin P$ and $\tau(x)$ otherwise.
 \item Return $\tau$ when the query phase terminates.
\end{enumerate}
For each partial function $\tau$ on $P$ with $|\dom\tau|<t$,
run $\Sim(u)$ with answers $\tau(x)$ on $\dom\tau$ and $C(x)$
elsewhere, and let $v_{u,\tau}$ be its output-one probability.
Put $v_{u,\bot}=0$. This defines $\tau_u$ with fewer than $t$ queries to $Y$.

Observe that conditioning on $C$ leaves the joint-inclusion bound for
$P$ unchanged. The message length
$m_{\rm msg}=s+t(\ceil{\log_2Q_{\mathrm{sim}}}+1)$ is polynomial in $\lambda$. Since
$\log_2Q_{\mathrm{sim}}\le\frac1{10}\log_2(1/\rho)+O(\log\lambda)$
and $t=s+\lambda^2$, for sufficiently large $\lambda$ we have
\[
 m_{\rm msg}+8\lambda^2\le t\log_2(1/\rho).
\]
The incompressibility bound (\cref{lem:simultaneous-incompressibility})
therefore gives $t_{\rm inc}(\lambda,m_{\rm msg})\le t$ for $P$,
as required by deterministic-classical-query simulation (\cref{thm:classical-simulation}).
On its good event,
$\Sim$ meets fewer than $t$ distinct points of $P$, so our algorithm
never returns $\bot$. Its returned $\tau$ reconstructs every answer
in this deterministic query phase, and hence the same final output
distribution. The error is therefore at most $\epsilon$, simultaneously
for every $u,Y$, while
\[
 \begin{aligned}
 Q_{\mathrm{sim}}&\le\poly(\lambda)\rho^{-1/10},\\
 \Pr_{P,C}\!\left[\begin{gathered}
 \exists u,Y:\quad\tau_u(Y)=\bot\text{ or}\\
 |v_{u,\tau_u(Y)}-\Pr[W_u^{C_P^Y}=1]|>\epsilon
 \end{gathered}\right]
 &\le2^{-8\lambda^2}+2^s(|\cX|+1)^t t e^{-\epsilon^2M_{\mathrm{trial}}/(16Q^2)}\\
 &\le\rho^4+2^{-4\lambda^2}\le2\rho^4.
 \end{aligned}
\]
Indeed, $\epsilon^2M_{\mathrm{trial}}/(16Q^2)=2^{\Omega(\lambda)}$, whereas
the logarithm of $2^s(|\cX|+1)^t t$ is polynomial in $\lambda$.
Although $\Sim$ may make more than polynomially many queries, those
outside $P$ use the fixed $C$, so only the fewer than $t$ queries in
Step~2 count toward access to $Y$.

\paragraph{Existential answers.}
For each $u$, enumerate the accepting transcripts
$\mathcal T_{u,w}$ of $M_u(w)$ and construct
\[
 f_u(\ora)=\bigvee_w\ \bigvee_{\tau\in\mathcal T_{u,w}}
       \bigwedge_{(x,b)\in\tau}[\ora(x)=b],\qquad
 T_u=\Switch_{f_u,t}(P,C|_{P^c}).
\]
The accepting-transcript identity (\cref{lem:accepting-dnf}) gives
$f_u(\ora)=\indicator[\exists w:M_u^\ora(w)=1]$, so running $T_u$
with answers from $Y$ will compute the desired predicate whenever
the switching guarantee holds. Choose a polynomial
width bound $k\ge1$, which exists because $M_u$ has polynomial query complexity.
The switching theorem (\cref{thm:biased-switching}), applied with
this $f_u$, bias parameter $1$, and independent uniform Boolean function $C$, gives fewer
than $t$ queries on every oracle and failure probability
$2(8k\rho)^t$. For sufficiently large $\lambda$,
\[
 k+t\le2^\lambda,\qquad \rho\le(16k)^{-1},\qquad
 2^{s+1}(8k\rho)^t\le\rho^4.
\]
A union bound over $u$ and the acceptance-estimation failures proves
the lemma.

\subsubsection{Constructing the recursive answers}
\label{sec:recursive-answer-construction}

We now prove \cref{lem:recursive-answer-construction} by constructing
each interval from the earlier answers, while keeping one independent
group unexposed in every oracle function.

Put
\[
 \rho_i=\rho_0^{(256m)^{-(i-1)}}\quad(i\in[d+1]),\qquad
 \cI_i=[\lambda^{2^i},\lambda^{2^{i+1}})\cap[0,L]\quad(i\in[d]),
\]
so $\rho_{d+1}=\nu$. Sample $\ora_{\rm pre}$, the $N,A$ answers
below $\lambda^2$, conditionally on the prefix specified by $I$,
and set $E_0=(I,\ora_{\rm pre})$. Since this prefix uses only
$\ora_{\rm fix}$ below $\lambda$ and independent coins, conditioning
on $E_0$ preserves the independent groups and uniform Boolean functions.

For each valid $N$ or $A$ query at address $u$, write
$\langle M_u,x_u\rangle$ for the padded encoding of its circuit and input.
The circuit $M_u$ is deterministic
for $N$ and quantum for $A$, and both query the full oracle
$\ora$. Address padding ensures that every such query reaches an earlier
interval when $u\in\cI_i$, so before constructing that interval we have
$E_{i-1}$ and an algorithm $\mathsf{Rec}_{<i}$ satisfying, on success,
\[
 \mathsf{Rec}_{<i}^{\mathbf T}(u)=\ora(u)\quad(|u|\le L,\ |u|<\lambda^{2^i}),
 \qquad
 \Pr[S\subseteq G\mid E_{i-1}]\le\rho_i^{|S|}\quad(|S|\le2^\lambda)
\]
for each of the $m-i+1$ independent groups $G$ still available in
every oracle function. All query counts are polynomial for every choice of
$\mathbf T$. The sampling of $B^{(1)},\ldots,B^{(m)},T$ above initializes these conditions.

\paragraph{Step 1: Construct $T_u$ and $\tau_u$.}
We first express the next interval's answers through classical queries
to the groups that remain after this step. Number the groups of $V_j$ by $(G_j^{(r)},B_j^{(r)})_{r\in[m-j]}$,
with $V_0=T$ and $(G_0^{(r)},B_0^{(r)})=(G^{(r)},B^{(r)})$.
Initialize $E_i=E_{i-1}$ and expose
$(G_j^{(i-j)},B_j^{(i-j)})$ for each $j<i$. Let $P$ contain
$(j,u)$ exactly when $u$ belongs to that exposed group, and set
$C(j,u)$ to the exposed function's value at $u$. Write $Y(j,u)$
for the value at $u$ of the Boolean function associated with the first
still unexposed group omitting $u$, or the final correction bit if every
remaining group contains $u$. The algorithm $\mathsf{Rec}_{<i}$ answers earlier exposed groups
using $E_{i-1}$, so its remaining queries use $C_P^Y(j,u)$.
Substitute $\mathsf{Rec}_{<i}$ into the circuits defining the $N,A$
answers in $\cI_i$, coherently for the quantum circuits
(\cref{lem:coherent-classical-query}), with their inputs $x_u$ fixed, then apply
the one-restriction lemma (\cref{lem:recursive-one-restriction}) with
\[
 \gamma_i=\rho_i^{1/128},\qquad
 \epsilon_i=b\gamma_i/1000\ge\rho_i^{1/40}.
\]
The resulting circuits are fixed conditional on $E_{i-1}$ with
polynomial query, index and witness bounds, and $(P,C)$ is independent
of them with
inclusion bound $\rho_i^{|S|}$. The displayed inequality follows from
$b\ge\nu^{1/256}\ge\rho_i^{1/256}$ for sufficiently large $\lambda$.
Thus the lemma supplies an existential algorithm $T_u$ at each $N$
address and an algorithm $\tau_u$ at each $A$ address, satisfying
\[
 \begin{gathered}
 T_u^Y=\indicator[\exists w:M_u^{\mathsf{Rec}_{<i}^{C_P^Y}}(x_u,w)=1],\\
 v_u(Y)=\Pr[M_u^{\mathsf{Rec}_{<i}^{C_P^Y}}(x_u)=1],\qquad
 \sup_Y|v_u(Y)-v_{u,\tau_u(Y)}|\le\epsilon_i.
 \end{gathered}
\]
Add $T_u,\tau_u$ to $E_i$. If this simultaneous
conclusion fails, set $\mathsf{Bad}=1$ and use the completion rule
below instead of proceeding to Step~2.

\paragraph{Step 2: Sample the $A$ answers.}
Set $a=b/4$ and sample independently at each $A$ address $u$, defining
$L_{u,\tau}$ for every possible output $\tau$, including $\bot$:
\[
 F_u\leftarrow\Bernoulli(2a),\qquad U_u\leftarrow[0,1],\qquad
 L_{u,\tau}=\indicator\!\left[
 U_u\le\frac{h_u(v_{u,\tau})-a}{1-2a}\right].
\]
Add $F_u,(L_{u,\tau})_\tau$ to $E_i$. Define the new Boolean function $V_i$ using $m-i$ groups, each
including every address independently with probability
$\gamma_i^{1/(m-i)}$, and independent uniform Boolean functions.
At each $u$, independently sample
\begin{equation}\label{eq:bounded-profile-correction}
 C_i(u)\leftarrow\Bernoulli(r_u),\qquad
 r_u=\frac12+
 \frac{h_u(v_u(Y))-h_u(v_{u,\tau_u(Y)})}{2a\gamma_i},
\end{equation}
where $Y$ gives the answers from the remaining groups in shorter intervals.
Set $V_i(u)$ to the value at $u$ of the uniform Boolean function
associated with the first group omitting $u$, or to $C_i(u)$ if every
group contains $u$. To answer $u$, if $F_u=0$, run $\tau_u$ with $Y$
and return $L_{u,\tau_u(Y)}$. If $F_u=1$, query $V_i(u)$ and return
its answer. At an $N$ address, run $T_u$.

Since $b\le h_u(v)\le1-b$, the thresholds defining $L_{u,\tau}$ lie in
$[0,1]$. Moreover, the estimate from Step~1 gives
\[
 |h_u(v_u(Y))-h_u(v_{u,\tau_u(Y)})|\le\epsilon_i,
 \qquad |r_u-1/2|\le\epsilon_i/(2a\gamma_i)=1/500.
\]
Conditional on $\orabase$, $Y$ and $(T_u,\tau_u)_u$ from Step~1,
the new coins give answer-one probability
\[
 h_u(v_{u,\tau_u(Y)})+2a\gamma_i(r_u-1/2)=h_u(v_u(Y)).
\]
Since distinct $u$ use independent coins and $v_u(Y)$ is determined
by the actual shorter answers, averaging over $Y$ and $(T_u,\tau_u)_u$ gives precisely the required product of Bernoulli distributions.

\paragraph{Step 3: Continue to the next interval.}
Keep the exposed groups in $E_i$ and continue with the $m-i$
unexposed groups in each function's sampling, including those just sampled. Their
inclusion bounds are at most $\rho_{i+1}^{|S|}$ because
\[
 \gamma_i^{1/(m-i)}
 =\rho_i^{1/(128(m-i))}\le\rho_{i+1}.
\]
We now have the required number of independent groups with the
next inclusion bound. Moreover, the algorithms and bits in $E_i$
were computed for every possible $Y$ before the actual shorter
answers were used, so conditioning on $E_i$ preserves the independence
of the unexposed groups and the uniform samples used to generate
$C_i(u)$ with their $Y$-dependent means $r_u$. The construction
can therefore continue to interval $i+1$.

\subsubsection{The final algorithm and its distribution}

Write $E=E_d$ and set
$(G_j,B_j)=(G_j^{(m-j)},B_j^{(m-j)})$. For $j\ge1$,
let $T_j$ equal $C_j$ on $G_j$ and $B_j$ elsewhere. The earlier groups belong to $E$,
so define $\mathsf{Read}_{j,r,E}^{T_j}(u)$ as follows:
\begin{enumerate}[nosep]
 \item For $\ell=r,\ldots,m-j-1$, if $u\notin G_j^{(\ell)}$,
 return $B_j^{(\ell)}(u)$.
 \item Otherwise query $T_j(u)$ and return its answer.
\end{enumerate}
This uses at most one oracle query. A query to $V_j$ uses $r=1$.
For $u\in\cI_i$, run $T_u$ or $\tau_u$ by answering each
$Y(j,v)$ query with $\mathsf{Read}_{j,i-j+1,E}^{T_j}(v)$.
Denote these algorithms by $T_u^E,\tau_u^E$, whose only oracles
are $(T_j)_j$. On input $u$, return zero if $|u|>L$, the recursive
query is invalid, or $\mathsf{Bad}(E)=1$. Otherwise return
\[
 \mathsf{Rec}_E^{\mathbf T}(u)=
 \begin{cases}
 \ora_{\rm fix}(x),&u\text{ queries }\orabase\text{ at }x\notin\cX,\\
 \mathsf{Read}_{0,1,E}^{T_0}(x),&u\text{ queries }\orabase\text{ at }x\in\cX,\\
 \ora_{\rm pre}(u),&u\text{ is an }N,A\text{ address below }\lambda^2,\\
 (T_u^E)^{\mathbf T},&u\text{ is an }N\text{ address in }\cI_i,\\
 L_{u,\tau_u^E(\mathbf T)},&u\text{ is an }A\text{ address in }\cI_i,\ F_u=0,\\
 \mathsf{Read}_{i,1,E}^{T_i}(u),&u\text{ is an }A\text{ address in }\cI_i,\ F_u=1,\\
 0,&\text{otherwise}.
 \end{cases}
\]
Every value other than a $T_j$ answer is specified by $E$, and the
algorithm evaluates $L_{u,\tau_u^E(\mathbf T)}$ without computing an
acceptance probability. Nested calls reach strictly shorter intervals,
so the algorithm terminates with polynomial query count for all
$\mathbf T$, since $d$ is constant. Coherent classical computation
(\cref{lem:coherent-classical-query}) implements its answer unitary
with at most twice as many queries.

On failure, expose any remaining groups belonging
to $E_{\mathsf{base}}$ and set $\mathsf{Bad}(E_d)=1$.
Unused pairs $(G_j,B_j)$ for $j\ge1$ are sampled independently with $T_j=B_j$,
while $(G_0,B_0)=(G^{(m)},B^{(m)})$. In either case, sample every
remaining recursive answer from its original conditional distribution. Thus the
construction preserves $(Z,\ora)$, with
$\Pr[\mathsf{Bad}]\le3\sum_{i=1}^d\rho_i^4\le3d\nu^4$.
The description $E_d$ uses only $E_{\mathsf{base}}$ and independent
coins. Conditional on successful $E_d$, the newly sampled sets
$(G_j)_{j\ge1}$ and functions $(B_j)_{j\ge1}$ are mutually independent
and independent of $(Z,B^{(m)})$, every $B_j$ is uniform,
$T_j|_{G_j^c}=B_j|_{G_j^c}$, and $\Pr[u\in G_j\mid E_d]\le\nu$.
Indeed, $T_u,\tau_u$ are constructed before the actual $Y$ is used,
and every $L_{u,\tau}$ is determined by the same $U_u$, which is
independent of the unused groups.

\clearpage\section{Preparing Quantum States from Boolean Oracles}
\label{sec:public-preparation}

We prepare real-amplitude states from Boolean oracles. The state-preparation bound
(\cref{lem:public-preparation}) also controls the change in sampling
probabilities when the target changes.

\begin{definition}[Preparation oracles]\label{def:public-preparation}
For an $n$-qubit real-amplitude state $\ket\Omega$, sample its
\emph{preparation oracles} $(R,S)\leftarrow\PrepTables(\ket\Omega)$ by
\begin{equation}\label{eq:preparation-probabilities}
\begin{gathered}
 R\leftarrow\{f:\zo^n\to\zo\},\qquad U_f\ket x=(-1)^{f(x)}\ket x,\\
 p_{\Omega,R}(x)=\frac12-\frac12\Clip_{[-1/2,1/2]}
 \left(\frac{2^{n/2}\bra x\Hadamard^{\otimes n}U_R\ket\Omega}{5n}\right)\in[1/4,3/4],\\
 S(x,y)\leftarrow\Bernoulli(p_{\Omega,R}(x))
 \quad\text{independently for }(x,y)\in\zo^n\times\zo^{4n}.
\end{gathered}
\end{equation}
Here $U_f$ is the phase oracle for $f$, and both $R$ and $S$ permit
quantum queries.
\end{definition}

We use the preparation error bound $\epsilon_{\mathrm{prep}}(n)=2^{-n}$.
\begin{lemma}[State preparation]\label{lem:public-preparation}
A uniform \QPT{} algorithm $\Prepare^{R,S}(1^n)$ outputs $\rho_{R,S}$
such that, for all $n$-qubit real-amplitude states
$\ket\Omega,\ket{\Omega^0}$ and sufficiently large $n$,
\[
\begin{gathered}
 \bbE_{(R,S)\leftarrow\PrepTables(\ket\Omega)}
 \trdist{\rho_{R,S}}{\ket\Omega\!\bra\Omega}\le\epsilon_{\mathrm{prep}}(n),\\
 \Pr_R\!\left[\max_x|p_{\Omega,R}(x)-p_{\Omega^0,R}(x)|>
 \tfrac14\|\ket\Omega-\ket{\Omega^0}\|_2\right]
 \le2^{-n^2}.
\end{gathered}
\]
\end{lemma}
The proof of \cref{lem:public-preparation} appears in
\cref{sec:state-preparation-proof}.

For $r$ preparations with the same $(R,S)$, changing one tensor
factor at a time gives
\begin{equation}\label{eq:preparation-joint-correctness}
 \bbE_{Z,R,S}\trdist{\rho_{R,S}^{\otimes r}}
 { (\ket{\Omega_Z}\!\bra{\Omega_Z})^{\otimes r}}\le r\epsilon_{\mathrm{prep}}(n),
 \qquad (R,S)\mid Z\sim\PrepTables(\ket{\Omega_Z}).
\end{equation}
Here $Z$ specifies the target. Tensoring both states with any jointly
sampled density matrix $\sigma_W$ preserves the bound, since
$\|\sigma_W\otimes X\|_1=\|X\|_1$ pointwise.

\subsection{Proof of state preparation}
\label{sec:state-preparation-proof}\label{app:quantum-evaluation}
\label{app:nearby-preparation-proof}\label{sec:nearby-preparation-proof}

We now prove \cref{lem:public-preparation} using the following
preparation algorithm. We identify its unnormalized output to bound
both the error and the success probability.

\label{sec:preparation-algorithm}
\begin{boxedalgorithm}{Preparing a state from Boolean oracles}
\noindent\underline{$\Prepare^{R,S}(1^n)$}\par
\begin{enumerate}[nosep]
 \item Repeat at most $4(5n)^3$ times, using fresh registers:
 \begin{enumerate}[nosep]
  \item Prepare $U_S\ket+^{\otimes5n}$, apply
  $\Hadamard^{\otimes4n}$ to the last $4n$ qubits, and measure them.
  \item On outcome $0^{4n}$, apply $U_R\Hadamard^{\otimes n}$ to
  the remaining register and return it.
 \end{enumerate}
 \item Return $\ket{0^n}$ if every attempt fails.
\end{enumerate}
\end{boxedalgorithm}

Put $m=5n$. The following coefficient bound controls both the
preparation error and a change of target: for every fixed real vector
$\ket v$, Hoeffding's
inequality and a union bound give
\begin{equation}\label{eq:preparation-no-clipping}
 \Pr_R\!\left[
 \max_x2^{n/2}|\bra x\Hadamard^{\otimes n}U_R\ket v|
 >\frac m2\|\ket v\|_2\right]\le2^{-n^2},
\end{equation}
because each term is a sum of independent signed coefficients whose
squares sum to $\|\ket v\|_2^2$.

\paragraph{Preparation error.}
Apply the coefficient bound (\cref{eq:preparation-no-clipping}) to
$\ket v=\ket\Omega$,
and let $\mathsf{Good}$ be the event that no coefficient is clipped.
For fixed $(R,S)$, define the unnormalized vector
\[
 \bar\sigma_x=2^{-4n}\sum_y(-1)^{S(x,y)},\qquad
 \ket\eta=U_R\Hadamard^{\otimes n}
       \left(2^{-n/2}\sum_x\bar\sigma_x\ket x\right).
\]
The zero measurement outcome followed by $U_R\Hadamard^{\otimes n}$
gives this unnormalized vector, so each attempt succeeds with probability $\|\ket\eta\|_2^2$ and
returns $\ket{\widehat\eta}=\ket\eta/\|\ket\eta\|_2$ on success.
For $R\in\mathsf{Good}$ the clipping is inactive, so the means
of the signs and independence of the bits of $S$ give
\begin{equation}\label{eq:preparation-average-error}
 \bbE_S\ket\eta=\ket\Omega/m,\qquad
 \bbE_S\|\ket\eta-\ket\Omega/m\|_2^2
 =2^{-n}\sum_x\operatorname{Var}(\bar\sigma_x)\le2^{-4n}.
\end{equation}
To turn this estimate into an output bound, put
$d=\|\ket\eta-\ket\Omega/m\|_2$. If $d\le1/(2m)$, normalization
and the triangle inequality bound the success probability and error,
while independent attempts give the failure bound:
\[
 \|\ket\eta\|_2^2\ge\frac1{4m^2},\qquad
 \trdist{\ket{\widehat\eta}\!\bra{\widehat\eta}}
 {\ket\Omega\!\bra\Omega}\le2md,
 \qquad
 \Pr[\text{all }4m^3\text{ attempts fail}]\le e^{-m}.
\]
Thus the output is within $2md+e^{-m}$ of the target when
$d\le1/(2m)$, and this bound also holds for larger $d$ because
trace distance is at most $1\le2md$. Averaging and applying
Cauchy--Schwarz to the second-moment estimate
(\cref{eq:preparation-average-error}) now gives
\[
 \bbE_{R,S}\trdist{\rho_{R,S}}{\ket\Omega\!\bra\Omega}
 \le2m2^{-2n}+e^{-m}+2^{-n^2}\le\epsilon_{\mathrm{prep}}(n).
\]
\paragraph{Changing the target.}
The same coefficient estimate also controls the change in the
Boolean sampling probabilities. Write $p_x=p_{\Omega,R}(x)$ and
$p_x^0=p_{\Omega^0,R}(x)$.
\label{lem:nearby-target-preparation}
Since $|\Clip_{[-1/2,1/2]}(a)-\Clip_{[-1/2,1/2]}(b)|\le|a-b|$, applying
the coefficient bound (\cref{eq:preparation-no-clipping}) to
$\ket\Delta=\ket\Omega-\ket{\Omega^0}$ gives
\begin{equation}\label{eq:nearby-mean-bound}
 \Pr_R[\mathsf{Good}_\Delta]\ge1-2^{-n^2},\qquad
 \mathsf{Good}_\Delta:=
 \left\{\max_x|p_x-p_x^0|\le\tfrac14\|\ket\Delta\|_2\right\}.
\end{equation}
This proves the claimed bound on the sampling probabilities.

\clearpage\section{Changing Preparation Targets}
\label{sec:preparation-replacement}

We show that exponentially close targets give indistinguishable
preparation oracles even with quantum access to their recursive
extension (\cref{thm:preparation-replacement}).

\begin{theorem}[Changing a preparation target]
\label{thm:preparation-replacement}\label{lem:common-profile-simulation}
\label{cor:preparation-ideal-input}
Fix polynomial bounds on query count and length, and $c_*>0$.
For sufficiently small $c>0$,
fix a Boolean oracle $\ora_{\rm fix}$, real-amplitude $n$-qubit states
$\ket{\Omega^0},\ket{\Omega^1}$, and functions $h_u$ satisfying
\[
\begin{gathered}
 n\ge\lambda,\qquad
 \|\ket{\Omega^0}-\ket{\Omega^1}\|_2\le2^{-c_*\lambda},\\
 2^{-c\lambda}\le h_u(v)\le1-2^{-c\lambda},\qquad
 |h_u(v)-h_u(w)|\le|v-w|\quad(v,w\in[0,1]),
\end{gathered}
\]
and sample the preparation oracles (\cref{def:public-preparation}) and
their recursive extension with means $h_u$ (\cref{def:recursive-qcma}):
\[
 (R^e,S^e)\leftarrow\PrepTables(\ket{\Omega^e}),\qquad
 \ora^e=(\ora_{\rm fix},R^e,S^e,N^e,A^e)\quad(e\in\zo).
\]
Any common supported $N,A$ prefix below $\lambda^2$ may be fixed.

For every quantum protocol $\cA$ obeying the query bounds, allow
unrestricted local computation and let its input density matrix
$\rho$ and local channels depend on both targets and $\ora_{\rm fix}$.
Uniformly over these choices, the averaged outputs satisfy
\begin{equation}\label{eq:nearby-preparation-oracles}
 \omega_e=\bbE_{\ora^e}\cA^{\ora^e}(\rho),\qquad
 \trdist{\omega_0}{\omega_1}\le2^{-\Omega(\lambda)}.
\end{equation}
\end{theorem}
The proof of \cref{thm:preparation-replacement} appears in
\cref{sec:preparation-replacement-proof}.

\subsection{Proof of the output bound}
\label{sec:preparation-replacement-proof}

We now prove \cref{thm:preparation-replacement}. We first sample the
two preparation oracles so that they can differ only on an
intersection of sparse sets. Recursive simulation then removes the
$N,A$ queries, after which a query hybrid bounds the effect of the
remaining difference.

Fix the choices in the statement, and let $Q,L$ bound the protocol's
query count and length. In each
experiment, set the sampled base answers above length $L$ to zero,
which changes neither the queried recursive answers nor the output.
Choose $m$ for this $L$
from classical recursive simulation (\cref{thm:split-recursive-preservation}), and put
\[
 0<c_0<c_*/m,\qquad
 \gamma=2^{-mc_0\lambda},\qquad \rho_0=\gamma^{1/m}.
\]
Each experiment returns the original protocol's full output and
changes only its oracle queries.

\paragraph{First hybrid: sample the preparation bits through sparse sets.}
Let $\mathsf H_{e,0}$ sample $\ora^e$ as in the target-change
experiments (\cref{thm:preparation-replacement}) and run
$\cA^{\ora^e}(\rho)$, so its averaged
output is $\omega_e$.
In $\mathsf H_{e,1}$, first sample a uniform function $R$,
then write $p_u^e=p_{\Omega^e,R}(x)$ for $u=(x,y)$ and sample
independently at each such input
\[
\begin{aligned}
 F_u&\leftarrow\Bernoulli(1/2),&
 C_u&\leftarrow\Bernoulli(2p_u^0-1/2),\\
 \indicator[u\in P_i]&\leftarrow\Bernoulli(\rho_0)\quad(i\in[m]),&
 B(u)&\leftarrow\Bernoulli(1/2),\\
 Z_u&\leftarrow\Bernoulli(q_u),&
 q_u&=\Clip_{[0,1]}\left(\frac12+
                  \frac{2(p_u^1-p_u^0)}\gamma\right).
\end{aligned}
\]
Write $P=\bigcap_{i=1}^mP_i$ and define
\[
 \begin{gathered}
 T^0=B,\qquad
 T^1(u)=\begin{cases}Z_u,&u\in P,\\B(u),&u\notin P,\end{cases}\\
 \widehat S^e(u)=\begin{cases}C_u,&F_u=0,\\T^e(u),&F_u=1.\end{cases}
 \end{gathered}
\]
Sample
\[
 \widehat\ora^e=(\ora_{\rm fix},R,\widehat S^e,
 \widehat N^e,\widehat A^e)
\]
with the fixed prefix and recursive means $h_u$, then run
$\cA^{\widehat\ora^e}(\rho)$ and return its output.

The state-preparation bound (\cref{lem:public-preparation}) gives
\[
 \Pr_R[\neg\mathsf{Good}]\le\eta:=2^{-n^2},\qquad
 \mathsf{Good}\Longrightarrow\max_{u,e}|p_u^e-p_u^0|
 \le\tfrac14\|\ket{\Omega^1}-\ket{\Omega^0}\|_2\le\gamma/4.
\]
$\widehat S^0$ has exactly the prescribed distribution for every $R$,
since $\Pr[\widehat S^0(u)=1\mid R]=p_u^0$.
On $\mathsf{Good}$, $q_u$ is unclipped and
\[
 \Pr[\widehat S^1(u)=1\mid R]
 =\frac12(2p_u^0-1/2)
  +\frac12\left(\frac{1-\gamma}{2}+\gamma q_u\right)=p_u^1.
\]
Using the same preparation answers and recursive coins whenever the
distributions agree therefore gives
\[
 \mathsf H_{0,0}=\mathsf H_{0,1},\qquad
 \trdist{\mathsf H_{1,0}}{\mathsf H_{1,1}}\le\eta.
\]

\paragraph{Second hybrid: simulate the recursive queries.}
We now simulate the $N,A$ queries. Choose $c>0$ from classical recursive simulation
(\cref{thm:split-recursive-preservation}) for $L,c_0$,
and condition on
\[
 I=(R,(F_u,C_u)_u,\text{fixed recursive prefix}).
\]
Restrict $T^e$ to the preparation inputs
$\cX=\{u=(x,y):|u|\le L,\ F_u=1\}$.
Conditional on $I$, $\widehat S^e(u)=C_u$ for $|u|\le L$ outside
$\cX$, and all other base answers are fixed. The sets $P_i$ have
inclusion bound $\rho_0^{|S|}$, and $B$ is uniform and independent
of them. Thus classical recursive simulation applies,
with an independent uniform choice
$e\leftarrow\zo$ included in its auxiliary variable $Z$.
It supplies a classical-query deterministic algorithm $\Answer_E^{T^e}$ with
polynomial query bound $t_{\mathsf{sim}}$ and
\[
 E=\Phi_I(P_1,\ldots,P_{m-1},B,\xi),\qquad
 \xi\perp(e,(P_i)_i,B,(Z_u)_u)\mid I.
\]
Let $\mathsf H_{e,2}$ replace every oracle query in
$\mathsf H_{e,1}$ by this algorithm, coherently evaluated and
uncomputed. For each $e$, the theorem gives
\[
 \trdist{\mathsf H_{e,1}}{\mathsf H_{e,2}}
 \le\epsilon_{\rm rec}=2^{-\Omega(\lambda)}.
\]
Indeed, its joint bound includes $e$, and each value has probability
$1/2$, so the conditional errors are at most twice that bound.

\paragraph{Third hybrid: remove the remaining sparse change.}
The first simulation already uses $\Answer_E^B$, since $T^0=B$.
Conditional on $I$, $(E,B)$ is independent of $(e,P_m)$, so we
compare $\mathsf H_{1,2}$ directly with $\mathsf H_{0,2}$.
The weight $w(u)$ of an internal query in $\mathsf H_{0,2}$ is
independent of $P_m$, and $T^1=B$ outside $P\subseteq P_m$. Thus
\[
 \bbE\sum_u w(u)\indicator[T^1(u)\ne B(u)]
 \le\rho_0\,\bbE\sum_u w(u)\le\rho_0.
\]
Use $\mathsf H_{0,2}$ for the prefix at each changed query.
Coherent evaluation and uncomputation use at most
$2Qt_{\mathsf{sim}}$ internal queries, so the quantum query bound
(\cref{lem:query-hybrid}) gives
\[
 \trdist{\mathsf H_{1,2}}{\mathsf H_{0,2}}
 \le4Qt_{\mathsf{sim}}\sqrt{\rho_0}.
\]
Combining the transitions now proves
\[
 \trdist{\mathsf H_{0,0}}{\mathsf H_{1,0}}
 \le\eta+2\epsilon_{\rm rec}
       +4Qt_{\mathsf{sim}}\sqrt{\rho_0}
 =2^{-\Omega(\lambda)}.
\]
Only oracle queries were changed, so each party still applies its
original channels, with no additional communication.

\clearpage\section{Encoded Functions with Quantum Keys}
\label{sec:quantum-evaluation}\label{sec:crypto-security}
\label{sec:common-key-reduction}

We evaluate encoded functions using predicate-preimage states and
simulate their recursive oracle extension using direct function queries
(\cref{thm:crypto-common-simulation}).
\label{sec:encoded-quantum-evaluation}
\begin{definition}[Encoded function and evaluation]\label{def:encoded-evaluation}
For $\mathfrak{F}:\zo^\kappa\to\zo$, a hash $H$ and
$B:\cX_\kappa\to\zo$, where $\cX_\kappa=\Sigma_\kappa^{q_\kappa}\times\zo^\kappa$,
define the encoded function on the code $C_\kappa\subseteq\Sigma_\kappa^{q_\kappa}$ by
\[
 T_{H,B,\mathfrak{F}}(v,y)=\begin{cases}
 \mathfrak{F}(y),&v\in C_\kappa\text{ and }H(v)=y\|0^{q_\kappa-\kappa},\\
 B(v,y),&\text{otherwise}.
 \end{cases}
\]
Its evaluation is $\EvalHash^T=U_{\Decode}^{-1}U_TU_{\Decode}$, where
the decoder uses $\ket{\Psi_H}$ and leaves $y$ unchanged
(\cref{def:hash-witness,lem:code-hash}).
Queries outside $\cX_\kappa$, including $v=\bot$, receive zero.
\label{def:hash-evaluation}
\end{definition}

For the encoded-function oracle, fix a quantum protocol $\cA$ with
polynomial query count and length,
$\ora_{\rm aux}:\zo^*\to\zo$,
$K\subseteq J\subseteq\zo^{\le\poly(\lambda)}$, and
$\lambda\le\kappa_j\le\poly(\lambda)$.
Sample a classical variable $\Theta$ containing
$(\mathfrak{F}_j)_{j\in K}$, and for each $j\in J$ independently sample
$H_j\leftarrow\cH_{\kappa_j}$ and a uniform Boolean function
$B_j:\cX_{\kappa_j}\to\zo$. Define
\begin{equation}\label{def:common-key-experiment}
 \ora=(\ora_{\rm aux},(T_{H_j,B_j,\mathfrak{F}_j})_{j\in J},N,A).
\end{equation}
Here the other $\mathfrak{F}_j$ may depend arbitrarily on
$\Theta,(H_j,B_j)_j$, and $N,A$ are the recursive extension
(\cref{def:recursive-qcma}). Queries $(j,v,y)$ return
$T_{H_j,B_j,\mathfrak{F}_j}(v,y)$.
A protocol opens $j\in K$ by a classical request for
$\ket{\Psi_{H_j}}$, using the same hash on repeated requests.
Initially no index is open. The querying parties know the current set
$K_t$ of opened indices. The input and response channels may depend on
$\Theta$, all $B_j$, and only the opened hashes.
\label{cond:common-function-dependence}\label{cond:common-supplied-input}

\begin{theorem}[Evaluation and simulation of encoded functions]
\label{thm:crypto-common-simulation}\label{thm:encoded-key-simulation}
\label{lem:quantitative-key-simulation}\label{lem:coherent-evaluation}
\leavevmode\par
\begin{enumerate}[leftmargin=*,itemsep=4pt]
\item For $T=T_{H,B,\mathfrak{F}}$ with $H\in\Good_\kappa$, let $\rho_q,\sigma_q$
be the joint outputs of $q$ calls to $\EvalHash^T$ or $\mathfrak{F}$,
starting with $\ket{\Psi_H}$ and any other input, with identical
intervening operations outside the key. Then
\begin{equation}\label{eq:repeated-hash-evaluation}
 \trdist{\rho_q}{\sigma_q}\le2q\sqrt{\delta_{\mathrm{dec}}(\kappa)},
 \qquad\delta_{\mathrm{dec}}(\kappa)=2^{1-\kappa}.
\end{equation}
\item There is a fixed coherent query algorithm
$\cS_\eta^{(\mathfrak{F}_j)_{j\in K}}(K_t,\cdot)$ making polynomially many
queries only to $\mathfrak{F}_j$ with $j\in K_t$, and answering unopened
encodings with $B_j$.
Its one classical sample $\eta\perp\Theta$ is jointly sampled with
$(H_j,B_j)_j$. Replace each party's oracle queries in $\cA$ by
$\cS_\eta(K_t,\cdot)$, keeping
the original input and response channels and using the same $H_j$ for
key requests. The averaged joint outputs $\rho_{\rm real},\rho_{\rm sim}$, including
$\Theta$, satisfy, uniformly for fixed query bounds,
\begin{equation}\label{eq:encoded-key-simulation}
 \trdist{\rho_{\rm real}}{\rho_{\rm sim}}\le2^{-\Omega(\lambda)}.
\end{equation}
\label{cor:crypto-online-simulation}
\end{enumerate}
\end{theorem}
Note that ideal evaluation leaves the key unchanged.

\Needspace{10\baselineskip}
\subsection{Proof of evaluation and simulation}
\label{app:encoded-key-reduction}\label{sec:common-simulation-proof}
\label{app:encoded-evaluation}\label{sec:evaluation-algorithms}

We now prove \cref{thm:crypto-common-simulation}. We first bound the
error in replacing an encoded query by a direct function query,
then construct one simulator for the entire protocol.
\paragraph{Evaluation.}
For $T=T_{H,B,\mathfrak{F}}$ and $H\in\Good_\kappa$, the decoder makes no
oracle queries and satisfies
\begin{equation}\label{eq:decoder-uniform-targets}
 \forall y:\qquad
 \Pr[(\Decode(y,\ket{\Psi_H}),y)\notin\cP_H]
 \le\delta_{\mathrm{dec}}(\kappa)
\end{equation}
by the definition of good hashes (\cref{def:good-hashes}).
Purify the other input as $\ket\phi$, apply $U_{\Decode}$ to
$\ket{\Psi_H}\ket\phi$, and let $\Pi$ project onto
$\cP_H$. Since decoding does not change $y$, the decoded state's weight outside
$\Pi$ is at most $\delta_{\mathrm{dec}}(\kappa)$. The queries to $T$
and $\mathfrak{F}$ agree on $\Pi$, and the ideal query
$U_{\mathfrak{F}}\ket{y,b}=\ket{y,b\oplus \mathfrak{F}(y)}$ commutes with $U_{\Decode}$.
Reversing the decoding therefore gives
\[
 \|\EvalHash^T(\ket{\Psi_H}\ket\phi)
       -\ket{\Psi_H}U_{\mathfrak{F}}\ket\phi\|_2
 \le2\sqrt{\delta_{\mathrm{dec}}(\kappa)}.
\]

The one-call estimate extends to repeated use by replacing calls
in chronological order with $U_{\mathfrak{F}}$.
Before the differing call, the ideal operations have left the key in
$\ket{\Psi_H}$, so the same estimate applies. The remaining channels
cannot increase trace distance, so the triangle inequality gives the stated
$2q\sqrt{\delta_{\mathrm{dec}}(\kappa)}$ joint error.

\paragraph{The query algorithm and experiments.}
We construct the simulator by first answering recursive queries
through the encodings, then removing the unopened encodings. Fix a
polynomial bound $L$ on the encoded addresses and all protocol queries. Choose the constant number $m$ of groups from classical recursive
simulation (\cref{thm:split-recursive-preservation}).
For each $j$, partition the suffix predicates into consecutive groups
$J_{j,1},\ldots,J_{j,m}$ whose sizes differ by at most one, and put
\[
 P_{j,i}=\{(v,y)\in C_{\kappa_j}\times\zo^{\kappa_j}:H_{j,[\kappa_j]}(v)=y,\ H_{j,J_{j,i}}(v)=0\}
 \qquad(i\in[m]).
\]
These are the suffix-constraint sets (\cref{def:suffix-constraints}).
Conditional on the prefix predicates, the codeword-hash bound
(\cref{lem:code-hash}, item~1) gives independent groups with, for distinct $u_1,\ldots,u_t$,
\[
 \Pr[u_1,\ldots,u_t\in P_{j,i}]
 \le2^{-\kappa_jt/(8m)}\le2^{-c_0\lambda t}
 \qquad(t\le2^\lambda)
\]
with $c_0=1/(8m)$, since $\kappa_j\ge\lambda$.
For the component addresses, put
\[
 P_i=\{(j,u):j\in J,\ u\in P_{j,i}\}.
\]
Independence across $j$ gives the same inclusion bound for $P_i$,
and every such address has length at least $\kappa_j\ge\lambda$.
Since each encoding $T_j=T_{H_j,B_j,\mathfrak{F}_j}$ equals
$B_j$ outside $\bigcap_iP_{j,i}$, apply classical recursive simulation
(\cref{thm:split-recursive-preservation}) with fixed oracle $\ora_{\rm aux}$ and
$h_u(v)=(1+v)/3$, whose range and difference bound are
$1/3\le h_u(v)\le2/3$ and $|h_u(v)-h_u(w)|=|v-w|/3$.
Include $\Theta$, every original hash and noise function, and the
prescribed channels in its auxiliary variable $Z$.
The theorem supplies a deterministic polynomial-query algorithm
$\Answer_E^{(T_j)_j}$, where $E$ uses only the prefix predicates, the
first $m-1$ suffix groups, the original $B_j$ and independent coins.
The algorithm answers queries to every original encoding $T_j$ exactly.
Set $\eta=(E,(H_j,B_j)_j)$, which is independent of $\Theta$.
We now define the simulator using this sample and direct access to
the opened functions.

\begin{boxedalgorithm}{$\cS_\eta^{(\mathfrak{F}_j)_{j\in K_t}}(K_t,u)$}
Run $\Answer_E(u)$ reversibly, answering its encoding queries
$(j,v,y)$, with $(v,y)\in\cX_{\kappa_j}$, by
\[
 \widetilde T_{j,K_t}(v,y)=
 \begin{cases}
 \mathfrak{F}_j(y),&j\in K_t\text{ and }(v,y)\in\cP_{H_j},\\
 B_j(v,y),&\text{otherwise}.
 \end{cases}
\]
XOR the output into the answer bit and reverse the computation,
using the same function queries.
\end{boxedalgorithm}

Membership in $\cP_{H_j}$ and the values of $B_j$ are determined by
$\eta$, so this algorithm makes polynomially many queries only to
the opened $\mathfrak{F}_j$. On a classical request $j$, set
$K_t=K_t\cup\{j\}$, apply the prescribed response channel, and
supply $\ket{\Psi_{H_j}}$ using the same $H_j$ on repeated requests.

We analyze the following experiments, with the original sample,
input and response channels.
\begin{enumerate}[leftmargin=*,itemsep=3pt]
 \item In $\mathsf H_0$, answer oracle queries with $\ora$.
 \item In $\mathsf H_1$, answer them by coherent evaluation of
 $\Answer_E^{(T_j)_j}$.
 \item In $\mathsf H_2$, answer them with $\cS_\eta(K_t,\cdot)$.
\end{enumerate}
Write $\mathsf H_b$ also for the averaged output density matrix,
including $\ket\Theta\bra\Theta$. The recursive simulation theorem
preserves the full original sampling and gives
\[
 \trdist{\mathsf H_0}{\mathsf H_1}\le2^{-\Omega(\lambda)}.
\]
The simulator in $\mathsf H_2$ makes no original $N,A$ queries.

\paragraph{Removing unopened encodings.}
\label{sec:adaptive-opening-simulation}\label{sec:adaptive-opening-proof}
It remains to justify replacing the unopened encodings by $B_j$.
For this purpose, condition on $\Theta$, the prefix predicates, all $B_j$, the first $m-1$
suffix groups, the coins defining $E$, and the opened hashes and request
history. In $\mathsf H_2$, the input and response channels use only
opened hashes, and every queried $\mathfrak{F}_j$ is contained in $\Theta$.
Conditional on the prefix predicates, induction on the classical requests
therefore leaves each unopened suffix $H_{j,J_{j,m}}$ with its original
distribution, independent of the remaining conditioned information.
The conditioned information determines the query density matrix.
If $w_t(j,u)$ is the weight of an internal query on address $(j,u)$,
the conditional inclusion bound for $P_{j,m}$ therefore gives
\begin{equation}\label{claim:unopened-query-weight}
 \bbE\sum_{j\notin K_t,u}w_t(j,u)\indicator[u\in P_{j,m}]
 \le2^{-c_0\lambda}\bbE\sum_{j\notin K_t,u}w_t(j,u)
 \le2^{-c_0\lambda}.
\end{equation}

This total query-weight bound controls the whole protocol. Replace
the internal queries of $\mathsf H_1$ by those of $\mathsf H_2$
in chronological order. Use $\mathsf H_2$ before the
differing query and the same $\mathsf H_1$ continuation afterward,
including future requests. The queried functions
differ only on unopened last groups, so the quantum query bound
(\cref{lem:query-hybrid}) and the estimate above bound each change
by $2\,2^{-c_0\lambda/2}$, including all quantum systems.
There are polynomially many internal queries, and their common
continuation cannot increase trace distance, so
\begin{equation}\label{eq:common-simulation-error}
 \trdist{\mathsf H_0}{\mathsf H_2}
 \le2^{-\Omega(\lambda)}
   +\poly(\lambda)2^{-c_0\lambda/2}
 =2^{-\Omega(\lambda)}.
\end{equation}
This proves the joint-output bound (\cref{eq:encoded-key-simulation}). The estimates are
uniform over the fixed choices for fixed query bounds, and each
party uses its known $K_t$ for local query substitutions, so the
simulation adds no communication.

\paperpart{Quantum Cryptography with $\PTIME=\QCMA$}
\label{part:cryptography}
\section{Reusable Keys for Coset Evaluation}
\label{sec:full-crypto-oracle}

We construct reusable quantum keys for coset evaluation and prove their
correctness (\cref{lem:function-key-correctness}), using separate key
families for PRF and encryption.

For a subspace $A\le\bbF_2^d$ and shifts $a,b\in\bbF_2^d$, write
\begin{equation}\label{eq:coset-state-definition}
 \ket{A_{a,b}}=|A|^{-1/2}\sum_{z\in A}(-1)^{\langle b,z\rangle}\ket{z+a}.
\end{equation}
For Boolean functions $f_0,f_1$, define the two-basis oracle
\begin{equation}\label{eq:two-basis-oracle}
 C_{A,a,b,f_0,f_1}(x,c,z)=
 \begin{cases}
 f_0(x),&c=0,\ z\in A+a,\\
 f_1(x),&c=1,\ z\in A^\perp+b,\\
 \bot,&\text{otherwise}.
 \end{cases}
\end{equation}
Its Boolean encoding is $\mathfrak{F}_C(x,c,z,j)=\operatorname{enc}(C(x,c,z))_j$ for $j\in\zo$,
where $\operatorname{enc}(\bot)=00$, $\operatorname{enc}(0)=10$ and
$\operatorname{enc}(1)=11$, so $\mathsf{val}(C(x,c,z))=\mathfrak{F}_C(x,c,z,1)$.

Throughout the constructions, put
\[
 \kappa=(n+4)^3,\qquad \kappa_M=(5a_\kappa+7n+20)^3,
\]
where $a_\kappa$ is the predicate-preimage state length, and use the
fixed prefix predicates of \cref{lem:code-hash}(2). Each oracle address
specifies $n$, a component listed in that oracle's definition, and
an input to that function, and the oracle returns its value. These addresses
have fixed efficient injective encodings, padded to a common strictly
increasing polynomial length $L_0(n)$, with zero on unused inputs.

The following definition specifies the key and its public encoding
for any fixed functions and cosets. The PRF and encryption sections
will give their sampling.

\begin{definition}[Functional keys and their encoding]
\label{def:functional-keys}\label{def:direct-key-oracle}\label{def:key-oracles}
\label{def:master-child-oracle}\label{def:coset-oracle-encoding}
For $s\in\zo^n$, let $F_{s,0},F_{s,1}:\zo^n\to\zo$,
$A_s\le\bbF_2^{4n}$ with $\dim A_s=2n$, and
$a_s,b_s\in\bbF_2^{4n}$. We define the functional keys by
\[
 \ket{K_s}=\ket{\Psi_{H_s}}\ket{\chi_s},\qquad
 \ket{\chi_s}=\ket{(A_s)_{a_s,b_s}},\qquad
 F_s=F_{s,0}\oplus F_{s,1},\quad
 C_s=C_{A_s,a_s,b_s,F_{s,0},F_{s,1}},
\]
where $H_s\in\cH_{\kappa_M}$ and
$B_s:\cX_{\kappa_M}\to\zo$ is a Boolean function. Their public
encodings and indexed
preparation target are
\begin{equation}\label{eq:shared-master-encoding}
 T_s=T_{H_s,B_s,\mathfrak{F}_{C_s}},
\end{equation}
\begin{equation}\label{eq:indexed-key-target}
 \ket{\Omega_K}=2^{-n/2}\sum_s\ket s\ket{K_s},\qquad
 n_K=5n+a_{\kappa_M},\qquad
 (R_K,S_K)\leftarrow\PrepTables(\ket{\Omega_K}).
\end{equation}
Here $T_{H,B,\mathfrak{F}}$ is the encoded function
(\cref{def:encoded-evaluation}) and $\PrepTables$ samples the
preparation oracles (\cref{def:public-preparation}).
\end{definition}

We use the preparation error bound $\epsilon_{\mathrm{prep}}(m)=2^{-m}$ and
the decoding failure bound $\delta_{\mathrm{dec}}(\kappa)=2^{1-\kappa}$.

\begin{lemma}[Functional-key correctness]\label{lem:function-key-correctness}
There are uniform \QPT{} algorithms $\genkey,\ceval$ with the
following guarantees. Fix the functions and keys $(F_s,T_s,\ket{K_s})_s$
above, and sample
$(R_K,S_K)\leftarrow\PrepTables(\ket{\Omega_K})$.
Writing $\rho_{\rm key}^{R_K,S_K}$
for the output density matrix of $\genkey^{R_K,S_K}(1^n)$, we have
\[
 \bbE_{R_K,S_K}\tfrac12\|\rho_{\rm key}^{R_K,S_K}
 -2^{-n}\sum_s\ket s\!\bra s\otimes\ket{K_s}\!\bra{K_s}\|_1
 \le\epsilon_{\mathrm{prep}}(n_K).
\]
For $H_s\in\Good_{\kappa_M}$, let $\rho_q,\sigma_q$ be the joint
outputs from $\ket{K_s}\!\bra{K_s}\otimes\sigma_{XBW}$ after $q$
calls to $\ceval$ or to
$\ket{x,b}\mapsto\ket{x,b\oplus F_s(x)}$, respectively, with the
same intervening operations on $XBW$. Then
\[
 \tfrac12\|\rho_q-\sigma_q\|_1\le4q\sqrt{\delta_{\mathrm{dec}}(\kappa_M)}.
\]
\end{lemma}
The proof of \cref{lem:function-key-correctness} appears in
\cref{sec:shared-key-correctness-proof}.
Observe that ideal evaluation leaves the key unchanged, and the
preparation bound also holds with conditional side information:
for every density matrix $\eta$, $\|\eta\otimes X\|_1=\|X\|_1$.

\subsection{Proof of functional-key correctness}
\label{sec:shared-key-correctness-proof}
\label{sec:crypto-key-construction}\label{sec:master-child-keys}
\label{sec:coset-evaluation}\label{sec:coset-evaluation-proof}

We specify the algorithms, then prove functional-key correctness
(\cref{lem:function-key-correctness}).
\begin{definition}[Encoded coset evaluation]\label{def:encoded-coset-evaluation}
We define encoded coset evaluation $\EvalKey^{T_s}$ by the four steps of
$\ceval$ below, using $\EvalHash$ and the predicate-preimage state register to implement
each value-bit query through $T_s$
(\cref{def:hash-evaluation}). Write $\EvalCoset^{C_s}$ for its direct
$C_s$ implementation.
\end{definition}

\begin{boxedalgorithm}{Functional-key generation and evaluation}
\noindent\underline{$\genkey(1^n)$}\par
Run $\Prepare^{R_K,S_K}(1^{n_K})$, measure the serial register, and
return its outcome $s$ with the remaining register $\regis{sk}$.

\smallskip
\noindent\underline{$\ceval(s,\regis{sk},x,b)$}\par
Write $\regis{sk}$ as a predicate-preimage state register and a coset register, and implement
each value-bit query below by $\EvalHash^{T_s}$.
Using the coset register as the query register $z$:
\begin{enumerate}[nosep]
 \item XOR $\mathsf{val}(C_s(x,0,z))$ into $b$.
 \item Apply $\Hadamard^{\otimes4n}$ to the coset register.
 \item XOR $\mathsf{val}(C_s(x,1,z))$ into $b$.
 \item Apply $\Hadamard^{\otimes4n}$ again and return all registers.
\end{enumerate}
If $b$ is omitted, initialize it to zero and measure the answer.
\end{boxedalgorithm}

\paragraph{Key generation.}
Let $\tau^{R_K,S_K}$ be the output of $\Prepare^{R_K,S_K}(1^{n_K})$
and let $\mathcal M$ measure its serial register, so the ideal output is
\[
 \mathcal M(\ket{\Omega_K}\!\bra{\Omega_K})
 =2^{-n}\sum_s\ket s\!\bra s\otimes\ket{K_s}\!\bra{K_s}.
\]
For these fixed keys, the target $\ket{\Omega_K}$ is fixed and
$(R_K,S_K)\leftarrow\PrepTables(\ket{\Omega_K})$, so the state-preparation
bound (\cref{lem:public-preparation}) gives error at most
$\epsilon_{\mathrm{prep}}(n_K)$. Since measurement cannot increase
trace distance, we obtain
\[
 \bbE_{R_K,S_K}\trdist{\mathcal M(\tau^{R_K,S_K})}
                    {\mathcal M(\ket{\Omega_K}\!\bra{\Omega_K})}
 \le\bbE_{R_K,S_K}\trdist{\tau^{R_K,S_K}}
                              {\ket{\Omega_K}\!\bra{\Omega_K}}
 \le\epsilon_{\mathrm{prep}}(n_K).
\]
The same estimate holds with conditional side information because
$\|\sigma_W\otimes X\|_1=\|X\|_1$ for every conditional
density matrix $\sigma_W$. Averaging these
conditional distances also gives the distance when the oracle and
all sampled functions and hashes are included as classical outputs.

\paragraph{Function evaluation.}
Character orthogonality gives
\begin{equation}\label{eq:coset-hadamard-transform}
 \Hadamard^{\otimes4n}\ket{\chi_s}
 =(-1)^{\langle a_s,b_s\rangle}\ket{(A_s^\perp)_{b_s,a_s}}.
\end{equation}
Let $U_{s,c}$ XOR $\mathsf{val}(C_s(x,c,z))$ into the answer register
and write $H=\Hadamard^{\otimes4n}$ on the coset register. The first
query returns $F_{s,0}(x)$ throughout the support of $\ket{\chi_s}$,
and the second returns $F_{s,1}(x)$ after the Hadamard transform, so
\[
 H U_{s,1} H U_{s,0}\ket{\chi_s}\ket x\ket b
 =\ket{\chi_s}\ket x\ket{b\oplus F_s(x)}.
\]
This proves exact correctness with direct queries to $C_s$.
For the encoded implementation, apply the repeated-evaluation bound
(\cref{lem:coherent-evaluation}) to its $2q$ calls to $\EvalHash^{T_s}$,
with ideal function $\mathfrak{F}_{C_s}$. Between these calls, the
Hadamard transforms and the prescribed channels on $XBW$ act outside
the predicate-preimage state register, so the theorem applies to the full joint input and gives
\begin{equation}\label{eq:repeated-coset-evaluation}
 \trdist{\rho_q}{\sigma_q}
 \le 2(2q)\sqrt{\delta_{\mathrm{dec}}(\kappa_M)}.
\end{equation}
The direct coset identity identifies $\sigma_q$ with the claimed
ideal evaluations of $F_s$, and the bound includes the returned key
and the auxiliary system $W$.

\clearpage\section{Copy-Protected Pseudorandom Functions}
\label{sec:function-keys}\label{sec:prf-construction}

We construct a copy-protected PRF from reusable functional keys
(\cref{thm:prf-security}), with the following public oracle.

\begin{definition}[PRF oracle]\label{def:prf-oracle}
Independently at every parameter $n$, and independently for each
$s\in\zo^n$, sample
\[
 \begin{gathered}
 F_{s,0},F_{s,1}\leftarrow\{f:\zo^n\to\zo\},\qquad
 A_s\leftarrow\{A\le\bbF_2^{4n}:\dim A=2n\},\\
 a_s,b_s\leftarrow\bbF_2^{4n},\qquad
 H_s\leftarrow\cH_{\kappa_M},\qquad
 B_s\leftarrow\zo^{\cX_{\kappa_M}}.
 \end{gathered}
\]
For the resulting keys and encodings
$\ket{K_s}=\ket{\Psi_{H_s}}\ket{(A_s)_{a_s,b_s}}$ and
$T_s=T_{H_s,B_s,\mathfrak{F}_{C_s}}$ (\cref{def:functional-keys}), sample
\[
 (R_K,S_K)\leftarrow\PrepTables\left(2^{-n/2}\sum_s\ket s\ket{K_s}\right).
\]
We define the PRF base oracle by
$\ora_{\mathsf{prf},n}=((T_s)_s,R_K,S_K)$, and let
$\ora_{\mathsf{prf}}(n,u)=\ora_{\mathsf{prf},n}(u)$,
where $u$ specifies a component above and its input.
\end{definition}

Use the full oracle
$\ora=(\orabase,N,A)$ with
$\orabase=(\ora_{\mathsf{prf}},\ora_{\rm rest})$, where
$\ora_{\rm rest}:\zo^*\to\zo$ is sampled independently on disjoint addresses, and
$N,A$ are sampled once over the entire base
(\cref{def:recursive-qcma}). All oracle access is coherent.
At $n=\lambda$, define
\begin{equation}\label{eq:prf-family}
 F_s^{\rm prf}(x)=F_s(x)=F_{s,0}(x)\oplus F_{s,1}(x),\qquad
 \ket{K_s^{\rm prf}}=\ket{K_s}.
\end{equation}
The public serial $s$ is the PRF function index (\cref{predef:prf}).

\subsection{Construction}\label{sec:prf-algorithms}

The PRF uses its functional key to evaluate $F_s$, with all calls
below using the PRF component of the oracle.

\begin{mdframed}[skipabove=5pt,skipbelow=6pt,innertopmargin=5pt,innerbottommargin=5pt,nobreak=false]
\noindent\underline{$\PRFGen^{\ora}(1^\lambda)$}\par\nopagebreak[4]
\begin{enumerate}[nosep]
 \item Return $(s,\regis{sk})\leftarrow\genkey(1^\lambda)$.
\end{enumerate}

\medskip
\noindent\underline{$\PRFEval^{\ora}(s,\regis{sk},x,b)$}\par\nopagebreak[4]
\begin{enumerate}[nosep]
 \item Run $\ceval(s,\regis{sk},x,b)$ with the caller's response register.
 \item Return all registers.
\end{enumerate}
\end{mdframed}

The following theorem gives correctness, pseudorandomness and
copy-protection for this construction.

\begin{theorem}[PRF security]\label{thm:prf-security}
For the PRF oracle $\ora$ above and every independent base component
$\ora_{\rm rest}$, the uniform \QPT{} construction $\prf^\ora$
satisfies the following guarantees, averaged over $\ora$:
\begin{itemize}
 \item \emph{Correctness.} Polynomially many coherent evaluations have
 negligible joint error, returning the key for reuse.
 \item \emph{Security.} The pseudorandomness and copy-protection
 guarantees in \cref{eq:prf-privacy-game,eq:function-piracy-game}
 hold with error $2^{-\Omega(\lambda)}$ against every adversary
 with polynomial query count and length and unrestricted local computation.
\end{itemize}
\end{theorem}
The proof of \cref{thm:prf-security} appears in
\cref{sec:prf-security-proof}.

\subsection{Proof of PRF security}\label{sec:prf-security-proof}

We now prove correctness, pseudorandomness and copy-protection in
\cref{thm:prf-security}.

\paragraph{Correctness.}
For $q$ coherent evaluations, compare the actual output with the ideal experiment
\[
 s\leftarrow\zo^\lambda,\qquad
 \regis{sk}\leftarrow\ket{K_s},\qquad
 \ket{x,b}\longmapsto\ket{x,b\oplus F_s(x)},
\]
which leaves $\regis{sk}$ unchanged after each query. Functional-key
correctness (\cref{lem:function-key-correctness}) applies to arbitrary intervening operations outside the key, so its joint
output, including the key, the sampled oracle and any entangled
auxiliary system, differs from the actual output by at most
\begin{equation}\label{eq:prf-correctness-error}
 \epsilon_{\mathrm{prep}}(n_K)+2^{-\kappa_M}+4q\sqrt{\delta_{\mathrm{dec}}(\kappa_M)}
 =\negl(\lambda).
\end{equation}
Here $2^{-\kappa_M}$ bounds the probability that $H_s$ is not good.

\paragraph{Pseudorandomness.}
Write $n=\lambda$ and fix the privacy adversary $\cA$. All bounds
below are uniform in the independently sampled additional base
$\ora_{\rm rest}$, which is included among the fixed choices in both security arguments.
In both branches $b\in\zo$, sample an independent uniform function
$R:\zo^n\to\zo$, and define $Q_0=F_s^{\rm prf}$ and $Q_1=R$. In experiment $\mathsf P_{i,b}$, write
$\ora^i=(\orabase^i,N^i,A^i)$ for the public oracle supplied to
$\cA$. Every experiment returns the bit output by $\cA(s)$,
with coherent challenge access to $Q_b$:
\begin{enumerate}[label=$\mathsf P_{\arabic*,b}$,start=0,itemsep=3pt]
 \item Sample $\ora^0=\ora$ and
 $(s,\regis{sk})\leftarrow\PRFGen^{\ora^0}(1^n)$,
 discard $\regis{sk}$, and output $\cA^{\ora^0,Q_b}(s)$.
 \item Sample $\ora^1=\ora$ and $s\leftarrow\zo^n$ independently,
 and output $\cA^{\ora^1,Q_b}(s)$.
 \item Sample all functions and hashes as above, but
 replace the key-preparation target by
 \[
  \ket{\Omega_K^0}=2^{-n/2}
  \left(\sum_{t\ne s}\ket t\ket{K_t}+\ket s\ket{0^{n_K-n}}\right).
 \]
 Sample $(R_K^0,S_K^0)\leftarrow\PrepTables(\ket{\Omega_K^0})$ and set
 \[
  \orabase^2=\orabase[(R_K,S_K)\leftarrow(R_K^0,S_K^0)],
  \qquad \ora^2=(\orabase^2,N^2,A^2),
 \]
 where $N^2,A^2$ have the recursive distribution conditional on
 $\orabase^2$. Output $\cA^{\ora^2,Q_b}(s)$.
 \item To construct one simulator for both branches, sample an independent
 analysis bit $\beta\leftarrow\zo$. Let $\orabase^{2,-s}$ denote
$\orabase^2$ with $T_s$ omitted, and put
\[
 \begin{aligned}
 \Theta_{\rm priv}&=(F_{s,0},F_{s,1},R,\beta),\\
 G_{\rm priv}&=(A_s,a_s,b_s,\orabase^{2,-s}).
 \end{aligned}
\]
Apply encoded-function simulation
(\cref{thm:crypto-common-simulation}) once to the protocol with challenge
oracle $Q_\beta$, using $J=\{s\}$ and $K=\varnothing$.
For each fixed $G_{\rm priv}$ this supplies $\cS_{G_{\rm priv}}$ and
one classical sample $\eta_{\rm priv}$ independent of $\Theta_{\rm priv}$.
In $\mathsf P_{3,b}$, condition on $\beta=b$ and output
$\cS_{G_{\rm priv}}^{Q_b}(s,\eta_{\rm priv})$.
 It runs $\cA$, forwards challenge queries to $Q_b$, and computes
 every $\ora^2$ answer by the coherent algorithm $\cS_{\eta_{\rm priv}}(\varnothing,\cdot)$
 with $T_s$ replaced by the original $B_s$. These answers use
 $\eta_{\rm priv}$ and no challenge-function query.
\end{enumerate}
Only the adversary's challenge queries are forwarded to $Q_b$ in
the last experiment.
Write $p_{i,b}=\Pr[\mathsf P_{i,b}=1]$.

Functional-key correctness (\cref{lem:function-key-correctness})
allows us to replace the generated serial by a uniform one, giving
$|p_{0,b}-p_{1,b}|\le\epsilon_{\mathrm{prep}}(n_K)$.
The following claim removes the remaining dependence of the public
oracle on the challenge function.
\begin{claim}\label{clm:prf-privacy-hybrids}
For $b\in\zo$, the privacy experiments satisfy
\[
 |p_{1,b}-p_{2,b}|+|p_{2,b}-p_{3,b}|\le2^{-\Omega(n)},
 \qquad p_{3,0}=p_{3,1}.
\]
\end{claim}
The proof of \cref{clm:prf-privacy-hybrids} appears in
\cref{sec:prf-oracle-hybrid-bounds}.

The two final experiments have equal acceptance probability by the
claim, so the triangle inequality gives
\[
 |p_{0,0}-p_{0,1}|
 \le\sum_{b=0}^1\sum_{i=0}^2|p_{i,b}-p_{i+1,b}|
 \le2\epsilon_{\mathrm{prep}}(n_K)+2^{-\Omega(n)}=2^{-\Omega(n)}.
\]

\paragraph{Copy-protection.}
Fix $(\cA_0,\cA_1,\cA_2)$. In $\mathsf V_0$, sample
$\ora^0=\ora$ and $(s,\regis{sk})\leftarrow\PRFGen^{\ora^0}(1^n)$,
then run
\[
 (\regi_1,\regi_2)\leftarrow\cA_0^{\ora^0}(s,\regis{sk}),\qquad
 u_1,u_2\leftarrow\zo^n,\qquad
 \widehat f_i\leftarrow\cA_i^{\ora^0}(u_i,\regi_i).
\]
The algorithms $\cA_1,\cA_2$ act on their possibly entangled
registers without communication. In $\mathsf V_1$, sample
$\ora^1=\ora$ and uniform $s$ independently, substitute
$\regis{sk}=\ket{K_s}$, and make the same calls with $\ora^1$.
Both experiments return
\[
 \indicator[\widehat f_1=F_s(u_1)\ \land\
             \widehat f_2=F_s(u_2)].
\]
Key-generation correctness gives
$|\Pr[\mathsf V_0=1]-\Pr[\mathsf V_1=1]|\le\epsilon_{\mathrm{prep}}(n_K)$.

We define the next experiments using an equivalent sampling for
$\mathsf V_1$. Privately sample the inputs before invoking $\cA_0$,
without giving them to any party, and sample
\[
 \begin{aligned}
 &u_1,u_2\leftarrow\zo^n,
 \quad G_0,G_1\leftarrow\{\zo^n\to\zo\},\\
 &b_{x,c}\leftarrow\zo\quad(x\in\{u_1,u_2\},\ c\in\zo),\qquad
 F_{s,c}=G_c[x\leftarrow b_{x,c}:x\in\{u_1,u_2\}].
 \end{aligned}
\]
There is one pair of bits per distinct input, so $F_{s,0},F_{s,1}$
are independent uniform functions even when $u_1=u_2$.
Sample every other component of $\ora$ with these functions.
The next two experiments use this sampling, the original issued
key and the same winning predicate:
\begin{enumerate}[label=$\mathsf V_{\arabic*}$,start=2,itemsep=3pt]
 \item Sample $(R_K^0,S_K^0)\leftarrow\PrepTables(\ket{\Omega_K^0})$,
 set $\orabase^2=\orabase[(R_K,S_K)\leftarrow(R_K^0,S_K^0)]$,
 and sample $(N^2,A^2)$ conditional on $\orabase^2$.
 Run
 \[
 (\regi_1,\regi_2)\leftarrow\cA_0^{(\orabase^2,N^2,A^2)}(s,\ket{K_s}),
 \qquad \widehat f_i\leftarrow\cA_i^{(\orabase^2,N^2,A^2)}(u_i,\regi_i).
 \]
 \item Let $\orabase^{2,-s}$ again omit $T_s$, and put
\[
 \Theta_{\rm pir}=(A_s,a_s,b_s,u_1,u_2,(b_{x,c})_{x,c}),\qquad
 G_{\rm pir}=(G_0,G_1,\orabase^{2,-s}).
\]
Apply encoded-function simulation (\cref{thm:crypto-common-simulation}) to
 $(\cA_0,\cA_1,\cA_2)$ for each fixed $G_{\rm pir}$, with
 $J=K=\{s\}$ and an initial request for $s$, obtaining local algorithms $(\cS_0,\cS_1,\cS_2)$.
 Draw their one classical sample $\eta_{\rm pir}$ and run
 \[
 (\regi'_1,\regi'_2)\leftarrow\cS_0^{\mathfrak{F}_{C_s}}(s,\ket{\chi_s},\eta_{\rm pir}),
 \qquad \widehat f_i\leftarrow\cS_i^{\mathfrak{F}_{C_s}}(u_i,\regi'_i).
 \]
 Here $\regi'_i$ includes the quantum state and classical information
 assigned to algorithm $i$. Each $\cS_i$ answers every
 $(\orabase^2,N^2,A^2)$ query by $\cS_{\eta_{\rm pir}}(\{s\},\cdot)$, using direct
 queries to $\mathfrak{F}_{C_s}$.
 The algorithms $\cS_1,\cS_2$ do not communicate.
\end{enumerate}
We use the following bound to reduce the original adversary to
one with direct coset-oracle access.
\begin{claim}\label{clm:prf-piracy-hybrids}
The piracy experiments satisfy
\[
 |\Pr[\mathsf V_1=1]-\Pr[\mathsf V_2=1]|
 +|\Pr[\mathsf V_2=1]-\Pr[\mathsf V_3=1]|
 \le2^{-\Omega(n)}.
\]
Moreover, $\mathsf V_3$ is a direct coset function-value experiment
(\cref{def:coset-experiments}) with $Q'=\poly(n)$ queries.
\end{claim}
The proof of \cref{clm:prf-piracy-hybrids} appears in
\cref{sec:prf-oracle-hybrid-bounds}.

Applying the coset security bound (\cref{lem:two-fresh-values})
and summing the three differences gives
\[
 \Pr[\mathsf V_0=1]
 \le\tfrac12+O((Q'+1)2^{-n/2})+\epsilon_{\mathrm{prep}}(n_K)+2^{-\Omega(n)}
 =\tfrac12+2^{-\Omega(n)}.
\]

\subsubsection{Bounds for the PRF hybrids}
\label{sec:prf-oracle-hybrid-bounds}

We now prove \cref{clm:prf-privacy-hybrids} and
\cref{clm:prf-piracy-hybrids}.
The privacy simulation opens no hash, whereas the piracy simulation
opens the issued hash before the adversary separates its registers.

\paragraph{Changing the key-preparation target.}
For both $\mathsf P_{1,b}\to\mathsf P_{2,b}$ and
$\mathsf V_1\to\mathsf V_2$, fix $s$, the functions $F_{t,c}$,
the hashes $H_t$, the coset descriptions $(A_t,a_t,b_t)$, and all
base components except $(R_K,S_K)$. Also fix $R$ in the privacy
experiment and $u_1,u_2$ in the piracy experiment. This determines the targets, the supplied inputs and
the challenge operations, including $Q_b$ in
the privacy experiment and the original $\ket{K_s},F_s,u_i$ in
the piracy experiment. Since
\[
 \|\ket{\Omega_K}-\ket{\Omega_K^0}\|_2\le2^{1-n/2},
\]
the target-change bound (\cref{thm:preparation-replacement}), with
dimension $n_K$ and recursive mean $h(v)=(1+v)/3$, applies to each
protocol. Averaging the fixed quantities gives
\[
 |p_{1,b}-p_{2,b}|\le2^{-\Omega(n)},\qquad
 |\Pr[\mathsf V_1=1]-\Pr[\mathsf V_2=1]|\le2^{-\Omega(n)}.
\]

\paragraph{Privacy.}
It remains to simulate $T_s$, the only base component still using
the challenge functions. Conditional on $G_{\rm priv}$, the variables $\Theta_{\rm priv},H_s,B_s$ are mutually
independent: the new target substitutes $\ket{0^{n_K-n}}$ for $\ket{K_s}$,
and all other primitive components were sampled independently. The single instance being simulated is
$T_s=T_{H_s,B_s,\mathfrak{F}_{C_s}}$, and no predicate-preimage state is supplied.
The challenge oracle depends on $\Theta_{\rm priv}$ but not $H_s$. Apply encoded-function simulation
(\cref{thm:crypto-common-simulation}) to the joint protocol, including
$\beta$ in its classical output. The averaged joint-output bound implies,
for either conditional branch,
\[
 |p_{2,b}-p_{3,b}|\le 2\cdot2^{-\Omega(n)}=2^{-\Omega(n)}.
\]
The one simulator sample and description are independent of
$(F_{s,0},F_{s,1},R,\beta)$ conditional on $G_{\rm priv}$.
Since the XOR of the two independent
functions is uniform, $\mathsf P_{3,0}$ and
$\mathsf P_{3,1}$ have the same distribution, completing
\cref{clm:prf-privacy-hybrids}.

\paragraph{Copy-protection.}
The target $\ket{\Omega_K^0}$ substitutes $\ket{0^{n_K-n}}$
for $\ket{K_s}$, and every other primitive base is independent, so $\Theta_{\rm pir},H_s,B_s$ are mutually independent
conditional on $G_{\rm pir}$. Moreover, $(G_{\rm pir},\Theta_{\rm pir})$ determines $\mathfrak{F}_{C_s}$, while the supplied
coset state depends only on $\Theta_{\rm pir}$. The serial $s$ is opened before separation and is known to all three
algorithms, so encoded-function simulation with $J=K=\{s\}$ gives
\[
 |\Pr[\mathsf V_2=1]-\Pr[\mathsf V_3=1]|\le2^{-\Omega(n)}.
\]
Its local simulations make $Q'=\poly(n)$ direct queries and use no
original $N,A$ oracle. After averaging
$\orabase^{2,-s}$, their classical sample and descriptions are
independent of $\Theta_{\rm pir}$ conditional on $(G_0,G_1)$. Hence $\mathsf V_3$ is the direct coset function-value experiment
(\cref{def:coset-experiments}), proving \cref{clm:prf-piracy-hybrids}.

\clearpage\section{Copy-Protected Public-Key Encryption}\label{sec:pke-construction}

We construct copy-protected encryption with classical public keys and
ciphertexts (\cref{thm:pke-security}). Its public oracle prepares
quantum decryption keys and graph states for encryption.

\begin{definition}[Encryption oracle]\label{def:pke-oracle}
Independently for every parameter $n$ and serial $s\in\zo^n$, sample
\[
 \begin{gathered}
 F_{s,0},F_{s,1}\leftarrow\{f:\zo^n\to\zo\},\qquad
 A_s\leftarrow\{A\le\bbF_2^{4n}:\dim A=2n\},\\
 a_s,b_s\leftarrow\bbF_2^{4n},\qquad
 H_s\leftarrow\cH_{\kappa_M},\qquad B_s\leftarrow\zo^{\cX_{\kappa_M}}.
 \end{gathered}
\]
Set $F_s=F_{s,0}\oplus F_{s,1}$ and
$C_s=C_{A_s,a_s,b_s,F_{s,0},F_{s,1}}$, with functional keys and encodings
\[
 \ket{K_s}=\ket{\Psi_{H_s}}\ket{(A_s)_{a_s,b_s}},\qquad
 T_s=T_{H_s,B_s,\mathfrak{F}_{C_s}}.
\]
Conditional on these variables, independently sample one indexed-key
preparation pair and one graph-preparation pair for each $s$:
\[
 (R_K,S_K)\leftarrow\PrepTables\left(2^{-n/2}\sum_s\ket s\ket{K_s}\right),
\]
\begin{gather}
 \ket{\Gamma_s}=2^{-n/2}\sum_{u\in\zo^n}\ket u\ket{F_s(u)}\ket{0^{4n-1}},
 \label{eq:pke-graph-state}\\
 (R_s^{\rm graph},S_s^{\rm graph})\leftarrow\PrepTables(\ket{\Gamma_s}).
 \label{eq:graph-preparation-target}
\end{gather}
We define the encryption base oracle by
\[
 \ora_{\mathsf{pke},n}=((T_s)_s,R_K,S_K,(R_s^{\rm graph},S_s^{\rm graph})_s),
\]
and let $\ora_{\mathsf{pke}}(n,u)=\ora_{\mathsf{pke},n}(u)$,
where $u$ specifies a component above and its input.
\end{definition}

Use the full oracle
$\ora=(\orabase,N,A)$ with
$\orabase=(\ora_{\mathsf{pke}},\ora_{\rm rest})$, where
$\ora_{\rm rest}:\zo^*\to\zo$ is sampled independently on disjoint addresses, and $N,A$ are sampled
once over the entire base (\cref{def:recursive-qcma}). The public
key is $s$, and measuring $\ket{\Gamma_s}$ produces a uniform input
and its pad without using the secret key $\ket{K_s}$.

\subsection{Construction}\label{sec:pke-algorithms}

Encryption samples an input and its pad from the graph state, while
decryption uses the functional key to recover that pad. All calls
below use the encryption component of the oracle.

\begin{mdframed}[skipabove=5pt,skipbelow=6pt,innertopmargin=5pt,innerbottommargin=5pt,nobreak=false]
\noindent\underline{$\PKEGen^{\ora}(1^\lambda)$}\par\nopagebreak[4]
\begin{enumerate}[nosep]
 \item Run $(s,\regis{sk})\leftarrow\genkey(1^\lambda)$.
 \item Return the public key $\pk=s$ and secret key $(s,\regis{sk})$.
\end{enumerate}

\medskip
\noindent\underline{$\PKEEnc^{\ora}(s,m)$}\par\nopagebreak[4]
\begin{enumerate}[nosep]
 \item Run $\Prepare^{R_s^{\rm graph},S_s^{\rm graph}}(1^{5\lambda})$.
 \item Measure the first $\lambda+1$ qubits to obtain $(u,b)$ and discard
       the remaining register.
 \item Return $(u,m\oplus b)$.
\end{enumerate}

\medskip
\noindent\begin{minipage}{\linewidth}
\noindent\underline{$\PKEDec^{\ora}((s,\regis{sk}),(u,c))$}\par\nopagebreak[4]
\begin{enumerate}[nosep]
 \item Compute $b\leftarrow\ceval(s,\regis{sk},u)$.
 \item Return $c\oplus b$ together with the updated key.
\end{enumerate}
\end{minipage}
\end{mdframed}

We now state correctness and security for this encryption scheme.

\begin{theorem}[Encryption security]\label{thm:pke-security}
For the encryption oracle $\ora$ above and every independent base
component $\ora_{\rm rest}$, the uniform \QPT{} construction
$\pke^\ora$ has classical public keys and ciphertexts and satisfies
the following guarantees, averaged over $\ora$:
\begin{itemize}
 \item \emph{Correctness.} Polynomially many encryptions and decryptions
 using one key have negligible joint error.
 \item \emph{Security.} Against every adversary with polynomial query
 count and length and unrestricted local computation, the scheme is
 chosen-plaintext secure and satisfies, for the independent ciphertext
 challenges of \cref{def:pke-piracy},
 \[
  \Pr_{\ora}[\Piracy^{\ora}_{\cA}(\lambda)=1]
  \le\tfrac12+2^{-\Omega(\lambda)}.
 \]
\end{itemize}
\end{theorem}
The proof of \cref{thm:pke-security} appears in
\cref{sec:pke-security-proof}.

\subsection{Proof of encryption security}\label{sec:pke-security-proof}

We now prove \cref{thm:pke-security} by reducing the encryption
experiment to the direct coset ciphertext experiment.

\paragraph{Correctness.}
The ideal generation, encryption and decryption operations are
\[
 \begin{aligned}
 &s\leftarrow\zo^\lambda,\qquad \regis{sk}\leftarrow\ket{K_s},\\
 &u\leftarrow\zo^\lambda,\qquad \ct=(u,m\oplus F_s(u)),\\
 &\widehat m=\ct_2\oplus F_s(\ct_1)=m.
 \end{aligned}
\]
Functional-key correctness (\cref{lem:function-key-correctness})
bounds the key error. For the graph preparations,
the state-preparation bound (\cref{lem:public-preparation}) gives
\[
 \bbE\max_s\trdist{\Prepare^{R_s^{\rm graph},S_s^{\rm graph}}(1^{5\lambda})}
 {\ket{\Gamma_s}\!\bra{\Gamma_s}}
 \le2^\lambda\epsilon_{\mathrm{prep}}(5\lambda)=2^{-4\lambda}.
\]
Applying these bounds sequentially shows that the joint output of $r$ encryptions and decryptions differs from the ideal
output by at most
\begin{equation}\label{eq:repeated-pke-correctness}
 \epsilon_{\mathrm{prep}}(n_K)+r2^{-4\lambda}+2^{-\kappa_M}
       +4r\sqrt{\delta_{\mathrm{dec}}(\kappa_M)}=\negl(\lambda).
\end{equation}
This includes the key, the sampled oracle and any entangled auxiliary
system, while $2^{-\kappa_M}$ bounds $\Pr[H_s\notin\Good_{\kappa_M}]$.

\paragraph{The piracy experiments.}
Fix $(\cA_0,\cA_1,\cA_2)$ and write $n=\lambda$. The estimates
are uniform in the independently sampled additional base
$\ora_{\rm rest}$, which is included in each $G$ below together with the other sampled
base components. In
$\mathsf H_0$, sample $\ora^0=\ora$ and run
\[
 \begin{aligned}
 (s,(s,\regis{sk}))&\leftarrow\PKEGen^{\ora^0}(1^n),\\
 ((m_{i,0},m_{i,1})_{i=1}^2,\regi_1,\regi_2)
   &\leftarrow\cA_0^{\ora^0}(s,\regis{sk}),\\
 \beta_1,\beta_2&\leftarrow\zo,\qquad
 \ct_i\leftarrow\PKEEnc^{\ora^0}(s,m_{i,\beta_i}),\\
 \widehat\beta_i&\leftarrow\cA_i^{\ora^0}(\ct_i,\regi_i).
 \end{aligned}
\]
The two challenge bits and encryption coins are independent, and
$\cA_1,\cA_2$ do not communicate. Return
$\indicator[\widehat\beta_1=\beta_1\land\widehat\beta_2=\beta_2]$.
To define $\mathsf H_1$, sample $s\leftarrow\zo^n$ and privately sample
\[
 \begin{aligned}
 &(u_1,u_2,\beta_1,\beta_2)\leftarrow(\zo^n)^2\times\zo^2,\\
 &G_0,G_1\leftarrow\{\zo^n\to\zo\},
       &&b_{x,c}\leftarrow\zo\quad(x\in\{u_1,u_2\},\ c\in\zo),\\
 &F_{s,c}=G_c[x\leftarrow b_{x,c}:x\in\{u_1,u_2\}],
       &&F_s=F_{s,0}\oplus F_{s,1}.
 \end{aligned}
\]
Use these functions to sample $\ora^1=(\orabase^1,N^1,A^1)$ and run
\[
 \begin{aligned}
 ((m_{i,0},m_{i,1})_{i=1}^2,\regi_1,\regi_2)
   &\leftarrow\cA_0^{\ora^1}(s,\ket{K_s}),\\
 \ct_i&=(u_i,m_{i,\beta_i}\oplus F_s(u_i)),\\
 \widehat\beta_i&\leftarrow\cA_i^{\ora^1}(\ct_i,\regi_i).
 \end{aligned}
\]
Return $\indicator[\widehat\beta_1=\beta_1\land
\widehat\beta_2=\beta_2]$. The algorithms $\cA_1,\cA_2$ run without communication and receive
$\ct_i$ only after $\cA_0$ has output its registers and message pairs.
Because there is one fresh bit pair per distinct $u_i$, this is the
ideal-key, ideal-encryption game even when $u_1=u_2$.

Define the next three experiments by changing only the public oracle
in this experiment. In each, the challenger uses the original
$\ket{K_s},F_s,u_i,\beta_i$ and the displayed winning predicate.
\begin{enumerate}[label=$\mathsf H_{\arabic*}$,start=2,itemsep=3pt]
 \item Replace the key target by
 \[
  \ket{\Omega_K^0}=2^{-n/2}
  \left(\sum_{t\ne s}\ket t\ket{K_t}+\ket s\ket{0^{n_K-n}}\right),
 \]
 sample $(R_K^0,S_K^0)\leftarrow\PrepTables(\ket{\Omega_K^0})$,
 set $\orabase^2=\orabase^1[(R_K,S_K)\leftarrow(R_K^0,S_K^0)]$,
 and sample $N^2,A^2$ conditional on $\orabase^2$.
 Run the displayed $\cA_0,\cA_1,\cA_2$ calls with
 $\ora^2=(\orabase^2,N^2,A^2)$ instead of $\ora^1$.
 \item In addition, replace only the $s$ graph target by
 \[
  \ket{\Gamma_s^0}=2^{-n/2}\sum_u\ket u
        \ket{G_0(u)\oplus G_1(u)}\ket{0^{4n-1}}.
 \]
 Sample $(R_s^0,S_s^0)\leftarrow\PrepTables(\ket{\Gamma_s^0})$,
 set
 \[
 \orabase^3=\orabase^2[(R_s^{\rm graph},S_s^{\rm graph})
                                  \leftarrow(R_s^0,S_s^0)],
 \]
 and sample $N^3,A^3$ conditional on $\orabase^3$.
 Run the same calls with $\ora^3=(\orabase^3,N^3,A^3)$.
 \item Let $\orabase^{3,-s}$ denote $\orabase^3$ with $T_s$ omitted, and set
\[
 \begin{aligned}
 \Theta&=(A_s,a_s,b_s,u_1,u_2,(b_{x,c})_{x,c},\beta_1,\beta_2),\\
 G&=(G_0,G_1,\orabase^{3,-s}).
 \end{aligned}
\]
Apply encoded-function simulation (\cref{thm:crypto-common-simulation}) to
 $(\cA_0,\cA_1,\cA_2)$ for each fixed $G$, with
 $J=K=\{s\}$ and an initial request for $s$, obtaining $(\cS_0,\cS_1,\cS_2)$. Draw their one classical sample $\eta_G$ and run
 \[
 \begin{aligned}
 ((m_{i,0},m_{i,1})_{i=1}^2,\regi'_1,\regi'_2)
 &\leftarrow\cS_0^{\mathfrak{F}_{C_s}}(s,\ket{\chi_s},\eta_G),\\
 \ct_i&=(u_i,m_{i,\beta_i}\oplus F_s(u_i)),\\
 \widehat\beta_i&\leftarrow\cS_i^{\mathfrak{F}_{C_s}}(\ct_i,\regi'_i).
 \end{aligned}
 \]
 Each $\regi'_i$ includes that algorithm's quantum state and classical
 information. The algorithms $\cS_1,\cS_2$ do not communicate.
\end{enumerate}
Write $p_i=\Pr[\mathsf H_i=1]$. Generation and two graph preparations,
with the same full oracle supplied to the adversary, give
\[
 |p_0-p_1|\le\epsilon_{\mathrm{prep}}(n_K)+2\cdot2^{-4n}
\]
by functional-key correctness and the graph-preparation bound above.

The following claim bounds the three changes to the public oracle.
\begin{claim}\label{clm:pke-oracle-hybrids}
For the experiments just defined,
\[
 |p_1-p_2|+|p_2-p_3|+|p_3-p_4|\le2^{-\Omega(n)}.
\]
In $\mathsf H_4$, the algorithms make $Q'=\poly(n)$ queries to
$\mathfrak{F}_{C_s}$, and their descriptions and classical sample
are independent of $\Theta$ conditional on $(G_0,G_1)$.
\end{claim}
The proof of \cref{clm:pke-oracle-hybrids} appears in
\cref{sec:pke-oracle-hybrid-bounds}.

We can now apply the direct coset bound. The final algorithms have
the coset state $\ket{\chi_s}$, the original ciphertext rule and
independent challenge bits $\beta_1,\beta_2$, so \cref{clm:pke-oracle-hybrids}
makes $\mathsf H_4$ a direct coset ciphertext experiment
(\cref{def:coset-experiments}). Its security bound
(\cref{lem:two-fresh-values}) gives
\[
 p_4\le\tfrac12+O((Q'+1)2^{-n/2}).
\]
Summing the four adjacent differences proves copy-protection:
\[
 p_0\le p_4+\sum_{i=0}^3|p_i-p_{i+1}|
 \le\tfrac12+O((Q'+1)2^{-n/2})+\epsilon_{\mathrm{prep}}(n_K)
       +2^{1-4n}+2^{-\Omega(n)}
 =\tfrac12+2^{-\Omega(n)}.
\]

\paragraph{Chosen-plaintext security.}
Given a CPA adversary $\mathcal B=(\mathcal B_0,\mathcal B_1)$,
define the following three algorithms, all with oracle $\ora$:
\[
 \begin{aligned}
 \cA_0(s,\regis{sk}):\quad
 &(m_0,m_1,\regi)\leftarrow\mathcal B_0(s),\\
 &\text{return }((m_0,m_1),(0,1),\regi,(s,\regis{sk})),\\
 \cA_1(\ct_1,\regi):\quad
 &\text{return }\mathcal B_1(\ct_1,\regi),\\
 \cA_2(\ct_2,(s,\regis{sk})):\quad
 &\text{return the plaintext from }\PKEDec((s,\regis{sk}),\ct_2).
 \end{aligned}
\]
Algorithm $\cA_1$ has exactly the CPA input distribution, while
$\cA_2$ decrypts an independent encryption of $\beta_2$ and fails
with probability at most $\negl(n)$ by correctness. Hence
\[
 \Pr[\mathsf{CPA}_{\mathcal B}^{\ora}(n)=1]
 \le\Pr[\mathsf H_0=1]+\negl(n)\le\tfrac12+\negl(n).
\]
Applying this construction also to the complement of
$\mathcal B_1$'s output bounds the absolute CPA advantage.

\subsubsection{Bounds for the encryption hybrids}
\label{sec:pke-oracle-hybrid-bounds}

We now prove \cref{clm:pke-oracle-hybrids}, keeping the issued key,
ciphertext maps and winning predicate fixed while changing the public oracle.

\paragraph{Changing the preparation targets.}
The two changes affect at most one key branch and two graph values,
so
\[
 \|\ket{\Omega_K}-\ket{\Omega_K^0}\|_2\le2^{1-n/2},\qquad
 \|\ket{\Gamma_s}-\ket{\Gamma_s^0}\|_2\le2\sqrt2\,2^{-n/2}.
\]
For the first change, fix $s,u_1,u_2,\beta_1,\beta_2$, all original
functions and hashes, and every preparation pair
except $(R_K,S_K)$. These choices determine both key targets, the
issued density matrix $\ket{K_s}\!\bra{K_s}$ and the maps producing
$\ct_i$ from the message pairs output by $\cA_0$. Apply the target-change bound (\cref{thm:preparation-replacement})
to these targets and the original three algorithms, with dimension
$n_K$ and recursive mean $h(v)=(1+v)/3$, to obtain
\[
 |p_1-p_2|\le2^{-\Omega(n)}.
\]
For the graph change, fix the first changed pair
$(R_K^0,S_K^0)$ as well, but sample the $s$ graph pair anew.
The same theorem now applies in dimension $5n$ and gives
\[
 |p_2-p_3|\le2^{-\Omega(n)}.
\]
Both estimates average over the fixed choices and use the original
ciphertext rule, including independent $\beta_1,\beta_2$.

\paragraph{Simulating the encoded functions.}
Since $\ket{\Omega_K^0}$ replaces $\ket{K_s}$ by
$\ket{0^{n_K-n}}$ and $\ket{\Gamma_s^0}$ uses only $G_0,G_1$, while all other primitive bases are independent, the variables $\Theta,H_s,B_s$ are mutually independent
conditional on $G$. Also, $(G,\Theta)$ determines $\mathfrak{F}_{C_s}$,
and $\ket{\chi_s}$ depends only on $\Theta$. The index $s$ is opened before the algorithms separate and remains known
to each of them. These are the hypotheses of encoded-function simulation (\cref{thm:crypto-common-simulation})
with $J=K=\{s\}$.

In $\mathsf H_4$, each $\cS_i$ substitutes the coherent algorithm $\cS_{\eta_G}(\{s\},\cdot)$ for every $\ora^3$ query of $\cA_i$,
using the one sample $\eta_G$ and direct queries to the Boolean
encoding $\mathfrak{F}_{C_s}$ of $C_s=C_{A_s,a_s,b_s,F_{s,0},F_{s,1}}$. All other answers and the master predicate-preimage state are computed from
$\eta_G$. Thus $\cS_i$ uses its assigned register $\regi'_i$ and these
same sampled functions, with no original $N,A$ access. The theorem applies to
the original ciphertext maps and winning predicate, and therefore
\[
 |p_3-p_4|\le2^{-\Omega(n)}.
\]
After averaging $\orabase^{3,-s}$, the simulator's classical sample
and descriptions are independent of $\Theta$ conditional on
$(G_0,G_1)$, and its number of direct queries is $Q'=\poly(n)$.
Together with the three displayed error bounds, this proves the claim.

\clearpage\section{Signatures with Quantum Keys}
\label{sec:signatures}

We construct two adaptively secure signature schemes with quantum signing
keys (\cref{thm:signature-security}): $\CSig$ has classical verification
keys and quantum signatures, whereas $\QSig$ has quantum verification
keys and classical signatures.

\subsection{Constructions}\label{sec:signature-algorithms}

Both schemes use one signature-specific family, sampled independently
of the PRF and encryption components. Put
\begin{equation}\label{eq:signature-leveraging-parameters}
 n=\lambda,\qquad m\in\zo^n,
\end{equation}
and use
\begin{equation}\label{eq:signature-encoding-parameters}
 \kappa=(n+4)^3,\qquad r=3n,\qquad
 \kappa_M=(5a_\kappa+7n+20)^3.
\end{equation}
For a hash $H=(H_i)_i$, let its preimage enumeration be
\begin{equation}\label{eq:hash-preimage-enumeration}
 E_H=\operatorname{enc}\bigl((|H_i^{-1}(1)|,\pi_i,\pi_i^{-1})_i\bigr),
\end{equation}
where $\pi_i$ lists $H_i^{-1}(1)$ followed by $H_i^{-1}(0)$,
each in canonical order. This function permits coherent preparation
of the predicate-preimage state, as shown in \cref{app:signature-correctness}.

\begin{definition}[Signature oracle]\label{def:signature-oracle}
\label{def:shared-signing-payload}
Independently for every parameter $n$ and serial $s\in\zo^n$, sample
\[
 \begin{gathered}
 H_s\leftarrow\cH_{\kappa_M},\quad B_s\leftarrow\zo^{\cX_{\kappa_M}},\quad
 F_s^{\rm sig}\leftarrow\{f:\zo^n\to\zo^n\},\\
 H_{s,t}\leftarrow\cH_\kappa,\quad B_{s,t}\leftarrow\zo^{\cX_\kappa}
 \quad\bigl(t\in\{(\mathsf C,m):m\in\zo^n\}\cup\{\mathsf V\}\bigr).
 \end{gathered}
\]
Write $H_{s,m}=H_{s,(\mathsf C,m)}$, $H_s^V=H_{s,\mathsf V}$,
and similarly for $B$. Define the signing and verification functions by
\begin{equation}\label{eq:shared-signing-payload}
 \begin{aligned}
 P_s(0,m,i)&=F_s^{\rm sig}(m)_i,&
 P_s(1,t,u)&=E_{H_{s,t}}(u),\\
 G_s(m,z)&=\indicator[z=F_s^{\rm sig}(m)].
 \end{aligned}
\end{equation}
Their encoded functions and the master signing state are
\begin{equation}\label{eq:signature-master-encoding}
 T_s=T_{H_s,B_s,P_s},\qquad \ket{K_s^M}=\ket{\Psi_{H_s}},
\end{equation}
\begin{equation}\label{eq:signature-verification-tables}
 T_{s,m}=T_{H_{s,m},B_{s,m},1},\qquad
 T_s^V=T_{H_s^V,B_s^V,G_s}.
\end{equation}
Conditional on these samples, define the master preparation target
and sample its preparation pair:
\[
 \ket{\Omega_M}=2^{-n/2}\sum_s\ket s\ket{\Psi_{H_s}},\qquad
 n_M=n+a_{\kappa_M},\qquad
 (R_M,S_M)\leftarrow\PrepTables(\ket{\Omega_M}).
\]
We define the signature base oracle by
\[
 \ora_{\mathsf{sig},n}=((T_s)_s,(T_{s,m})_{s,m},(T_s^V)_s,R_M,S_M),
\]
and let $\ora_{\mathsf{sig}}(n,u)=\ora_{\mathsf{sig},n}(u)$,
where $u$ specifies a component above and its input.
\end{definition}

Use $\ora=(\orabase,N,A)$ with
$\orabase=(\ora_{\mathsf{sig}},\ora_{\rm rest})$, where
$\ora_{\rm rest}:\zo^*\to\zo$ is sampled independently on disjoint addresses, and one $N,A$ extension
is sampled over the entire base (\cref{def:recursive-qcma}).
All oracle access is coherent.\footnote{The displayed tuples have fixed injective binary encodings.
We pad the oracle addresses to one polynomial length $L_0(n)$,
as for the functional-key construction (\cref{def:functional-keys}).}

The subroutine $\mathsf{Read}(s,\regis{sk},u,b)$ runs
$\EvalHash^{T_s}$ on those registers. Using these encoded queries to
$E_{H_{s,t}}$, $\mathsf{Issue}(s,\regis{sk},t)$ prepares
$\ket{\Psi_{H_{s,t}}}$ and returns the signing register for reuse
(\cref{sec:shared-key-algorithms}).

\subsubsection{Classical verification keys}
\label{sec:classical-vk-signatures}

We define $\CSig$, whose signature is a predicate-preimage state reused by
all verification checks.

\begin{definition}[Classical-verification-key signature oracle]
\label{def:classical-vk-signature-oracle}
The classical-verification-key signature oracle and its signature states are
\begin{equation}\label{eq:classical-vk-signature-oracle}
 \ket{\Sigma_{s,m}}=\ket{\Psi_{H_{s,m}}},\qquad
 \ora_{{\rm csig},n}=(T_{s,m})_{s,m}.
\end{equation}
Here $T_{s,m}$ encodes the constant-one function
(\cref{eq:signature-verification-tables}). The signing function $P_s$
supplies $E_{H_{s,m}}$, and no public preparation oracle supplies
$\ket{\Sigma_{s,m}}$.
\end{definition}

\begin{mdframed}[skipabove=4pt,skipbelow=4pt,innertopmargin=3pt,innerbottommargin=3pt,nobreak=false]
\noindent\underline{$\CSigGen(1^\lambda)$}\par\nopagebreak[4]
\begin{enumerate}[nosep]
 \item Run $\Prepare^{R_M,S_M}(1^{n_M})$ and measure its serial register
       to obtain $s$, leaving the predicate-preimage state register $\regis{sk}$.
 \item Return the verification key $\vk=s$ and signing key
       $(s,\regis{sk})$.
\end{enumerate}

\smallskip
\noindent\underline{$\CSigSign((s,\regis{sk}),m)$}\par\nopagebreak[4]
\begin{enumerate}[nosep]
 \item Return $\regis{sig}\leftarrow\mathsf{Issue}(s,\regis{sk},(\mathsf C,m))$,
       together with the updated signing register.
\end{enumerate}

\smallskip
\noindent\underline{$\CSigVer(s,m,\regis{sig})$}\par\nopagebreak[4]
\begin{enumerate}[nosep]
 \item Reject incorrect lengths.
 \item For $j=1,\ldots,r$, choose fresh $y_j\leftarrow\zo^\kappa$ and run
       $\EvalHash^{T_{s,m}}$ on $\regis{sig}$, input $y_j$,
       and a zero response bit. Measure the response and reject unless
       it equals one. Discard the fresh private workspace and use the
       updated signature register for the next check.
 \item Accept if all checks pass.
\end{enumerate}
\end{mdframed}

\subsubsection{Quantum verification keys}
\label{sec:quantum-vk-signatures}

We define $\QSig$, whose classical signatures are checked by
encoded equality queries using a quantum verification key.

\begin{definition}[Quantum-verification-key signature oracle]
\label{def:quantum-vk-signature-oracle}
The quantum-verification-key signature oracle and its verification states are
\begin{equation}\label{eq:quantum-vk-signature-oracle}
 \ket{K_s^V}=\ket{\Psi_{H_s^V}},\qquad
 \ora_{{\rm qsig},n}=(T_s^V)_s.
\end{equation}
Here $T_s^V$ encodes the equality test $G_s(m,z)$
(\cref{eq:signature-verification-tables}), and both schemes use
the same signing function $P_s$.
\end{definition}

\begin{mdframed}[skipabove=5pt,skipbelow=6pt,innertopmargin=5pt,innerbottommargin=5pt,nobreak=false]
\noindent\underline{$\QSigGen(1^\lambda)$}\par\nopagebreak[4]
\begin{enumerate}[nosep]
 \item Run $\Prepare^{R_M,S_M}(1^{n_M})$ and measure its serial register
       to obtain $s$, leaving the predicate-preimage state register $\regis{sk}$.
 \item Generate $(s,\regis{vk})\leftarrow\QSigVKGen((s,\regis{sk}))$.
 \item Return the verification key $(s,\regis{vk})$ and signing key
       $(s,\regis{sk})$.
\end{enumerate}

\medskip
\noindent\underline{$\QSigVKGen((s,\regis{sk}))$}\par\nopagebreak[4]
\begin{enumerate}[nosep]
 \item Run $\regis{vk}\leftarrow\mathsf{Issue}(s,\regis{sk},\mathsf V)$.
 \item Return $(s,\regis{vk})$, together with the updated signing register.
\end{enumerate}

\medskip
\noindent\underline{$\QSigSign((s,\regis{sk}),m)$}\par\nopagebreak[4]
\begin{enumerate}[nosep]
 \item For $i\in[n]$, run $\mathsf{Read}(s,\regis{sk},(0,m,i),0)$ and measure
       the response bit to obtain $\sigma_i$.
 \item Return $\sigma$ together with the updated signing register.
\end{enumerate}

\medskip
\noindent\underline{$\QSigVer((s,\regis{vk}),m,\sigma)$}\par\nopagebreak[4]
\begin{enumerate}[nosep]
 \item Reject incorrect lengths.
 \item Run $\EvalHash^{T_s^V}$ on $\regis{vk}$, input $(m,\sigma)$,
       and a zero response bit.
 \item Measure that bit and accept iff it is one.
\end{enumerate}
\end{mdframed}

\subsection{Security}

We now state the common security guarantee for both signature schemes.

\begin{theorem}[Adaptive signature security]\label{thm:signature-security}
For the signature oracle $\ora$ above and every independent base
component $\ora_{\rm rest}$, both $\CSig^\ora$ and $\QSig^\ora$
are uniform \QPT{} and satisfy the following guarantees, averaged over $\ora$:
\begin{itemize}
 \item \emph{Correctness.} Polynomially many adaptive honest signing,
 verification-key, and verification calls have negligible joint error.
 \item \emph{Adaptive unforgeability.} For either
 $\SigScheme\in\{\CSig,\QSig\}$ and every adversary with
 polynomial query count, query length and number of classical signing
 and verification-key requests, and unrestricted local computation,
 \[
  \Pr_{\ora}[\Forge^{\ora}_{\SigScheme,\cA}(\lambda)=1]
  \le2^{-\Omega_{\cA}(\lambda)}.
 \]
\end{itemize}
\end{theorem}
The proof of \cref{thm:signature-security} appears in
\cref{sec:adaptive-signature-proof}.

\Needspace{6\baselineskip}
Combining these signatures with the encryption construction gives
public-key quantum money.

\begin{corollary}[Public-key quantum money]\label{cor:public-key-money}
For almost every cryptographic oracle $\ora$ (\cref{def:full-crypto-oracle}),
there is a uniform \QPT{} public-key quantum money scheme with
classical verification keys, a reusable quantum minting key and
reusable banknotes, secure against polynomial-size quantum
counterfeiters with classical descriptions depending on $\ora$
and no quantum advice for every polynomial number of adaptive
minting requests.
\end{corollary}
The proof of \cref{cor:public-key-money} appears in
\cref{app:public-key-money-proof}.

\subsection{Proof of adaptive signature security}
\label{sec:adaptive-signature-proof}

We prove \cref{thm:signature-security}, using the following two
bounds for verification with a random Boolean function and for
predicting an unsigned function value. The joint output guarantee in
\cref{lem:signature-correctness} includes the full oracle and all
responses to adaptive classical requests.

\begin{lemma}[Verification with a random function]\label{lem:signature-random-table}
Put $\kappa=(n+4)^3$ and
$\tau_{\mathrm{sig}}(\kappa)=6\sqrt{(a_\kappa+1)2^{-\kappa}}$.
Let $V_B^r$ be the $r$ checks of $\CSigVer$ with $T_{s,m}$
replaced by $B$, setting $B(\bot,y)=0$. For a uniform random Boolean function
$B:\Sigma_\kappa^{q_\kappa}\times\zo^\kappa\to\zo$, with the maximum over
all $a_\kappa$-qubit density matrices,
\[
 \Pr_B\!\left[
  \max_\rho\Pr[V_B^r(\rho)=1]>(2/3)^r
 \right]\le\tau_{\mathrm{sig}}(\kappa).
\]
\end{lemma}
The proof of \cref{lem:signature-random-table} appears in
\cref{app:signature-security}.

\begin{lemma}[Adaptive equality queries]\label{lem:signature-equality-queries}
Let $F:\zo^n\to\zo^n$ be a uniform random function and
$\EqTest_F(m,z)=\indicator[z=F(m)]$.
An algorithm with unrestricted local computation and an arbitrary
auxiliary state independent of $F$ may make classical signing
requests $m\mapsto F(m)$ and at most
$Q$ quantum queries to $\EqTest_F$. If $S$ is the set of signed messages
and the algorithm outputs $(m,z)$, then
\[
 \Pr_F[m\notin S\ \land\ z=F(m)]\le2(Q+1)2^{-n/2}.
\]
\end{lemma}
The proof of \cref{lem:signature-equality-queries} appears in
\cref{app:signature-security}.
We now prove adaptive security for both signature schemes.

\Needspace{10\baselineskip}
\paragraph{The forgery experiments.}
Fix a forger $\cA$ and write $n=\lambda$. The bounds below are
uniform in the independently sampled additional base
$\ora_{\rm rest}$. Use $D=C,Q$ for the
two schemes, writing $(\mathsf{Gen}_D,\mathsf{Sign}_D,\mathsf{Ver}_D)$
for their generation, signing and verification algorithms. In
$\mathsf S_0^D$, sample
$\ora^0=(\orabase^0,N^0,A^0)=\ora$ and run
$(\regis{vk},\regis{sk})\leftarrow\mathsf{Gen}_D^{\ora^0}(1^n)$.
Initialize $S=\varnothing$ and run $\cA^{\ora^0}(\regis{vk})$,
answering each classical signing request $m$ by
$\mathsf{Sign}_D^{\ora^0}(\regis{sk},m)$, updating the
signing register and setting $S=S\cup\{m\}$.
For $D=Q$, answer verification-key requests by
$\QSigVKGen^{\ora^0}(\regis{sk})$ with the updated signing key.
When $\cA$ outputs $(m,\regis{sig})$, return zero if $m\in S$,
and otherwise return $\mathsf{Ver}_D^{\ora^0}(\regis{vk},m,\regis{sig})$,
using a newly generated verification key for the final $Q$ test.

For $\mathsf S_1^D$, sample $\ora^1=\ora$ and $s\leftarrow\zo^n$
independently, initialize $S=\varnothing$, and run $\cA$ as follows:
\begin{enumerate}[nosep]
 \item Give $s$ to $\cA$ in scheme $C$, or
 $(s,\ket{\Psi_{H_s^V}})$ in scheme $Q$, with coherent access to
 $\ora^1$ in either case.
 \item On a classical signing request $m$, set
 $S=S\cup\{m\}$ and return $\ket{\Psi_{H_{s,m}}}$ in
 scheme $C$, or $F_s^{\rm sig}(m)$ in scheme $Q$. In scheme $Q$,
 answer each verification-key request with a fresh
 $\ket{\Psi_{H_s^V}}$ using the same sampled hash $H_s^V$.
 \item On output $(m,\regis{sig})$, reject if $m\in S$. Otherwise
 run $\CSigVer(s,m,\regis{sig})$ in scheme $C$, or run
 $\QSigVer((s,\ket{\Psi_{H_s^V}}),m,\regis{sig})$ with a fresh
 verification key in scheme $Q$, and return its acceptance bit.
\end{enumerate}
Signature correctness (\cref{lem:signature-correctness}) applies
jointly to all adaptively requested outputs and the fresh final
verification key, including their correlations with $\ora$. Thus,
writing $p_i^D=\Pr[\mathsf S_i^D=1]$, we have
\[
 |p_0^D-p_1^D|\le2^{-\Omega(n)}.
\]

In $\mathsf S_2^D$, use exactly this challenger, but replace the
signing-state target and sample its oracle as follows:
\[
 \begin{gathered}
 \ket{\Omega_M^0}=2^{-n/2}
  \left(\sum_{t\ne s}\ket t\ket{\Psi_{H_t}}+\ket s\ket{0^{n_M-n}}\right),\\
 (R_M^0,S_M^0)\leftarrow\PrepTables(\ket{\Omega_M^0}),\qquad
 \ora^2=(\orabase^2,N^2,A^2).
 \end{gathered}
\]
Set
$\orabase^2=\orabase^1[(R_M,S_M)\leftarrow(R_M^0,S_M^0)]$
and sample $N^2,A^2$ conditional on $\orabase^2$.
Run $\cA^{\ora^2}$ with the same initial input and classical
request rules as in $\mathsf S_1^D$, and give the final verifier
$\ora^2$ as well. Thus the signing and verification-key responses
use the original functions and hashes in Steps 1--2.

\paragraph{Quantum verification keys.}
In $\mathsf S_3^Q$, keep $\ora^3=\ora^2$ and run the same
$\cA^{\ora^3}$ with the same signing and verification-key channels,
replacing only its final test by
\[
 \indicator[m\notin S]\,G_s(m,\sigma),\qquad
 G_s(m,z)=\indicator[z=F_s^{\rm sig}(m)].
\]
The final verifier in $\mathsf S_2^Q$ uses a fresh
$\ket{\Psi_{H_s^V}}$, so encoded evaluation
(\cref{lem:coherent-evaluation}) on the good-hash event, followed by
averaging its complement, gives
\[
 |p_2^Q-p_3^Q|\le2\sqrt{\delta_{\mathrm{dec}}(\kappa)}+2^{-\kappa}.
\]
We replace this test before simulating the public encodings.

Let $G_Q$ consist of the message hashes $(H_{s,m})_m$ and
$\orabase^2$ with $T_s,T_s^V$ omitted.
For each fixed $G_Q$, encoded-function simulation
(\cref{thm:crypto-common-simulation}) supplies the algorithm
$\cS_{G_Q}$ used in the next experiment, with only the verification
instance opened. Its hypotheses are checked below.
Experiment $\mathsf S_4^Q$ runs it as follows:
\begin{enumerate}[nosep]
 \item Draw one classical sample $\eta_Q$ conditional on $G_Q$,
 initialize $S=\varnothing$, and run
 $\cS_{G_Q}^{G_s}(s,\eta_Q)$.
 Inside this call, $\cS_{G_Q}$ initializes $\cA$ on
 $(s,\ket{\Psi_{H_s^V}})$ using its sampled $H_s^V$.
 \item On a signing request $m$, query the classical signing oracle
 for $F_s^{\rm sig}(m)$ to obtain $z$, set $S=S\cup\{m\}$,
 and return $z$. On a verification-key request, prepare another
 copy of $\ket{\Psi_{H_s^V}}$ from the same sampled $H_s^V$.
 \item Answer the simulator's function queries by $G_s$. On the
 final $(m,\sigma)$, reject if $m\in S$ and otherwise return
 $G_s(m,\sigma)$.
\end{enumerate}
\paragraph{Classical verification keys.}
Let $\orabase^{2,-\mathsf C}$ denote $\orabase^2$ with $T_s$ and all
$T_{s,m}$ omitted, and set
\[
 G_C=(F_s^{\rm sig},H_s^V,\orabase^{2,-\mathsf C}).
\]
For each fixed $G_C$, encoded-function simulation
(\cref{thm:crypto-common-simulation}) supplies $\cS_{G_C}$, opening a message
index precisely when it is requested for signing. The proof below
checks this adaptive application.
In $\mathsf S_3^C$, run this algorithm as follows:
\begin{enumerate}[nosep]
 \item Draw one classical sample $\eta_C$ conditional on $G_C$,
 initialize $S=\varnothing$, and run
 $\cS_{G_C}(s,\eta_C)$, which initializes $\cA$ on $s$.
 \item On a signing request $m$, set $S=S\cup\{m\}$
 and supply $\ket{\Psi_{H_{s,m}}}$ using the sampled $H_{s,m}$.
 Between requests, queries to the encodings use
 \[
  T_s\mapsto B_s,\qquad
  T_{s,m}\mapsto
  \begin{cases}
    T_{H_{s,m},B_{s,m},1},&m\in S,\\
    B_{s,m},&m\notin S.
  \end{cases}
 \]
 Every $\ora^2$ query is answered by the query algorithm $\cS_{\eta_C}(S,\cdot)$ using this one classical sample. Repeated signing requests use the same $H_{s,m},B_{s,m}$.
 \item On $(m,\regis{sig})$, reject if $m\in S$. Otherwise run
 the $3n$ checks of $\CSigVer$ with $B_{s,m}$ and return their
 joint acceptance bit.
\end{enumerate}
The following claim justifies the three public-oracle changes used
by these experiments.
\begin{claim}\label{clm:signature-oracle-hybrids}
For the fixed forger and either $D\in\{C,Q\}$,
\[
 |p_1^D-p_2^D|+|p_3^Q-p_4^Q|+|p_2^C-p_3^C|
 \le2^{-\Omega(n)}.
\]
The simulator in $\mathsf S_4^Q$ makes $Q'=\poly(n)$ queries only
to $G_s$, in addition to the classical signing requests, and its
classical sample and description are independent of $F_s^{\rm sig}$.
In $\mathsf S_3^C$, every unsigned message is tested with its original
uniform Boolean function $B_{s,m}$.
\end{claim}
The proof of \cref{clm:signature-oracle-hybrids} appears in
\cref{sec:signature-oracle-hybrid-bounds}.

\paragraph{Bounding the forgery probabilities.}
In $\mathsf S_4^Q$, the simulator has classical signing access to
$F_s^{\rm sig}$ and $Q'=\poly(n)$ equality queries, with an independent
classical sample. Adaptive equality-query security
(\cref{lem:signature-equality-queries}) therefore gives
\[
 p_4^Q\le2(Q'+1)2^{-n/2}.
\]
Combining this with the adjacent bounds proves
\[
 p_0^Q\le p_4^Q+\sum_{i=0}^3|p_i^Q-p_{i+1}^Q|
 \le2(Q'+1)2^{-n/2}+2\sqrt{\delta_{\mathrm{dec}}(\kappa)}
       +2^{-\kappa}+2^{-\Omega(n)}.
\]

Define the event
\[
 \mathsf{Bad}_{\rm sig}=\left\{\exists m\in\zo^n:
       \max_\rho\Pr[V_{B_{s,m}}^{3n}(\rho)=1]>(2/3)^{3n}\right\}.
\]
Verification with a random function (\cref{lem:signature-random-table}) applies
to each $B_{s,m}$, and a union bound gives
\[
 \Pr[\mathsf{Bad}_{\rm sig}]\le2^n\tau_{\mathrm{sig}}(\kappa),\qquad
 p_3^C\le(2/3)^{3n}+2^n\tau_{\mathrm{sig}}(\kappa).
\]
The maximum ranges over every state depending on the selected function,
so this estimate applies to the adaptively chosen forged message
without conditioning on that choice. Therefore
\[
 p_0^C\le p_3^C+\sum_{i=0}^2|p_i^C-p_{i+1}^C|
 \le(2/3)^{3n}+2^n\tau_{\mathrm{sig}}(\kappa)+2^{-\Omega(n)}.
\]
Both forgery bounds are $2^{-\Omega(n)}$.

\subsubsection{Bounds for the signature hybrids}
\label{sec:signature-oracle-hybrid-bounds}

We now prove \cref{clm:signature-oracle-hybrids}. The preparation
change keeps the original signing responses, and the two encoded
simulations open precisely the hashes used in those responses.

\paragraph{Changing the signing-state target.}
Fix $s$, all sampled functions and hashes, and
the preparation pairs other than $(R_M,S_M)$. They determine both
targets and every request-response rule before the changed pair and
recursive coins are sampled. The targets satisfy
\[
 \|\ket{\Omega_M}-\ket{\Omega_M^0}\|_2\le2^{1-n/2}.
\]
Apply the target-change bound (\cref{thm:preparation-replacement})
with dimension $n_M$, the initial state in Step 1, the response
channels in Steps 2--3, and recursive mean $h(v)=(1+v)/3$, obtaining
\[
 |p_1^D-p_2^D|\le2^{-\Omega(n)}.
\]
The original signing and verification-key rules are the same in
both experiments, so this bound includes adaptive requests.

\paragraph{The quantum verification key.}
The target $\ket{\Omega_M^0}$ replaces $\ket{\Psi_{H_s}}$
by $\ket{0^{n_M-n}}$, and all other primitive bases are independent, so the families
\[
 F_s^{\rm sig},\quad(H_s,H_s^V),\quad(B_s,B_s^V)
 \qquad\text{are mutually independent conditional on }G_Q.
\]
For each fixed $G_Q$, apply encoded-function simulation
(\cref{thm:crypto-common-simulation}) with $J=\{\mathsf M,\mathsf V\}$ and $K=\{\mathsf V\}$
for $T_s,T_s^V$, opening only the verification instance at the start and on later
verification-key requests, with direct access only to $G_s$. The function
$G_s$ depends on $F_s^{\rm sig}$, whereas the discarded master
function may also use $H_s^V$ through $P_s(1,\mathsf V,u)=E_{H_s^V}(u)$,
as allowed by the theorem. The opened verification index is known throughout this single-adversary
interaction. Every $\ora^2$ query is answered by the coherent query
algorithm with this opened set and the same classical sample $\eta_Q$.
Its only direct queries are to $G_s$, and it has no original $N,A$
access. The unchanged request channels use $F_s^{\rm sig}$ and the
opened verification hash, so the theorem gives
\[
 |p_3^Q-p_4^Q|\le2^{-\Omega(n)}.
\]
The sample and description are independent of $F_s^{\rm sig}$
conditional on $G_Q$, which is itself independent of that function.
Averaging over $G_Q$ therefore establishes the equality-query experiment in the claim.

\paragraph{Classical verification keys.}
Conditional on $G_C$, the hashes are independent draws
$H_s\leftarrow\cH_{\kappa_M}$ and $H_{s,m}\leftarrow\cH_\kappa$
for each $m$. The functions $B_s,(B_{s,m})_m$ are independent uniform
Boolean functions, independent of those hashes. A signing response uses only its requested hash
$H_{s,m}$, and these hashes occur elsewhere only in the discarded
master function, through $P_s(1,(\mathsf C,m),u)=E_{H_{s,m}}(u)$.
There is no public preparation pair for a message hash.

Apply encoded-function simulation (\cref{thm:crypto-common-simulation})
with $J=\{\mathsf M\}\cup\zo^n$ and $K=\zo^n$, where $\mathsf M$
indexes $T_s$ and each message $m$ indexes $T_{s,m}$. Each signing
request supplies a state using only its opened hash, so the adaptive
input hypothesis holds and gives
$|p_2^C-p_3^C|\le2^{-\Omega(n)}$. In particular, the final test uses
the original uniformly random Boolean function $B_{s,m}$ for every unsigned $m$.

The forger knows its signed-message set $S$, which is exactly the
opened set used in $\cS_{\eta_C}(S,\cdot)$. Thus no additional
request information or oracle access is required.

\clearpage\section{Pseudorandom States}\label{sec:pseudorandom-states}

We construct PRS with classical keys and polynomial-copy security
(\cref{thm:pseudorandom-states}), using binary-phase states as the
preparation targets.

\begin{definition}[PRS oracle]\label{def:prs-preparation-oracle}
Independently for every $n$ and $k\in\zo^n$, sample a uniform
function $\Theta_k:\zo^{5n}\to\zo$ and then sample
\begin{gather}
 \ket{\Omega_{\Theta_k}}=2^{-5n/2}\sum_{x\in\zo^{5n}}(-1)^{\Theta_k(x)}\ket x,
 \label{eq:prs-binary-phase-state}\\
 \ket{\Omega_k}=\ket{\Omega_{\Theta_k}},\qquad
 (R_k^{\mathsf{prs}},S_k^{\mathsf{prs}})
 \leftarrow\PrepTables(\ket{\Omega_k}).
 \label{eq:prs-prescribed-state}
\end{gather}
We define the PRS base oracle by
$\ora_{\mathsf{prs},n}=((R_k^{\mathsf{prs}},S_k^{\mathsf{prs}}))_k$,
and let $\ora_{\mathsf{prs}}(n,u)=\ora_{\mathsf{prs},n}(u)$,
where $u$ specifies a component above and its input.
Only the preparation pairs are oracle components.
\end{definition}

Use $\ora=(\orabase,N,A)$ with
$\orabase=(\ora_{\mathsf{prs}},\ora_{\rm rest})$, where
$\ora_{\rm rest}:\zo^*\to\zo$ is sampled independently on disjoint addresses, and one $N,A$ extension
is sampled over the entire base (\cref{def:recursive-qcma}).
In particular, direct access to the phase functions is unavailable.

\subsection{Construction}\label{sec:prs-algorithms}

For fixed $\ora,k$, write $\rho_k^\ora$ for the output density
matrix of the following state-generation algorithm. Its $r$ calls
use independent internal coins, so their joint output is
$(\rho_k^\ora)^{\otimes r}$.

\begin{mdframed}[skipabove=5pt,skipbelow=6pt,innertopmargin=5pt,innerbottommargin=5pt,nobreak=false]
\noindent\underline{$\KeyGen(1^\lambda)$}\par\nopagebreak[4]
\begin{enumerate}[nosep]
 \item Return $k\leftarrow\zo^\lambda$.
\end{enumerate}

\medskip
\noindent\underline{$\StateGen^{\ora}(1^\lambda,k)$}\par\nopagebreak[4]
\begin{enumerate}[nosep]
 \item Run $\Prepare^{R_k^{\mathsf{prs}},S_k^{\mathsf{prs}}}(1^{5\lambda})$
       and return the entire output.
\end{enumerate}
\end{mdframed}

All calls use the same hidden classical key $k$, and the following
theorem establishes pseudorandomness for polynomially many copies.

\begin{theorem}[PRS security]\label{thm:pseudorandom-states}
Use the PRS oracle $\ora$ above with any independent base component
$\ora_{\rm rest}$. The generator is uniform \QPT{}. Sample
$k\leftarrow\zo^\lambda$ and put
$\rho_{k,r}^\ora=(\rho_k^\ora)^{\otimes r}$.
For every polynomial $r$ and quantum algorithm $\cA$ with polynomial
query count and length and unrestricted local computation,
\begin{equation}\label{eq:prs-fixed-security}
 \left|\Pr_{\ora,k}[\cA^{\ora}(\rho_{k,r}^{\ora})=1]
 -\Pr_{\ora,\psi\leftarrow\HaarMeasure_{5\lambda}}
       [\cA^{\ora}(\ket\psi^{\otimes r})=1]\right|
 \le2^{-\Omega_{\cA}(\lambda)}.
\end{equation}
\end{theorem}
The proof of \cref{thm:pseudorandom-states} appears in
\cref{sec:prs-security-proof}.

\subsection{Proof of PRS security}\label{sec:prs-security-proof}

We now prove \cref{thm:pseudorandom-states} by changing the supplied
state first to an independent binary-phase state and then to a Haar state.

Fix the additional base component. Each experiment below samples
the full oracle $\ora=(\orabase,N,A)$ using the PRS preparation pairs
(\cref{def:prs-preparation-oracle}), and changes only the input
supplied to $\cA^\ora$:
\begin{enumerate}[label=$\mathsf H_{\arabic*}$,start=0,itemsep=3pt]
 \item Sample one $k\leftarrow\zo^\lambda$, run
 $\StateGen^{\ora}(1^\lambda,k)$ independently $r$ times, and output
 \[
  \cA^{\ora}((\rho_k^{\ora})^{\otimes r}).
 \]
 \item Sample $\Theta\leftarrow\{\zo^{5\lambda}\to\zo\}$ independently
 of $\ora$, and output
 $\cA^{\ora}((\ket{\Omega_\Theta}\!\bra{\Omega_\Theta})^{\otimes r})$.
 \item Sample $\psi\leftarrow\HaarMeasure_{5\lambda}$ independently
 of $\ora$, and output
 $\cA^{\ora}((\ket\psi\!\bra\psi)^{\otimes r})$.
\end{enumerate}
All $r$ outputs in $\mathsf H_0$ use the same key and preparation
pair, while the other experiments use $r$ copies of one independent
state. To justify the first change, apply the independent-input bound (\cref{thm:crypto-family-replacement}) to the family
$(\Theta_k)_{k\in\zo^\lambda}$, its preparation pairs, and the input
$(\rho_k^\ora)^{\otimes r}$. Its hypotheses hold because these
phase functions are independent and occur in $\orabase$ only
through their preparation pairs, while all other base components
are independent. Its joint-output guarantee includes all $r$ copies and the full
quantum oracle access, so
\[
 |\Pr[\mathsf H_0=1]-\Pr[\mathsf H_1=1]|
 \le2^{-\Omega_{\cA}(\lambda)}.
\]
For the next step, the binary-phase moment bound~\cite{BS19BinaryPhase} gives
\begin{equation}\label{lem:binary-phase-haar}
 \trdist{\bbE_\Theta[(\ket{\Omega_\Theta}\!\bra{\Omega_\Theta})^{\otimes r}]}
 {\bbE_{\psi\leftarrow\HaarMeasure_{5\lambda}}
             [(\ket\psi\!\bra\psi)^{\otimes r}]}
 \le O(r^2 2^{-5\lambda}).
\end{equation}
The input state in $\mathsf H_1$ and $\mathsf H_2$ is independent
of $\ora$, so running $\cA^\ora$ and measuring its output is the
same quantum channel applied to those two averaged states. Since
quantum operations cannot increase trace distance,
\[
 |\Pr[\mathsf H_1=1]-\Pr[\mathsf H_2=1]|
 \le O(r^2 2^{-5\lambda}).
\]
The triangle inequality now gives
\[
 \bigl|\Pr[\mathsf H_0=1]-\Pr[\mathsf H_2=1]\bigr|
 \le2^{-\Omega_{\cA}(\lambda)}+O(r^2 2^{-5\lambda})
 =2^{-\Omega_{\cA}(\lambda)}.
\]
The estimate is uniform in the fixed additional base component, so
averaging over that component proves the theorem.

\cref{thm:fixed-oracle-crypto} gives the PRS guarantee
(\cref{predef:prs}) and bounds the preparation error
(\cref{eq:prs-preparation-error}) simultaneously for every key.

\clearpage\section{Quantum Cryptography Relative to an Oracle with $\PTIME=\QCMA$}\label{sec:fixed-crypto}

In this section, we establish all cryptographic guarantees relative
to one fixed oracle. \cref{thm:fixed-oracle-crypto} includes security
against nonuniform classical circuit descriptions depending on that
oracle, with no quantum advice.

The four independently sampled base oracles can now be combined
before adding the recursive extension.

\begin{definition}[The cryptographic oracle]\label{def:full-crypto-oracle}
Independently sample the PRF, encryption, signature and PRS base
oracles (\cref{def:prf-oracle,def:pke-oracle,def:signature-oracle,def:prs-preparation-oracle}).
We define the cryptographic oracle by
\begin{equation}\label{eq:cryptographic-base-oracle}
 \orabase=(\ora_{\mathsf{prf}},\ora_{\mathsf{pke}},
             \ora_{\mathsf{sig}},\ora_{\mathsf{prs}}),
 \qquad \ora=(\orabase,N,A),
\end{equation}
where $\orabase(c,u)=\ora_c(u)$ for
$c\in\{\mathsf{prf},\mathsf{pke},\mathsf{sig},\mathsf{prs}\}$,
with zero on unused addresses. One pair $N,A$ is sampled from this
entire base (\cref{def:recursive-qcma}).
\end{definition}

The following theorem makes the four guarantees simultaneous for a
single fixed oracle.

\begin{theorem}[Cryptography for one fixed oracle]
\label{thm:fixed-oracle-crypto}\label{cor:fixed-oracle-signatures}
With probability one over the cryptographic oracle
$\ora=(\orabase,N,A)$ (\cref{def:full-crypto-oracle}), the following hold relative to $\ora$. Every security guarantee is
against polynomial-size quantum oracle circuit families with classical
descriptions, allowing dependence on $\ora$ but no quantum advice.
\begin{itemize}
 \item The promise classes satisfy
 \[
  \PROMISEP^{\ora}=\PROMISEBQP^{\ora}
        =\PROMISEQCMA^{\ora}=\PROMISEQCMAH^{\ora}.
 \]

 \item The PRF $\prf^{\ora}$ (\cref{sec:prf-construction})
 satisfies correctness, pseudorandomness and copy-protection
 as in \cref{predef:prf,def:prf-security}.

 \item The encryption scheme $\pke^{\ora}$ (\cref{sec:pke-construction})
 satisfies correctness, chosen-plaintext security and copy-protection
 under independent ciphertext challenges, as in
 \cref{predef:pke,def:pke-piracy}.

 \item The signature schemes $\CSig^{\ora}$ and $\QSig^{\ora}$ (\cref{sec:signatures})
 satisfy correctness and adaptive unforgeability as in
 \cref{def:signature-security}.

 \item The PRS scheme $\prs^{\ora}$ (\cref{sec:pseudorandom-states})
 satisfies the PRS definition (\cref{predef:prs}), with
 $\lambda$-bit keys and $5\lambda$-qubit outputs. Its preparation
 error (\cref{eq:prs-preparation-error}) is negligible uniformly over
 its keys, and its security holds for polynomially many same-key outputs.
\end{itemize}
\end{theorem}
The proof of \cref{thm:fixed-oracle-crypto} appears in
\cref{app:fixed-crypto}.
Observe that the uniform preparation bound also gives PRS security
for polynomially many copies of the prescribed state $\ket{\Omega_k}$
(\cref{eq:prs-prescribed-state}), by the tensor-product trace-distance bound.

Note that this also (trivially) implies that these primitives exist relative to a fixed oracle $\ora$ such that $ \PROMISEP^{\ora}=\PROMISEQCMAH^{\ora}.$

The encryption scheme also yields average-case hardness in \QMA.

\begin{corollary}[Average-case $\QCMAH$ hardness in \QMA{}]\label{cor:pke-average-qma}
Fix an oracle $\ora$ satisfying \cref{thm:fixed-oracle-crypto}.
There are a \QMA{} promise problem $\Pi=(\Pi_0,\Pi_1)$\footnote{We state
the promise-problem version here. A total-language version can also be
proved by adapting our lemmas.} and a uniform \QPT{}
sampler $\mathsf D^{\ora}(1^\lambda)\to(X,b)$ with classical
output and uniform $b\in\zo$ such that, for every constant $d$
and every uniform $\QCMAH_d$ machine $M_d$, counting $\bot$ answers
as incorrect,
\[
 \Pr_{X,b}[X\in\Pi_b]\ge1-\negl(\lambda),\qquad
 \Pr_{X,b}[M_d^{\ora}(X)=b]\le\tfrac12+\negl_{M_d}(\lambda).
\]
\end{corollary}
The proof of \cref{cor:pke-average-qma} appears in
\cref{app:pke-average-qma-proof}.

Through known black-box constructions, we obtain the following corollary.

\begin{corollary}[Further cryptographic primitives]\label{cor:crypto-consequences}
For almost every cryptographic oracle $\ora$
(\cref{def:full-crypto-oracle}), the following primitives exist relative
to $\ora$ and satisfy correctness and security against polynomial-size
quantum oracle circuit families with classical descriptions depending
on $\ora$ and no quantum advice:
\begin{itemize}
 \item Collusion-resistant public-key quantum money with classical
 verification keys.
 \item For every polynomial $q$, $q$-bounded collusion-resistant
 copy-protected PKE and PRF schemes.
 \item Pseudorandom unitaries with a quantum key, secure under
 polynomially many forward oracle queries.
 \item One-way puzzles with possibly inefficient verification.
 \item One-way state generators with classical keys.
 \item For every polynomial $q$, signatures with quantum verification
 keys and classical signing keys and signatures, secure against
 $q(\lambda)$ adaptive signing queries.
\end{itemize}
\end{corollary}

\clearpage\newcommand{\etalchar}[1]{$^{#1}$}

\paperappendices
\paperappendixpart{Deferred Proofs}
\section{Proofs of the Quantum Query and Encoding Bounds}
\label{app:query-bounds}

In this appendix, we prove the coherent implementation bound and the
random-function bound used in \QMAbb{} soundness.

\subsection{Proof of coherent classical computation (\texorpdfstring{\cref{lem:coherent-classical-query}}{Lemma \getrefnumber{lem:coherent-classical-query}})}

We now prove the coherent implementation guarantee of
\cref{lem:coherent-classical-query}.

\begin{proof}[Proof of \cref{lem:coherent-classical-query}]
Run $T$ reversibly while storing its intermediate computation, XOR
its output into the response bit, and run the computation backward.
The working registers return to their initial state, so the operation
is the required coherent map. Each of the at most $h$ oracle queries
is made once in each direction, giving the bound $2h$.
\end{proof}

\subsection{Proof of the random-function bound (\texorpdfstring{\cref{lem:random-table}}{Lemma \getrefnumber{lem:random-table}})}

We prove \cref{lem:random-table}, which bounds the probability
that a random Boolean function admits a successful quantum witness,
using the following random-access-code bound.

\begin{lemma}[Random access codes~\cite{Nayak}]
\label{lem:random-access-code}
Let $x\mapsto\sigma_x$ encode $x\in\zo^N$ into $m$ qubits, and
let $M_1,\ldots,M_N$ be binary measurements, where $N\ge1$ and
$m\ge0$ are integers. Then
\[
 \Pr_{x\leftarrow\zo^N,\,i\leftarrow[N]}
     [M_i(\sigma_x)=x_i]\le\frac12+\sqrt{\frac mN}.
\]
\end{lemma}
The proof of \cref{lem:random-access-code} appears in
\cref{app:query-bounds}, after the proof of \cref{lem:random-table}.

We now prove \cref{lem:random-table} by constructing a random access code
from a witness accepted on a random Boolean function.

\paragraph{Constructing the code.}
Let $N=2^\lambda$ and define
\[
 \cG=\left\{B:\ \exists\sigma,\quad
 \Pr_{y\leftarrow\zo^\lambda,\,v\leftarrow A(\sigma,y)}
       [v\ne\bot\land B(v,y)=1]>\frac23\right\},
 \qquad \gamma=\Pr_B[B\in\cG].
\]
For each $B\in\cG$, choose an $a$-qubit state $\sigma_B$
whose acceptance probability exceeds $2/3$. To encode a uniform
string using these states, independently sample
\[
 Z:\Sigma_\lambda^{q_\lambda}\times\zo^\lambda\to\zo\ \text{uniformly},
 \qquad x=(x_y)_{y\in\zo^\lambda}\leftarrow\zo^N,
\]
and put
\[
 B(v,y)=Z(v,y)\oplus x_y.
\]
For each fixed $x$, $B$ is uniform. Both the encoder and the recovery
algorithm receive $Z$.
\begin{boxedalgorithm}{Encoding $x$ and recovering $x_y$}
\textbf{Encoding.} Output $(1,\sigma_B)$ if $B\in\cG$,
and $(0,\ket{0^a})$ otherwise.

\smallskip
\textbf{Recovery.} On input $y$ and an encoding $(e,\sigma)$:
\begin{enumerate}
\item If $e=0$, output a uniform bit and terminate.
\item Run $v\leftarrow A(\sigma,y)$, then output a uniform bit if
      $v=\bot$ and $1\oplus Z(v,y)$ otherwise.
\end{enumerate}
\end{boxedalgorithm}

\paragraph{Bounding the recovery probability.}
For $B\in\cG$, every pair with $B(v,y)=1$ recovers
$x_y=1\oplus Z(v,y)$, giving success at least $2/3$ over $y$ and $A$,
whereas for $B\notin\cG$ recovery succeeds with probability $1/2$.
Thus uniformity of $B$ gives average success over $Z,x$ at least
\[
 \frac{1-\gamma}{2}+\frac{2\gamma}{3}
 =\frac12+\frac\gamma6.
\]
Some fixed Boolean function $Z$ attains at least this average success. Fixing it
makes the construction
an $(a+1)$-qubit random access code for a uniform $N$-bit string, so
\cref{lem:random-access-code} gives
\[
 \frac12+\frac\gamma6
 \le\Pr[\text{recovery is correct}]
 \le\frac12+\sqrt{\frac{a+1}{N}}.
\]
Therefore $\gamma\le6\sqrt{(a+1)2^{-\lambda}}$, as required.

For completeness, we now derive the random access code bound from
mutual information.

\begin{proof}[Proof of \cref{lem:random-access-code}]
Let $X$ be uniform on $\zo^N$, let $Q$ contain $\sigma_X$, and put
$p_i=\Pr[M_i(\sigma_X)=X_i]$ and $p=N^{-1}\sum_i p_i$.
Write $Y_i$ for the outcome of measurement $M_i$. Since measuring
$Q$ cannot increase mutual information, Fano's inequality gives
$I(X_i:Q)\ge I(X_i:Y_i)\ge1-h_2(p_i)$. Combining this with
the $m$-qubit entropy bound and independence of the $X_i$, we obtain
\[
 m\ge I(X:Q)\ge\sum_{i=1}^N I(X_i:Q)
 \ge\sum_{i=1}^N(1-h_2(p_i))
 \ge N(1-h_2(p)),
\]
where $I$ denotes quantum mutual information and $h_2$ is binary entropy,
with the last inequality following from concavity of $h_2$. Since
$1-h_2(p)\ge(p-1/2)^2$, we conclude that
$p\le1/2+\sqrt{m/N}$.
\end{proof}

\clearpage\section{Proof of the Codeword-Hash Guarantees (\texorpdfstring{\cref{lem:code-hash}}{Lemma \getrefnumber{lem:code-hash}})}
\label{app:code-export}

In this appendix, we prove \cref{lem:code-hash} using the multiplicity
code of~\cite{BHV}.

\subsection{The multiplicity code}\label{app:code-construction}

We first specify the code family whose properties are stated in
\cref{lem:code-hash}. Let $q_\lambda$ be the smallest prime greater than
$\lambda^5$, put $\Sigma_\lambda=\bbF_{q_\lambda}^\lambda$, and fix an
ordering $\alpha_1,\ldots,\alpha_{q_\lambda}$ of $\bbF_{q_\lambda}$. For
$f\in\bbF_{q_\lambda}[T]$ with $\deg f<\lambda^3$, define
\[
 v(f)_i=\bigl([T^j]f(\alpha_i+T)\bigr)_{j=0}^{\lambda-1},
 \qquad
 C_\lambda=\{(v(f)_1,\ldots,v(f)_{q_\lambda}):\deg f<\lambda^3\}.
\]
This is the multiplicity code
$\mathsf{Mult}_{\lambda,\bbF_{q_\lambda},\lambda^3}$ of~\cite{BHV}.
Its generator matrix and a left inverse are computable in polynomial
time, giving efficient encoding and membership testing. Write
$\ket{C_\lambda}=|C_\lambda|^{-1/2}\sum_{v\in C_\lambda}\ket v$
for the uniform codeword state. Preparing
uniform field coefficients, applying the linear encoder, and uncomputing
the coefficients with the left inverse prepares the uniform codeword
state, with the approximation guarantee in \cref{lem:code-hash}.
Encode each field element using $\lceil\log_2q_\lambda\rceil$ bits, so
$b_\lambda=\lambda\lceil\log_2q_\lambda\rceil$ and
$a_\lambda=2q_\lambda\lambda\lceil\log_2q_\lambda\rceil$.

\subsection{Expansion and decoding bounds}

We use the following expansion and decoding bounds to establish the
properties of this code family.

\begin{lemma}[Multiplicity-code expansion~\cite{BHV}]
\label{lem:bhv-expansion}
For all sufficiently large $\lambda$, every set $W\subseteq C_\lambda$
with $|W|\le2^\lambda$ satisfies
\[
 \bigl|\{(i,v_i):i\in[q_\lambda],\ v\in W\}\bigr|
 \ge(q_\lambda-4\lambda^4)|W|.
\]
\end{lemma}
\begin{proof}
Apply the expansion bound of~\cite{BHV} with degree bound
$k=\lambda^3$, multiplicity $s=\lambda$ and set-size bound
$K=2^{\lambda+1}$. Since $\lambda^5<q_\lambda\le2\lambda^5$, its hypotheses
$15\le\lambda+1\le\lambda^3\le q_\lambda$ hold, and its expansion factor is
\[
 q_\lambda-\frac{\lambda^3(\lambda+1)}2
       (q_\lambda2^{\lambda+1})^{1/(\lambda+1)}
 =q_\lambda-\lambda^3(\lambda+1)q_\lambda^{1/(\lambda+1)}
 \ge q_\lambda-4\lambda^4.
\]
The formal derivatives in that bound are nonzero scalar multiples
of the Hasse derivatives defining $C_\lambda$, so they count the
same distinct coordinate--symbol pairs.
\end{proof}

The decoding bound below holds simultaneously for every target.

\begin{lemma}[Simultaneous biased-hash decoding]
\label{lem:bhv-decoding}
For the code $C_\lambda$ and predicate-preimage state $\ket{\Psi_H}$, independently
sample $H_i(u)\leftarrow\Bernoulli(\lambda^{-4})$ for every
$i\in[q_\lambda]$ and $u\in\Sigma_\lambda$.
There is a uniform polynomial-time oracle-free quantum decoder
$\BiasYZ$ such that, for all sufficiently large $\lambda$,
with probability at least $1-2^{-\lambda}$ over $H$, every
$H_i^{-1}(b)$ is nonempty and
\[
 \forall y\in\zo^\lambda:\qquad
 \Pr_{v\leftarrow\BiasYZ(y,\ket{\Psi_H}\otimes\ket{C_\lambda})}
      [(v,y)\in\cP_H]\ge1-2^{-\lambda}.
\]
\end{lemma}
\begin{proof}
For this code and hash distribution, \cite[Corollary~5.24]{BHV} gives
the displayed simultaneous decoding guarantee. Its successful event
excludes empty preimages, since the decoder returns $\bot$ if any preimage is empty
(\cite[Definition~5.21 and Figure~1]{BHV}). On this event its advice
state is exactly $\ket{\Psi_H}\otimes\ket{C_\lambda}$, proving the claim.
\end{proof}

On input $(y,\sigma)$, the decoder $\Decode$ first prepares the
public state $\ket{C_\lambda}$ to trace-distance error $2^{-\lambda}$
using the linear encoder above. It then runs $\BiasYZ$ on $y$ and the
combined state, returning its output $v$ if $v\in C_\lambda$ and
$\bot$ otherwise. The preparation error changes any acceptance
probability by at most $2^{-\lambda}$, so the decoding error for
$\ket{\Psi_H}$ is at most $2^{1-\lambda}$. This is the bound used
in the definition of good hashes (\cref{def:good-hashes}).

\subsection{Proof of the codeword-hash guarantees (\texorpdfstring{\cref{lem:code-hash}}{Lemma \getrefnumber{lem:code-hash}})}

We now prove the codeword-hash guarantees (\cref{lem:code-hash}).

\paragraph{Bounding joint inclusion.}
Fix a constant $d\ge1$ and predicates $H_1,\ldots,H_\lambda$, and let $J_j$ and $P_j(H)$
be the suffix-coordinate groups and their constraint sets
(\cref{def:suffix-constraints}). For the joint-inclusion bound,
fix $j\in[d]$ and distinct oracle inputs
$u_r=(v_r,y_r)$, $r\in[t]$.
If some $v_r\notin C_\lambda$, the joint-inclusion probability is zero,
so we may assume that every $v_r$ is a codeword.
If two inputs have the same codeword, their targets differ, so they
cannot both belong to $P_j(H)$. We may therefore assume that the
codewords are distinct. By \cref{lem:bhv-expansion}, they contain
at least $(q_\lambda-4\lambda^4)t$ distinct coordinate--symbol pairs.
We need only those in $J_j$, so write
\[
 S_j=\{(i,(v_r)_i):i\in J_j,\ r\in[t]\}.
\]
Removing coordinates outside $J_j$ deletes at most $(q_\lambda-|J_j|)t$
pairs from the expansion bound. Since the groups $J_1,\ldots,J_d$
partition the suffix coordinates and differ in size by at most one,
this gives
\[
 |S_j|\ge(|J_j|-4\lambda^4)t
 \ge\left(\frac{q_\lambda-\lambda}{d}-1-4\lambda^4\right)t
 \ge\frac{\lambda^5t}{2d}.
\]
Simultaneous membership requires all predicate values indexed by $S_j$ to
be zero. Under $H\leftarrow\cH_\lambda$, these are
independent bits of mean $\lambda^{-4}$, giving
\[
 \Pr[u_1,\ldots,u_t\in P_j(H)]
 \le(1-\lambda^{-4})^{|S_j|}
 \le\exp\left(-\frac{\lambda t}{2d}\right)
 \le2^{-\lambda t/(8d)}.
\]
With the prefix fixed, each $P_j(H)$ depends on a disjoint group of
suffix predicates, giving mutual independence, while their intersection
imposes every suffix constraint and is therefore $\cP_H$.

\paragraph{Fixing the prefix predicates.}
When all predicate values are sampled independently with mean $\lambda^{-4}$,
\cref{lem:bhv-decoding} and the definition of $\Decode$ give
$\Pr_H[H\in\Good_\lambda]\ge1-2^{-\lambda}$. Averaging over
the first $\lambda$ predicates gives a fixed choice of
$H_1,\ldots,H_\lambda$ for which
\[
 \Pr_{H\leftarrow\cH_\lambda}
    [H\in\Good_\lambda]\ge1-2^{-\lambda}.
\]
The decoder is uniform and runs in polynomial time by
\cref{lem:bhv-decoding}, and neither it nor the chosen prefix depends
on the partition into $d$ groups. Together with the code-state
preparation in \cref{app:code-construction}, this proves all three
parts of the lemma.

\clearpage\section{Correctness of the Signature Constructions}
\label{app:signature-correctness}

We prove joint correctness for the adaptive honest operations of
$\CSig$ and $\QSig$ (\cref{lem:signature-correctness}), including
the returned keys and signatures. Use the signature oracle
$\ora=(\orabase,N,A)$ (\cref{def:signature-oracle}) with any
independent additional base components. The ideal honest experiment
starts with uniform $s\in\zo^n$ and signing state
$\ket{\Psi_{H_s}}$, and returns
\[
 \begin{array}{c|c|c}
 &\CSig&\QSig\\ \hline
 \text{signature on }m&\ket{\Psi_{H_{s,m}}}&F_s^{\rm sig}(m)\\
 \text{verification-key copy}&s&(s,\ket{\Psi_{H_s^V}}).
 \end{array}
\]
Each ideal operation leaves its predicate-preimage state unchanged, and verification
uses an exact query to the encoded function in place of $\EvalHash$.
Let $\Pi$ be generation followed by at most $q(n)$ adaptive honest
signing, verification-key-generation and verification calls. For any conditional density matrix $\sigma_W^\ora$, let
$\Pi_{\rm act}^\ora(\sigma_W^\ora)$ and
$\Pi_{\rm id}^\ora(\sigma_W^\ora)$ be the joint output density
matrices obtained by these actual and ideal operations, and put
\[
 \Delta_\Pi(\ora)=\tfrac12\|
 \Pi_{\rm act}^\ora(\sigma_W^\ora)
       -\Pi_{\rm id}^\ora(\sigma_W^\ora)\|_1.
\]
Both outputs include $W$ and all keys and signatures. Thus averaging
$\Delta_\Pi(\ora)$ also bounds the joint distance with the sampled
oracle included as a classical output.
The following lemma bounds this error simultaneously for all such
honest executions.

\begin{lemma}[Signature correctness]\label{lem:signature-correctness}
There are a polynomial $p$ and sets $\cG_n^{\rm sig}$ such that, for
every polynomial $q\ge1$, every honest execution $\Pi$ above and sufficiently
large $n$,
\[
 \begin{aligned}
 \Pr_\ora[\ora\notin\cG_n^{\rm sig}]&\le2^{-n},\\
 \Delta_\Pi(\ora)&\le q(n)p(n)2^{-n}
             &&(\ora\in\cG_n^{\rm sig}).
 \end{aligned}
\]
\end{lemma}
The proof of \cref{lem:signature-correctness} appears in
\cref{sec:signature-correctness-proof}.
Observe that the ideal experiment always accepts, so the same error
bounds honest rejection, and averaging the good event gives
$\bbE_\ora\Delta_\Pi(\ora)\le q(n)p(n)2^{-n}+2^{-n}=2^{-\Omega(n)}$.

\subsection{Proof of signature correctness}
\label{sec:shared-key-algorithms}\label{sec:signature-correctness-proof}

We now prove \cref{lem:signature-correctness}. The signing state
$\ket{\Psi_{H_s}}$ evaluates $P_s$ through
$T_s=T_{H_s,B_s,P_s}$. For an issuance label $t$, the function
$P_s(1,t,u)=E_{H_{s,t}}(u)$ supplies the counts and permutations
\[
 D=|\Sigma_\kappa|,\qquad k_i=|H_i^{-1}(1)|,\qquad
 I_{i,1}=[0,k_i),\quad I_{i,0}=[k_i,D),
\]
where $H=H_{s,t}$ and $\pi_i:[0,D)\to\Sigma_\kappa$ lists
$H_i^{-1}(1)$ before $H_i^{-1}(0)$
(\cref{eq:hash-preimage-enumeration}). Thus $\pi_i$ maps a uniform
rank in $I_{i,b}$ to a uniform preimage of $b$, while $\pi_i^{-1}$
erases the rank coherently. The following algorithms evaluate these
functions and issue the resulting state.

\begin{boxedalgorithm}{Reading signing values and issuing predicate-preimage states}
\noindent\underline{$\mathsf{Read}(s,\regis{sk},u,b)$}\par
Run $\EvalHash^{T_s}$ with the predicate-preimage state register, input $u$ and
response $b$, and return all registers.

\smallskip
\noindent\underline{$\mathsf{Issue}(s,\regis{sk},t)$}\par
\begin{enumerate}[nosep]
 \item Use $\mathsf{Read}$ with zero response bits and measure those
 responses to obtain the counts $k_i$ for $H_{s,t}$.
 If a count is outside $[0,D]$ or either preimage is empty, return
 the default state $\ket{0^{a_\kappa}}$ and the signing register.
 \item For each interval $I_{i,b}=[a,a+M)$, put $\ell=\lceil\log_2 M\rceil$.
 Prepare $2^{-\ell/2}\sum_{j=0}^{2^\ell-1}\ket j$, measure whether $j<M$,
 and add $a$ on success. Retry a failed test polynomially many times,
 using gate accuracy that makes the total failure and approximation
 error at most $2^{-\kappa-1}$ for the intervals and encodings.
 Return the default state if an interval preparation fails.
 \item Use $\mathsf{Read}$ to query $\pi_i$ and $\pi_i^{-1}$:
 \[
  \sum_{j\in I_{i,b}}\ket j\ket0
  \longmapsto\sum_{j\in I_{i,b}}\ket j\ket{\pi_i(j)}
  \longmapsto\sum_{j\in I_{i,b}}\ket0\ket{\pi_i(j)}.
 \]
 \item Return the prepared preimage registers and the signing register,
 discarding workspace.
\end{enumerate}
\end{boxedalgorithm}

We prove the joint guarantee by bounding issuance and verification
at each fixed good oracle, then averaging over the oracle sample.

Issuance requires a joint bound for several encoded calls. We use
the following bound for that operation.
\begin{claim}\label{clm:signature-issuance-error}
Fix $H_s\in\Good_{\kappa_M}$ and an issuance label $t$.
Starting with $\ket{\Psi_{H_s}}$ and any auxiliary system $W$, let
$\rho_{\rm iss}$ be the joint output of $\mathsf{Issue}(s,\ket{\Psi_{H_s}},t)$,
and let $\rho_{\rm id}$ instead append $\ket{\Psi_{H_{s,t}}}$ while
leaving the signing state and $W$ unchanged. Then
\[
 \trdist{\rho_{\rm iss}}{\rho_{\rm id}}\le2^{-\kappa}.
\]
\end{claim}
The proof of \cref{clm:signature-issuance-error} appears in
\cref{sec:signature-issuance-error}.

We first use this bound to prove \cref{lem:signature-correctness}.
\paragraph{A simultaneous good-oracle event.}
Let $\tau^\ora$ be the output of $\Prepare^{R_M,S_M}(1^{n_M})$
and put
\[
 e_K(\ora)=\trdist{\tau^\ora}{\ket{\Omega_M}\!\bra{\Omega_M}}.
\]
We now require all signing hashes and this preparation to be good:
\[
 \begin{aligned}
 \cH_n^{\rm sig}
   &=\{H_s\in\Good_{\kappa_M},\ H_{s,t}\in\Good_\kappa
                                      \text{ for every }s,t\},\\
 \cG_n^{\rm sig}
   &=\cH_n^{\rm sig}\cap\{e_K(\ora)\le2^{-n}\}.
 \end{aligned}
\]
The state-preparation bound (\cref{lem:public-preparation}) gives
$\bbE_\ora e_K\le2^{-n_M}$. There are $2^n$ master hashes and
$2^{2n}+2^n$ child hashes, so the good-hash bound
(\cref{eq:good-hash-probability}) and Markov's inequality imply
\[
 \Pr_\ora[\ora\notin\cG_n^{\rm sig}]
 \le2^{n-\kappa_M}+2^{2n+1-\kappa}+2^{n-n_M}
 \le2^{-n}.
\]

\paragraph{The honest-operation hybrids.}
Fix $\ora\in\cG_n^{\rm sig}$ and let $\mathsf H_0$ run the actual
execution $\Pi$. In $\mathsf H_1$, change only indexed-key preparation
in generation: prepare
$\ket{\Omega_M}$, measure its serial, yielding $s$ and $\ket{\Psi_{H_s}}$.
During the remaining execution, list in chronological order each
issuance, signing-bit $\mathsf{Read}$ and verification $\EvalHash$
call, including the initial verification-key issuance in $\QSigGen$.
Since verification uses at most $r=3n$ checks, there are at most
$T=(q+1)+qn+qr$ such operations. Append identity operations when
necessary to use this common bound.

For $j=1,\ldots,T$, define $\mathsf H_{j+1}$ from $\mathsf H_1$ by running the first
$j$ operations ideally and all later operations as in the actual
algorithms, with these ideal operations:
\begin{enumerate}[nosep]
 \item On issuance label $t$, return $\ket{\Psi_{H_{s,t}}}$,
 leaving the signing state unchanged.
 \item On a signing-bit request $(m,i)$, return
 $F_s^{\rm sig}(m)_i$, leaving the signing state unchanged.
 \item On a verification call for $T_{H,B,\mathfrak{F}}$, query $\mathfrak{F}$ directly
 and leave $\ket{\Psi_H}$ unchanged.
\end{enumerate}
Each experiment uses the fixed $\ora$, chooses its adaptive requests
from its own preceding outputs, and returns all registers produced
by $\Pi$ jointly with $W$. Write $\rho_j^\ora$ for this output.
The preparation error on $\cG_n^{\rm sig}$ gives
\[
 \trdist{\rho_0^\ora}{\rho_1^\ora}\le2^{-n}.
\]
Before the operation changed between $\mathsf H_j$ and
$\mathsf H_{j+1}$, every honest operation is ideal. The master state
therefore equals $\ket{\Psi_{H_s}}$, and each issued child state
equals $\ket{\Psi_{H_{s,t}}}$ when used by that operation. The
issuance bound (\cref{clm:signature-issuance-error}) and encoded evaluation
(\cref{lem:coherent-evaluation}) thus apply with respective errors
\[
 \epsilon(\mathsf{op})=
 \begin{cases}
  2^{-\kappa},&\mathsf{op}\text{ is issuance},\\
  2\sqrt{\delta_{\mathrm{dec}}(\kappa_M)},&\mathsf{op}\text{ is a signing-bit query},\\
  2\sqrt{\delta_{\mathrm{dec}}(\kappa)},&\mathsf{op}\text{ is a verification query},\\
  0,&\mathsf{op}\text{ is an identity operation}.
 \end{cases}
\]
The common ideal prefix has the same distribution as in the fully
ideal experiment. The identical remaining operations cannot increase
trace distance, so averaging the next operation $\mathsf{op}_j$ gives
\[
 \trdist{\rho_j^\ora}{\rho_{j+1}^\ora}
 \le\bbE_{\rm ideal}\epsilon(\mathsf{op}_j)
 \qquad(1\le j\le T).
\]
In $\mathsf H_{T+1}$, every $\CSig$ check queries the constant-one
function, while every $\QSig$ check satisfies
\[
 G_s(m,F_s^{\rm sig}(m))=1.
\]
Thus $\mathsf H_{T+1}$ is the ideal honest experiment and always
accepts. Every execution contains at most $q+1$ issuances, $qn$
signing-bit queries and $qr$ verification queries, so summing the bounds gives
\[
 \begin{aligned}
 \Delta_\Pi(\ora)
 &\le\sum_{j=0}^T\trdist{\rho_j^\ora}{\rho_{j+1}^\ora}\\
 &\le2^{-n}+(q+1)2^{-\kappa}
       +2qn\sqrt{\delta_{\mathrm{dec}}(\kappa_M)}+2qr\sqrt{\delta_{\mathrm{dec}}(\kappa)}
 \le q(n)p(n)2^{-n}.
 \end{aligned}
\]
The distance also bounds the probability of any honest rejection.
Averaging the good-oracle event and its complement proves
\[
 \bbE_\ora\Delta_\Pi(\ora)
 \le q(n)p(n)2^{-n}+\Pr[\ora\notin\cG_n^{\rm sig}]
 =2^{-\Omega(n)}.
\]

\subsubsection{The issuance error}
\label{sec:signature-issuance-error}

We now prove \cref{clm:signature-issuance-error}. Replacing the encoded
calls by exact queries to $P_s$ leaves only the interval preparation
and permutation computations in $\mathsf{Issue}$ to analyze.

For a nonempty preimage, write
\[
 \ket{I_{i,b}}=|I_{i,b}|^{-1/2}\sum_{j\in I_{i,b}}\ket j,
 \qquad
 \ket{H_i^{-1}(b)}=|H_i^{-1}(b)|^{-1/2}
                         \sum_{z\in H_i^{-1}(b)}\ket z.
\]
Because $\pi_i:I_{i,b}\to H_i^{-1}(b)$ is a bijection, the forward
and inverse permutation queries implement
\[
 \ket{I_{i,b}}\ket0
 \longmapsto |I_{i,b}|^{-1/2}\sum_{j\in I_{i,b}}\ket j\ket{\pi_i(j)}
 \longmapsto\ket0\ket{H_i^{-1}(b)}.
\]
Taking the tensor product over all $i,b$ therefore gives exactly
$\ket{\Psi_{H_{s,t}}}$, with the prescribed default state when a
preimage is empty (\cref{def:hash-witness}).

Fix $H_s\in\Good_{\kappa_M}$, and let $q_0(n)$ bound the encoded
bit queries in one issuance. Define three experiments starting with
the exact signing state $\ket{\Psi_{H_s}}$ and the same issuance label:
\begin{enumerate}[label=$\mathsf I_{\arabic*}$,start=0,nosep]
 \item Run $\mathsf{Issue}(s,\ket{\Psi_{H_s}},t)$ with $T_s$.
 \item Run the same algorithm, replacing each
 $\mathsf{Read}(s,\regis{sk},u,b)$ by
 $\ket{u,b}\mapsto\ket{u,b\oplus P_s(u)}$ and leaving
 $\regis{sk}$ unchanged.
 \item Return $\ket{\Psi_{H_{s,t}}}\!\bra{\Psi_{H_{s,t}}}$
 together with the unchanged signing state and $W$.
\end{enumerate}
Let $\rho_i$ be their joint outputs with $W$. To pass from
$\mathsf I_0$ to $\mathsf I_1$, replace the $\mathsf{Read}$ calls
in execution order. Every earlier replacement leaves
$\ket{\Psi_{H_s}}$ unchanged, so encoded evaluation
(\cref{lem:coherent-evaluation}) gives the first estimate below.
For the second, each interval trial succeeds with probability
$M/2^\ell\ge1/2$, and the chosen trial count and gate accuracy bound
the total error in interval preparation and encoding:
\[
 \trdist{\rho_0}{\rho_1}
 \le2q_0(n)\sqrt{\delta_{\mathrm{dec}}(\kappa_M)}\le2^{-\kappa-1},\qquad
 \trdist{\rho_1}{\rho_2}\le2^{-\kappa-1}.
\]
The default output on failure is included in $\rho_0,\rho_1$, so no
conditioning on success is needed. Thus one issuance starting with
$\ket{\Psi_{H_s}}$ differs from its ideal operation by at most
$2^{-\kappa}$, including the signing state and $W$.

\clearpage\section{Proof of the Ordered Switching Lemma (\texorpdfstring{\cref{lem:ordered-switching}}{Lemma \getrefnumber{lem:ordered-switching}})}
\label{app:ordered-switching}

We prove a switching lemma that returns the label of the first matching
database and permits independent answers with different biases.

\begin{lemma}[Ordered switching]\label{lem:ordered-switching}
Fix positive integers $k,h$, a nonnegative integer $m$, a finite
domain $\cX\subseteq\zo^*$, and an ordered collection
$\cC=(D_1,\ldots,D_m)$ of databases $D_i\subseteq\cX\times\zo$,
each with at most $k$ entries and at most one entry for each input.
For each $D_i$,
fix an order of its entries and an arbitrary label $l_i$.
For fixed parameters $0<\beta\le1/2$ and $0\le\rho\le\beta/(8k)$,
let $P\subseteq\cX$ be random, with
\[
 \Pr_P[S\subseteq P]\le\rho^{|S|}
 \quad\text{for every }S\subseteq\cX
 \text{ of size at most }h+k-1.
\]
Independently of $P$, sample
\[
 B(u)\overset{\mathrm{ind}}\leftarrow\Bernoulli(p_u)\quad(u\in\cX),
 \qquad p_u\in[\beta,1-\beta].
\]
To identify the first matching label, define
\[
 j(\ora)=\min\bigl(\{i\in[m]:\ora\text{ agrees with }D_i\}\cup\{m+1\}\bigr),
 \qquad l_{m+1}=\bot.
\]
For $U=B|_{\cX\setminus P}$, there is a fixed mapping
$(P,U)\mapsto T_{P,U}$ with unrestricted local computation.
Its output makes fewer than $h$ classical queries on every oracle,
all in $P$, and satisfies
\[
 \Pr_{P,B}\left[
  \forall \ora\text{ agreeing with }U:\quad T_{P,U}^\ora=l_{j(\ora)}
 \right]
 \ge1-2(4k\rho/\beta)^h.
\]
\end{lemma}
The proof of \cref{lem:ordered-switching} appears in \cref{sec:ordered-switching-proof}.
Observe that the correctness event depends only on $P,U$, since it
requires agreement for every oracle extending $U$.

\subsection{Proof of ordered switching}\label{sec:ordered-switching-proof}

We now prove \cref{lem:ordered-switching}. We construct an algorithm that
tests the databases in order, stopping before its $h$th query, and bound
the probability that this stopping rule changes any output.

\begin{boxedalgorithm}{$T_{P,U}^{\ora}$: first matching database}
Given $(P,U)$ and the ordered, labelled collection $\cC$, the
algorithm has classical access to $\ora$. If it would make query $h$,
it returns $\bot$ instead.
\begin{enumerate}
 \item Initialize the answer database
 $D=\{(u,U(u)):u\in\cX\setminus P\}$.
 \item Delete every $D_i$ containing an entry $(u,a)$ with
 $u\in\dom D$ and $D(u)\ne a$. If none remains, output $\bot$
 and terminate. Otherwise, let $D_i$ be the first remaining database.
 \item Set $Q=\dom D_i\setminus\dom D$. If $Q=\varnothing$,
 output $l_i$ and terminate. Otherwise, for each $u\in Q$ in the
 fixed entry order of $D_i$, query $\ora(u)$ to obtain $a$ and
 set $D(u)=a$. Subject to the stopping rule, complete
 this batch even if an earlier answer contradicts $D_i$.
 \item Repeat from Step~2.
\end{enumerate}
\end{boxedalgorithm}

To bound the probability of a long execution, we encode its queried
positions and answers in a tuple.

\begin{definition}[Encoding a first-match execution]
\label{def:ordered-switching-record}
In an untruncated execution of $T_{P,U}$ that reaches query $h$,
stop after the first batch containing that query. If the execution
has $r$ batches and $s$ distinct queries, we define its \emph{execution encoding} to be
\[
 a=\bigl((s_i)_{i\le r},(Q_i)_{i\le r},(\alpha_i)_{i<r}\bigr),
 \qquad h\le s=\sum_{i=1}^r s_i\le h+k-1,
\]
where $s_i$ is the positive length of batch $i$, $Q_i$ lists its
queried positions in the selected database, and $\alpha_i$ lists
its answers. The final batch's answers are omitted.
\end{definition}

For fixed $s$, there are at most $2^s k^s2^s=(4k)^s$
encodings, accounting respectively for compositions of $s$, queried
positions, and answers. The following bound does not include the
database indices.

\begin{lemma}[Probability of a fixed execution encoding]\label{lem:ordered-switching-record}
Under the hypotheses of \cref{lem:ordered-switching}, fix
$h\le s\le h+k-1$ and an execution encoding $a$ of $s$ queries
(\cref{def:ordered-switching-record}).
Then
\[
 \Pr_{P,B}\!\left[
  \exists \ora\text{ agreeing with }U:
  \text{the untruncated execution has encoding }a
 \right]\le(\rho/\beta)^s.
\]
\end{lemma}
The proof of \cref{lem:ordered-switching-record} appears in
\cref{sec:ordered-execution-proof}.

We now sum over the possible encodings to prove \cref{lem:ordered-switching}.

\begin{proof}[Proof of \cref{lem:ordered-switching}]
The untruncated algorithm tests databases in order and returns
$l_{j(\ora)}$, so truncation can fail only if some assignment to $P$
reaches query $h$. Counting encodings and applying
\cref{lem:ordered-switching-record} gives
\[
 \Pr[\text{some untruncated execution reaches query }h]
 \le\sum_{s=h}^{h+k-1}(4k\rho/\beta)^s
 \le2(4k\rho/\beta)^h.
\]
Outside this event, truncation changes no execution, proving
simultaneous correctness on an event determined by $P,U$.
\end{proof}

\subsection{Bounding the probability of a long execution}
\label{sec:ordered-execution-proof}

We now prove \cref{lem:ordered-switching-record} by recovering the
queried inputs from an independently sampled Boolean function.
Fix an execution encoding $a$ (\cref{def:ordered-switching-record}).
We first define a decoder that uses a Boolean function $W$ to
recover database indices, following Kelley's canonical switching
argument~\cite{Kelley}. For a tuple
$\ell=(\ell_1,\ldots,\ell_r)$, let $S_{a,\ell}$ be the set of
inputs at the positions $Q_i$ of $D_{\ell_i}$ specified by $a$.

\begin{boxedalgorithm}{$J_a(W)$}
\begin{enumerate}[leftmargin=*,itemsep=3pt]
 \item Enumerate the increasing tuples
 $1\le\ell_1<\cdots<\ell_r\le m$ in lexicographic order.
 Discard a tuple if a position in $Q_i$ is invalid for $D_{\ell_i}$
 or if $|S_{a,\ell}|\ne s$.
 \item For each remaining tuple, set $W'=W$ and process $i=1,\ldots,r$:
 \begin{enumerate}[leftmargin=*,itemsep=2pt]
  \item If $W'(u)\ne b$ for some $(u,b)\in D_{\ell_i}$,
  discard this tuple and continue the enumeration in Step~1.
  \item If $i<r$, replace $W'(u)$ by the corresponding answer
  in $\alpha_i$ at each position of $Q_i$ in $D_{\ell_i}$.
 \end{enumerate}
 \item Return the first tuple passing every test. If no tuple
 passes, return $\bot$.
\end{enumerate}
\end{boxedalgorithm}
The decoder uses only $W$ and the answers in $a$, so its selected
inputs will be independent of $P$ whenever $W$ is. To obtain such a
function while preserving the answers outside $P$, independently sample
$X(u)\leftarrow\Bernoulli(p_u)$ at each $u\in\cX$ and set
\[
 W(u)=\begin{cases}X(u),&u\in P,\\U(u),&u\notin P.\end{cases}
\]
The next claim relates a long execution to the decoder's output.

\begin{claim}
For fixed $P,U$, if some execution has encoding $a$, then
\[
 \Pr_X[J_a(W)\ne\bot\ \land\ S_{a,J_a(W)}\subseteq P]\ge\beta^s.
\]
\end{claim}
\begin{proof}
Choose such an execution and write $\ell$ for its database indices.
At batch $i$, the selected database $D_{\ell_i}$ agrees with all
known answers, and its unknown entries are exactly those in $Q_i$.
Consequently, the tests in Step~2 for $\ell$ require $s$ consistent
literals, one at each input in $S_{a,\ell}\subseteq P$. Each literal
holds independently with probability at least $\beta$.

When all these literals hold, $J_a(W)=\ell$. Indeed, a
lexicographically earlier tuple first selects a database that the
execution rejected using $U(u)$ or an answer from an earlier batch.
The function $W'$ supplies that same answer when testing the earlier
database, so the decoder rejects the earlier tuple as well. Thus all
$s$ literals suffice to recover $S_{a,\ell}$, proving the claim.
\end{proof}

Now average the claim over $P,U$. Conditional on $P$, the bits of $W$ have
the product distribution of $B$, so $W$ is independent of $P$.
Whenever $J_a(W)\ne\bot$, its selected set $S_{a,J_a(W)}$ has
size $s$ and is fixed by $W$, and hence the joint-inclusion hypothesis gives
\[
 \Pr_{P,W}[J_a(W)\ne\bot\ \land\ S_{a,J_a(W)}\subseteq P]
 \le\rho^s.
\]
Therefore the claim and the joint-inclusion bound give
\[
 \Pr_{P,B}[\text{some execution has encoding }a]
 \le\beta^{-s}
 \Pr_{P,W}[J_a(W)\ne\bot\ \land\ S_{a,J_a(W)}\subseteq P]
 \le(\rho/\beta)^s,
\]
which proves \cref{lem:ordered-switching-record}.

\clearpage\section{Estimating Acceptance and Searching Classical Inputs}\label{sec:recursive-qcma}

In this section, we use the recursive oracle $\ora=(\orabase,N,A)$
(\cref{def:recursive-qcma}) to maximize acceptance over classical
inputs and decide promise QCMA (\cref{lem:helper-estimates}).

For a valid length-$n$ description $a=\langle M,x\rangle$, where $M$
is a witnessless quantum circuit with classical input $x$, define
\begin{equation}\label{eq:helper-estimates}
 \begin{gathered}
 v_a=\Pr[M^\ora(x)=1],\qquad
 \widehat v_a=\frac3{n^4}\sum_{i=1}^{n^4}A(a,i)-1,\\
 \cE_n=\{\forall\text{ valid }a\in\zo^n:
                  |\widehat v_a-v_a|\le3/n\}.
 \end{gathered}
\end{equation}
A test family is a polynomial-time map
$a:\zo^\ell\to\zo^n$ to such padded pairs, with $n$, $\ell$ and
evaluation time polynomial in the explicit input length.

\begin{lemma}[Estimation, search and collapse]\label{lem:helper-estimates}
Over any base distribution, there is a deterministic polynomial-time
oracle algorithm $\mathsf{Search}$ such that, for every test family $a$,
\[
 \begin{aligned}
 \Pr_\ora[\neg\cE_n]&\le2^{n+1}e^{-2n^2},\\
 \cE_n\quad&\Longrightarrow\quad
 v_{a(\mathsf{Search}^{\ora}(a))}
       \ge\max_{z\in\zo^\ell}v_{a(z)}-6/n.
 \end{aligned}
\]
Moreover, with probability one over $\ora$,
\[
 \PTIME^\ora=\BQP^\ora=\QCMA^\ora
 \qquad\text{for languages and promise problems}.
\]
\end{lemma}
The proof of \cref{lem:helper-estimates} appears in
\cref{sec:helper-estimates-proof}.

\subsection{Proof}\label{sec:helper-estimates-proof}

We now prove \cref{lem:helper-estimates}. The recursive $N$ oracle
lets a classical algorithm maximize the estimate $\widehat v_a$.
A simultaneous bound on that estimate will then turn maximization
into both approximate search and a decision procedure for promise QCMA.
\paragraph{Step 1: Maximize the acceptance estimate.}
Define
\[
 \begin{gathered}
 s(z)=\sum_{i=1}^{n^4}A(a(z),i),\qquad
 H(p,t)=\indicator[\exists z\in\zo^\ell:
                         p\text{ is a prefix of }z,\ s(z)\ge t].
 \end{gathered}
\]
To compute $H(p,t)$, form the deterministic verifier $M_{\rm search}$
with $a,n,\ell$ in its description. On ordinary input $(p,t)$ and
witness $z\in\zo^\ell$, it rejects unless $p$ is a prefix of $z$,
queries $A(a(z),i)$ for $i\in[n^4]$, and accepts exactly when their
sum is at least $t$. Pad $\langle M_{\rm search},(p,t)\rangle$ to a
polynomial length that bounds the verifier's size, witness length and
query lengths, then query $N(\langle M_{\rm search},(p,t)\rangle)$.
The recursive oracle (\cref{def:recursive-qcma}) returns $H(p,t)$,
since $M_{\rm search}$ uses the same fixed $A$ as the definition of
$s(z)$. The algorithm $\mathsf{Search}^{\ora}(a)$ is then:
\begin{enumerate}[leftmargin=*,itemsep=2pt]
 \item Binary-search $t\in\{0,\ldots,n^4\}$ to find
 $s_* =\max\{t:H(\varnothing,t)=1\}$.
 \item Set $p=\varnothing$. While $|p|<\ell$, append $0$ if
 $H(p0,s_*)=1$, and $1$ otherwise.
 \item Return $z_*=p$.
\end{enumerate}
By the definition of $s_*$, the initial prefix satisfies
$H(\varnothing,s_*)=1$. Each prefix step preserves this condition,
since at least one of $p0,p1$ extends to a string whose sum is at least
$s_*$. At termination the prefix itself is the returned string, so
\[
 s(z_*)=s_* =\max_{z\in\zo^\ell}s(z).
\]
Thus $\mathsf{Search}$ uses $O(\log n+\ell)$ queries to $N$, each of
polynomial length.

\paragraph{Step 2: Bound the estimation error.}
We now compare the maximized estimate with actual acceptance.
Conditional on the shorter oracle determining $v_a$, the defining
sampling of $A$ is
\[
 A(a,i)\overset{\mathrm{ind}}\leftarrow
 \Bernoulli((1+v_a)/3),\qquad i\in[n^4].
\]
Hoeffding's inequality followed by a union bound over the at most
$2^n$ valid descriptions gives
\[
 \begin{aligned}
 \Pr_\ora[|\widehat v_a-v_a|>3/n]&\le2e^{-2n^2},\\
 \Pr_\ora[\neg\cE_n]&\le2^{n+1}e^{-2n^2}.
 \end{aligned}
\]
On $\cE_n$, let
$z_{\rm opt}\in\arg\max_zv_{a(z)}$. Since
$\widehat v_{a(z)}=3s(z)/n^4-1$, Step~1 implies
\[
 \begin{aligned}
 v_{a(z_*)}
 &\ge\widehat v_{a(z_*)}-3/n\\
 &\ge\widehat v_{a(z_{\rm opt})}-3/n
 \ge v_{a(z_{\rm opt})}-6/n.
 \end{aligned}
\]
This proves the search bound.

\paragraph{Step 3: Decide promise QCMA.}
To obtain a decision procedure for every sufficiently padded
description, observe that the failure bound in Step~2 is summable.
The Borel--Cantelli lemma therefore gives
$\cE_n$ for all $n\ge n_0(\ora)$ with probability one. Fix such an
oracle and a QCMA verifier $V$, viewed as a witnessless quantum
circuit on ordinary input $(x,w)$. On input $x$, form the descriptions
$a(x,w)=\langle V,(x,w)\rangle$ for classical witnesses $w$, padding
all pairs to the same polynomial length
\[
 n\ge\max\{72,n_0(\ora)\}.
\]
The finite constant $n_0(\ora)$ is hardwired into the resulting
classical machine. On input $x$, use the $N$ query from Step~1
with the test family $a(x,\cdot)$ to output
\[
 H(\varnothing,\lfloor n^4/2\rfloor+1)
 =\indicator[\exists w:\widehat v_{a(x,w)}>1/2].
\]
On $\cE_n$, the estimation bound gives
\[
 \begin{aligned}
 x\in L_{\mathrm{yes}}&\ \Longrightarrow\
  \exists w:\widehat v_{a(x,w)}\ge2/3-3/n>1/2,\\
 x\in L_{\mathrm{no}}&\ \Longrightarrow\
  \forall w:\widehat v_{a(x,w)}\le1/3+3/n<1/2.
 \end{aligned}
\]
Thus this deterministic polynomial-time algorithm is correct on
both sides of the promise and returns a bit on every input. Therefore
$\QCMA^\ora\subseteq\PTIME^\ora$ for promise problems, and the
inclusions $\PTIME^\ora\subseteq\BQP^\ora\subseteq\QCMA^\ora$
complete the proof.

\paperappendixpart{Cryptographic Security Proofs}
\section{Copy-Protection Definitions}\label{app:copy-protection-definitions}
\label{sec:crypto-security-games}

We define copy-protection for one supplied key and two independent
challenges. Every party has quantum access to the same oracle $\ora$.
After the split, the two recipients do not communicate, although
their registers may be entangled.

\begin{definition}[Copy-protected PRFs]\label{def:prf-security}
For a one-bit-output PRF (\cref{predef:prf}),
consider the following experiment.
\begin{boxedalgorithm}{PRF copy-protection}
\begin{enumerate}[nosep]
 \item Generate $(j,\regis{sk})\leftarrow\gen^{\ora}(1^\lambda)$ with function index $k$, and run
 $(\regi_1,\regi_2)\leftarrow\cA_0^{\ora}(j,\regis{sk})$.
 \item Independently sample $x_1,x_2\leftarrow\zo^{m(\lambda)}$ and run
 $b_i\leftarrow\cA_i^{\ora}(x_i,\regi_i)$ for $i=1,2$.
 \item Return $\indicator[b_1=F_k(x_1)\land b_2=F_k(x_2)]$.
\end{enumerate}
\end{boxedalgorithm}
The PRF is copy-protected if, for every \QPT{} $\cA$,
\begin{equation}\label{eq:function-piracy-game}
 \Pr[b_1=F_k(x_1)\land b_2=F_k(x_2)]\le\tfrac12+\negl(\lambda).
\end{equation}
\end{definition}

\begin{definition}[Copy-protected PKE]\label{def:pke-piracy}
For a bit-encryption scheme (\cref{predef:pke}), define the following experiment:
\begin{boxedalgorithm}{$\Piracy^{\ora}_{\cA}(\lambda)$}
\begin{enumerate}[nosep]
 \item Generate $(\pk,\regis{sk})\leftarrow\gen^{\ora}(1^\lambda)$
 and run
 \[
 ((m_{i,0},m_{i,1})_{i=1}^2,\regi_1,\regi_2)
 \leftarrow\cA_0^\ora(\pk,\regis{sk}),\qquad m_{i,b}\in\zo.
 \]
 \item Independently for $i=1,2$, sample $b_i\leftarrow\zo$ and
 $\ct_i\leftarrow\enc^{\ora}(\pk,m_{i,b_i})$.
 Run $\widehat b_i\leftarrow\cA_i^\ora(\ct_i,\regi_i)$.
 \item Return $\indicator[\widehat b_1=b_1\land\widehat b_2=b_2]$.
\end{enumerate}
\end{boxedalgorithm}
We call the scheme \emph{copy-protected} if, for every \QPT{} $\cA$,
\[
 \Pr[\Piracy^{\ora}_{\cA}(\lambda)=1]
 \le\tfrac12+\negl(\lambda).
\]
\end{definition}

\clearpage\section{Security of Direct Coset Evaluation}
\label{app:crypto-security}\label{app:coset-security}

In this section, we prove \cref{lem:two-fresh-values}, the direct
coset bound used in the PRF and encryption proofs
(\cref{sec:prf-security-proof,sec:pke-security-proof}). The bound
applies to independent function-value and ciphertext challenges
with access to the two-basis oracle (\cref{eq:two-basis-oracle}).
We also prove the public-key quantum money guarantee
(\cref{cor:public-key-money}) in \cref{app:public-key-money-proof}.

We first define the two challenge experiments. The functions are
sampled by replacing their values at the challenge inputs, which
leaves them uniform while separating the fresh values from any
auxiliary information held before the challenges.

\begin{definition}[Coset experiments]\label{def:coset-experiments}
Fix $n\ge1$. We define the \emph{direct coset experiments} as follows.
\begin{enumerate}[leftmargin=*,itemsep=5pt]
\item Independently sample uniform Boolean functions
$G_0,G_1:\zo^n\to\zo$ and
\[
 A\leftarrow\{V\le\bbF_2^{4n}:\dim V=2n\},\quad
 a,b\leftarrow\bbF_2^{4n},\quad x_1,x_2\leftarrow\zo^n.
\]
Sample independent fair bits
$(b_{i,0},b_{i,1})$ for each distinct $x_i$, using the same pair
when $x_1=x_2$. Set
\[
 F_c=G_c[x_i\leftarrow b_{i,c}:i\in\{1,2\}],\qquad
 F=F_0\oplus F_1,\qquad C_F=C_{A,a,b,F_0,F_1}.
\]
Its query $(x,c,z)$ returns $F_c(x)$ when $z\in A+a$ for $c=0$
or $z\in A^\perp+b$ for $c=1$, and returns $\bot$ otherwise.
Write $G=(G_0,G_1)$.
\item Fix algorithms $(\cA_0,\cA_1,\cA_2)$ whose descriptions
may depend on $G$, and give $\cA_0$ the input density matrix
$\ket{A_{a,b}}\!\bra{A_{a,b}}\otimes\sigma_J^G$, where
$\sigma_J^G$ is an arbitrary classical--quantum state depending
only on $G$. For function-value challenges, run
\[
 (\regi_1,\regi_2)\leftarrow\cA_0^{C_F}(\ket{A_{a,b}},\sigma_J^G),
 \qquad \widehat f_i\leftarrow\cA_i^{C_F}(x_i,\regi_i),
\]
and return $\indicator[\widehat f_1=F(x_1)\land\widehat f_2=F(x_2)]$.
\item For ciphertext challenges, instead run
\[
 \begin{aligned}
 ((m_{i,0},m_{i,1})_{i=1}^2,\regi_1,\regi_2)
 &\leftarrow\cA_0^{C_F}(\ket{A_{a,b}},\sigma_J^G),\\
 \beta_1,\beta_2&\leftarrow\zo,\qquad
 \ct_i=(x_i,m_{i,\beta_i}\oplus F(x_i)),\\
 \widehat\beta_i&\leftarrow\cA_i^{C_F}(\ct_i,\regi_i),
 \end{aligned}
\]
and return $\indicator[\widehat\beta_1=\beta_1\land\widehat\beta_2=\beta_2]$.
The challenge bits are independent, $\regi_1,\regi_2$ may be entangled,
and $\cA_1,\cA_2$ act without communication after $\cA_0$ terminates.
All three algorithms have unrestricted local computation.
\end{enumerate}
\end{definition}

The following bound applies to both challenges.

\begin{lemma}[Security of the coset experiments]\label{lem:two-fresh-values}
In either direct coset experiment (\cref{def:coset-experiments}), every
adversary making at most $Q$ total quantum queries to $C_F$ satisfies
\begin{equation}\label{eq:coset-function-security}
 \Pr[\text{both challenges are answered correctly}]
 \le\tfrac12+O\bigl((Q+1)2^{-n/2}\bigr).
\end{equation}
\end{lemma}
The proof of \cref{lem:two-fresh-values} appears in
\cref{sec:direct-coset-security-proof}.

\subsection{Proof of the direct coset bound}
\label{sec:direct-coset-security-proof}

We prove \cref{lem:two-fresh-values} using two supporting lemmas:
decisional coset monogamy with membership queries, and a reduction
from fresh function values and ciphertexts to that game.

\begin{lemma}[Decisional monogamy with oracle access]
\label{lem:query-decisional-moe}
Fix $n\ge1$ and $Q\ge0$, and independently sample
\[
 A\leftarrow\{V\le\bbF_2^{4n}:\dim V=2n\},\qquad
 a,b\leftarrow\bbF_2^{4n}.
\]
Let $M_0$ and $M_1$ test membership in $A+a$ and $A^\perp+b$.
The algorithms have unrestricted local computation and may receive
auxiliary information independent of $(A,a,b)$. Run $(\regi_1,\regi_2)\leftarrow\cA_0^{M_0,M_1}(\ket{A_{a,b}})$, then
independently sample
\[
 u\leftarrow A\setminus\{0\},\qquad
 v\leftarrow A^\perp\setminus\{0\},\qquad
 e=\langle u,b\rangle\oplus\langle v,a\rangle,
\]
and run $e_i\leftarrow\cA_i^{M_0,M_1}(u,v,\regi_i)$ without
communication, optionally giving $A$ to $\cA_1,\cA_2$ after
$\cA_0$ terminates. Their registers may be entangled. If all three
algorithms make at most $Q$ total queries, then
\[
 \Pr_{A,a,b,u,v}[e_1=e_2=e]
 \le\tfrac12+O\bigl((Q+1)2^{-n/2}\bigr).
\]
\end{lemma}

The proof of \cref{lem:query-decisional-moe} appears in
\cref{sec:decisional-coset-proof}.

The next lemma changes only the oracle used by $\cA_0$, so that
its queries reveal no fresh challenge values.

\begin{lemma}[Reduction from fresh values to coset monogamy]
\label{lem:fresh-value-reduction}
In the direct coset experiments (\cref{def:coset-experiments}),
condition on $x_1\ne x_2$, using the input state
$\ket{A_{a,b}}\!\bra{A_{a,b}}\otimes\sigma_J^G$.
Run $\cA_0$ with $C_G=C_{A,a,b,G_0,G_1}$ and $\cA_1,\cA_2$ with $C_F$.
For each fixed $G=(G_0,G_1)$, any $Q$-query adversary induces a
strategy in \cref{lem:query-decisional-moe} using at most $4Q$
membership queries, with success equal to the adversary's in the
function-value experiment and at least its success in the ciphertext
experiment.
\end{lemma}
The proof of \cref{lem:fresh-value-reduction} appears in
\cref{sec:fresh-value-reduction-proof}.

\paragraph{Combining the two bounds.}
We now prove \cref{lem:two-fresh-values} by a query hybrid.
Put $N=2^n$ and first condition on $x_1\ne x_2$. Define
\[
 \begin{array}{c|c|c}
 &\text{oracle for }\cA_0&\text{oracle for }\cA_1,\cA_2\\\hline
 \mathsf H_0&C_F&C_F\\
 \mathsf H_1&C_G&C_F.
 \end{array}
\]
Both experiments use the conditional sampling and the displayed
calls to $\cA_0,\cA_1,\cA_2$ in the coset experiment
(\cref{def:coset-experiments}), with the same input density matrix
and winning predicate. Thus only $\cA_0$'s oracle changes in
$\mathsf H_1$, while $\cA_1,\cA_2$ still query $C_F$.
Write $p_i=\Pr[\mathsf H_i=1\mid x_1\ne x_2]$. For any query by $\cA_0$ in $\mathsf H_1$, let $w_x$ be its weight on input $x$.
Conditional on $(G,A,a,b)$, the query state, including $\sigma_J^G$, is independent of the uniform distinct inputs
$x_i$, so
\[
 \bbE_{x_1\ne x_2}[w_{x_1}+w_{x_2}]
 =\frac2N\sum_{x\in\zo^n}w_x\le\frac2N.
\]
Since the functions differ only at $x_1,x_2$, the query hybrid bound
(\cref{lem:query-hybrid}) applies with expected query weight at most
$2/N$ and gives
\[
 |p_0-p_1|\le2Q\sqrt{2/N}.
\]
Now the fresh-value reduction and decisional monogamy
(\cref{lem:fresh-value-reduction,lem:query-decisional-moe}) give
$p_1\le1/2+O((Q+1)2^{-n/2})$. Finally,
\[
 \Pr[\text{both win}]
 \le p_0+\Pr[x_1=x_2]
 \le\tfrac12+O((Q+1)2^{-n/2})+N^{-1},
\]
which proves the claimed bound for both challenge rules.

\subsection{Proof of decisional coset monogamy}
\label{sec:decisional-coset-proof}

We now prove \cref{lem:query-decisional-moe}, a quantitative
adaptation of the oracle formulation in \cite[footnote~15]{BCG26}.

\paragraph{Replacing membership by larger sets.}
Conditional on $A$, independently sample
\[
 L\leftarrow\{V\le A:\dim V=n\},\qquad
 U\leftarrow\{V\ge A:\dim V=3n\}.
\]
Consider the three games
\[
 \begin{array}{c|c|c}
 &M_0&M_1\\\hline
 \mathsf H_0&\indicator[A+a]&\indicator[A^\perp+b]\\
 \mathsf H_1&\indicator[U+a]&\indicator[A^\perp+b]\\
 \mathsf H_2&\indicator[U+a]&\indicator[L^\perp+b].
 \end{array}
\]
Writing $(M_0^{(j)},M_1^{(j)})$ for the displayed pair, game
$\mathsf H_j$ runs
\[
 (\regi_1,\regi_2)\leftarrow\cA_0^{M_0^{(j)},M_1^{(j)}}(\ket{A_{a,b}}),
 \qquad e_i\leftarrow\cA_i^{M_0^{(j)},M_1^{(j)}}(u,v,\regi_i),
\]
where $(u,v)$ is sampled after the first call exactly as in the
lemma. It returns $\indicator[e_1=e_2=e]$, using the original
$e=\langle u,b\rangle\oplus\langle v,a\rangle$.
For any point outside its original coset, the probability of entering
the corresponding larger set is
\[
 \beta_n=\frac1{2^n+1}\le2^{-n}.
\]
In $\mathsf H_0$, the query state is independent of $U$ conditional
on $(A,a,b)$, so the query hybrid bound (\cref{lem:query-hybrid})
applied to the added points of $U+a$ gives
\[
 |\Pr[\mathsf H_0=1]-\Pr[\mathsf H_1=1]|\le2Q\sqrt{\beta_n}.
\]
Similarly, queries in $\mathsf H_1$ are independent of $L$
conditional on $(A,a,b,U)$, and the added points of $L^\perp+b$
therefore give
\[
 |\Pr[\mathsf H_1=1]-\Pr[\mathsf H_2=1]|\le2Q\sqrt{\beta_n}.
\]

The next claim bounds the game with both larger-set membership oracles.
\begin{claim}\label{clm:larger-coset-game}
For $\mathsf H_2$ defined above,
\[
 \Pr[\mathsf H_2=1]\le\tfrac12+O(2^{-n}).
\]
The bound still holds when $A$ is given to the two later algorithms.
\end{claim}
The proof of \cref{clm:larger-coset-game} appears in
\cref{sec:larger-coset-game}.

Combining this bound with the two query hybrids gives
\[
 \begin{aligned}
 \Pr[\mathsf H_0=1]
 &\le\Pr[\mathsf H_2=1]
   +\sum_{i=0}^1|\Pr[\mathsf H_i=1]-\Pr[\mathsf H_{i+1}=1]|\\
 &\le\tfrac12+4Q\sqrt{\beta_n}+O(2^{-n})
 =\tfrac12+O((Q+1)2^{-n/2}).
 \end{aligned}
\]

\subsubsection{The larger-set game}
\label{sec:larger-coset-game}

We now prove \cref{clm:larger-coset-game} by expressing the remaining
hidden coset in $2n$ coordinates. Revealing the other coordinates
makes both larger-set membership oracles computable.

Sample $\sigma\leftarrow U+a$ and $\tau\leftarrow L^\perp+b$,
and give $(L,U,\sigma,\tau)$ to $\cA_0$ and both later algorithms. Both larger-set
membership oracles can now be evaluated from
$U+a=U+\sigma$ and $L^\perp+b=L^\perp+\tau$. Choose a linear
isomorphism $T$ determined by $(L,U)$ such that
\[
 \begin{aligned}
 T(L)&=\bbF_2^n\oplus0\oplus0,\\
 T(A)&=\bbF_2^n\oplus\bar A\oplus0,\\
 T(U)&=\bbF_2^n\oplus\bbF_2^{2n}\oplus0,
 \end{aligned}
 \qquad
 \bar A\leftarrow\{V\le\bbF_2^{2n}:\dim V=n\},
\]
where the second equality defines $\bar A$. Write $U_T\ket z=\ket{Tz}$.
Applying $U_T X^\sigma Z^\tau$ to the supplied state gives, up to
global phase,
\[
 \ket+^{\otimes n}\otimes\ket{\bar A_{\bar a,\bar b}}
       \otimes\ket0^{\otimes n},
 \qquad \bar a,\bar b\leftarrow\bbF_2^{2n}\text{ independently}.
\]
Writing $\bar u,\bar v$ for the middle coordinates of
$Tu,T^{-\mathsf T}v$, respectively, decomposes the challenge as
\[
 e=\underbrace{\langle u,\tau\rangle\oplus\langle v,\sigma\rangle}
                  _{\text{known after the challenge}}
   \oplus\langle\bar u,\bar b\rangle
   \oplus\langle\bar v,\bar a\rangle.
\]
To apply the decisional bound, exclude the event $\mathsf{Bad}=\{\bar u=0\text{ or }\bar v=0\}$,
which has probability at most $2\beta_n$. Outside it, $\bar u,\bar v$
are independent uniform nonzero vectors of $\bar A,\bar A^\perp$.
Define $(\mathcal B_0,\mathcal B_1,\mathcal B_2)$ for the residual
game by the following calls:
\begin{enumerate}[nosep]
 \item Algorithm $\mathcal B_0$, on input
       $\ket{\bar A_{\bar a,\bar b}}$, forms the
       three-factor state above and applies
       $(U_T X^\sigma Z^\tau)^{-1}$. Sample independent uniform
       $r,t\in\zo^n$ for the two local simulators. Give the state to
       $\cA_0$, answering its queries by membership in
       $U+\sigma$ and $L^\perp+\tau$. If $\cA_0$ returns
       $(\regi_1,\regi_2)$, return $(\regi_1,r,t)$ to $\mathcal B_1$ and
       $(\regi_2,r,t)$ to $\mathcal B_2$.
 \item Algorithm $\mathcal B_i$, on input
       $(\bar u,\bar v,\regi_i,r,t)$, sets
       $u=T^{-1}(r,\bar u,0)$ and
       $v=T^{\mathsf T}(0,\bar v,t)$, and runs
       $e_i\leftarrow\cA_i^{\indicator[U+\sigma],\indicator[L^\perp+\tau]}
       (u,v,\regi_i)$, then outputs
       $e_i\oplus\langle u,\tau\rangle\oplus\langle v,\sigma\rangle$.
\end{enumerate}
Conditional on $(L,U,\sigma,\tau)$ and $\neg\mathsf{Bad}$, these
operations give the original state and challenge distribution, as
in the change-of-basis argument of \cite[proof of Theorem~22]{BCG26}.
The outputs both equal the residual challenge bit exactly when
both original outputs equal $e$. Thus \cite[Corollary~3]{BCG26}
bounds the conditional success probability by $1/2+O(2^{-n})$.
The bound permits giving $A$ to $\mathcal B_1,\mathcal B_2$, since they
can reconstruct it from $\bar A,L,U,T$ without either residual shift.
Finally, $\Pr[\mathsf{Bad}]\le2\beta_n\le2^{1-n}$, so
\[
 \Pr[\mathsf H_2=1]
 \le\Pr[\mathsf H_2=1\mid\neg\mathsf{Bad}]+\Pr[\mathsf{Bad}]
 \le\tfrac12+O(2^{-n}).
\]

\subsection{Proof of the fresh-value reduction}
\label{sec:fresh-value-reduction-proof}

We now prove \cref{lem:fresh-value-reduction} for the direct coset
experiments (\cref{def:coset-experiments}), by expressing the fresh
function bits using the monogamy challenge and independent fair bits.

\paragraph{Algorithm $\mathcal B_0$.}
Fix $G$. On input $\ket{A_{a,b}}$, algorithm $\mathcal B_0^{M_0,M_1}$
prepares $\sigma_J^G$ and samples
\[
 (x_1,x_2)\leftarrow\{(x_1,x_2)\in(\zo^n)^2:x_1\ne x_2\},\qquad
 (r_1,r_2,\zeta_1,\zeta_2)\leftarrow\zo^4.
\]
Run $\cA_0^{C_G}(\ket{A_{a,b}},\sigma_J^G)$, implementing
$C_G(x,c,z)$ by returning $G_c(x)$ if $M_c(z)=1$ and $\bot$
otherwise, then uncomputing the membership test. If $\cA_0$
outputs registers $\regi_1,\regi_2$ and, in the ciphertext experiment,
message pairs, give $\mathcal B_i$ its register $\regi_i$, both message
pairs and $(x_1,x_2,r_1,r_2,\zeta_1,\zeta_2)$.
These classical values specify the oracle answers below and are
not given to $\cA_i$.

\paragraph{Algorithms $\mathcal B_1,\mathcal B_2$.}
On receiving the monogamy challenge $(u,v)$, each $\mathcal B_i$
implements the oracle below for $\cA_i$. To verify its distribution, define
\begin{equation}\label{eq:moe-branch-values}
 b_{i,0}=\langle v,a\rangle\oplus r_i,
 \qquad
 b_{i,1}=\langle u,b\rangle\oplus r_i\oplus\zeta_i.
\end{equation}
These values describe the simulated function rather than a
computation of the unknown shifts $a,b$. The algorithm $\mathcal B_i^{M_0,M_1}$ answers each query of
$\cA_i$ by the following coherent computation.

\begin{boxedalgorithm}{Answering a query of $\cA_i$}
On query $(x,c,z)$, perform these steps coherently.
\begin{enumerate}
 \item Test $z\in A+a$ when $c=0$, or $z\in A^\perp+b$ when $c=1$,
 returning $\bot$ if the test fails.
 \item For $x\notin\{x_1,x_2\}$, return $G_c(x)$.
 \item For $x=x_j$, return $\langle v,z\rangle\oplus r_j$ when $c=0$,
 or $\langle u,z\rangle\oplus r_j\oplus\zeta_j$ when $c=1$.
\end{enumerate}
Uncompute the membership test after copying the answer.
\end{boxedalgorithm}

This implements $C_F$ exactly because
\[
 \begin{aligned}
 z\in A+a&\Longrightarrow\langle v,z\rangle=\langle v,a\rangle,\\
 z\in A^\perp+b&\Longrightarrow\langle u,z\rangle=\langle u,b\rangle.
 \end{aligned}
\]
Including uncomputation, each simulated query uses at most four
membership queries. For every fixed coset and monogamy challenge,
the independent fair bits $(r_i,\zeta_i)$ make the pairs
$(b_{i,0},b_{i,1})$ independent and uniform, with
\[
 F(x_i)=e\oplus\zeta_i,
 \qquad e=\langle v,a\rangle\oplus\langle u,b\rangle.
\]

\paragraph{Function-value challenges.}
Algorithm $\mathcal B_i$ runs
$\widehat f_i\leftarrow\cA_i^{C_F}(x_i,\regi_i)$ using the displayed oracle answers and outputs $\widehat e_i=\widehat f_i\oplus\zeta_i$. Since
$F(x_i)=e\oplus\zeta_i$, the winning events agree:
\[
 \widehat f_1=F(x_1)\land\widehat f_2=F(x_2)
 \quad\Longleftrightarrow\quad
 \widehat e_1=\widehat e_2=e.
\]

\paragraph{Ciphertext challenges.}
Each $\mathcal B_i$ has both message pairs output by $\cA_0$.
Suppose first that $m_{i,0}\ne m_{i,1}$ for both $i$.
Independently, $\mathcal B_i$ samples $c_i\leftarrow\zo$ and runs
$\widehat\beta_i\leftarrow\cA_i^{C_F}((x_i,c_i),\regi_i)$ using the
displayed oracle answers. The implied bits
\[
 \beta_i=c_i\oplus m_{i,0}\oplus F(x_i)
\]
are independent and uniform conditional on the messages and $F$,
and satisfy $c_i=m_{i,\beta_i}\oplus F(x_i)$. Thus these are exactly
the required independent ciphertext challenges. Translate the outputs by
\[
 \widehat e_i=
 \widehat\beta_i\oplus m_{i,0}\oplus c_i\oplus\zeta_i,
\]
so both translated outputs equal $e$ exactly when both ciphertext
guesses are correct.

If either message pair is equal, $\mathcal B_1$ and $\mathcal B_2$ instead
output the saved bit $r_1$. This bit is uniform and independent of
$\cA_0$'s output and the monogamy challenge, so the branch succeeds
with probability $1/2$. In the original ciphertext experiment, the
corresponding challenge bit is independent of the joint input and outputs of $\cA_1,\cA_2$, making their joint success at most $1/2$. The overall strategy
therefore succeeds at least as often as the ciphertext adversary,
using at most $4Q$ membership queries.

\subsection{Public-key quantum money}
\label{app:public-key-money-proof}

We prove \cref{cor:public-key-money} by authenticating a coset
mini-scheme with $\CSig$. Repeating the encoded validity test is
necessary because its off-preimage answers are random bits.
Use the cryptographic oracle
$\ora=(\orabase,N,A)$ with
$\orabase=(\ora_{\mathsf{prf}},\ora_{\mathsf{pke}},
\ora_{\mathsf{sig}},\ora_{\mathsf{prs}})$
(\cref{def:full-crypto-oracle}). The mini-scheme uses the encryption
encoding $T_s=T_{H_s,B_s,\mathfrak{F}_{C_s}}$ and key
$\ket{K_s}=\ket{\Psi_{H_s}}\ket{\chi_s}$
(\cref{def:pke-oracle}).

\begin{boxedalgorithm}{The coset mini-scheme}
\noindent\underline{Generation on input $1^n$}\par
Run $(s,(s,\regis{note}))\leftarrow\PKEGen(1^n)$ and return
$(s,\regis{note})$.

\smallskip
\noindent\underline{Verification of $(s,\regis{note})$}\par
Reject incorrect lengths and write $\regis{note}=\regis{hash}\regis{coset}$
for its predicate-preimage state register and $4n$-qubit coset register. Put $r=3n$.
\begin{enumerate}[nosep]
 \item For $c=0,1$, first apply $\Hadamard^{\otimes4n}$ to $\regis{coset}$
 when $c=1$, then run $r$ checks. In each check, independently sample
 $x\leftarrow\zo^n$, use $\EvalHash^{T_s}$ with key $\regis{hash}$,
 input $(x,c,\regis{coset},0)$ and a zero response bit, then measure
 that bit.
 Reject unless the bit is one.
 \item Apply $\Hadamard^{\otimes4n}$ to $\regis{coset}$. Accept and
 return the updated note.
\end{enumerate}
Each check uses fresh private workspace, which is discarded after
uncomputing the decoder. Here $j=0$ is the validity bit of the
Boolean encoding $\mathfrak{F}_{C_s}$ (\cref{eq:two-basis-oracle}).
\end{boxedalgorithm}

The mini-scheme counterfeiting game gives $(s,\regis{note})$ to
an adversary, which returns two registers. It wins if both pass
verification under the same serial $s$. The adversary has full
access to $\ora$, polynomial query count and length, and unrestricted
local computation. We use the following search
bound for a single coset state.

\begin{lemma}[Coset search hardness, \cite{CLLZ}]
\label{lem:coset-search-hardness}
Independently sample
$A\leftarrow\{V\le\bbF_2^{4n}:\dim V=2n\}$ and
$a,b\leftarrow\bbF_2^{4n}$, and let $M_0,M_1$ test membership in
$A+a,A^\perp+b$. Every algorithm with unrestricted local computation and auxiliary
information independent of $(A,a,b)$, given one copy of
$\ket{A_{a,b}}$ and polynomially many quantum queries to
$M_0,M_1$, satisfies
\[
 \Pr_{A,a,b}[(v,w)\leftarrow\cA^{M_0,M_1}(\ket{A_{a,b}}):
                    v\in A+a,\ w\in A^\perp+b]=\negl(n).
\]
\end{lemma}

By the direct-product theorem of~\cite{CLLZ}, polynomially many
queries have negligible success.
Independent auxiliary information can be fixed before applying this
uniform query bound.

The following claim supplies the mini-scheme guarantee needed by
the authentication construction.
\begin{claim}\label{clm:coset-mini-scheme}
Averaged over the full oracle $\ora$, the coset mini-scheme has
negligible joint correctness error for polynomially many verifications.
Every counterfeiter with full access to $\ora$, polynomial query count
and length, and unrestricted local computation satisfies
\[
 \Pr_\ora[\text{both output notes pass for the issued serial}]
 =\negl(n).
\]
\end{claim}
The proof of \cref{clm:coset-mini-scheme} appears in
\cref{sec:coset-mini-scheme-security}.

We now prove the public-key money guarantee
(\cref{cor:public-key-money}) using this mini-scheme and $\CSig$.
\paragraph{Authenticating the serial.}
Generate a $\CSig$ signing key as the quantum minting key and use
its classical verification key as the money verification key.
To mint, generate a mini-scheme note $(s,\regis{note})$, sign $s$
with the current minting key, and return $(s,\regis{note},\regis{sig})$,
using the updated minting key for subsequent calls. Verification runs $\CSigVer$
and the mini-scheme verifier on their respective registers,
accepting only if both accept and returning the updated registers.
Signature correctness (\cref{thm:signature-security}) and
\cref{clm:coset-mini-scheme} give reusable minting and banknotes.

The mini-scheme compiler of~\cite{AC13Money} applies by forwarding
quantum signatures unchanged. Pad an adversary making at most $q$
adaptive minting requests to exactly $q$ requests, appending the
extra honest notes to its output, and let $\ell=\poly(n)$ bound its
number of outputs. Write $\mathsf{Count}$ for the number accepted
in the joint verification experiment. Use the following reductions.
\begin{enumerate}[itemsep=3pt]
 \item \emph{Fresh serial.} Guess one of the $\ell$ output positions.
 Generate the mini-scheme notes honestly and obtain their signatures
 from the signing challenger. Return the guessed serial and its
 untouched signature register as the forgery.
 \item \emph{Repeated issued serial.} Guess one of the $q$ minting
 requests and a pair of output positions. Embed the mini-scheme
 challenge at that request, generate all other notes honestly, and
 sign with an independently generated $\CSig$ key. Return the two
 guessed mini-scheme registers without first verifying any output.
\end{enumerate}
Let $\operatorname{Adv}_{\mathsf{sig}}$ and
$\operatorname{Adv}_{\mathsf{mini}}$ be these reductions' success
probabilities. Outside an issued-serial collision, the event
$\mathsf{Count}>q$ contains either an accepted fresh serial or two
accepted outputs with the same issued serial. Verification acts
separately on each note, so the reductions' untouched outputs have
the corresponding verification marginals. A union bound gives
\[
 \Pr[\mathsf{Count}>q]
 \le\ell\operatorname{Adv}_{\mathsf{sig}}
   +q\binom\ell2\operatorname{Adv}_{\mathsf{mini}}
   +q^2 2^{-n}+\negl(n).
\]
Here $q^2 2^{-n}$ bounds collisions among the ideal uniform serials,
and the negligible term accounts for replacing the honest generated
notes by ideal notes and for errors on the appended honest notes.
Both advantages are negligible by signature security
(\cref{thm:signature-security}) and \cref{clm:coset-mini-scheme}.

\paragraph{One fixed oracle.}
For a polynomial circuit-size bound $q$, let $p_{D,\ora}$ be the
mini-scheme success probability of description $D$, and define the
uniform quantum test $T_{\rm mini}^{\ora}(1^n,D)$: generate one note,
reject an invalid description, otherwise run $D$, verify both outputs,
and return one iff both accept.
This test has acceptance probability exactly $p_{D,\ora}$.

Fix an inverse-polynomial $\epsilon>0$. Pad the descriptions $D$
to one polynomial length, interpreting invalid strings as rejection,
and encode each test by
$a(D)=\langle T_{\rm mini},(1^n,D)\rangle$ of common polynomial
length $m\ge\max\{n,6/\epsilon\}$. The adversary $\cA_{\rm sel}$
first runs $\mathsf{Search}^{\ora}(a)$, obtaining $D_*$. It then
receives the challenge note, runs $D_*$ on it, and returns its outputs.
The estimation and search bound (\cref{lem:helper-estimates}) gives
\[
 \bbE_\ora\max_{|D|\le q(n)}p_{D,\ora}
 \le\Pr_\ora[\cA_{\rm sel}\text{ wins}]+\epsilon+2^{m+1}e^{-2m^2}
 \le\epsilon+\negl(n),
\]
since $\cA_{\rm sel}$ makes polynomially many queries and is covered
by \cref{clm:coset-mini-scheme}. Since $\epsilon$ is arbitrary,
$\delta_q(n):=\bbE_\ora\max_{|D|\le q(n)}p_{D,\ora}$ is negligible.
Markov's inequality gives
\[
 \Pr_\ora\!\left[\max_{|D|\le q(n)}p_{D,\ora}>\sqrt{\delta_q(n)}\right]
 \le\sqrt{\delta_q(n)}.
\]
This bound is summable, so Borel--Cantelli and a countable intersection
over polynomial bounds $q$ give security for every polynomial-size
classical description, including one depending on $\ora$.
Intersect this probability-one event with the fixed-oracle guarantees
(\cref{thm:fixed-oracle-crypto}). The authentication construction above
then proves \cref{cor:public-key-money} for the same oracle, against
nonuniform counterfeiters with no quantum advice.

\subsubsection{Correctness and security of the coset mini-scheme}
\label{sec:coset-mini-scheme-security}

We now prove \cref{clm:coset-mini-scheme}. The key is to turn two
accepted notes into a primal and a dual coset vector without
conditioning the counterfeiter's input on acceptance.

\paragraph{Correctness and reuse.}
On an ideal note $\ket{K_s}$, direct validity queries return one
in both bases and preserve the state. For $H_s\in\Good_{\kappa_M}$,
replace the $2r$ encoded calls one at a time by those direct calls.
The coherent evaluation bound (\cref{lem:coherent-evaluation})
gives joint trace-distance error at most
$4r\sqrt{\delta_{\mathrm{dec}}(\kappa_M)}$, including the acceptance bits and
returned note. Averaging the bad-hash event, whose probability is at
most $2^{-\kappa_M}$, and adding the functional-key preparation error
(\cref{lem:function-key-correctness}) gives negligible joint error
for any polynomial number of verifications.

\paragraph{Rejecting invalid coset components.}
Let $p_{\rm real}$ be the probability that both forged notes pass.
Replace key generation by $s\leftarrow\zo^n$ and the ideal note
$\ket{K_s}$, and let $p_{\rm ideal}$ denote the resulting probability.
Functional-key correctness (\cref{lem:function-key-correctness}) gives
$|p_{\rm real}-p_{\rm ideal}|\le\epsilon_{\mathrm{prep}}(n_K)$.
We analyze this ideal experiment, in which $s$ is uniform and
independent of the sampled oracle.
Fix a serial $s$ and basis $c$, and let $\Pi_c$ project the coset
register onto $A_s+a_s$ for $c=0$ or $A_s^\perp+b_s$ for $c=1$.
Let $E_c$ be the acceptance effect of the $r=3n$ checks after the
indicated change of basis,
so their acceptance probability on $\rho$ is $\operatorname{Tr}(E_c\rho)$.
We need the following bound on acceptance outside the correct coset.
\begin{claim}\label{clm:money-invalid-cosets}
Put $\alpha=(2/3)^r$ and
$\tau_{\mathrm{money}}(n)=6\sqrt{(a_{\kappa_M}+4n+1)2^{-n}}$.
Except for an event of probability at most $2\tau_{\mathrm{money}}(n)$
over $B_s$, both acceptance effects satisfy
\begin{equation}\label{eq:money-coset-effect}
 E_c\le\Pi_c+\alpha I\qquad(c\in\zo).
\end{equation}
Each accepting channel preserves the ranges of $\Pi_c$ and $I-\Pi_c$.
\end{claim}
The proof of \cref{clm:money-invalid-cosets} appears in
\cref{sec:money-invalid-cosets}.
These operator bounds hold for every forged state, including states
entangled with other notes and depending on the sampled functions.

\paragraph{From two accepted notes to coset search.}
Consider the following search algorithm.
\begin{boxedalgorithm}{Search from a mini-scheme counterfeiter}
On input $(s,\ket{K_s})$, run the counterfeiter and the primal
checks on both output notes. Abort if either group rejects.
Otherwise measure the first coset register in the computational
basis and the second in the Hadamard basis, and return their
outcomes $(v,w)$.
\end{boxedalgorithm}
No conditional state is supplied to the search algorithm.

Let $\rho$ be the joint output of a counterfeiter given one note.
Run the first, primal-basis group of checks on both output notes,
and write $\sigma$ for the unnormalized state when both groups
accept. The preceding bound gives
\[
 \operatorname{Tr}[((I-\Pi_0)\otimes I)\sigma]\le\alpha.
\]
The two notes' operations commute with each other. Thus, writing
$\widehat\Pi_1=\Hadamard^{\otimes4n}\Pi_1\Hadamard^{\otimes4n}$
and $\widehat E_1$ for the dual acceptance effect in the original
basis, their joint verification probability is at most
\begin{equation}\label{eq:money-to-search}
 \begin{aligned}
 \operatorname{Tr}[(\widehat E_1\otimes\widehat E_1)\sigma]
 &\le\operatorname{Tr}[(I\otimes\widehat E_1)\sigma]\\
 &\le\operatorname{Tr}[(I\otimes\widehat\Pi_1)\sigma]+\alpha\\
 &\le\operatorname{Tr}[(\Pi_0\otimes\widehat\Pi_1)\sigma]+2\alpha.
 \end{aligned}
\end{equation}
The last trace is exactly the unconditional success probability of
the displayed search algorithm. Let $q_0$ be this success probability
in the ideal experiment, averaged over the original oracle sample.
The two exceptional noise events therefore give
\[
 p_{\rm ideal}\le q_0+2\alpha+2\tau_{\mathrm{money}}(n).
\]

\paragraph{Removing the other oracle access.}
We show that this search algorithm has negligible success even
with the full cryptographic oracle. Use the ideal generation
experiment fixed above and define the following experiments,
each running the displayed search algorithm.
\begin{enumerate}[label=$\mathsf H_{\arabic*}$,start=0,itemsep=3pt]
 \item Use the original oracle and ideal note.
 \item Replace only the indexed encryption-key target by
 \[
 \ket{\Omega_K^0}=2^{-n/2}
 \left(\sum_{t\ne s}\ket t\ket{K_t}
                         +\ket s\ket{0^{n_K-n}}\right),
 \]
 sample its preparation pair from $\PrepTables(\ket{\Omega_K^0})$,
 and sample the one recursive extension conditional on the resulting
 full base. Continue to supply the original $\ket{K_s}$.
 \item Apply encoded-function simulation (\cref{thm:crypto-common-simulation})
 to the entire preceding experiment for each fixed $G$ below, with
 $J=K=\{s\}$, an initial request for $s$, and direct oracle $\mathfrak{F}_{C_s}$.
\end{enumerate}
Write $q_i$ for the search success probability in $\mathsf H_i$.
For the first change, fix all functions, hashes and preparation
pairs except the indexed encryption-key pair. Both targets and
the initial density matrix are then fixed, and
$\|\ket{\Omega_K}-\ket{\Omega_K^0}\|_2\le2^{1-n/2}$.
The target-change bound
(\cref{thm:preparation-replacement}) applies in dimension $n_K$
to the displayed targets. Since $h(v)=(1+v)/3$ and the query
bounds are polynomial, it gives
\[
 |q_0-q_1|\le2^{-\Omega(n)}.
\]

Let $\orabase^{1,-s}$ be the base in $\mathsf H_1$ with $T_s$
omitted, and put
\[
 G=(F_{s,0},F_{s,1},\orabase^{1,-s}),\qquad
 \Theta=(A_s,a_s,b_s).
\]
The indexed target replaces $\ket{K_s}$ by $\ket{0^{n_K-n}}$,
and the unchanged graph pair depends only on $F_{s,0},F_{s,1}$. All other primitive components were sampled
independently. Hence $\Theta,H_s,B_s$ are independent conditional
on $G$. The direct function $\mathfrak{F}_{C_s}$ is determined by $(G,\Theta)$,
and the supplied coset state depends only on $\Theta$.
For each fixed $G$, these are the sampling and request hypotheses of
encoded-function simulation (\cref{thm:crypto-common-simulation}). Averaging its joint-output bound,
including $\Theta$, over $G$ gives
\[
 |q_1-q_2|\le2^{-\Omega(n)},
\]
and a polynomial-query algorithm whose one classical sample $\eta_G$
is independent of $\Theta$ conditional on $G$.
Each query to $\mathfrak{F}_{C_s}$ is implemented with two membership queries
and the fixed functions $F_{s,0},F_{s,1}$. Applying
coset search hardness (\cref{lem:coset-search-hardness}) for each fixed $G$ therefore gives
the bound $q_2=\negl(n)$. Combining this with
\cref{eq:money-to-search} and the probability $2\tau_{\mathrm{money}}(n)$ of bad noise
gives
\[
 p_{\rm real}\le\epsilon_{\mathrm{prep}}(n_K)+2\tau_{\mathrm{money}}(n)+2\alpha
 +q_2+|q_0-q_1|+|q_1-q_2|
 =\negl(n).
\]
Thus the mini-scheme has negligible counterfeiting probability.

\subsubsection{Rejecting invalid coset components}
\label{sec:money-invalid-cosets}

We now prove \cref{clm:money-invalid-cosets}. Fix the serial $s$ and
a basis $c\in\zo$. We first bound a single random-function check
for every input state, then apply that bound after each accepted check.

For a random Boolean function $B_s$, put
\[
 p_c(B_s)=\max_{\rho\text{ on }a_{\kappa_M}+4n\text{ qubits}}
 \Pr_{x\leftarrow\zo^n,\,(z,v)\leftarrow D_c(x,\rho)}
             [B_s(v,(x,c,z,0))=1],
\]
where the oracle-free channel $D_c$ measures the coset register in
the computational basis to obtain $z$, then runs $\Decode$ on the
predicate-preimage state register with input $(x,c,z,0)$.
Decoder failure is assigned answer zero. The random access encoding
argument gives
\begin{equation}\label{eq:money-noise-bound}
 \Pr_{B_s}[p_c(B_s)>2/3]
 \le \tau_{\mathrm{money}}(n).
\end{equation}
To prove this estimate, sample independent uniform bits $Z_{x,z,v}$
and $w\leftarrow\zo^{2^n}$, and set
\[
 B_w(v,(x,c,z,0))=Z_{x,z,v}\oplus w_x.
\]
For each $w$, encode a state maximizing $p_c(B_w)$ together with the
flag $\indicator[p_c(B_w)>2/3]$. To recover $w_x$ when the flag is
one, run $D_c$ and output $1\oplus Z_{x,z,v}$ on decoder success.
Output an independent fair bit if the flag is zero or decoding fails.
Writing $\gamma=\Pr[p_c(B_s)>2/3]$, the average recovery probability
is at least $1/2+\gamma/6$. Fixing $Z$ to attain this average and
applying Nayak's bound~\cite{Nayak} to $a_{\kappa_M}+4n+1$ qubits
gives $\gamma\le\tau_{\mathrm{money}}(n)$, proving
\cref{eq:money-noise-bound}.

Fix $B_s$ outside the two exceptional events in
\cref{eq:money-noise-bound}. On the range of $I-\Pi_c$, the encoded
validity bit is zero on a correct preimage and equals $B_s$ otherwise.
It is therefore pointwise at most the bit used to define $p_c(B_s)$.
Each check has acceptance probability at most $2/3$ on this range.
Moreover, each check leaves the coset basis value $z$ unchanged, because
$\EvalHash$ uses it as an unchanged input. The accepting channel
for $r$ checks consequently preserves the ranges of $\Pi_c$ and
$I-\Pi_c$, and the operator $E_c$ giving acceptance probability
$\operatorname{Tr}(E_c\rho)$ obeys
\[
 E_c\le\Pi_c+\alpha I,\qquad \alpha=(2/3)^r.
\]
These inequalities hold for every forged state, including states
entangled with other notes and depending on the sampled functions.

\section{Proofs of the Signature Verification Bounds}
\label{app:signature-security}

In this section, we prove
\cref{lem:signature-random-table,lem:signature-equality-queries},
the two verification bounds used in \cref{sec:adaptive-signature-proof}
after removing the encoded signature oracles.

\subsection{Repeated verification with a random function}

We now prove \cref{lem:signature-random-table} using the random-function
bound. Once $B$ is fixed outside its exceptional event, the same bound
applies after every accepted check.

Set $B(\bot,y)=0$ and define
\[
 p(B)=\max_{\rho\text{ on }a_\kappa\text{ qubits}}
       \Pr_{y\leftarrow\zo^\kappa,\,v\leftarrow\Decode(y,\rho)}
                                 [B(v,y)=1].
\]
The decoder is oracle-free and returns an element of
$C_\kappa\cup\{\bot\}$. Applying the random-function bound
(\cref{lem:random-table}) with $\lambda=\kappa$ and $a=a_\kappa$ gives
\[
 \Pr_B[p(B)>2/3]
 \le6\sqrt{(a_\kappa+1)2^{-\kappa}}=\tau_{\mathrm{sig}}(\kappa).
\]

\paragraph{Repeated verification with fixed $B$.}
Fix $B$ with $p(B)\le2/3$. Uncomputing the decoder in
$\EvalHash^B$ leaves its response bit unchanged, so each check, with its uniformly sampled $y$,
has acceptance probability at most $2/3$ for every input state.
This includes the state conditioned on earlier checks accepting,
even for states entangled with an auxiliary system and depending on $B$.
Therefore
\[
 \Pr[V_B^r(\rho)=1]
 =\prod_{j=1}^r
   \Pr[\text{check }j\text{ accepts}\mid B,
                            \text{checks }1,\ldots,j-1\text{ accept}]
 \le(2/3)^r,
\]
which proves the claimed bound outside an event of probability
$\tau_{\mathrm{sig}}(\kappa)$.

\subsection{Predicting an unsigned value from equality queries}

We now prove \cref{lem:signature-equality-queries} by removing
equality queries at messages that have not yet been
signed. This must be done during the interaction, since fixing the
final forged message could reveal information about the function.
Recall that $F:\zo^n\to\zo^n$ is uniform and
$\EqTest_F(m,z)=\indicator[z=F(m)]$.

\Needspace{6\baselineskip}
\paragraph{The experiments.}
For $0\le j\le Q+1$, define $\mathsf H_j$ as follows. Sample
$F\leftarrow\{\zo^n\to\zo^n\}$ and the adversary's auxiliary state
independently. Write $\sigma_J$ for that auxiliary density matrix,
initialize an empty database $D$, and run $\cA(\sigma_J)$ with
classical signing requests and coherent equality queries answered
as follows:
\begin{enumerate}[nosep]
 \item On a classical signing request $m$, set $D(m)=F(m)$
 if $m\notin\operatorname{dom}D$, then return $D(m)$.
 \item Number equality queries consecutively and answer the first
 $j$ with $E_D$, and the rest with $\EqTest_F$, where
 \[
 E_D(m,z)=\begin{cases}
   \indicator[z=D(m)],&m\in\operatorname{dom}D,\\
   0,&m\notin\operatorname{dom}D.
 \end{cases}
 \]
 \item On output $(m,z)$, reject if $m\in\operatorname{dom}D$.
 Otherwise make one more equality query under the same rule and
 return its answer.
\end{enumerate}
The signed set is always $S=\operatorname{dom}D$. Thus
$\mathsf H_0$ returns the forgery bit and $\mathsf H_{Q+1}$ always
returns zero, while every experiment uses the same classical signing
oracle and updates its database from its own requests.

\paragraph{One equality query.}
Consider the next query after a prefix that has used only $E_D$,
and condition on its classical history $h$, including $D$.
During this prefix, $E_D$ reveals no function value outside $D$,
and a signing request adds only its requested value to $D$. Inducting
on the interaction therefore shows that, conditional on $h$, the unused values
$(F(m))_{m\notin\operatorname{dom}D}$ are still independent and
uniform. If $w_{m,z}(h)$ is the next query's squared weight, then
\[
 \begin{aligned}
 \bbE_{F\mid h}
   \sum_{m\notin S,z}w_{m,z}(h)\indicator[F(m)=z]
 &=2^{-n}\sum_{m\notin S,z}w_{m,z}(h)\\
 &\le2^{-n}.
 \end{aligned}
\]
This includes the final test in Step 3. At the query where
$\mathsf H_{j-1}$ and $\mathsf H_j$ differ, both have used only
$E_D$ so far, and afterwards both use $\EqTest_F$ with the same
signing rule. That common continuation cannot increase trace distance,
so the quantum query bound (\cref{lem:query-hybrid}) bounds each
adjacent difference by $2\sqrt{2^{-n}}$. Summing over at most
$Q+1$ queries gives
\[
 \Pr_F[m\notin S\land z=F(m)]
 \le\sum_{j=1}^{Q+1}|\Pr[\mathsf H_{j-1}=1]-\Pr[\mathsf H_j=1]|
 \le2(Q+1)2^{-n/2}.
\]

\clearpage\section{Changing the Oracle at a Hidden Key}
\label{sec:coherent-key-sampling}

We prove that polynomially many preparations at one hidden key can
be replaced by copies of an independent binary-phase state, even
with quantum access to the full recursive oracle
(\cref{thm:crypto-family-replacement}).

The PRS preparation oracle (\cref{def:prs-preparation-oracle}) samples
\[
 \Theta_k\leftarrow\{\zo^{5\lambda}\to\zo\},\qquad
 \ket{\Omega_{\Theta_k}}
 =2^{-5\lambda/2}\sum_{x\in\zo^{5\lambda}}(-1)^{\Theta_k(x)}\ket x,
 \qquad
 (R_k,S_k)\leftarrow\PrepTables(\ket{\Omega_{\Theta_k}}).
\]
These samples are independent over $k\in\zo^\lambda$, and only
$(R_k,S_k)$ are oracle components. The other base components form a Boolean oracle
$\ora_{\rm aux}:\zo^*\to\zo$, independent of this family. The
cryptographic oracle (\cref{def:full-crypto-oracle}) is
$\ora=(\ora_{\rm aux},(R_k,S_k)_k,N,A)$. Queries
$(\mathrm{aux},u)$, $(R,k,x)$ and $(S,k,x,y)$ return
$\ora_{\rm aux}(u)$, $R_k(x)$ and $S_k(x,y)$, respectively.
Sample one recursive extension $(N,A)$ of these functions
(\cref{def:recursive-qcma}).
For fixed $\ora$, write $\rho_{R_k,S_k}$ for the output density matrix
of $\Prepare^{R_k,S_k}(1^{5\lambda})$, so $r$ calls on separate input
registers produce $\rho_{k,r}^{\ora}=\rho_{R_k,S_k}^{\otimes r}$.

\begin{theorem}[Independent binary-phase input]
\label{thm:coherent-family-replacement}\label{thm:crypto-family-replacement}
Fix a family of density matrices $(\zeta_\theta)_\theta$ indexed only
by the phase function. For polynomial $r$ and every
quantum algorithm $\cA$ with polynomial query count and length and
unrestricted local computation,
\begin{equation}\label{eq:coherent-family-replacement}
 \trdist{\bbE_{\ora,k}[\cA^{\ora}
       (\rho_{k,r}^{\ora}\otimes\zeta_{\Theta_k})]}
 {\bbE_{\ora,\Theta}[\cA^{\ora}
       ((\ket{\Omega_\Theta}\!\bra{\Omega_\Theta})^{\otimes r}
         \otimes\zeta_\Theta)]}
 \le2^{-\Omega_{\cA}(\lambda)},
\end{equation}
where $k\leftarrow\zo^\lambda$ and
$\Theta\leftarrow\{\zo^{5\lambda}\to\zo\}$ are independent of $\ora$.
\end{theorem}
The proof of \cref{thm:crypto-family-replacement} appears in
\cref{app:pseudorandom-states}.

\subsection{Proof of the independent-input bound}\label{app:pseudorandom-states}
\label{sec:family-replacement-from-coupling}

We prove the independent-input bound (\cref{thm:crypto-family-replacement}).
Fix polynomial query bounds $Q,L$ with $L\ge\lambda$, and put
\[
 \sigma_\Theta=(\ket{\Omega_\Theta}\!\bra{\Omega_\Theta})^{\otimes r}
                  \otimes\zeta_\Theta.
\]
Every experiment below samples its oracle and input as specified,
runs $\cA$ with quantum access to the entire oracle, and returns
$\cA$'s output density matrix.
\begin{enumerate}[label=$\mathsf H_{\arabic*}$,start=0,leftmargin=*]
 \item Sample $\ora$ and $k\leftarrow\zo^\lambda$, and supply
 $\rho_{R_k,S_k}^{\otimes r}\otimes\zeta_{\Theta_k}$, using
 $\Prepare^{R_k,S_k}$ for each of the $r$ output factors.
 \item Sample $\ora,k$ as above and supply
 $\sigma_{\Theta_k}$ instead of
 $\rho_{R_k,S_k}^{\otimes r}\otimes\zeta_{\Theta_k}$.
 \item Sample $\ora$ and an independent uniform
 $\Theta:\zo^{5\lambda}\to\zo$, and supply $\sigma_\Theta$.
\end{enumerate}
For a joint sample $(\ora,\ora',\Theta)$ and a supported prefix
$o=\ora_{\le L}$, write
\begin{equation}\label{eq:prs-conditional-disagreement}
 g_u(o,\theta)
 =\Pr[\ora(u)\ne\ora'(u)\mid\ora_{\le L}=o,\Theta=\theta].
\end{equation}
The following coupling bounds this disagreement even after the old
oracle and the supplied phase function are known.
\begin{lemma}[Conditional oracle disagreement]
\label{lem:prs-conditional-disagreement}
There is a joint sample $(\ora,\ora',\Theta)$ such that
$(\ora,\Theta)$ has the independent sampling in $\mathsf H_2$,
$(\ora',\Theta)$ has the sampling $(\ora,\Theta_k)$ in $\mathsf H_1$, and
\[
 \bbE_{o,\Theta}\max_{|u|\le L}g_u(o,\Theta)
 \le2^{-\Omega(\lambda)}.
\]
The bound is uniform in the auxiliary oracle $\ora_{\rm aux}$.
\end{lemma}
The proof of \cref{lem:prs-conditional-disagreement} appears in
\cref{sec:prs-conditional-disagreement-proof}.

We now use conditional oracle disagreement to finish the main proof.
For each fixed oracle and key, changing one tensor factor at a time gives
\[
 \trdist{\rho_{R_k,S_k}^{\otimes r}\otimes\zeta_{\Theta_k}}
 {\sigma_{\Theta_k}}
 \le r\trdist{\rho_{R_k,S_k}}
                  {\ket{\Omega_{\Theta_k}}\!\bra{\Omega_{\Theta_k}}}.
\]
Conditional on $k,\Theta_k$, the preparation pair has distribution
$\PrepTables(\ket{\Omega_{\Theta_k}})$. Applying $\cA^\ora$ cannot
increase trace distance, so the state-preparation bound
(\cref{lem:public-preparation}) gives
\[
 \trdist{\mathsf H_0}{\mathsf H_1}\le r2^{-5\lambda}.
\]
For the second transition, use the coupling from conditional oracle
disagreement (\cref{lem:prs-conditional-disagreement}). Conditional on
$(o,\Theta)$, the query states of $\cA^o(\sigma_\Theta)$ are fixed,
so each query has expected weight at most
$\max_{|u|\le L}g_u(o,\Theta)$ on inputs where the two oracles differ.
The quantum query bound (\cref{lem:query-hybrid}) therefore gives
\begin{equation}\label{eq:prs-hybrid-target}
 \trdist{\mathsf H_1}{\mathsf H_2}
 \le2Q\sqrt{\bbE_{o,\Theta}\max_{|u|\le L}g_u(o,\Theta)}
 =2^{-\Omega(\lambda)}.
\end{equation}
The maximum allows queries to depend on every supplied copy and on
$\zeta_\Theta$. Averaging $\ora_{\rm aux}$ and adding the two errors
proves the theorem, including its allowed auxiliary states.

\subsection{Proof of conditional oracle disagreement}
\label{sec:prs-conditional-disagreement-proof}

We now prove \cref{lem:prs-conditional-disagreement} by fixing the
auxiliary oracle $\ora_{\rm aux}$ and constructing the joint sample
whose conditional disagreement we will bound.
\subsubsection{Joint sampling of the recursive oracles}

We replace the preparation pair at one hidden key and sample the
new recursive answers jointly with the old answers. The extra
randomness in this joint sampling will let us change the target
while conditioning on the entire old oracle.
Choose a constant $d\ge1$ with $L<\lambda^{2^{d+1}}$, put
$m=d+1$, and process interval $i\in[d]$ only at query lengths
\[
 \lambda^{2^i}\le |u|<\min\{L+1,\lambda^{2^{i+1}}\}.
\]
Set
\[
 \rho_{P,i}=2^{-c_P\lambda(256m)^{-(i-1)}},\qquad
 \gamma_{P,i}=\rho_{P,i}^{1/128},\qquad
 t_i=\gamma_{P,i}/12\quad(i\in[m]).
\]
The constant $c_P>0$ will be chosen below. The following definition
specifies the distribution before we construct an implementation.

\begin{definition}[Joint oracle sample at a hidden key]
\label{def:hidden-component-coupling}
We define the \emph{joint oracle sample} $(\ora,\ora',\Theta)$ as follows.
\begin{enumerate}[leftmargin=*,itemsep=3pt]
 \item Sample independent
 $C_k=(\Theta_k,R_k^{\mathsf{prs}},S_k^{\mathsf{prs}})$ using the
 phase and preparation sampling above. Choose $k\leftarrow\zo^\lambda$
 and an independent $C'_k$, and set $\Theta$ to its phase function.
 The old base uses the preparation pairs in $(C_j)_j$, and the new
 base replaces the pair at $k$ by that in $C'_k$.
 \item Use the same recursive prefix below $\lambda^2$, then process
 query inputs in increasing length, computing each $N$ answer
 exactly from its own shorter oracle.
 \item At an $A$ input in interval $i$ with old and new means $p_0,p_1$,
 sample the pair of bits from the matrix
 \begin{equation}\label{eq:canonical-recursive-pair}
 \begin{gathered}
 K_{t_i}(p_0,p_1)=M(p_0,p_1)+t_iJ,\qquad
 J=\begin{pmatrix}-1&1\\1&-1\end{pmatrix},\\
 M(p_0,p_1)=
 \begin{pmatrix}
 1-\max\{p_0,p_1\}&(p_1-p_0)_+\\
 (p_0-p_1)_+&\min\{p_0,p_1\}
 \end{pmatrix},
 \end{gathered}
 \end{equation}
 with rows indexed by the old bit and columns by the new bit.
 Beyond length $L$, use any causal joint sampling with the prescribed
 marginals.
\end{enumerate}
\end{definition}
Observe that the shared prefix uses only base inputs shorter than $\lambda$:
a reserved input of length $(n+2)^2<\lambda^2$ describes queries
of length at most $n<\lambda$. At every later $A$ input,
$p_0,p_1\in[1/3,2/3]$ and $0<t_i\le1/6$ make $K_{t_i}$ a positive
probability matrix. Since each row and column of $J$ sums to zero,
its marginals are Bernoulli with means $p_0,p_1$, so induction on
query length gives the original recursive distribution for each oracle.

\subsubsection{An implementation that leaves the key hidden}
\label{sec:hidden-key-comparison-proof}

We implement the joint sample using classical queries, so disagreement
on any input can be charged to a query at the changed key or at a
sparse set of recursive inputs.

For each of the $N_*=2^\lambda$ keys, sample two independent components
$C_{k,0},C_{k,1}$. To select a component, independently sample $m$
sets $P_1,\ldots,P_m$, including each key with probability
$\rho_{P,1}$, and independent fair bits $B_1(k),\ldots,B_m(k),Z(k)$.
Set
\[
 X_k=\begin{cases}B_{j(k)}(k),&j(k)=\min\{i:k\notin P_i\}<\infty,\\
 Z(k),&k\in P:=\bigcap_iP_i,
 \end{cases}
 \qquad \Pr[k\in P]=p:=\rho_{P,1}^m,
\]
and use the preparation pair in $C_{k,X_k}$ in the old base.
Conditional on the sets, the selectors are independent fair bits, so the selected and unused
component arrays are independent of each other and of the sets.
A preparation-oracle query at key $k$ reads $X_k$ and returns the
requested bit of the pair in $C_{k,X_k}$. Its coherent implementation
uses two selector queries, including uncomputation. This substitution
preserves deterministic verification, polynomial query and witness
bounds, and the shorter-query order because selector inputs have
length $\lambda$.

Let $Z_{\mathsf{base}}$ denote the component arrays, selectors and
the values of $\ora_{\rm aux}$, and put $\nu_P=\rho_{P,m}$. We use the
following implementation before selecting which $k\in P$ to replace,
writing $(\ora_0,\ora_1)=(\ora,\ora')$.

\begin{lemma}[Classical implementation of coupled oracles]
\label{lem:prs-paired-implementation}
The joint oracle sample $(\ora,\ora',\Theta)$
(\cref{def:hidden-component-coupling}) has an implementation with a classical description
$E$ and a deterministic query algorithm $\Answer_E^{\mathbf T_e}$,
where $\mathbf T_e=(T_{0,e},\ldots,T_{d,e})$ gives the unexposed
selector answers and the successive intervals' $A$ answers. Averaging its
coins, including failed stages, gives the original distribution of the old oracle and
\[
 \ora\perp\bigl((C_{k,1-X_k})_k,(P_i)_{i\in[m]}\bigr)
       \mid(C_{k,X_k})_k.
\]
When any $k\in P$ is replaced by its unused component,
\[
 \ora_e(u)=\Answer_E^{\mathbf T_e}(u)\quad(e\in\{0,1\},\ |u|\le L)
\]
outside an $E$-determined event $\mathsf{Fail}$ of probability at most
$3d\nu_P^4$. Each execution uses at most $h=\poly(\lambda)$ queries
for every choice of Boolean oracle functions, and changing that component does
not change $E$. Write $T_j=T_{j,0}$ for the old oracle. On success,
conditional on $E,Z_{\mathsf{base}}$ and the earlier functions $T_{<j}$, each remaining
Boolean function $T_j$, $j\ge1$, is sampled independently at each input by
\[
\begin{aligned}
 G_j(u)&\leftarrow\Bernoulli(\pi_j),&B_j(u)&\leftarrow\Bernoulli(1/2),\\
 C_j(u)&\leftarrow\Bernoulli(c_{j,u}(T_{<j})),&\pi_j&\le\nu_P,\\
 T_j(u)&=(1-G_j(u))B_j(u)+G_j(u)C_j(u).
\end{aligned}
\]
\end{lemma}

The proof of \cref{lem:prs-paired-implementation} appears in
\cref{sec:prs-paired-implementation-proof}.

\subsubsection{Disagreement conditional on the old oracle}

We now select the changed key and bound each answer's probability
of changing after all old answers have been fixed. Generate the
old oracle using its classical implementation
(\cref{lem:prs-paired-implementation}), choose $k$
uniformly from $P$, and complete that implementation with the unused
component at $k$ to obtain the new oracle. If $P=\varnothing$, choose $k\leftarrow\zo^\lambda$
and use the conditional rows of $K_{t_i}$ directly.

After averaging implementation coins, the old oracle is independent
of the unused components and of the selector sets, while the selected
and unused arrays are independent. Symmetry of the keys consequently
gives
\[
 k\leftarrow\zo^\lambda,\qquad
 C'_k\perp(k,\orabase,\ora_{\le L}),\qquad \Theta\perp\ora.
\]
At each later input the pair has matrix $K_{t_i}$, so this sampling
gives the joint oracle sample
$(\ora,\ora',\Theta)$ (\cref{def:hidden-component-coupling}).

To bound a query chosen from the old oracle, we first check that
conditioning on its actual answers does not reveal the remaining
recursive sets.

\begin{claim}[Conditioning on the old answers]\label{clm:prs-conditioned-sets}
On success, for every $j\ge1$ and input $w$,
\[
 \Pr[G_j(w)=1\mid E,Z_{\mathsf{base}},T_0,\ldots,T_d]
 \le4\nu_P.
\]
The bound also holds after averaging $Z_{\mathsf{base}}$, for an input
chosen from $E$ and $T_0,\ldots,T_d$.
\end{claim}
\begin{proof}
Condition on $E,Z_{\mathsf{base}},T_{<j}$. The classical implementation
(\cref{lem:prs-paired-implementation}) gives $\Pr[G_j(w)=1]\le\nu_P$. On $G_j(w)=0$, the bit $T_j(w)$
is fair, so either value has probability at least $(1-\nu_P)/2$.
Bayes' rule therefore gives
\[
 \Pr[G_j(w)=1\mid E,Z_{\mathsf{base}},T_{<j},T_j(w)=x]
 \le\frac{2\nu_P}{1-\nu_P}\le4\nu_P.
\]
Other inputs use independent pairs of bits, so conditioning on the
whole function $T_j$ does not change this bound. Later correction
probabilities depend on $G_j$ only through $T_j$, and hence conditioning
on $T_{j+1},\ldots,T_d$ also leaves the bound unchanged. It holds
for every $w$ with these functions fixed, which allows $w$ to depend
on them and permits averaging $Z_{\mathsf{base}}$.
\end{proof}

Since $|P|$ is binomial with mean $N_*p$, Chernoff's bound gives
\[
 \Pr[|P|<N_*p/2]\le e^{-N_*p/8}.
\]
Put $\mathsf{Bad}=\mathsf{Fail}\cup\{|P|<N_*p/2\}$ and
$\xi=\Pr[\mathsf{Fail}]\le3d\nu_P^4$. Fix $E,T_0,\ldots,T_d$
and follow $\Answer_E(u)$ with the old answers. For a selector query,
condition also on $P$: the changed key is uniform in $P$, so its
probability of being queried is at most $1/|P|\le2/(N_*p)$ when
$|P|\ge N_*p/2$. For a query to $T_j$, $j\ge1$, instead use the
conditional sparse-set bound (\cref{clm:prs-conditioned-sets}), which
bounds its probability of lying in $G_j$ by $4\nu_P$ conditional
on $E,T_0,\ldots,T_d$, without fixing $P$.

Outside $\mathsf{Bad}$, an answer can change only if one of the at
most $h$ queries encounters one of these two events. Averaging the
respective conditional bounds and taking a union bound therefore
gives, for every supported $o=\ora_{\le L}$,
\begin{equation}\label{eq:shared-answer-change}
\begin{aligned}
 \Pr[\ora(u)\ne\ora'(u)\mid\ora_{\le L}=o]
 &\le b(o):=\frac{2h}{N_*p}+4h\nu_P+
               \Pr[\mathsf{Bad}\mid\ora_{\le L}=o],\\
 \bbE_o b(o)&\le\frac{2h}{N_*p}+4h\nu_P+\xi+e^{-N_*p/8}.
\end{aligned}
\end{equation}
Choosing $mc_P<1/2$ makes $\eta:=\bbE_o b(o)=2^{-\Omega(\lambda)}$.
The same $b(o)$ works for every input $u$, and the last term covers
the empty-set case as well.

\subsubsection{Queries chosen from copies of the target}
\label{sec:hidden-family-dependence}

Because $\Theta\perp\ora$, the preceding estimate gives
$\bbE_\Theta g_u(o,\Theta)\le b(o)$ for every $o,u$. We must make
this bound simultaneous over query inputs that may depend on
$\sigma_\Theta$. Put $D=2^{5\lambda}$ and $H=\lfloor D^{3/4}\rfloor$.
The following claim lets us pass from an average over phases to that
simultaneous bound.

\begin{claim}[Changing the phase function]\label{clm:prs-phase-change}
For sufficiently small $c_P>0$, there is
$\epsilon=2^{-\Omega(\lambda)}$ such that, for every supported old
prefix $o$, every $|u|\le L$, and all phase functions $\theta,\theta'$,
\begin{equation}\label{eq:prs-answer-modulus}
 d_H(\theta,\theta')\le H
 \quad\Longrightarrow\quad
 |g_u(o,\theta)-g_u(o,\theta')|\le\epsilon.
\end{equation}
\end{claim}
The proof of \cref{clm:prs-phase-change} appears in
\cref{sec:prs-phase-change-proof}.

\paragraph{Concentration over the phases.}
For fixed $o,u$, Markov's inequality gives
\[
 S=\{\theta:g_u(o,\theta)\le2b(o)\},\qquad
 \Pr_\Theta[\Theta\in S]\ge\tfrac12.
\]
When $b(o)=0$, nonnegativity gives $g_u(o,\Theta)=0$ almost surely,
so this conclusion also holds. The function $d_H(\Theta,S)$ changes
by at most one when one bit changes. Bounded differences therefore
implies
\[
 \frac12\le\Pr[d_H(\Theta,S)=0]
 \le\exp\!\left(-\frac{2(\bbE d_H(\Theta,S))^2}{D}\right),
 \qquad \bbE d_H(\Theta,S)\le\sqrt{D\ln2/2}.
\]
For sufficiently large $\lambda$, $H\ge2\sqrt{D\ln2/2}$, so a
second application gives
\[
 \Pr[d_H(\Theta,S)>H]\le\delta:=\exp(-H^2/(2D)).
\]
By the phase-change bound (\cref{clm:prs-phase-change}), every phase
function within distance $H$ of $S$ satisfies
$g_u(o,\Theta)\le2b(o)+\epsilon$.
There are fewer than $2^{L+1}$ query inputs of length at most $L$,
so a union bound and $g_u\le1$ give
\begin{equation}\label{eq:prs-query-concentration}
 \bbE_{o,\Theta}\max_{|u|\le L}g_u(o,\Theta)
 \le2\eta+\epsilon+2^{L+1}\delta
 =2^{-\Omega(\lambda)}.
\end{equation}
The doubly exponential bound on $\delta$ allows every polynomial $L$.

The exact marginals of the coupled oracles and the bound on the
conditional maximum prove the lemma.

\subsection{Proof of the phase-change bound}
\label{sec:prs-phase-change-proof}

We now prove \cref{clm:prs-phase-change}. We identify the distribution
of the new oracle conditional on the old oracle, so that the target-change
bound applies with the old answers fixed.

\paragraph{Conditional sampling of the new oracle.}
Fix a supported old prefix $o$. For an $A$ input in interval $i$,
let $x$ be its old answer and $p\in[1/3,2/3]$ its old mean, both
determined by $o$. For a new acceptance probability $v$, put
$q(v)=(1+v)/3$. Dividing row $x$ of $K_{t_i}(p,q(v))$ by its sum
gives the new conditional mean
\begin{equation}\label{eq:prs-conditional-profiles}
\begin{aligned}
 h_1(v)&=\min\{1,q(v)/p\}-t_i/p,\\
 h_0(v)&=\max\{0,(q(v)-p)/(1-p)\}+t_i/(1-p).
\end{aligned}
\end{equation}
Thus, conditional on $o$ and the new base, generate the new oracle
in increasing query length, answering $N$ exactly from its shorter
oracle and sampling each $A$ bit with mean $h_x(v)$.
To verify this whole conditional distribution, factor each paired
answer in a finite prefix as
\[
 K_{t_i}(p,q(v))_{x,y}
 =p^x(1-p)^{1-x}h_x(v)^y(1-h_x(v))^{1-y}.
\]
Conditioning on $o$ cancels the old factors, whose product is its
marginal probability. Old exact $N$ answers contribute one on the
supported prefix, leaving precisely the new exact $N$ constraints.
Since $\ora$ is independent of the new component $C'_k$, its
preparation pair still has distribution
$(R,S)\leftarrow\PrepTables(\ket{\Omega_\Theta})$ given $\Theta$.
The shared prefix below $\lambda^2$ queries only base inputs shorter
than $\lambda$, so it is the same for both oracles.

Since $p,1-p\ge1/3$ and $t_i\le1/12$, the conditional means satisfy
\[
 |h_x(v)-h_x(w)|\le|v-w|,\qquad
 t_i\le h_x(v)\le1-t_i\quad(v,w\in[0,1]).
\]
These are exactly the mean and margin hypotheses needed to change
the new preparation target while keeping $o$ fixed.

\paragraph{Changing the phase function.}
Let $D=2^{5\lambda}$ and $H=\lfloor D^{3/4}\rfloor$. For functions
$\theta,\theta'$ with $d_H(\theta,\theta')\le H$,
\[
 \|\ket{\Omega_\theta}-\ket{\Omega_{\theta'}}\|_2
 =2\sqrt{d_H(\theta,\theta')/D}
 \le2\sqrt{H/D}\le2^{-\lambda/9}
\]
for sufficiently large $\lambda$. Fix $o$, the changed key $k$, and all base answers outside $k$. Define the test $\mathsf T_e$, for $e\in\zo$,
by the following operations:
\begin{enumerate}[leftmargin=*,itemsep=2pt]
 \item Sample the changed component with target
 $\ket{\Omega_\theta}$ if $e=0$ and $\ket{\Omega_{\theta'}}$ if $e=1$.
 \item Sample its recursive extension with exact $N$ answers,
 the fixed shared prefix, and conditional means $h_x(v)$
 (\cref{eq:prs-conditional-profiles}).
 \item Query the new oracle at $u$, obtain $a$, and return
 $\indicator[a\ne o(u)]$.
\end{enumerate}
Apply the target-change bound (\cref{thm:preparation-replacement})
to the oracles sampled by
$\mathsf T_0,\mathsf T_1$ and the one-query algorithm in Step~3,
with $c_*=1/9$ and $n=5\lambda$. The conditional means satisfy
the displayed difference and range bounds. If $c$ is the theorem's
margin constant, choose $c_P$ so that $c_P/128<c$ and $mc_P<1/2$.
Then
\[
 \min_i t_i=\frac1{12}2^{-c_P\lambda/128}\ge2^{-c\lambda}
\]
for sufficiently large $\lambda$. The test uses the fixed bit
$o(u)$ after its query. The theorem gives
$|\Pr[\mathsf T_0=1]-\Pr[\mathsf T_1=1]|\le\epsilon$ for
$\epsilon=2^{-\Omega(\lambda)}$, uniformly over the fixed quantities.
The changed key and the base answers outside it, conditioned on $o$, are
independent of the new phase function. Averaging these fixed choices therefore proves the phase-change bound
(\cref{eq:prs-answer-modulus}).

\subsection{Proof of the classical simulation of coupled oracles}
\label{sec:prs-paired-implementation-proof}

We now prove \cref{lem:prs-paired-implementation} by constructing
the classical query algorithm $\Answer_E$ and its common description $E$.
At the beginning of interval $i$, the representation of each existing
oracle function uses $m-i+1$ independent sets with joint inclusion
bound $\rho_{P,i}^{|S|}$.
The description $E$ specifies exposed sets and finite functions of
query transcripts, and implements every earlier answer for arbitrary
oracles for shorter inputs. The selector sampling establishes this
invariant at the first interval. The following stages extend it.

\paragraph{Classical query algorithms for the next interval.}
For each valid $N$ or $A$ query at address $u$, write
$\langle M_u,x_u\rangle$ for the padded encoding of its circuit and input,
so $M_u$ is deterministic for $N$ and witnessless quantum for $A$.
Address padding restricts both
types of circuit to shorter queries to the full oracle.
For each existing oracle function $j$, expose its next set $P_{i,j}$
and uniform Boolean function $B_{i,j}$, and define
\[
 P_i=\{(j,x):x\in P_{i,j}\},\qquad B_i(j,x)=B_{i,j}(x).
\]
Independence of the exposed sets gives $P_i$ inclusion bound
$\rho_{P,i}^{|S|}$. For each $A$ query, substitute the earlier answer
algorithms and component selection into its defining circuit, giving
\[
 v_u(Y)=\Pr[M_u^{\Answer_{<i}^{(B_i)_{P_i}^{Y}}}(x_u)=1].
\]
For the sampled oracle, $Y(j,x)$ returns the bit of the first
not-yet-exposed Boolean function whose corresponding set omits $x$.
If every remaining set contains $x$, it returns the correction bit.
The function $(B_i)_{P_i}^{Y}$ equals $Y$ on $P_i$ and $B_i$ elsewhere. The following query algorithms will be stored in $E$ before any
actual shorter answers are used.

\begin{claim}[Query algorithms for one interval]\label{clm:prs-interval-queries}
Conditional on the preceding description, the exposed $(P_i,B_i)$
determine classical algorithms $T_u,\tau_u$ and numbers
$\widehat v_{u,\tau}\in[0,1]$. The algorithms have polynomial query
bounds on every oracle, and except with probability $3\rho_{P,i}^4$
over $(P_i,B_i)$, simultaneously
for every address $u$ and every $Y$, $T_u^Y$ gives the exact existential
answer at each $N$ address, while at each $A$ address
\[
 |v_u(Y)-\widehat v_{u,\tau_u(Y)}|\le\gamma_{P,i}/4000.
\]
\end{claim}
The proof of \cref{clm:prs-interval-queries} appears in
\cref{sec:prs-interval-queries-proof}.

Add these algorithms and numbers to $E$. If the simultaneous guarantee
fails, use the completion rule below. Otherwise it applies to both
sets of shorter answers $Y_0,Y_1$, allowing us to construct the paired
$A$ answers next.

\paragraph{Sampling the next $A$ answers.}
For each $A$ input $u$ set
\[
 q_{u,\tau}=\frac{1+\widehat v_{u,\tau}}3,\qquad
 p_e=\frac{1+v_u(Y_e)}3,\qquad q_e=q_{u,\tau_u(Y_e)},\qquad
 \delta_i=\frac{2\gamma_{P,i}}3.
\]
The estimate gives $q_e\in[1/3,2/3]$ and
$\max_e|p_e-q_e|\le\delta_i/1000$. Independently at each $u$, sample
$F_u\leftarrow\Bernoulli(2/3)$ and $V_u\leftarrow[0,1]$, and put
in $E$ the bit $F_u$ and the finite function
$L_u(\tau)=\indicator[V_u\le\widehat v_{u,\tau}]$.
For the new oracle function, sample $m-i$ independent sets of
density $\gamma_{P,i}^{1/(m-i)}$ and independent uniform Boolean
functions. Independently at each input $u$, sample a correction pair
$(C_{i,0}(u),C_{i,1}(u))$ from
\begin{equation}\label{eq:paired-residual-repair}
 R_i=D_{\rm fair}+J/8+
       \frac{M(p_0,p_1)-M(q_0,q_1)}{\delta_i},
 \qquad D_{\rm fair}=\begin{pmatrix}1/2&0\\0&1/2\end{pmatrix}.
\end{equation}
To sample the old oracle before choosing the changed key, first
sample only the old bit with mean
$1/2+(p_0-q_0)/\delta_i$. Once the key and new shorter answers are
chosen, sample the new bit from the row of $R_i$ indexed by the
old bit, divided by that row's sum. Sample these correction pairs
also where they will not be used. If $F_u=0$, run $\tau_u$ with oracle $Y_e$ and output
$L_u(\tau_u(Y_e))$. Otherwise, find the
first new set that omits $u$ and return the value at $u$ of its
corresponding uniform Boolean function, or return $C_{i,e}(u)$ if all
new sets contain $u$. Compute each exact $N$ answer separately from
its own shorter oracle.

To check the sampling, observe that using the same threshold $V_u$
for the two transcript estimates gives
\[
 \tfrac13 M(\widehat v_{u,\tau_u(Y_0)},
                 \widehat v_{u,\tau_u(Y_1)})
       +\tfrac23D_{\rm fair}=M(q_0,q_1).
\]
The event requiring a correction has probability $\delta_i$ and
would contribute $\delta_iD_{\rm fair}$ to this matrix. Replacing
that contribution by $\delta_iR_i$ is valid because
\[
 D_{\rm fair}+J/8=\begin{pmatrix}3/8&1/8\\1/8&3/8\end{pmatrix},
 \qquad
 \|M(p_0,p_1)-M(q_0,q_1)\|_{\max}\le\delta_i/500,
\]
so $R_i$ is a positive probability matrix. Its old marginal is the
one used above, and $t_i/\delta_i=1/8$ gives the required distribution
\[
 M(q_0,q_1)-\delta_iD_{\rm fair}+\delta_iR_i
 =K_{t_i}(p_0,p_1).
\]
In particular, sampling the old answer does not require the new
component or its shorter oracle.

\paragraph{Completing the construction and answering queries.}
After this stage the representation of each earlier function has one
fewer set, and the new function uses $m-i$ sets whose density satisfies
\[
 \gamma_{P,i}^{1/(m-i)}
 =\rho_{P,i}^{1/(128(m-i))}\le\rho_{P,i+1}.
\]
The functions $L_u$ and estimates $\widehat v_{u,\tau}$ are defined
for every possible partial function $\tau$ and for $\tau=\bot$, without using the
remaining sets or correction bits. This remains true at later stages, whose algorithms use
arbitrary oracle functions on shorter inputs. Conditional on $E$, the
selected base and the earlier actual functions $T_{<j}$, the last sets
and their fair and correction bits therefore have exactly the product distribution
in the statement.

If a stage fails, set $\mathsf{Fail}$ in $E$, sample subsequent old
answers with their prescribed means, and, after choosing the key,
sample the new answers from the appropriate rows of $K_{t_i}$.
Thus failure changes the implementation but never the prescribed
oracle distribution. The union of the three failure events per
interval has probability at most $3d\nu_P^4$.

For completeness, the final algorithm uses the following explicit
oracle access to the function introduced at stage $j$. Number its
sets and uniform Boolean functions by
$(G_j^{(r)},B_j^{(r)})_{r\in[m-j]}$, with those for $r<m-j$ in $E$.
For $j\ge1$, put $G_j=G_j^{(m-j)}$, $B_j=B_j^{(m-j)}$ and
\[
 T_{j,e}(u)=\begin{cases}C_{j,e}(u),&u\in G_j,\\B_j(u),&u\notin G_j.\end{cases}
\]
For $j=0$, use $B_m$ outside $P_m$ and $Z$ inside $P_m$, flipping
only the changed key for the new selector function $T_{0,1}$.
Define $\mathsf{Read}_{j,r,E}^{T_{j,e}}(u)$ by two steps:
\begin{enumerate}[leftmargin=*,itemsep=2pt]
 \item For $\ell=r,\ldots,m-j-1$, return $B_j^{(\ell)}(u)$
 at the first $\ell$ with $u\notin G_j^{(\ell)}$.
 \item If every tested set contains $u$, query $T_{j,e}(u)$ and return its answer.
\end{enumerate}
The first step uses only functions in $E$, and the second makes one
oracle query. In an algorithm constructed for interval $i$, answer $Y(j,v)$ with
$\mathsf{Read}_{j,i-j+1,E}^{T_{j,e}}(v)$. Let $T_u^E$ and $\tau_u^E$ be the resulting
existential and transcript algorithms, whose only oracles are
$(T_{j,e})_j$.

\Needspace{11\baselineskip}
On input $u$, $\Answer_E^{\mathbf T_e}$ returns zero if
$|u|>L$ or $\mathsf{Fail}$ is set. Otherwise it performs the applicable step:
\begin{enumerate}[leftmargin=*,itemsep=2pt]
 \item For an $\ora_{\rm aux}$ or shared-prefix input, return its
 bit in the corresponding fixed function in $E$.
 \item For a preparation-oracle input with key $k$, compute
 $x=\mathsf{Read}_{0,1,E}^{T_{0,e}}(k)$ and return the requested bit
 of the pair in $C_{k,x}$.
 \item For an $N$ input, execute $(T_u^E)^{\mathbf T_e}$ and return
 its output.
 \item For an $A$ input in interval $i$, return
 $L_u(\tau_u^E(\mathbf T_e))$ if $F_u=0$, and otherwise return
 $\mathsf{Read}_{i,1,E}^{T_{i,e}}(u)$.
\end{enumerate}
Unused inputs and invalid $N,A$ encodings return zero. Strictly shorter recursive calls and
constantly many intervals give termination and the polynomial query
bound $h$ for every choice of its Boolean oracle functions.

Averaging the fresh implementation coins at each input gives old
mean $p_0$ independently of unused components and the initial sets.
Induction over query length proves the original conditional
distribution of the entire old oracle, including failed stages.
Finally, no exposed base set reveals a selector in $P$, and all
later functions in $E$ are specified for every possible $\tau$, including $\bot$.
Changing a selected component at any $k\in P$ therefore leaves $E$
unchanged, as required.

\subsubsection{Proof of the interval query algorithms}
\label{sec:prs-interval-queries-proof}

We now prove \cref{clm:prs-interval-queries}, fixing the preceding
description and the substituted circuits. The exposed restriction
$(P_i,B_i)$ is independent of these circuits, with inclusion bound
$\rho_{P,i}^{|S|}$ and independent uniform $B_i$.

Set $\epsilon=\gamma_{P,i}/4000$. Let $Q_i\ge1$ bound the substituted
query count, let $a_i$ bound the logarithm of the number of circuit
indices, and let $\cX_i$ be their query domain. Choose
\[
 s=a_i+\lambda^2,\qquad
 M_{\mathrm{trial}}=\lceil\rho_{P,i}^{-1/10}\rceil,\qquad
 Q_{\mathrm{sim}}=s^2M_{\mathrm{trial}}.
\]
Let $\Sim_u$ be the classical-query simulator
(\cref{thm:classical-simulation}) for the circuit at $u$, with fixed
function $B_i$, random set $P_i$, no witness and accuracy $\epsilon$.
Implement its deterministic query phase using the partial function
$\tau=\varnothing$ on $P_i$:
\begin{enumerate}[leftmargin=*,itemsep=2pt]
 \item At a requested input $x\notin P_i$, supply $B_i(x)$.
 For $x\in\dom\tau$, supply $\tau(x)$.
 \item At any other input, return $\bot$ if $|\dom\tau|=s-1$.
 Otherwise query $Y(x)$, set $\tau(x)$ to its answer, and continue.
 \item Return $\tau$ when the query phase ends.
\end{enumerate}
For each partial function $\tau$ on $P_i$ with fewer than $s$ inputs,
run the same $\Sim_u$ with answers $\tau(x)$ on its domain and $B_i(x)$
elsewhere, and define $\widehat v_{u,\tau}$ to be its output-one
probability. Put $\widehat v_{u,\bot}=0$. This defines $\tau_u$ with
fewer than $s$ queries to $Y$ on every execution.

Conditional on $B_i$, the joint-inclusion bound for $P_i$ is unchanged.
The message length is
\[
 m_{{\rm msg},i}=a_i+s(\ceil{\log_2Q_{\mathrm{sim}}}+1)=\poly(\lambda).
\]
Since
$\log_2Q_{\mathrm{sim}}\le\frac1{10}\log_2(1/\rho_{P,i})+O(\log\lambda)$
and $s=a_i+\lambda^2$, for sufficiently large $\lambda$,
\[
 m_{{\rm msg},i}+8\lambda^2\le s\log_2(1/\rho_{P,i}).
\]
Contractive simultaneous incompressibility
(\cref{lem:simultaneous-incompressibility})
therefore gives $t_{\rm inc}(\lambda,m_{{\rm msg},i})\le s$ for $P_i$.
Deterministic-classical-query simulation (\cref{thm:classical-simulation}), with index
length $a_i$ and no witness, bounds the
probability of any truncated or inaccurate execution by
\[
 2^{-8\lambda^2}+
 2^{a_i}(|\cX_i|+1)^s s
 e^{-\epsilon^2M_{\mathrm{trial}}/(16Q_i^2)}
 \le\rho_{P,i}^4+2^{-4\lambda^2}\le2\rho_{P,i}^4.
\]
The second term is at most $2^{-4\lambda^2}$ because
$\epsilon^2M_{\mathrm{trial}}/(16Q_i^2)=2^{\Omega(\lambda)}$,
whereas the logarithm of its prefactor is polynomial.
Outside this event, the returned $\tau$ determines every answer in
the simulator's deterministic query phase, so its output probability
is $\widehat v_{u,\tau_u(Y)}$. Consequently,
\[
 \sup_Y|v_u(Y)-\widehat v_{u,\tau_u(Y)}|
 \le\gamma_{P,i}/4000.
\]
For an $N$ input $u$, let $f_u$ be the accepting-transcript DNF of
$M_u(x_u,w)$ after the same substitutions (\cref{lem:accepting-dnf}),
whose width $k$ is polynomial. Define $T_u=\Switch_{f_u,s}(P_i,B_i|_{P_i^c})$,
and evaluate it by forwarding its queries to $Y$. The switching
theorem (\cref{thm:biased-switching}) applies with bias parameter $1$
because $B_i$ is independent and uniform. The same cutoff
$s=a_i+\lambda^2$ satisfies, for sufficiently large $\lambda$,
\[
 k+s\le2^\lambda,\qquad \rho_{P,i}\le(16k)^{-1},\qquad
 2^{a_i}\,2(8k\rho_{P,i})^s\le\rho_{P,i}^4.
\]
It therefore supplies exact $N$ answers simultaneously for every
$u,Y$, with the stated failure probability. Both algorithms depend
only on the exposed restriction and have polynomial query bounds
also when they fail.

\paperappendixpart{Proofs of Fixed-Oracle Separations}
\section[Proof of the Fixed-Oracle Collapse and Separation (\texorpdfstring{\cref{thm:boolean-collapse}}{Theorem \getrefnumber{thm:boolean-collapse}})]{Proof of the Fixed-Oracle Collapse\\and Separation (\texorpdfstring{\cref{thm:boolean-collapse}}{Theorem \getrefnumber{thm:boolean-collapse}})}\label{app:fixed-oracles}

We prove the \QMA{} separation and \QCMA{} collapse relative
to one fixed classical oracle.

\subsection{\texorpdfstring{Proof of \cref{thm:boolean-collapse}}{Proof of Theorem \getrefnumber{thm:boolean-collapse}}}\label{app:boolean-collapse}

We prove the fixed-oracle collapse and separation of
\cref{thm:boolean-collapse}.

The oracle $\ora_{\mathsf{sep}}$ has inputs $\mathsf{base}(\lambda,v,y)$ of length
$\ell_\lambda=(\lambda+c_0)^{10}$ and inputs $\mathsf{help}(a)$ of
length $(|a|+2)^2$, for a sufficiently large constant $c_0$.
Here the query pairs range over
$(v,y)\in\cX_\lambda=\Sigma_\lambda^{q_\lambda}\times\zo^\lambda$ for the linear code
$C_\lambda\subseteq\Sigma_\lambda^{q_\lambda}$ (\cref{def:code-pairs}), and
$a=\langle V,x\rangle$ is a padded description of a quantum verifier
$V$ with ordinary input $x$ and classical witness $w$, using at most
$|a|$ gates, witness bits and query length. Use injective polynomial-time encodings and zero on
malformed inputs. Put
\begin{equation}\label{eq:boolean-qma-value}
 \ora_{\mathsf{sep},\lambda}(v,y)=\ora_{\mathsf{sep}}(\mathsf{base}(\lambda,v,y)),\qquad
 \alpha(\ora)=\max_\sigma\Pr[V_{\rm \QMA{}}^{\ora}(1^\lambda,\sigma)=1],
\end{equation}
where $V_{\rm \QMA{}}$ is the \QMA{} verifier
(\cref{thm:qma-verification}). Require
\begin{equation}\label{eq:boolean-helper}
 \ora_{\mathsf{sep}}(\mathsf{help}(\langle V,x\rangle))=
 \begin{cases}
  1,&\max_w\Pr[V^{\ora_{\mathsf{sep}}}(x,w)=1]\ge2/3,\\
  0,&\max_w\Pr[V^{\ora_{\mathsf{sep}}}(x,w)=1]\le1/3,
 \end{cases}
\end{equation}
with either answer allowed in the gap. These requirements depend
only on shorter oracle inputs.

A finite prefix $\ora_{\mathsf{sep},\le N}$ is \emph{admissible} if its \QCMA{}
answers satisfy~\cref{eq:boolean-helper} and
$\alpha(\ora_{\mathsf{sep},\lambda})\in[0,1/3]\cup[2/3,1]$ whenever
$\ell_\lambda\le N$.

The following extension lemma lets us diagonalize against a classical
algorithm while preserving admissibility.

\begin{lemma}[An extension preserving \QCMA{} answers]
\label{lem:boolean-extension}
Let $\ora_{\mathsf{sep},\le N}$ be admissible and let $M$ be a deterministic
polynomial-time oracle algorithm. For every sufficiently large
$\lambda$ with $\ell_\lambda>N$, there is an admissible extension
$\ora'_{\mathsf{sep},\le N'}$, with $N'\ge\ell_\lambda$ bounding $M$'s query lengths on
$1^\lambda$ and all newly assigned base components other than
$\lambda$ zero, such that
\[
 M^{\ora_{\mathsf{sep}}'}(1^\lambda)\ne\indicator[\alpha(\ora'_{\mathsf{sep},\lambda})\ge2/3].
\]
\end{lemma}
The proof of \cref{lem:boolean-extension} appears in
\cref{app:boolean-collapse}, after the proof of \cref{thm:boolean-collapse}.

We now apply the extension lemma to prove \cref{thm:boolean-collapse}.

\paragraph{Constructing the oracle.}
Enumerate the deterministic oracle algorithms with explicit polynomial
time bounds as $M_1,M_2,\ldots$, and start with the empty admissible
prefix $\ora_{\mathsf{sep}}^{(0)}$. At stage $i$, choose $\lambda_i$ large
enough for \cref{lem:boolean-extension} and with
$\ell_{\lambda_i}$ beyond the domain of $\ora_{\mathsf{sep}}^{(i-1)}$, then let
$\ora_{\mathsf{sep}}^{(i)}$ be an extension supplied by that lemma against $M_i$.
Define the final oracle by
\[
 \ora_{\mathsf{sep}}(u)=\ora_{\mathsf{sep}}^{(i)}(u)\quad\text{for any stage }i
 \text{ whose prefix contains }u.
\]
The prefixes extend one another and their lengths tend to infinity,
so this defines every answer of $\ora_{\mathsf{sep}}$. Each helper requirement depends
only on shorter inputs, and hence remains satisfied after the stage
at which its answer is fixed.

\paragraph{Separating the language.}
Admissibility gives
\[
 L=\{1^\lambda:\alpha(\ora_{\mathsf{sep},\lambda})\ge2/3\}
 \in\QMA^{\ora_{\mathsf{sep}}},
\]
because every component has acceptance probability at most $1/3$
or at least $2/3$. On the other hand, the answers
queried by $M_i(1^{\lambda_i})$ never change after stage $i$.
Thus $M_i^{\ora_{\mathsf{sep}}}$ errs on $1^{\lambda_i}$ and $L\notin\PTIME^{\ora_{\mathsf{sep}}}$.

\paragraph{Collapsing the hierarchy.}
For any \QCMA{} verifier $V$ on input $x$, pad $\langle V,x\rangle$
to a polynomial length satisfying the bounds above, then query
$\mathsf{help}(\langle V,x\rangle)$ and return its answer.
The helper requirement (\cref{eq:boolean-helper}) makes this a
deterministic polynomial-time solver for the verifier's promise problem. Hence
$\PTIME^{\ora_{\mathsf{sep}}}=\QCMA^{\ora_{\mathsf{sep}}}$ also for promise problems.
Coherently substituting each lower level's deterministic solver
into its parent gives $\QCMAH^{\ora_{\mathsf{sep}}}=\PTIME^{\ora_{\mathsf{sep}}}$ by induction.

\subsubsection{Extending an admissible oracle}

We now prove the extension lemma (\cref{lem:boolean-extension}).

\paragraph{Assigning the shorter answers.}
Choose a constant $d\ge0$ so that $M$'s running time on $1^\lambda$
is less than $b_{d+1}$, where
\[
 b_i=\ell_\lambda^{2^i}\quad(0\le i\le d+1),\qquad N'=b_{d+1}-1.
\]
Keep the existing prefix and set every new base component through
$N'$ to zero except component $\lambda$. A helper input of length
less than $b_1$ describes a circuit querying only inputs shorter
than $b_0$, so these helper answers can be fixed in increasing
length order using~\cref{eq:boolean-helper}, choosing zero in the
gap and leaving every answer in the existing prefix unchanged. Let $T_0$ output these fixed answers without making oracle queries.

Independently sample $H\leftarrow\cH_\lambda$ and
$B\leftarrow\Bernoulli(2^{-\lceil\sqrt\lambda\rceil})^{\cX_\lambda}$,
and form the coupled target oracles
\[
 \oranoise=B,\qquad \oraplant=B\vee\indicator_{\cP_H}
\]
from \cref{def:oracle-distributions}. Use $d+1$ suffix groups
(\cref{def:suffix-constraints}) and put
\[
 X_0=\cX_\lambda,\qquad X_i=\bigcap_{j\le i}P_j(H),\qquad
 e_i=(P_1,\ldots,P_i,B|_{\cX_\lambda\setminus X_i}).
\]
Here $e_0$ is empty. We next construct total answer algorithms
$T_i$ for helper inputs shorter than $b_{i+1}$, whose descriptions
depend only on $e_i$ and whose target-oracle queries lie in $X_i$.

\paragraph{Constructing the helper-answer algorithms.}
For $i=1,\ldots,d+1$, perform the following steps, with $F$
denoting the function $F:X_{i-1}\to\zo$ used in Step~1.
\begin{enumerate}[leftmargin=*,itemsep=4pt]
 \item If $i\le d$, decode $a=\langle V,x\rangle$ and set
 \[
 W_i^F(\mathsf{help}(a),w)=V^{F,T_{i-1}^F}(x,w),\qquad
 \cZ_i=\{z:b_i\le|z|<b_{i+1}\},
 \]
 rejecting malformed inputs. If $i=d+1$, instead take
 $\cZ_{d+1}=\{1^\lambda\}$ and
 $W_{d+1}^F=M^{F,T_d^F}$, with no witness.
 In either case, answer each target query outside $X_{i-1}$
 by the stored value $B(u)$ in $e_{i-1}$ and leave only queries
 in $X_{i-1}$ for the target oracle. All other base answers
 are the fixed answers chosen above.
 \item Put $h_i=\lceil\log_2(2|\cZ_i|)\rceil$. On the domain
 $X_{i-1}$, apply the mapping $\Replace$ of
 \cref{thm:qcma-replacement} to $W_i$, the remaining set
 $P_i\cap X_{i-1}$, and the fixed answers
 $B|_{X_{i-1}\setminus X_i}$, using the conditional distribution
 given $e_{i-1}$. Call its output $S_i$.
 If $X_{i-1}=\varnothing$, use the oracle-free completion that
 returns the non-$\bot$ answer of the \QCMAbb{} machine for $W_i$
 and returns zero in its gap.
 \item For $i\le d$, extend the earlier answers by setting, for
 every function $F:X_i\to\zo$ and valid helper input $z$,
 \[
 T_i^F(z)=
 \begin{cases}
 T_{i-1}^{\,B|_{X_{i-1}\setminus X_i}\,\cup F}(z),&|z|<b_i,\\
 S_i^F(z),&b_i\le|z|<b_{i+1}.
 \end{cases}
 \]
 Set all other inputs to zero. In the first branch, answer each query $u\in X_{i-1}\setminus X_i$
 with $B(u)$ and forward queries in $X_i$ to $F$.
 For $i=d+1$, set the output bit to
 $c=S_{d+1}(1^\lambda)$ and end the construction.
\end{enumerate}

\paragraph{Verifying each interval.}
At interval $i\le d$, the padding gives $|a|<b_i$, so every
helper query made by $V(x,w)$ is already covered by $T_{i-1}$.
Coherent substitution (\cref{lem:coherent-classical-query}) therefore
makes $W_i$ a quantum query verifier with polynomial resource
bounds on every sample. The returned $S_i$ has fewer than $h_i$
queries, all in $X_i$. Substitution across constantly many intervals
gives polynomial query and length bounds for every $T_i$, while
Step~3 shows directly that its description depends only
on $e_i$. Thus we can apply the same resource bound at every interval,
including after an incorrect answer in an earlier interval.

For $i\in[d+1]$, let $N_{W_i}$ be the partial \QCMAbb{} machine
associated with $W_i$. We need the event
\[
 \cE_i=\left\{
 \forall z\in\cZ_i,\ F\in\{\oranoise,\oraplant\}:\quad
 N_{W_i}^{F|_{X_{i-1}}}(z)
 \in\{S_i^{F|_{X_i}}(z),\bot\}
 \right\}.
\]
The following short claim supplies its conditional probability bound.

\begin{claim}
For every $i\in[d+1]$,
$\Pr[\neg\cE_i\mid e_{i-1}]\le2^{-\Omega(\sqrt\lambda)}$,
uniformly in $e_{i-1}$.
\end{claim}
\begin{proof}
Conditional on $e_{i-1}$, $W_i$ is fixed and
\[
 B|_{X_{i-1}}\samp
 \Bernoulli(2^{-\lceil\sqrt\lambda\rceil})^{X_{i-1}}
 \quad\text{independently of }P_i\cap X_{i-1}.
\]
The suffix-group bound (\cref{lem:code-hash}(1)) gives joint
inclusion at most $2^{-\lambda t/(8(d+1))}$ for this remaining
set, while $\oranoise,\oraplant$ agree outside it on $X_{i-1}$.
Removing classical witnesses (\cref{thm:qcma-replacement}) therefore
applies in Step~2 with the displayed $h_i$, giving
$\Pr[\neg\cE_i\mid e_{i-1}]\le2^{-\Omega(\sqrt\lambda)}$.
For an empty domain, $\cE_i$ holds exactly.
\end{proof}

The claim holds without conditioning on earlier correctness, so averaging
and summing over the constantly many intervals gives
\[
 \Pr\left[\bigcap_{i=1}^{d+1}\cE_i\right]
 \ge1-\sum_{i=1}^{d+1}\bbE_{e_{i-1}}
                 \Pr[\neg\cE_i\mid e_{i-1}]
 \ge1-2^{-\Omega(\sqrt\lambda)}.
\]
On this event, Step~3 preserves each earlier helper answer under either
target oracle $F$, because $F=B$ on $X_{i-1}\setminus X_i$.
A verifier described by a helper input in $[b_i,b_{i+1})$ queries
only those earlier helpers, so $\cE_i$ makes its new answer satisfy
\cref{eq:boolean-helper}. Induction on $i$ therefore shows that
$T_d$ satisfies every helper requirement through $N'$ under both
target oracles.
For the terminal interval, $h_{d+1}=1$, so $S_{d+1}$ makes no
queries and both executions of $M$ output its bit $c$.

\paragraph{Choosing the component on which $M$ errs.}
The acceptance bounds for \QMA{} verification (\cref{thm:qma-verification}) give
\[
 \Pr_{H,B}[\alpha(\oraplant)\ge2/3\land\alpha(\oranoise)\le1/3]
 \ge1-2^{-\Omega(\lambda)}.
\]
For sufficiently large $\lambda$, there is therefore a sample
satisfying both promises and all replacement events. Fix such a
sample, set
\[
 \ora'_{\mathsf{sep},\lambda}=
 \begin{cases}\oraplant,&c=0,\\\oranoise,&c=1,\end{cases}
\]
and assign every new helper answer through $N'$ using
$T_d^{\ora'_{\mathsf{sep},\lambda}|_{X_d}}$, with the fixed answers outside $X_d$
stored in its description. The construction agrees with the old prefix,
satisfies every new helper requirement and gives the promised
\QMA{} value on every new base component, since the components
other than $\lambda$ are zero. Hence $\ora'_{\mathsf{sep},\le N'}$ is admissible,
and $M^{\ora_{\mathsf{sep}}'}(1^\lambda)=c$ is the wrong answer for its target
component, proving the lemma.

\clearpage\section[Proof of the Fixed-Oracle Cryptography Theorem (\texorpdfstring{\cref{thm:fixed-oracle-crypto}}{Theorem \getrefnumber{thm:fixed-oracle-crypto}})]{Proof of the Fixed-Oracle\\Cryptography Theorem (\texorpdfstring{\cref{thm:fixed-oracle-crypto}}{Theorem \getrefnumber{thm:fixed-oracle-crypto}})}\label{app:fixed-crypto}

We now prove \cref{thm:fixed-oracle-crypto} by one common
Markov--Borel--Cantelli argument.

\paragraph{Correctness.}
The cryptographic oracle $\ora=(\orabase,N,A)$ has independent
PRF, encryption, signature and PRS base components, with one
recursive extension (\cref{def:full-crypto-oracle}). Together they have at most
$2^{2\lambda+2}$ hashes and $2^{\lambda+2}$ preparation targets,
each on at least $5\lambda$ qubits. By the good-hash and state-preparation
bounds (\cref{lem:code-hash,lem:public-preparation}), Markov's
inequality, and a union bound, the probability that some hash is bad
or some preparation has trace-distance error greater than $2^{-\lambda}$
is at most
\[
 2^{2\lambda+2-\kappa}
 +2^{\lambda+2}2^\lambda\epsilon_{\mathrm{prep}}(5\lambda)
 =2^{2\lambda+2-\kappa}+2^{-3\lambda+2},
 \qquad \kappa=(\lambda+4)^3.
\]
Since this bound is summable, Borel--Cantelli implies that all
hashes and preparations are good at sufficiently large parameters
for almost every oracle. Functional-key and signature correctness
(\cref{lem:function-key-correctness,lem:signature-correctness})
then give correctness for all four constructions. The same preparation bound holds uniformly
over the PRS keys, and replacing $r$ actual outputs by prescribed
copies costs at most $r2^{-\lambda}$.

\paragraph{Security for all classical descriptions.}
For each primitive, condition on the other three independent base
components and apply its security theorem with those components as
$\ora_{\rm rest}$. The theorem includes the single global $N,A$
extension, so averaging gives its security bound for the combined
oracle.
Fix one of the security games in \cref{thm:prf-security,thm:pke-security,thm:signature-security,thm:pseudorandom-states}, and a polynomial size bound $q$.
For distinguishing games, use balanced guessing and allow an output
complement. Write $p_{A,\ora}$ for success probability and
$p_0\in\{0,1/2\}$ for the security threshold, and put
\[
 \Delta_q(\ora)
 :=\left(\max_{|A|\le q(\lambda)}p_{A,\ora}-p_0\right)_+.
\]
The next claim replaces a best oracle-dependent description by a
single adversary that selects its description using the oracle.
\begin{claim}\label{clm:crypto-description-selection}
For every inverse-polynomial $\epsilon>0$, there is a uniform
polynomial-query adversary $A_{\rm sel}$ such that
\[
 \bbE_\ora\Delta_q(\ora)
 \le\bbE_\ora(p_{A_{\rm sel},\ora}-p_0)+\epsilon+\negl(\lambda).
\]
Selection takes place before using the game's inputs, and separated
algorithms receive only their selected descriptions and assigned registers.
\end{claim}
The proof of \cref{clm:crypto-description-selection} appears in
\cref{sec:crypto-description-selection}.

The averaged security theorem applies to $A_{\rm sel}$ because its
description is uniform and its queries have polynomial count and
length. Thus $\bbE_\ora p_{A_{\rm sel},\ora}\le p_0+\negl(\lambda)$,
and \cref{clm:crypto-description-selection} gives
\[
 \bbE_\ora\Delta_q(\ora)\le\epsilon+\negl(\lambda).
\]
Since $\epsilon$ is any inverse polynomial, this expectation is a
negligible function $\delta_q(\lambda)$. As in~\cite{KQT}, Markov's
inequality gives
\[
 \Pr_{\ora}[\Delta_q(\ora)>\sqrt{\delta_q(\lambda)}]
 \le\sqrt{\delta_q(\lambda)},\qquad
 \sum_\lambda\sqrt{\delta_q(\lambda)}<\infty.
\]
Borel--Cantelli and a countable intersection over integer polynomial
bounds $q$, including copy and request bounds, give security for
all nonuniform classical descriptions, even those depending on $\ora$.

\paragraph{The promise collapse.}
Finally, the estimation, search and collapse lemma
(\cref{lem:helper-estimates}) applies to this entire base distribution and gives
\[
 \PROMISEP^{\ora}=\PROMISEBQP^{\ora}=\PROMISEQCMA^{\ora}.
\]
For a hierarchy machine, start at the lowest level and choose the
total deterministic solver supplied for its promise problem.
This solver is a Boolean completion of that level's partial
function, so substitute it at every query of the next verifier and
apply the collapse to the resulting QCMA promise problem. Repeat
up to the top level. The completion convention ensures that each
resulting solver agrees with every non-$\bot$ answer of the
corresponding hierarchy machine, giving
$\PROMISEQCMAH^{\ora}=\PROMISEP^{\ora}$.
Intersect these probability-one events.

\subsection{Selecting an adversary description}
\label{sec:crypto-description-selection}

We now prove \cref{clm:crypto-description-selection}. We first implement
the chosen security game as a quantum test, then use the recursive
oracle to select a description with nearly maximal test acceptance.

\paragraph{The game test.}
We construct a uniform \QPT{} test $T_{\rm game}^\ora(1^\lambda,A)$, which
runs the described adversary with full access to $\ora$ and returns
the game's success bit. The test samples the honest keys and independent randomness as
specified below, and every adversarial algorithm has coherent
access to the same $\ora$:
\begin{itemize}[itemsep=3pt]
 \item For PRF privacy, sample $\beta\leftarrow\zo$ and
 $(s,\regis{sk})\leftarrow\PRFGen^\ora(1^\lambda)$. Give $s$ to
 the adversary and answer its challenge queries with
 $\PRFEval^\ora$ when $\beta=0$, or an independent limited-independence
 function when $\beta=1$. Return $\indicator[\widehat\beta=\beta]$.
 \item For PRF piracy, generate $(s,\regis{sk})$, privately sample
 $u_1,u_2\leftarrow\zo^\lambda$, and obtain $f_i$ by measuring
 the response of $\PRFEval^\ora(s,\regis{sk},u_i,0)$ sequentially.
 Run
 $(\regi_1,\regi_2)\leftarrow\cA_0^\ora(s,\regis{sk})$ on the resulting
 key, then $\widehat f_i\leftarrow\cA_i^\ora(u_i,\regi_i)$ without
 communication, and return
 $\indicator[\widehat f_1=f_1\land\widehat f_2=f_2]$.
 \item For encryption piracy, generate $(\pk,\regis{sk})$ and run
 \[
 ((m_{i,0},m_{i,1})_{i=1}^2,\regi_1,\regi_2)
       \leftarrow\cA_0^\ora(\pk,\regis{sk}).
 \]
 Independently sample $\beta_i\leftarrow\zo$ and
 $\ct_i\leftarrow\PKEEnc^\ora(\pk,m_{i,\beta_i})$, then run
 $\widehat\beta_i\leftarrow\cA_i^\ora(\ct_i,\regi_i)$ without
 communication and return
 $\indicator[\widehat\beta_1=\beta_1\land\widehat\beta_2=\beta_2]$.
 For CPA, instead run $(m_0,m_1,\regi)\leftarrow\mathcal B_0^\ora(\pk)$,
 sample $\beta\leftarrow\zo$ and
 $\ct\leftarrow\PKEEnc^\ora(\pk,m_\beta)$, then run
 $\widehat\beta\leftarrow\mathcal B_1^\ora(\ct,\regi)$ and return
 $\indicator[\widehat\beta=\beta]$.
 \item For signatures, generate $(\regis{vk},\regis{sk})$, give
 $\regis{vk}$ to the forger, and initialize $S=\varnothing$.
 On a classical request $m$, run $\sign^\ora(\regis{sk},m)$, return its signature
 and set $S=S\cup\{m\}$. Answer quantum verification-key
 requests with $\QSigVKGen$. On a forgery $(m,\sigma)$, return
 $\indicator[m\notin S]\ver^\ora(\regis{vk},m,\sigma)$, generating a
 fresh verification key for the final $\QSig$ test.
 \item For PRS, sample $\beta\leftarrow\zo$. If $\beta=0$, sample
 one key $k$ and supply $r$ fresh outputs of
 $\StateGen^\ora(1^\lambda,k)$. Otherwise sample one
 limited-independence Boolean function $f$ and supply $r$ copies of
 $2^{-5\lambda/2}\sum_x(-1)^{f(x)}\ket x$. Return
 $\indicator[\widehat\beta=\beta]$.
\end{itemize}
The independent functions and binary-phase states can be generated
efficiently.\footnote{For $q$ quantum queries, use a fresh
$2q$-wise independent function~\cite{FOCS:Zhandry12}. For $r$ Haar
copies on $5\lambda$ qubits, prepare the normalized state
$\sum_x(-1)^{f(x)}\ket x$ with one $2r$-wise independent Boolean $f$.
The binary-phase moment bound~\cite{BS19BinaryPhase} gives
trace-distance error $O(r^2 2^{-5\lambda})$.}
Replacing the two preliminary PRF evaluations by ideal evaluations
leaves the key unchanged, so $u_1,u_2$ can then be sampled after
$\cA_0$ runs. Reusable correctness bounds this change jointly with
the key, $f_1,f_2$ and subsequent coherent queries. Thus each test
has the original winning predicate, up to the joint correctness
and moment errors, giving uniformly over the descriptions,
\[
 \bbE_{\ora}\max_{|A|\le q(\lambda)}
 \left|\Pr[T_{\rm game}^{\ora}(1^\lambda,A)=1]-p_{A,\ora}\right|
 =\negl(\lambda).
\]
Increase the size bound to include the following fixed strategy: abort
when $p_0=0$, guess a fair bit in a distinguishing game, or give the
key to one honest recipient while the other guesses in a piracy game.
Write $A_{\rm triv}$ for this strategy. Correctness gives
$\bbE_{\ora}(p_0-p_{A_{\rm triv},\ora})_+\le\negl(\lambda)$.

\paragraph{Selecting and running a description.}
Fix an inverse-polynomial $\epsilon>0$ and pad adversary descriptions
$z$ to length $\ell=\poly(\lambda)$. With the test $T_{\rm game}$ interpreting
invalid $z$ as $A_{\rm triv}$, let
$a(z)=\langle T_{\rm game},(1^\lambda,z)\rangle$ encode the test and its ordinary
input, padding all pairs to one polynomial length
$n\ge\max\{\lambda,6/\epsilon\}$. Define the following uniform
adversary $A_{\rm sel}$, using $\mathsf{Search}^{\ora}$ from the estimation, search and collapse lemma
(\cref{lem:helper-estimates}):
\begin{enumerate}[nosep]
 \item Before using the game's inputs, run
       $\mathsf{Search}^{\ora}(a)$ and denote its output by $z_*$.
 \item Run the adversary described by $z_*$ on those inputs, forwarding
       its oracle queries and external responses. In a piracy game,
       perform selection before invoking $\cA_0$, then give
       algorithm $i$ only its selected description and register.
 \item Return the selected adversary's outputs at the prescribed times.
\end{enumerate}
The selection algorithm queries the fixed oracles $N,A$, uses
polynomially many queries of polynomial length, and never uses the
current game's hidden randomness. Except with negligible probability,
its selected test has acceptance probability within $6/n\le\epsilon$
of the best test. The test approximation and the bound for $A_{\rm triv}$
above then transfer this estimate to the true success probabilities.
To finish the claim, write
\[
 e_{\rm test}(\ora)=\max_{|A|\le q(\lambda)}
 \left|\Pr[T_{\rm game}^{\ora}(1^\lambda,A)=1]-p_{A,\ora}\right|.
\]
Except when selection fails, its additive test error is at most
$\epsilon$, so
\[
 \max_{|A|\le q(\lambda)}p_{A,\ora}
 \le p_{z_*,\ora}+\epsilon+2e_{\rm test}(\ora).
\]
Moreover,
$(p_0-\max_A p_{A,\ora})_+\le(p_0-p_{A_{\rm triv},\ora})_+$.
Selection precedes the game inputs, so averaging $p_{z_*,\ora}$
gives $p_{A_{\rm sel},\ora}$. The expected test error, the expected
deficit of $A_{\rm triv}$ and the selection-failure probability are negligible. Adding these
three errors proves the displayed bound for $\bbE_\ora\Delta_q(\ora)$.

\subsection{Proof of average-case \QCMAH{} hardness in \QMA{} (\texorpdfstring{\cref{cor:pke-average-qma}}{Corollary \getrefnumber{cor:pke-average-qma}})}
\label{app:pke-average-qma-proof}
We now prove \cref{cor:pke-average-qma} using the encryption scheme
$\pke^\ora$ (\cref{sec:pke-construction}) for a fixed cryptographic
oracle satisfying \cref{thm:fixed-oracle-crypto}.

\paragraph{The sampler and its promise problem.}
Let $k(\lambda)$ bound the quantum secret-key length, including any
classical secret information, and put $t=k+\lambda$.
All algorithms below use the fixed oracle $\ora$. The sampler chooses
\[
 \begin{gathered}
  (\pk,\regis{sk})\leftarrow\PKEGen(1^\lambda),
  \qquad m\leftarrow\zo^t,\quad b\leftarrow\zo,\\
  c_i\leftarrow\PKEEnc(\pk,bm_i)\quad(i\in[t]),
  \qquad X=(1^\lambda,\pk,(c_i)_{i\in[t]},m).
 \end{gathered}
\]
It discards the secret key and returns $(X,b)$. On input $X$, the
verifier initializes a classical output string $z$ and uses its
$k$-qubit witness as the decryption key. For $i=1,\ldots,t$, it runs
\mbox{$z_i\leftarrow\PKEDec(\regis{sk},c_i)$}, using the updated key
for the next call. After all decryptions, it accepts exactly when $z=m$.
Let $a(X)$ be its maximum acceptance probability and set
$\Pi_1=\{X:a(X)\ge2/3\}$ and $\Pi_0=\{X:a(X)\le1/3\}$.
Reusable correctness (\cref{thm:pke-security} and
\cref{thm:fixed-oracle-crypto}) gives
$\Pr[X\notin\Pi_1\mid b=1]\le\negl(\lambda)$.

For $b=0$, condition on $\pk,(c_i)_i$. Decryption produces
its entire output before testing equality with the independent uniform
$m$. For a witness density matrix $\rho$, write
$\Pr[z=m]=\operatorname{Tr}(E_m\rho)$, where $E_m\ge0$ and
$\sum_mE_m\le I$. Markov's inequality and
$\|E_m\|_\infty\le\operatorname{Tr}E_m$ give the quantum-witness
guessing bound~\cite{AaronsonGuessing12}:
\[
 \Pr[X\notin\Pi_0\mid b=0]
 \le3\,2^{-t}\sum_m\|E_m\|_\infty
 \le3\,2^{-t}\operatorname{Tr}I
 =3\,2^{-\lambda}.
\]
\paragraph{Reducing prediction to chosen-plaintext security.}
To prove hardness, let $\mathsf H_j$ sample a fresh key and uniform
$m\in\zo^t$, encrypt $m_i$ for $i\le j$ and $0$ for $i>j$ with
independent encryption randomness, and return $X$. Thus
$\mathsf H_0$ and $\mathsf H_t$ are the conditional distributions
of the sampler's $X$ given $b=0$ and $b=1$.
For any polynomial-time bit predictor $D$, define a CPA adversary
$(\mathcal B_0,\mathcal B_1)$ by the following algorithms:
\begin{enumerate}[nosep]
 \item Algorithm $\mathcal B_0^\ora(\pk)$ samples $j\leftarrow[t]$ and
       $m\leftarrow\zo^t$, submits the message pair $(0,m_j)$,
       and stores $(\pk,j,m)$.
 \item Algorithm $\mathcal B_1^\ora(c,\pk,j,m)$ sets $c_j=c$, independently encrypts
       $m_i$ for $i<j$ and $0$ for $i>j$, forms
       $X=(1^\lambda,\pk,(c_i)_i,m)$, and outputs $D^\ora(X)$.
\end{enumerate}
For each $j$, the two CPA challenge bits give $\mathsf H_{j-1}$
and $\mathsf H_j$, so its success minus $1/2$ is
\[
 \frac{1}{2t}\sum_{j=1}^t
  \bigl(\Pr[D^\ora(\mathsf H_j)=1]
          -\Pr[D^\ora(\mathsf H_{j-1})=1]\bigr)
 =\frac{\Pr_{X,b}[D^\ora(X)=b]-1/2}{t}.
\]
Chosen-plaintext security in \cref{thm:fixed-oracle-crypto}
therefore bounds the predictor's advantage by a negligible function.
For a uniform $\QCMAH_d$ machine $M_d$, the promise collapse in
that theorem supplies a total deterministic polynomial-time solver
$D$ agreeing with $M_d$ wherever $M_d$ outputs a bit. Since
$\bot$ answers count as incorrect, replacing $M_d$ by $D$ cannot
decrease its prediction success, which proves the stated hardness.

\bibliographystyle{alpha}

\begin{thebibliography}{GMMY24}

\bibitem[AA11]{ITCS:AarAmb11}
Scott Aaronson and Andris Ambainis.
\newblock The need for structure in quantum speedups.
\newblock In Bernard Chazelle, editor, {\em ICS 2011: 2nd Innovations in
  Computer Science}, pages 338--352, Tsinghua University, Beijing, China,
  January~7--9, 2011. Tsinghua University Press.

\bibitem[AA15]{STOC:AarAmb15}
Scott Aaronson and Andris Ambainis.
\newblock Forrelation: {A} problem that optimally separates quantum from
  classical computing.
\newblock In Rocco~A. Servedio and Ronitt Rubinfeld, editors, {\em 47th Annual
  {ACM} Symposium on Theory of Computing}, pages 307--316, Portland, OR, USA,
  June~14--17, 2015. {ACM} Press.

\bibitem[Aar09]{AaronsonCP}
Scott Aaronson.
\newblock Quantum copy-protection and quantum money, 2009.
\newblock CCC 2009. arXiv:1110.5353, version 1.

\bibitem[Aar10]{AaronsonPH}
Scott Aaronson.
\newblock {BQP} and the polynomial hierarchy, 2010.
\newblock STOC 2010. Full version: arXiv:0910.4698, version 1.

\bibitem[Aar12]{AaronsonGuessing12}
Scott Aaronson.
\newblock Impossibility of succinct quantum proofs for collision-freeness.
\newblock {\em Quantum Information and Computation}, 12(1--2):21--28, 2012.

\bibitem[Aar14]{AaronsonSampling14}
Scott Aaronson.
\newblock The equivalence of sampling and searching.
\newblock {\em Theory of Computing Systems}, 55(2):281--298, 2014.

\bibitem[AB24]{C:AnaBeh24}
Prabhanjan Ananth and Amit Behera.
\newblock A modular approach to unclonable cryptography.
\newblock In Leonid Reyzin and Douglas Stebila, editors, {\em Advances in
  Cryptology -- {CRYPTO}~2024, Part~VII}, volume 14926 of {\em Lecture Notes in
  Computer Science}, pages 3--37, Santa Barbara, CA, USA, August~18--22, 2024.
  Springer, Cham, Switzerland.

\bibitem[ABD26]{AB}
Avantika Agarwal and Shalev Ben-David.
\newblock Oracle separations for the quantum-classical polynomial hierarchy.
\newblock \url{https://arxiv.org/abs/2410.19062v2}, 2026.
\newblock ITCS 2026. arXiv:2410.19062, version 2, 3 December 2025.

\bibitem[ABH{\etalchar{+}}26]{ABHKY26UPO}
Prabhanjan Ananth, Amit Behera, Zikuan Huang, Fuyuki Kitagawa, and Takashi
  Yamakawa.
\newblock Copy-protection from unclonable puncturable obfuscation, revisited.
\newblock In {\em Advances in Cryptology---{EUROCRYPT} 2026}, volume 16541 of
  {\em Lecture Notes in Computer Science}, pages 541--570. Springer, 2026.

\bibitem[AC13]{AC13Money}
Scott Aaronson and Paul Christiano.
\newblock Quantum money from hidden subspaces.
\newblock {\em Theory of Computing}, 9(9):349--401, 2013.

\bibitem[AIK22]{AIK}
Scott Aaronson, DeVon Ingram, and William Kretschmer.
\newblock The acrobatics of {BQP}.
\newblock In {\em 37th Computational Complexity Conference (CCC 2022)}, volume
  234 of {\em Leibniz International Proceedings in Informatics}, pages
  20:1--20:17, 2022.
\newblock Full version: arXiv:2111.10409, version 4.

\bibitem[AK07]{AK07}
Scott Aaronson and Greg Kuperberg.
\newblock Quantum versus classical proofs and advice.
\newblock {\em Theory of Computing}, 3(7):129--157, 2007.

\bibitem[ALL{\etalchar{+}}20]{ALLLZ}
Scott Aaronson, Jiahui Liu, Qipeng Liu, Mark Zhandry, and Ruizhe Zhang.
\newblock New approaches for quantum copy-protection, 2020.
\newblock arXiv:2004.09674, version 7, 16 October 2020.

\bibitem[AN02]{AN02}
Dorit Aharonov and Tomer Naveh.
\newblock Quantum {NP}---a survey, 2002.

\bibitem[AQY22]{C:AnaQiaYue22}
Prabhanjan Ananth, Luowen Qian, and Henry Yuen.
\newblock Cryptography from pseudorandom quantum states.
\newblock In Yevgeniy Dodis and Thomas Shrimpton, editors, {\em Advances in
  Cryptology -- {CRYPTO}~2022, Part~I}, volume 13507 of {\em Lecture Notes in
  Computer Science}, pages 208--236, Santa Barbara, CA, USA, August~15--18,
  2022. Springer, Cham, Switzerland.

\bibitem[BBBV97]{BBBV}
Charles~H. Bennett, Ethan Bernstein, Gilles Brassard, and Umesh Vazirani.
\newblock Strengths and weaknesses of quantum computing.
\newblock {\em SIAM Journal on Computing}, 26(5):1510--1523, 1997.

\bibitem[BBC{\etalchar{+}}98]{FOCS:BBCMW98}
Robert Beals, Harry Buhrman, Richard Cleve, Michele Mosca, and Ronald de~Wolf.
\newblock Quantum lower bounds by polynomials.
\newblock In {\em 39th Annual Symposium on Foundations of Computer Science},
  pages 352--361, Palo Alto, CA, USA, November~8--11, 1998. {IEEE} Computer
  Society Press.

\bibitem[BBF13]{AC:BaeBrzFis13}
Paul Baecher, Christina Brzuska, and Marc Fischlin.
\newblock Notions of black-box reductions, revisited.
\newblock In Kazue Sako and Palash Sarkar, editors, {\em Advances in Cryptology
  -- {ASIACRYPT}~2013, Part~I}, volume 8269 of {\em Lecture Notes in Computer
  Science}, pages 296--315, Bengalore, India, December~1--5, 2013. Springer
  Berlin Heidelberg, Germany.

\bibitem[BCG26]{BCG26}
Amit Behera, Alper Cakan, and Vipul Goyal.
\newblock Decisional monogamy-of-entanglement for coset states and applications
  to unclonable cryptography with correlated challenges, 2026.
\newblock arXiv:2608.18841, version 2, 1 October 2026.

\bibitem[BCN25]{BCN25}
John Bostanci, Boyang Chen, and Barak Nehoran.
\newblock Oracle separation between quantum commitments and quantum
  one-wayness, 2025.
\newblock arXiv:2410.03358, version 3, 11 April 2025.

\bibitem[BCQ23]{ITCS:BraCanQia23}
Zvika Brakerski, Ran Canetti, and Luowen Qian.
\newblock On the computational hardness needed for quantum cryptography.
\newblock In Yael~Tauman Kalai, editor, {\em ITCS 2023: 14th Innovations in
  Theoretical Computer Science Conference}, volume 251, pages 24:1--24:21,
  Cambridge, MA, USA, January~10--13, 2023. Leibniz International Proceedings
  in Informatics (LIPIcs).

\bibitem[BDK24]{BK24}
Shalev Ben-David and Srijita Kundu.
\newblock Oracle separation of {QMA} and {QCMA} with bounded adaptivity.
\newblock In {\em 51st International Colloquium on Automata, Languages, and
  Programming (ICALP 2024)}, volume 297 of {\em Leibniz International
  Proceedings in Informatics}, pages 21:1--21:18, 2024.

\bibitem[BG26]{BG26UE}
James Bartusek and Eli Goldin.
\newblock Unclonable encryption in the {Haar} random oracle model, 2026.
\newblock CRYPTO 2026. Full version: arXiv:2603.11437, version 2, 15 June 2026.

\bibitem[BHNZ26]{BHNZ}
John Bostanci, Jonas Haferkamp, Chinmay Nirkhe, and Mark Zhandry.
\newblock Separating {QMA} from {QCMA} with a classical oracle.
\newblock \url{https://arxiv.org/abs/2511.09551v2}, 2026.
\newblock STOC 2026. arXiv:2511.09551, version 2.

\bibitem[BHV26]{BHV}
John Bostanci, Andrew Huang, and Vinod Vaikuntanathan.
\newblock Separating quantum and classical advice with good codes.
\newblock \url{https://arxiv.org/abs/2602.09385v2}, 2026.
\newblock arXiv:2602.09385, version 2.

\bibitem[BJ24]{FOCS:BatJai24}
Rishabh Batra and Rahul Jain.
\newblock Commitments are equivalent to statistically-verifiable one-way state
  generators.
\newblock In {\em 65th Annual Symposium on Foundations of Computer Science},
  pages 1178--1192, Chicago, IL, USA, October~27--30, 2024. {IEEE} Computer
  Society Press.

\bibitem[BMM{\etalchar{+}}25]{BMMMY25}
Amit Behera, Giulio Malavolta, Tomoyuki Morimae, Tamer Mour, and Takashi
  Yamakawa.
\newblock A new world in the depths of microcrypt: Separating {OWSGs} and
  quantum money from {QEFID}, 2025.
\newblock EUROCRYPT 2025. Full version: arXiv:2410.03453, version 3, 5 February
  2025.

\bibitem[BS19]{BS19BinaryPhase}
Zvika Brakerski and Omri Shmueli.
\newblock (pseudo) random quantum states with binary phase, 2019.
\newblock Version 2, 26 June 2019.

\bibitem[BS20]{C:BraShm20}
Zvika Brakerski and Omri Shmueli.
\newblock Scalable pseudorandom quantum states.
\newblock In Daniele Micciancio and Thomas Ristenpart, editors, {\em Advances
  in Cryptology -- {CRYPTO}~2020, Part~II}, volume 12171 of {\em Lecture Notes
  in Computer Science}, pages 417--440, Santa Barbara, CA, USA, August~17--21,
  2020. Springer, Cham, Switzerland.

\bibitem[Cak26]{CakanQCIP26}
Alper Cakan.
\newblock Separating {QMA} from {QCIP} with a classical oracle, or, the power
  of quantum proofs over classical interaction for quantum verifiers, 2026.

\bibitem[CCHL22]{FOCS:CCHL21}
Sitan Chen, Jordan Cotler, Hsin-Yuan Huang, and Jerry Li.
\newblock Exponential separations between learning with and without quantum
  memory.
\newblock In {\em 62nd Annual Symposium on Foundations of Computer Science},
  pages 574--585, Denver, CO, USA, February~7--10, 2022. {IEEE} Computer
  Society Press.

\bibitem[CCHY26]{ClonableQMA}
Alper Cakan, Kai-Min Chung, Wei-Hsiang Hung, and Tzu-Yi Yang.
\newblock Separating {ClonableQMA} and {QCMA} relative to a classical oracle,
  2026.
\newblock arXiv:2610.06740, version 1, 5 October 2026.

\bibitem[CG24a]{TCC:CakGoy24}
Alper Cakan and Vipul Goyal.
\newblock Unclonable cryptography with unbounded collusions and impossibility
  of hyperefficient shadow tomography.
\newblock In Elette Boyle and Mohammad Mahmoody, editors, {\em TCC~2024: 22nd
  Theory of Cryptography Conference, Part~III}, volume 15366 of {\em Lecture
  Notes in Computer Science}, pages 225--256, Milan, Italy, December~2--6,
  2024. Springer, Cham, Switzerland.

\bibitem[CG24b]{STOC:ColGun24}
Andrea Coladangelo and Sam Gunn.
\newblock How to use quantum indistinguishability obfuscation.
\newblock In Bojan Mohar, Igor Shinkar, and Ryan {O'Donnell}, editors, {\em
  56th Annual {ACM} Symposium on Theory of Computing}, pages 1003--1008,
  Vancouver, BC, Canada, June~24--28, 2024. {ACM} Press.

\bibitem[CG25]{CG25CopyProtection}
Alper Cakan and Vipul Goyal.
\newblock How to copy-protect all puncturable functionalities without
  conjectures: A unified solution to quantum protection.
\newblock Cryptology {ePrint} Archive, Paper 2025/1197, 2025.

\bibitem[CGG24]{C:ChuGolGra24}
Kai-Min Chung, Eli Goldin, and Matthew Gray.
\newblock On central primitives for quantum cryptography with classical
  communication.
\newblock In Leonid Reyzin and Douglas Stebila, editors, {\em Advances in
  Cryptology -- {CRYPTO}~2024, Part~VII}, volume 14926 of {\em Lecture Notes in
  Computer Science}, pages 215--248, Santa Barbara, CA, USA, August~18--22,
  2024. Springer, Cham, Switzerland.

\bibitem[CGK{\etalchar{+}}26]{CGKNY26MultiCopy}
Alper Cakan, Vipul Goyal, Fuyuki Kitagawa, Ryo Nishimaki, and Takashi Yamakawa.
\newblock Multi-copy security in quantum cryptography and more.
\newblock \url{https://arxiv.org/abs/2510.12626v2}, 2026.

\bibitem[CGL{\etalchar{+}}20]{STOC:CGLLTW20}
Nai-Hui Chia, Andr{\'a}s Gily{\'e}n, Tongyang Li, Han-Hsuan Lin, Ewin Tang, and
  Chunhao Wang.
\newblock Sampling-based sublinear low-rank matrix arithmetic framework for
  dequantizing quantum machine learning.
\newblock In Konstantin Makarychev, Yury Makarychev, Madhur Tulsiani, Gautam
  Kamath, and Julia Chuzhoy, editors, {\em 52nd Annual {ACM} Symposium on
  Theory of Computing}, pages 387--400, Chicago, IL, USA, June~22--26, 2020.
  {ACM} Press.

\bibitem[CGS26]{QFire}
Alper Cakan, Vipul Goyal, and Omri Shmueli.
\newblock Public-key quantum fire and key-fire from classical oracles.
\newblock \url{https://arxiv.org/abs/2504.16407v2}, 2026.
\newblock arXiv:2504.16407, version 2, 14 March 2026.

\bibitem[CLLZ22]{CLLZ}
Andrea Coladangelo, Jiahui Liu, Qipeng Liu, and Mark Zhandry.
\newblock Hidden cosets and applications to unclonable cryptography, 2022.
\newblock arXiv:2107.05692, version 3, 14 July 2022.

\bibitem[FK18]{FK18}
Bill Fefferman and Shelby Kimmel.
\newblock Quantum vs. classical proofs and subset verification.
\newblock In {\em 43rd International Symposium on Mathematical Foundations of
  Computer Science (MFCS 2018)}, volume 117 of {\em Leibniz International
  Proceedings in Informatics}, pages 22:1--22:23, 2018.

\bibitem[GGJL25]{STOC:GGJL25}
Mika G{\"o}{\"o}s, Tom Gur, Siddhartha Jain, and Jiawei Li.
\newblock Quantum communication advantage in {TFNP}.
\newblock In Michal Kouck\'{y} and Nikhil Bansal, editors, {\em 57th Annual
  {ACM} Symposium on Theory of Computing}, pages 1465--1475, Prague, Czechia,
  June~23--27, 2025. {ACM} Press.

\bibitem[GKK{\etalchar{+}}07]{STOC:GKKRW07}
Dmitry Gavinsky, Julia Kempe, Iordanis Kerenidis, Ran Raz, and Ronald de~Wolf.
\newblock Exponential separations for one-way quantum communication complexity,
  with applications to cryptography.
\newblock In David~S. Johnson and Uriel Feige, editors, {\em 39th Annual {ACM}
  Symposium on Theory of Computing}, pages 516--525, San Diego, CA, USA,
  June~11--13, 2007. {ACM} Press.

\bibitem[GKT26]{GKT26PRU}
Aditya Gulati, Dakshita Khurana, and Kabir Tomer.
\newblock How to build pseudorandom unitaries in microcrypt, 2026.

\bibitem[GKW26]{GKW26}
David Gosset, Robin Kothari, and Kewen Wu.
\newblock Quantum state preparation with optimal {T}-count.
\newblock {\em Quantum}, 10:2168, 2026.
\newblock Full version: arXiv:2411.04790, version 3.

\bibitem[GMMY24]{CountCrypt}
Eli Goldin, Tomoyuki Morimae, Saachi Mutreja, and Takashi Yamakawa.
\newblock {CountCrypt}: Quantum cryptography between {QCMA} and {PP}, 2024.
\newblock arXiv:2410.14792, version 3, 6 October 2025.

\bibitem[Gol01]{Goldreich_2001}
Oded Goldreich.
\newblock {\em Foundations of Cryptography}.
\newblock Cambridge University Press, 2001.

\bibitem[Gol04]{Goldreich_2004}
Oded Goldreich.
\newblock {\em Foundations of Cryptography}.
\newblock Cambridge University Press, 2004.

\bibitem[GZ20]{GZ20}
Marios Georgiou and Mark Zhandry.
\newblock Unclonable decryption keys.
\newblock Cryptology ePrint Archive, Paper 2020/877, 2020.
\newblock Revised 29 July 2020.

\bibitem[H{\aa}s86]{Hastad86}
Johan H{\aa}stad.
\newblock Computational limitations for small depth circuits, 1986.
\newblock Book manuscript based on the author's MIT Ph.D. thesis.

\bibitem[HN22]{HiraharaNanashima22}
Shuichi Hirahara and Mikito Nanashima.
\newblock Finding errorless pessiland in error-prone heuristica.
\newblock In {\em 37th Computational Complexity Conference (CCC 2022)}, volume
  234 of {\em Leibniz International Proceedings in Informatics}, pages
  25:1--25:28, 2022.

\bibitem[Imp11]{Impagliazzo11}
Russell Impagliazzo.
\newblock Relativized separations of worst-case and average-case complexities
  for {NP}.
\newblock In {\em 2011 IEEE 26th Annual Conference on Computational
  Complexity}, pages 104--114, 2011.

\bibitem[JLS18]{C:JiLiuSon18}
Zhengfeng Ji, Yi-Kai Liu, and Fang Song.
\newblock Pseudorandom quantum states.
\newblock In Hovav Shacham and Alexandra Boldyreva, editors, {\em Advances in
  Cryptology -- {CRYPTO}~2018, Part~III}, volume 10993 of {\em Lecture Notes in
  Computer Science}, pages 126--152, Santa Barbara, CA, USA, August~19--23,
  2018. Springer, Cham, Switzerland.

\bibitem[Kel21]{Kelley}
Zander Kelley.
\newblock An improved derandomization of the switching lemma.
\newblock
  \url{https://eccc.weizmann.ac.il/report/2020/182/revision/1/download}, 2021.
\newblock Electronic Colloquium on Computational Complexity, TR20-182, revision
  1.

\bibitem[KNSY26]{KNSY26Trapdoor}
Fuyuki Kitagawa, Ryo Nishimaki, Yuki Shirakawa, and Takashi Yamakawa.
\newblock Trapdoor functions with secure key leasing and copy protection, 2026.

\bibitem[KOW26]{KOW26}
Robin Kothari, Ryan O'Donnell, and Kewen Wu.
\newblock No exponential quantum speedup for {$\mathrm{SIS}^{\infty}$} anymore.
\newblock In {\em Proceedings of the 58th Annual ACM Symposium on Theory of
  Computing}, pages 101--105. ACM, 2026.

\bibitem[KQST23]{KQST}
William Kretschmer, Luowen Qian, Makrand Sinha, and Avishay Tal.
\newblock Quantum cryptography in algorithmica, 2023.
\newblock arXiv:2212.00879, version 2, 2 May 2023.

\bibitem[KQT24]{KQT}
William Kretschmer, Luowen Qian, and Avishay Tal.
\newblock Quantum-computable one-way functions without one-way functions, 2024.
\newblock arXiv:2411.02554, version 1, 4 November 2024.

\bibitem[Kre21]{KretschmerHaar24}
William Kretschmer.
\newblock Quantum pseudorandomness and classical complexity.
\newblock In {\em 16th Conference on the Theory of Quantum Computation,
  Communication and Cryptography (TQC 2021)}, volume 197 of {\em Leibniz
  International Proceedings in Informatics}, pages 2:1--2:20, 2021.
\newblock Full version: arXiv:2103.09320, version 5, 18 September 2024.

\bibitem[KT24]{KT}
Dakshita Khurana and Kabir Tomer.
\newblock Commitments from quantum one-wayness.
\newblock In {\em Proceedings of the 56th Annual ACM Symposium on Theory of
  Computing}, 2024.
\newblock Full version: arXiv:2310.11526, version 4.

\bibitem[Liu23]{EC:Liu23}
Qipeng Liu.
\newblock Non-uniformity and quantum advice in the quantum random oracle model.
\newblock In Carmit Hazay and Martijn Stam, editors, {\em Advances in
  Cryptology -- {EUROCRYPT}~2023, Part~I}, volume 14004 of {\em Lecture Notes
  in Computer Science}, pages 117--143, Lyon, France, April~23--27, 2023.
  Springer, Cham, Switzerland.

\bibitem[LLPY24]{ITCS:LLPY24}
Xingjian Li, Qipeng Liu, Angelos Pelecanos, and Takashi Yamakawa.
\newblock Classical vs quantum advice and proofs under classically-accessible
  oracle.
\newblock In Venkatesan Guruswami, editor, {\em ITCS 2024: 15th Innovations in
  Theoretical Computer Science Conference}, volume 287, pages 72:1--72:19,
  Berkeley, CA, USA, January~30~--~February~2, 2024. Leibniz International
  Proceedings in Informatics (LIPIcs).

\bibitem[LLQZ22]{TCC:LLQZ22}
Jiahui Liu, Qipeng Liu, Luowen Qian, and Mark Zhandry.
\newblock Collusion resistant copy-protection for watermarkable
  functionalities.
\newblock In Eike Kiltz and Vinod Vaikuntanathan, editors, {\em TCC~2022: 20th
  Theory of Cryptography Conference, Part~I}, volume 13747 of {\em Lecture
  Notes in Computer Science}, pages 294--323, Chicago, IL, USA, November~7--10,
  2022. Springer, Cham, Switzerland.

\bibitem[LMY25]{STOC:LiuMutYue25}
Jiahui Liu, Saachi Mutreja, and Henry Yuen.
\newblock {QMA} vs {QCMA} and pseudorandomness.
\newblock In Michal Kouck\'{y} and Nikhil Bansal, editors, {\em 57th Annual
  {ACM} Symposium on Theory of Computing}, pages 1520--1531, Prague, Czechia,
  June~23--27, 2025. {ACM} Press.

\bibitem[Mah18]{FOCS:Mahadev18a}
Urmila Mahadev.
\newblock Classical verification of quantum computations.
\newblock In Mikkel Thorup, editor, {\em 59th Annual Symposium on Foundations
  of Computer Science}, pages 259--267, Paris, France, October~7--9, 2018.
  {IEEE} Computer Society Press.

\bibitem[MH25]{MH25}
Fermi Ma and Hsin-Yuan Huang.
\newblock How to construct random unitaries.
\newblock In {\em Proceedings of the 57th Annual ACM Symposium on Theory of
  Computing}, pages 806--809. ACM, 2025.

\bibitem[MSY25]{STOC:MorShiYam25}
Tomoyuki Morimae, Yuki Shirakawa, and Takashi Yamakawa.
\newblock Cryptographic characterization of quantum advantage.
\newblock In Michal Kouck\'{y} and Nikhil Bansal, editors, {\em 57th Annual
  {ACM} Symposium on Theory of Computing}, pages 1863--1874, Prague, Czechia,
  June~23--27, 2025. {ACM} Press.

\bibitem[MY22]{MY22}
Tomoyuki Morimae and Takashi Yamakawa.
\newblock Quantum commitments and signatures without one-way functions.
\newblock In {\em Advances in Cryptology -- CRYPTO 2022}, volume 13507 of {\em
  Lecture Notes in Computer Science}, pages 269--295. Springer, 2022.

\bibitem[MY24]{MY24OneWayness}
Tomoyuki Morimae and Takashi Yamakawa.
\newblock One-wayness in quantum cryptography.
\newblock In {\em 19th Conference on the Theory of Quantum Computation,
  Communication and Cryptography (TQC 2024)}, volume 310 of {\em Leibniz
  International Proceedings in Informatics}, pages 4:1--4:21, 2024.
\newblock Full version: arXiv:2210.03394, version 3.

\bibitem[Nay99]{Nayak}
Ashwin Nayak.
\newblock Optimal lower bounds for quantum automata and random access codes.
\newblock In {\em 40th Annual Symposium on Foundations of Computer Science},
  pages 369--376, 1999.
\newblock Full version: arXiv:quant-ph/9904093, version 3.

\bibitem[NC10]{NC10}
Michael~A. Nielsen and Isaac~L. Chuang.
\newblock {\em Quantum Computation and Quantum Information: 10th Anniversary
  Edition}.
\newblock Cambridge University Press, 2010.

\bibitem[NN24]{NN24}
Anand Natarajan and Chinmay Nirkhe.
\newblock A distribution testing oracle separation between {QMA} and {QCMA}.
\newblock {\em Quantum}, 8:1377, 2024.

\bibitem[RK11]{STOC:RegKla11}
Oded Regev and Bo'az Klartag.
\newblock Quantum one-way communication can be exponentially stronger than
  classical communication.
\newblock In Lance Fortnow and Salil~P. Vadhan, editors, {\em 43rd Annual {ACM}
  Symposium on Theory of Computing}, pages 31--40, San Jose, CA, USA,
  June~6--8, 2011. {ACM} Press.

\bibitem[RST15]{FOCS:RosSerTan15}
Benjamin Rossman, Rocco~A. Servedio, and Li-Yang Tan.
\newblock An average-case depth hierarchy theorem for {Boolean} circuits.
\newblock In Venkatesan Guruswami, editor, {\em 56th Annual Symposium on
  Foundations of Computer Science}, pages 1030--1048, Berkeley, CA, USA,
  October~17--20, 2015. {IEEE} Computer Society Press.

\bibitem[RT22]{RT}
Ran Raz and Avishay Tal.
\newblock Oracle separation of {BQP} and {PH}.
\newblock {\em Journal of the ACM}, 69(4), 2022.

\bibitem[Sat]{MicrocryptZoo}
Or~Sattath.
\newblock {Microcrypt Zoo}.
\newblock \url{https://sattath.github.io/microcrypt-zoo/}.
\newblock Accessed 23 October 2026.

\bibitem[SLSB19]{SLSB19}
Yuval~R. Sanders, Guang~Hao Low, Artur Scherer, and Dominic~W. Berry.
\newblock Black-box quantum state preparation without arithmetic.
\newblock {\em Physical Review Letters}, 122(2):020502, 2019.
\newblock Full version: arXiv:1807.03206, version 2.

\bibitem[Sto83]{STOC:Stockmeyer83}
Larry~J. Stockmeyer.
\newblock The complexity of approximate counting (preliminary version).
\newblock In {\em 15th Annual {ACM} Symposium on Theory of Computing}, pages
  118--126, Boston, MA, USA, April~25--27, 1983. {ACM} Press.

\bibitem[Tan19]{STOC:Tang19}
Ewin Tang.
\newblock A quantum-inspired classical algorithm for recommendation systems.
\newblock In Moses Charikar and Edith Cohen, editors, {\em 51st Annual {ACM}
  Symposium on Theory of Computing}, pages 217--228, Phoenix, AZ, USA,
  June~23--26, 2019. {ACM} Press.

\bibitem[Wat00]{FOCS:Watrous00}
John Watrous.
\newblock Succinct quantum proofs for properties of finite groups.
\newblock In {\em 41st Annual Symposium on Foundations of Computer Science},
  pages 537--546, Redondo Beach, CA, USA, November~12--14, 2000. {IEEE}
  Computer Society Press.

\bibitem[YZ24]{YZ}
Takashi Yamakawa and Mark Zhandry.
\newblock Verifiable quantum advantage without structure.
\newblock {\em Journal of the ACM}, 71(3), 2024.

\bibitem[Zha12a]{ZhandryQRF12}
Mark Zhandry.
\newblock How to construct quantum random functions, 2012.
\newblock Electronic Colloquium on Computational Complexity, TR12-140, revision
  2.

\bibitem[Zha12b]{FOCS:Zhandry12}
Mark Zhandry.
\newblock How to construct quantum random functions.
\newblock In {\em 53rd Annual Symposium on Foundations of Computer Science},
  pages 679--687, New Brunswick, NJ, USA, October~20--23, 2012. {IEEE} Computer
  Society Press.

\bibitem[Zha19]{EC:Zhandry19b}
Mark Zhandry.
\newblock Quantum lightning never strikes the same state twice.
\newblock In Yuval Ishai and Vincent Rijmen, editors, {\em Advances in
  Cryptology -- {EUROCRYPT}~2019, Part~III}, volume 11478 of {\em Lecture Notes
  in Computer Science}, pages 408--438, Darmstadt, Germany, May~19--23, 2019.
  Springer, Cham, Switzerland.

\bibitem[Zha25]{ITCS:Zhandry25}
Mark Zhandry.
\newblock Toward separating {QMA} from {QCMA} with a classical oracle.
\newblock In Raghu Meka, editor, {\em ITCS 2025: 16th Innovations in
  Theoretical Computer Science Conference}, volume 325, pages 95:1--95:19, New
  York, NY, USA, January~7--10, 2025. Leibniz International Proceedings in
  Informatics (LIPIcs).

\end{thebibliography}
\end{document}